\documentclass[article]{aastex631}

\usepackage[utf8]{inputenc}

\usepackage{booktabs}
\usepackage{newtxtext,newtxmath}
\usepackage{ulem}
\usepackage{amssymb}
\usepackage{graphicx}

\begin{document}

\title{Reducing the uncertainty on the Hubble constant up to 35\% with an improved statistical analysis: different best-fit likelihoods for Supernovae Ia, Baryon Acoustic Oscillations, Quasars, and Gamma-Ray Bursts}

\author{Maria Giovanna Dainotti}
\altaffiliation{First and second authors share the same contribution.}
\altaffiliation{Corresponding author, maria.dainotti@nao.ac.jp}
\affiliation{Division of Science, National Astronomical Observatory of Japan, 2 Chome-21-1 Osawa, Mitaka, Tokyo 181-8588, Japan}
\affiliation{The Graduate University for Advanced Studies, SOKENDAI, Shonankokusaimura, Hayama, Miura District, Kanagawa 240-0193, Japan}
\affiliation{Space Science Institute, Boulder, CO, USA}
\author{Giada Bargiacchi}
\affiliation{Scuola Superiore Meridionale,
                Largo S. Marcellino 10,
                80138, Napoli, Italy}
\affiliation{Istituto Nazionale di Fisica Nucleare (INFN), Sez. di Napoli,  Complesso Univ. Monte S. Angelo, Via Cinthia 9, 80126, Napoli, Italy}
\author{Malgorzata Bogdan}
\affiliation{ University of Wroclaw, plac Grunwaldzki 2/4, Wroclaw, Lower Silesia Province, 50-384, Poland}
\affiliation{Department of Statistics, Lund University, Box 117, SE-221 00 Lund, Sweden}
\author{Aleksander Łukasz Lenart}
\affiliation{Astronomical Observatory, Jagiellonian University, ul. Orla 171, 31-501 Kraków, Poland}
\author{Kazunari Iwasaki}
\affiliation{The Graduate University for Advanced Studies, SOKENDAI, Shonankokusaimura, Hayama, Miura District, Kanagawa 240-0193, Japan}
\affiliation{Center for Computational Astrophysics, National Astronomical Observatory of Japan, 2 Chome-21-1 Osawa, Mitaka, Tokyo 181-8588, Japan}

\author{Salvatore Capozziello}
\affiliation{Dipartimento di Fisica "E. Pancini" , Universit\'a degli Studi di  Napoli "Federico II" Complesso Univ. Monte S. Angelo, Via Cinthia 9, 80126, Napoli, Italy}
\affiliation{Scuola Superiore Meridionale, Largo S. Marcellino 10, 80138, Napoli, Italy}
\affiliation{Istituto Nazionale di Fisica Nucleare (INFN), Sez. di Napoli, Complesso Univ. Monte S. Angelo, Via Cinthia 9, 80126, Napoli, Italy}
\author{Bing Zhang}
\affiliation{Nevada Center for Astrophysics, University of Las Vegas, NV 89154, USA}
\affiliation{Department of Physics and Astrophysics, University of Las Vegas, NV 89154, USA}
\author{Nissim Fraija}
\affiliation{Instituto de Astronomía, Universidad Nacional Autónoma de México Circuito Exterior, C.U., A. Postal 70-264, 04510 México D.F., México}
\begin{abstract}
Cosmological models and their parameters are widely debated, especially about whether the current discrepancy between the values of the Hubble constant, $H_{0}$, obtained by type Ia supernovae (SNe Ia), and the Planck data from the Cosmic Microwave Background Radiation could be alleviated when alternative cosmological models are considered. 
Thus, combining high-redshift probes, such as Gamma-Ray Bursts (GRBs) and Quasars (QSOs), together with Baryon Acoustic Oscillations (BAO) and SNe Ia is important to assess the viability of these alternative models and if they can cast further light on the Hubble tension. In this work, for GRBs, we use a 3-dimensional relation between the peak prompt luminosity, the rest-frame time at the end of the X-ray plateau, and its corresponding luminosity in X-rays: the 3D Dainotti fundamental plane relation. 
Regarding QSOs, we use the Risaliti-Lusso relation among the UV and X-ray luminosities for a sample of 2421 sources.
We correct both the QSO and GRB relations by accounting for selection and evolutionary effects with a reliable statistical method. 
We here use both the traditional Gaussian likelihoods ($\cal L_G$) and the new best-fit likelihoods ($\cal L_N$) to infer cosmological parameters of a non-flat $\Lambda$CDM and flat $w$CDM models.
We obtain for all the parameters reduced uncertainties, up to $35\%$ for $H_{0}$, when applying the new $\cal L_N$ likelihoods in place of the Gaussian ones.
Our results remain consistent with a flat $\Lambda$CDM model, although with a shift of the dark energy parameter $w$ toward $w<-1$ and a curvature density parameter toward $\Omega_k<0$.
\end{abstract}


\section{Introduction}
Although the flat $\Lambda$CDM model \citep{peebles1984} remains the most suitable cosmological model to parameterize the Universe, discussions about the possibility of an open and closed Universe are obtaining increasing attention.
Undoubtedly, the flat $\Lambda$CDM model, with the cold dark matter (CDM) component and a cosmological constant $\Lambda$ \citep{2001LRR.....4....1C}, as suggested by the current accelerated expansion of the Universe \citep{riess1998,perlmutter1999}, allows a series of observations with cosmic microwave background \citep[CMB; ][]{planck2018} and the baryon acoustic oscillations \citep[BAO;][]{eboss2021}, which agree with the accelerated expansion based on type Ia supernovae (SNe Ia). Nevertheless, there is a current crisis of the $\Lambda$CDM model, which suffers from theoretical and observational shortcomings. Among all the issues, we can enumerate the cosmological constant problem \citep{1989RvMP...61....1W}, the fine-tuning, the origin and properties of dark energy, and the Hubble constant ($H_0$) tension, the discrepancy between the $H_0$ value obtained from the Planck data of the CMB under the assumption of a flat $\Lambda$CDM model ($H_0 = 67.4 \pm 0.5  \, \mathrm{km} \, \mathrm{s}^{-1} \, \mathrm{Mpc}^{-1}$, \citealt{planck2018}), and the local $H_0$ from SNe Ia and Cepheids ($H_0 = 73.04 \pm 1.04  \, \mathrm{km} \, \mathrm{s}^{-1} \, \mathrm{Mpc}^{-1}$, \citealt{2022ApJ...934L...7R}). This discrepancy ranges from 4.4 to 6 $\sigma$ when different samples are considered \citep{2019ApJ...876...85R,2020PhRvR...2a3028C,2020MNRAS.498.1420W}.
Time-delay and strong lensing measurements from QSOs \citep{2019ApJ...886L..23L} provide a value close to the one of SNe Ia, while cosmic chronometers favor the value of $H_0$ derived from the CMB \citep{2018JCAP...04..051G}. Other probes, such as QSOs \citep{rl19,2020A&A...642A.150L,biasfreeQSO2022}, the Tip of the Red-Giant Branch (TRGB) \citep{2021ApJ...919...16F}, and Gamma-ray bursts (GRBs) \citep{Cardone2009MNRAS.400..775C,cardone10,Dainotti2013a, Postnikov14, Dainotti2021ApJ...912..150D,Dainotti2022Galax..10...24D,Dainotti2022MNRAS.tmp.2639D,Dainotti2022PASJ}, hint at a $H_0$ value halfway between these two. Nevertheless, the $H_0$ tension remains one of the most challenging open questions in astrophysics and cosmology. 
So far, probes like QSOs and GRBs have only been used separately with SNe Ia and BAO \citep{Dainotti2013b,Postnikov14,Dainotti2022MNRAS.tmp.2639D,Dainotti2022PASJ,Dainotti2022MNRAS.514.1828D,Cao2021arXiv211014840C,Cao2022MNRAS.510.2928C,Cao2022arXiv220105245C} or, even if used together, selection biases or redshift evolution have not been taken into account, except for the case of \citet{Bargiacchi2023arXiv230307076B}, and thus the resulting cosmological parameters could have been affected by biases and induced by these effects rather than the intrinsic physics.
Indeed, the evolution of the cosmological parameters is an extremely important topic, and it has been discussed especially in relation to ${H_{0}}$. It has been highlighted even for the SNe Ia by \citet{Dainotti2021ApJ...912..150D} and \citet{snelikelihood2022} that there is an evolutionary trend on $H_0$ as a function of the redshift, which can possibly be explained either by selection biases or by a new physics (i.e., invoking the so-called ${f(R)}$-gravity theory as in \citet{Schiavone:2022wvq}). For a general review of the Hubble constant tension, see \citet{2022JHEAp..34...49A}.
In addition, we here use a platinum sample of GRBs \citep{Dainotti2020a} and the most suitable sample of QSOs \citep{2020A&A...642A.150L}. This choice is dictated by the fact that tighter correlations, if grounded on fundamental physics, naturally produce tighter and more reliable constraints on cosmological parameters.
The GRB relation we here use as a cosmological tool is the GRB fundamental plane relation,\footnote{We note that we are referring to the fundamental plane correlation related to GRBs, and not to other definitions of fundamental planes used in astronomy, such as the fundamental plane of elliptical galaxies \citep{1987ApJ...313...59D}.} namely the three-dimensional Dainotti relation, which links the time at the end of the plateau emission measured in the rest frame, $T^{*}_{X}$, with the corresponding X-ray luminosity of the light curve, $L_{a,X}$ \citep{Dainotti2008,Dainotti2010ApJ...722L.215D,dainotti11a,Dainotti2017A&A...600A..98D}, with the prompt peak luminosity, $L_{peak}$ \citep{Dainotti2016ApJ...825L..20D,Dainotti2017ApJ...848...88D,Dainotti2020a,Srinivasaragavan2020}. This relation is theoretically supported by the magnetar model \citep{2001ApJ...552L..35Z,2011A&A...526A.121D, 2012MNRAS.425.1199B,Rowlinson2014MNRAS.443.1779R,Rea2015ApJ...813...92R} where the equation of the magnetars reproduces this anti-correlation.
The application of the two-dimensional Dainotti relation and the fundamental plane relation as a cosmological probe has been presented in \citet{cardone09}, \citet{cardone10}, \citet{Postnikov14}, 
 \citet{Dainotti2013a}, \citet{Dainotti2022PASJ}, \citet{Dainotti2022MNRAS.514.1828D}, and \citet{Dainotti2022MNRAS.tmp.2639D}.
We here remind that the two-dimensional Dainotti relation has also been discovered in optical \citep{Dainotti2020b, Dainotticlosureoptical2022ApJ...940..169D} and radio
\citep{Levine2022ApJ...925...15L}, while the three-dimensional (3D) correlation has also been found in high energy $\gamma$-rays \citep{Dainotti2021ApJS..255...13D} and in optical \citep{Dainotti2022ApJS..261...25D}. For a more extensive discussion on the prompt, prompt-afterglow relations, their selection biases and the application as cosmological tools, see \citet{Dainotti2017NewAR..77...23D}, \citet{Dainotti2018PASP..130e1001D}, and  \citet{Dainotti2018AdAst2018E...1D}. 
We also here remind that for the study of GRB relation, it is important to distinguish among GRB classes since each class of GRB may depend on a different progenitor or the same progenitor, but with a different environment. In the current analysis, we focus on the long-duration GRBs (LGRBs) which arise from the collapse of a massive star. 
For QSOs we use the Risali-Lusso relation (RL), between the ultraviolet (UV) and X-ray luminosities proposed after the observations of the first X-ray surveys
\citep{1979ApJ...234L...9T,1981ApJ...245..357Z,1986ApJ...305...83A}, and then validated by using several different QSO samples \citep[e.g.][]{steffen06,just07,2010A&A...512A..34L,lr16,2021A&A...655A.109B}. This relation well reproduces the commonly accepted picture where QSOs are powered by the accretion on a central supermassive black hole, which efficiently converts mass into energy \citep{1998A&A...334...39S,1999RvMPS..71..180H,qsophysics,2020MNRAS.498.5652K}. 

Encouraged by our previous results \citep{biasfreeQSO2022,Bargiacchi2023arXiv230307076B} in which QSOs and GRBs have been used alone or in combination with other probes, including the correction of selection biases and redshift evolution, we are finally ready to combine QSOs and GRBs and we use different assumptions of the likelihood traditionally used for SNe Ia as we have proved in a previous paper \citep{snelikelihood2022}. The common practice of constraining cosmological parameters with a Gaussian likelihood ($\cal L_G$) based on distance moduli relies on the implicit assumption that the difference between the distance moduli measured and expected from a cosmological model normalized over the uncertainties is normally distributed. Since we have already explored in previous papers the assumptions about the most appropriate likelihoods, we use the assumption of the logistic distribution for the \textit{Pantheon} \citep{scolnic2018} and the QSO sample, of the T-student for the \textit{Pantheon +} \citep{pantheon+} and BAO, and the Gaussian distribution for the GRB platinum sample. 

Section \ref{data} introduces the SNe Ia, GRB, QSO and BAO samples. Section \ref{meth} presents the treatment of the physical quantities for each probe, the likelihood assumptions, and the cosmological fits in the non-flat and $w$CDM model. In Section \ref{results}, we describe the results. In Section \ref{sec5}, we show the goodness of the fits of the results. In Section \ref{sec6}, we introduce the cosmological analysis with QSOs alone. In Section \ref{conclusions}, we draw our conclusions. 

\section{Data sets}
\label{data}
We use a combination of four cosmological probes: SNe Ia, GRBs, QSOs, and BAO. We here summarize the data set, and for further details, we refer to the original papers.
We employ the two most recent SNe Ia collections of ``\textit{Pantheon}" \citep{scolnic2018} and ``\textit{Pantheon +}" \citep{pantheon+} with the full covariance matrix that includes both statistical and systematic uncertainties. The ``\textit{Pantheon}" sample is composed of 1048 sources in the redshift range between $z=0.01$ and $z=2.26$ gathered from CfA1-4, \textit{Carnegie Supernova Project}, \textit{Pan-STARRS1},  \textit{Sloan Digital Sky Survey} (SDSS), \textit{Supernova Legacy Survey}, and \textit{Hubble Space Telescope}, while ``\textit{Pantheon +}" consists of 1701 SNe Ia collected from 18 different surveys in the range $z=0.001 - 2.26$. The use of both samples in this work allows us to reveal if and how a combined analysis leads to different results with the use of \citet{scolnic2018} or \citet{pantheon+} samples.

For GRBs, we use the ``Platinum" sample first defined in \citet{Dainotti2020ApJ...904...97D}, which is an improvement of the golden sample \citep{Dainotti2016ApJ...825L..20D,Dainotti2017ApJ...848...88D, Dainotti2020ApJ...904...97D}, and it is composed of 50 X-ray LGRBs with observed redshift between $z=0.055$ and $z=5$ that have been selected according to the following criteria: a) the plateau must present an inclination $<$41°. The angle is computed as $\Delta F / \delta T = F_i - F_{a,X}/T_X - T_i$ by the use of trigonometry where $F_{a,X}$ indicates the flux at the end of the plateau emission, $i$ indicates the time of the beginning of the plateau, b) it should have no flares during or at the end of the plateau emission, c) it should have at least 5 points in its beginning to better determine its onset; d) and a duration $> 500$s \citep[see also][]{Dainotti2022MNRAS.tmp.2639D,Dainotti2022MNRAS.514.1828D}.
This sample is selected from initial 372 GRBs with a known redshift observed by the {\it Neil Gehrels Swift Observatory} (Swift) from January 2005 to August 2019 from both the Burst Alert Telescope (BAT) and X-Ray Telescope (XRT) repositories \citep{Evans2009}, which are then reduced to 222 GRBs by retaining only the one with a plateau fittable using the \cite{2007ApJ...662.1093W} model, to which the above-defined criteria are finally applied to obtain the Platinum sample.

The complete description of the QSO data set used in this work is presented in \citet{2020A&A...642A.150L}. It consists of 2421 sources from $z=0.009$ to $z =7.54$ \citep{banados2018}, and it is composed starting from eight samples taken from the
literature and archives. 
The QSO community has been striving to fine-tune and choose sources to be suitable for cosmological studies by carefully removing as much as possible observational biases, as described in \citet{rl15}, \citet{lr16}, \citet{rl19}, \citet{salvestrini2019} and \citet{2020A&A...642A.150L}. 
As in \citet{DainottiQSO}, \citet{biasfreeQSO2022},\citet{Bargiacchi2023arXiv230307076B}, and \citet{DainottiGoldQSOApJ2023}, we use this final cleaned sample, however without imposing any cut in redshift, such as the one at $z=0.7$ investigated in \citet{2020A&A...642A.150L}, to avoid introducing additional truncation and that the sample is biased against high-redshift sources only.

The BAO collection we use is described in \citet{2018arXiv180707323S}, composed of 26 data points for which the covariance matrix is detailed in \citet{2016JCAP...06..023S}. 
This data set has already been used in \citet{Dainotti2022Galax..10...24D} and \citet{Dainotti2022MNRAS.tmp.2639D} in combination with \textit{Pantheon} SNe Ia and GRBs.
\section{Methodology}
\label{meth}

\subsection{Cosmological models}

We here describe the two cosmological models analyzed in this work: the non-flat $\Lambda$CDM and the flat $w$CDM. These models are the ones most frequently studied as extensions of the standard flat $\Lambda$CDM model. Indeed, they can be easily built by modifying the equation for the background evolution assumed in the flat $\Lambda$CDM model, which reads as
\begin{equation}
\label{lcdm}
E(z) = \frac{H(z)}{H_0} = \Bigg[\Omega_{M}\,(1+z)^{3} +
\Omega_{\Lambda}\Bigg]^{\frac{1}{2}}\,.
\end{equation}
Here, $\Omega_{\Lambda} = 1-\Omega_{M}$ from the flatness condition and we have omitted the negligible contribution of the current relativistic density parameter.

\subsubsection{Non-flat $\Lambda$CDM model}

The non-flat $\Lambda$CDM model allows for the curvature of the Universe to be different from zero. Thus, Eq. \eqref{lcdm} becomes
\begin{equation}\label{lcdmnonflat}
E(z) =\Bigg[\Omega_{M}\,(1+z)^{3} + \Omega_{k}\,(1+z)^{2} + \Omega_{\Lambda}\Bigg]^{\frac{1}{2}}\,,
\end{equation}
where the parameters are linked via the relation $ 1= \Omega_{k} + \Omega_{M} + \Omega_{\Lambda}$. By using this formula, we can compute the luminosity distance within this model as
\begin{equation}
    d_{l} = (1+z) \frac{c}{H_{0}} 
    \begin{cases}
  \frac{1}{\sqrt{|\Omega_{k}|}} sin(\sqrt{|\Omega_{k}|} \int^{z}_{0} \frac{d \zeta}{E(\zeta)} ) & \text{ if } \Omega_{k} < 0,\\ 
 \int^{z}_{0} \frac{d \zeta}{E(\zeta)} & \text{ if } \Omega_{k} = 0, \\ 
 \frac{1}{\sqrt{\Omega_{k}}} sinh(\sqrt{\Omega_{k}} \int^{z}_{0} \frac{d \zeta}{E(\zeta)} ) & \text{ if } \Omega_{k} > 0 .
\end{cases}
\label{nonflatdl}
\end{equation}
The case with $\Omega_k = 0$ corresponds to the flat $\Lambda$CDM model, while the cases with $\Omega_k > 0$ and $\Omega_k <0$ describe an open and closed Universe, respectively.

\subsubsection{Flat $w$CDM model}

In the flat $w$CDM model, the equation of state of dark energy can assume any constant value, also different from -1, the value required by the flat $\Lambda$CDM model. Hence, in this case, we consider a generic equation of state parameter $w= P_{\Lambda}/\rho_{\Lambda}$, where $P_{\Lambda}$ and $\rho_{\Lambda}$ are the pressure and energy density of dark energy, respectively. Within this model, Eq. \eqref{lcdm} changes to
\begin{equation}\label{wcdm}
E(z) = \Bigg[\Omega_{M}\,(1+z)^{3} +
\Omega_{\Lambda}\, (1+z)^{3(1+w)}\Bigg]^{\frac{1}{2}}\,.
\end{equation}
Accordingly, $d_l$ is obtained from Eq. \eqref{nonflatdl} in the case of $\Omega_k=0$.
 The two distinguished regimes for $w>-1$ and $w<-1$ are referred to as ``quintessence'' and ``phantom'', respectively. In the latter scenario, the Universe ends with a ``Big Rip'' in which the extreme acceleration of the expansion rips the matter apart.

\subsection{The definition of the physical quantities for the various probes}
\label{quantities}
\subsubsection{The distance moduli of SNe Ia}
The observed distance moduli of SNe Ia $\mu$ is 
\begin{equation}
\label{muobs}
\mu_{obs, \mathrm{SNe \, Ia}} = m_B - M + \alpha x_1 - \beta c_1 + \Delta_M + \Delta_B\,,
\end{equation}
where $m_B$ is the B-band overall amplitude, $x_1$ the stretch parameter, $c_1$ the color, $\alpha$ and $\beta$ the coefficients of the relation of luminosity with $x_1$ and $c_1$, respectively, $M$ the fiducial B-band absolute magnitude of a SNe with $x_1 = 0$ and $c_1=0$, $\Delta_M$ and $\Delta_B$ the corrections on the distance that account for the mass of the host-galaxy and biases predicted through simulations, respectively \citep{scolnic2018,pantheon+}. $M$ is degenerate with $H_0$, hence $H_0$ cannot be determined by SNe Ia alone
\citep{1998A&A...331..815T,scolnic2018}, but it can be derived if $M$ is fixed.
Indeed, the first part of the distance ladder, composed of Cepheids and TRGB, is traditionally used to calibrate $M$ from which $\mu$ is then computed \citep{2022ApJ...938..110B,2022ApJ...934L...7R}. \citet{scolnic2018} fix $M=-19.35$
corresponding to $H_0 = 70 \, \mathrm{km} \, \mathrm{s}^{-1} \, \mathrm{Mpc}^{-1}$ \citep[][]{Dainotti2021ApJ...912..150D,Dainotti2022Galax..10...24D}. \citet{2022ApJ...934L...7R} find $M = -19.253 \pm 0.027$ and $H_0 = 73.04 \pm 1.04  \, \mathrm{km} \, \mathrm{s}^{-1} \, \mathrm{Mpc}^{-1}$ using 42 SNe Ia combined with Cepheids hosted in the same galaxies of these SNe Ia. The $\mu$ provided for the \textit{Pantheon +} sample are computed assuming this value of $M$. In our analysis, we directly use the $\mu_{obs, \mathrm{SNe \, Ia}}$ supplied by \textit{Pantheon}\footnote{\url{https://github.com/dscolnic/Pantheon}} and \textit{Pantheon +}\footnote{\url{https://github.com/PantheonPlusSH0ES}} releases. 
The theoretical $\mu$ is
\begin{equation}
\label{dmlcdm}
\mu_{th} =  25+5 \, \mathrm{log_{10}} \, d_l (\Omega_M, H_0)\,, 
\end{equation}
where $d_l$ is the luminosity distance defined in Eq. \eqref{nonflatdl} in Megaparsec (Mpc) and $c$ the speed of light.
In the case of SNe Ia, we slightly modify Eq. \eqref{dmlcdm} using the more precise formula provided by \citet{2019ApJ...875..145K}:
\begin{equation}
\label{dmlcdm_corr}
\mu_{th} =25+ 5 \, (1+z_{hel}) \frac{c}{H_{0}} \,  \begin{cases}
  \frac{1}{\sqrt{|\Omega_{k}|}} sin(\sqrt{|\Omega_{k}|} \int^{z_{HD}}_{0} \frac{d \zeta}{E(\zeta)} ) & \text{ if } \Omega_{k} < 0,\\ 
 \int^{z_{HD}}_{0} \frac{d \zeta}{E(\zeta)} & \text{ if } \Omega_{k} = 0, \\ 
 \frac{1}{\sqrt{\Omega_{k}}} sinh(\sqrt{\Omega_{k}} \int^{z_{HD}}_{0} \frac{d \zeta}{E(\zeta)} ) & \text{ if } \Omega_{k} > 0\,,
\end{cases}
\end{equation}
where $z_{HD}$, the ``Hubble-Lema\^itre diagram” redshift, accounts for peculiar velocity and CMB corrections, $z_{hel}$ is the heliocentric redshift, and $E(z)$ is generally given by Eq. \eqref{wcdm}.\\

\subsubsection{The distance moduli for GRBs}
For GRBs, the 3D X-ray fundamental plane relation employed to use them as cosmological tools have the following form:
\begin{equation}
\log_{10} L_{a,X} =  a \, \log_{10} T^{*}_{X} + b \, \log_{10} L_{peak} + c_2,
\label{3drelation}
\end{equation}
where $a$, $b$, and $c_2$ are the fitting parameters of the relation in a given cosmological model. 
The luminosities are computed from the measured fluxes by applying the relationship between fluxes $F$ and luminosities $L$, $L= 4 \pi d_{l}^{2}\, \cdot  F \cdot  \, K$, where $d_l$ is in units of cm and $K$ is K-correction accounting for the cosmic expansion \citep{2001AJ....121.2879B}. 
The values of $a$, $b$, and $c_2$ parameters in Eq. \eqref{3drelation} are determined by fitting the 3D relation with the Bayesian technique, \cite{Kelly2007}, which is based on the Markov Chain Monte Carlo (MCMC) approach and allows us to consider error bars on all quantities and also an intrinsic dispersion $sv$ of the relation. The following method accounts for heteroskedastic errors on all random variables considered when performing fitting.
As explained in Section \ref{EP correction}, $L_{a,X}$, $T^{*}_{X}$, and $L_{peak}$ in Eq. \eqref{3drelation} could also be corrected to account for their redshift evolution through the application of the EP method \citep{Dainotti2022MNRAS.tmp.2639D}.
As for SNe Ia, the physical quantity we consider for GRBs is $\mu_{obs, \mathrm{GRBs}}$ computed assuming the reliability of the 3D relation. Converting luminosities into fluxes in Eq. \eqref{3drelation} through the above-defined relation between these two quantities \citep[see][for the mathematical derivation]{Dainotti2022MNRAS.tmp.2639D} and using $\mu = 5 \, \mathrm{log_{10}} \, d_l + 25$, we obtain 
\begin{small}
\begin{equation}
\label{dmGRBs}
\mu_{obs, \mathrm{GRBs}} =  5 \Bigg[ -\frac{\log_{10} F_{a,X}}{2 (1-b)}+\frac{b \cdot \log_{10} F_{peak}}{2 (1-b)} - \frac{(1-b)\log_{10}(4\pi)+c_2}{2 (1-b)} + \frac{a \log_{10} T^{*}_{X}}{2 (1-b)} \Bigg] + 25
\end{equation}
\end{small}
where we consider the K-correction already applied to all quantities. 
\citet{Dainotti2022MNRAS.tmp.2639D} have proved that the cosmological constraints do not depend on the choice of using distance moduli or logarithmic luminosities in the likelihood since the values of cosmological parameters obtained in the two approaches are consistent within 1 $\sigma$. Thus, we prefer to use the distance moduli to remove one degree of freedom, as the parameter $c_2$ is fixed in Eq. \eqref{dmGRBs} following \citet{Dainotti2022MNRAS.tmp.2639D}.
Under the same cosmological assumption applied to SNe Ia, $\mu_{th}$ for GRBs is the one already defined in Eq. \eqref{dmlcdm}.\\

\subsubsection{The luminosities for QSOs}
The  X-ray and UV relation is defined as
\begin{equation}\label{RL}
\mathrm{log_{10}} \, L_{X} = g_1 \, \mathrm{log_{10}} \, L_{UV} + n_1\,,
\end{equation}
where $L_X$ and $L_{UV}$ are the luminosities (in units of $\mathrm{erg \, s^{-1} \, Hz^{-1}}$) at 2 keV and 2500 \AA, respectively. 
As for GRBs, we compute $L_{X}$ and $L_{UV}$ in Eq. \eqref{RL} from the observed flux densities $F_{X}$ and $F_{UV}$ (in $\mathrm{erg \, s^{-1} \, cm^{-2} \, Hz^{-1}}$), respectively, using $L_{X,UV}= 4 \pi d_{l}^{2}\ K F_{X, UV}$, where $K=1$ \citep{2020A&A...642A.150L} since the QSO spectral index is assumed to be 1.
One can argue (see \citet{2022ApJ...935L..19P}) that given that the X-UV relation is based on both luminosities the relation is independent from cosmology. However, this is not the case because if we subtract from both members of the Eq. \eqref{RL} the distance luminosity, $d_l$ (Eq. \eqref{nonflatdl}), the parameter $n_1$ still should be subtracted by $d_l$ and thus (unless $g_1=1$) the parameters $g_1$ and $n_1$ change when varying the cosmological parameters.

As detailed in Section \ref{EP correction}, we here also correct $L_X$ and $L_{UV}$ for evolutionary effects by applying the EP method \citep{DainottiQSO,biasfreeQSO2022}. The parameters $g_1$ and $n_1$ can be fitted, along with the intrinsic dispersion $sv_1$ of the relation, through the Kelly method.
Inserting $\mathrm{log_{10}}L_{UV} =  \mathrm{log_{10}}(4 \, \pi \, d_l^2) + \mathrm{log_{10}}F_{UV}$ in Eq. \eqref{RL} provides us the observed physical quantity $\mathrm{log_{10}}L_{X,obs}$ under the assumption of the RL relation.
The theoretical quantity is computed according to $\mathrm{log_{10}}L_{X,th} = \mathrm{log_{10}}(4 \, \pi \, d_l^2) + \mathrm{log_{10}}F_X$, where $d_l$ is the luminosity distance defined in Eq. \eqref{nonflatdl}. 
As a matter of fact, we could also consider as physical quantities the logarithmic luminosities for GRBs and the distance moduli for QSOs, since the approaches with luminosities and with distance moduli are equivalent (i.e. they are related through $d_l$). Nevertheless, we have here described only the case of distance moduli for GRBs and the case of luminosities for QSOs to be consistent with the fitting procedure applied in our previous analysis. 
We prefer to construct the cosmological likelihood for QSOs with the logarithmic luminosities as these are the quantities intrinsic to the RL relation (see Eq. \eqref{RL}) and also the ones that are commonly used for cosmological analyses in the literature \citep[e.g.][]{2020MNRAS.492.4456K,2020MNRAS.497..263K,2020A&A...642A.150L,2021MNRAS.502.6140K,Colgain2022arXiv220611447C,2022MNRAS.510.2753K,Colgain2022arXiv220310558C,2022MNRAS.515.1795B,biasfreeQSO2022,2022arXiv221014432W,2022MNRAS.517.1901L,Cao2023,Khadka2023}. Thus, this approach allows an easier and more immediate comparison with results from other studies, without the need for taking into account possible differences in the respective methodologies. 
\subsubsection{The distance quantity for BAO}
For BAO, the investigated physical quantity is $d_{z} = r_s(z_d)/D_V(z)$ \citep[e.g.][]{2005ApJ...633..560E}, where $r_s(z_d)$ is the sound horizon at the baryon drag epoch $z_d$ and $D_V(z)$ is the volume averaged distance. The observed $d_{z,obs}$ provided in \citet{2018arXiv180707323S} are obtained from the measured $D_V(z)$ and assuming the fiducial $(r_s(z_d) \, \cdot h)_{\mathrm{fid}} = 104.57$ Mpc, where $h$ is the dimensionless Hubble constant $\displaystyle h= {H_{0}}/{100 \, \mathrm{km \,s^{-1} \, Mpc^{-1}}}$, which corresponds to the best-fit of a $\Lambda$CDM model \citep[see][]{2016JCAP...06..023S}.
The theoretical $d_{z,th}$ are instead computed as follows. Since an exact computation of $r_s(z_d)$ would require the use of Boltzmann codes, we estimate it with the following numerical approximation \citep{2015PhRvD..92l3516A,2019JCAP...10..044C}:

\begin{equation}
\label{rs}
r_s(z_d) \sim \frac{55.154 \, e^{-72.3  (\Omega_{\nu}\, h^2 + 0.0006)^{2}}}{(\Omega_M \, h^2)^{0.25351} \, (\Omega_b \, h^2)^{0.12807}} \, \mathrm{Mpc}\,,
\end{equation}
where $\Omega_{b}$ and $\Omega_{\nu}$ are the baryon and neutrino density parameters. In this formula, we fix $\Omega_{\nu} \, h^2 = 0.00064$ and $\Omega_{b} \, h^2 = 0.002237 $ (according to \citealt{Hinshaw_2013} and \citealt{planck2018}).
The theoretical distance $D_V(z)$ required to compute $d_{z,th}$ is defined as \cite{2005ApJ...633..560E}
\begin{equation} \label{DVBAO}
D_{V}(z) = \left[ \frac{cz}{H(z)} \frac{d_l^{2}(z)}{(1+z)^{2}} \right]^{\frac{1}{3}}.
\end{equation}

\subsection{The likelihood statistical assumption}
\label{bestfitlikelihood}

As proved by \citet{Bargiacchi2023arXiv230307076B} and \citet{snelikelihood2022}, only GRBs, independently whether or not the variables are corrected for redshift evolution, can be well approximated by the Gaussian probability distribution function (PDF)
\begin{equation}
\label{gaussianlf}
\displaystyle
\mathrm{PDF_{Gaussian}} = \frac{1}{\sqrt{2 \, \pi} \, \sigma} \, e^{-\frac{1}{2} \left(\frac{x- \hat{x}}{\sigma}\right)^2}\,,
\end{equation}
where, in our case, $x = \Delta \mu_{ \mathrm{GRBs}} = \mu_{obs, \mathrm{GRBs}} - \mu_{th}$. 
Investigating \textit{Pantheon} and \textit{Pantheon +} SNe Ia, \citet{snelikelihood2022} have determined that the best-fit $\Delta \mu_{ \mathrm{SNe \, Ia}}= \mu_{obs, \mathrm{SNe \, Ia}} - \mu_{th}$ distributions are a logistic and a Student's t distribution, respectively (see their Fig. 1). The logistic PDF reads as follows:
\begin{equation}
\label{logistic}
\mathrm{PDF_{logistic}} = \frac{e^{-\frac{(x-\hat{x})}{s}}}{s \, \left(1+ e^{\frac{-(x-\hat{x})}{s}}\right)^2}\,,
\end{equation}
where $s$ is the scale parameter and the variance $\sigma^2$ is given by $\sigma^2 = (s^2 \, \pi^2)/3$. The best-fit parameters of \citet{snelikelihood2022} are $\hat{x}= -0.004$ and $s=0.08$. The generalized Student's t PDF is defined as 
\begin{equation}
\label{student}
\mathrm{PDF_{student}} = \frac{\Gamma\left(\frac{\nu +1}{2}\right)}{\sqrt{\nu \, \pi} \, s \, \Gamma \left(\frac{\nu}{2}\right)} \, \left[1 + \frac{((x- \hat{x})/s)^2}{\nu}\right]^{-\frac{\nu +1}{2}}\,,
\end{equation}
where $\Gamma$ is the $\gamma$ function, $\nu$ are the degrees of freedom ($>0$), and the variance is $\sigma^2 = (s^2 \, \nu) / (\nu -2)$. The corresponding best-fit parameters in \citet{snelikelihood2022} are $\hat{x}= 0.1$, $s=0.12$, and $\nu = 3.8$. 

\citet{Bargiacchi2023arXiv230307076B} have proved that, for all evolutionary cases, the best-fit distribution of $\Delta \mathrm{log_{10}}L_X = \mathrm{log_{10}} \, L_{X,obs} -\mathrm{log_{10}} \, L_{X,}$ is the logistic distribution, whose PDF is the one in Eq. \eqref{logistic}. 
The best-fit logistic curve has $\hat{x}= 0.009$ and $s=0.13$ for the case with evolution and $\hat{x}= 0.0006$ and $s=0.13$ for the case with no evolution.

For BAO, the best fit of the $\Delta d_z=d_{z,obs} - d_{z,th}$ histogram is the Student's t distribution in Eq. \eqref{student}.
The best-fit Student's t has $\hat{x} = 0.0003$, $s = 0.002$, and $\nu = 2.23$.

\subsection{Correction for redshift evolution and selection biases}\label{EP correction}

GRBs and QSOs are invaluable probes of the evolution of the Universe as observed at early epochs. Nevertheless, at high redshifts, selection biases and effects of evolution could deform or generate a correlation between the physical quantities of a source. Hence, using such a relation to test cosmological models would lead to underestimating or overestimating the cosmological parameters \citep{Dainotti2013a}. The Efron \& Petrosian (EP) method \citep{1992ApJ...399..345E} is a statistical technique applied to correct relations for these effects. It has already been employed for GRBs \citep{Dainotti2013a,Dainotti2015b,Dainotti2017A&A...600A..98D,Dainotti2021Galax...9...95D,Dainotti2022MNRAS.tmp.2639D} and QSOs \citep{DainottiQSO,biasfreeQSO2022,Bargiacchi2023arXiv230307076B}. Here, we apply this method by using the results obtained by \citet{Dainotti2022MNRAS.tmp.2639D} for GRBs and by \citet{DainottiQSO} and \citet{biasfreeQSO2022} for QSOs, as already detailed in \citet{Bargiacchi2023arXiv230307076B} and \citet{DainottiGoldQSOApJ2023}. 

In the above-mentioned studies on GRBs and QSOs the variables of luminosities (and time for GRBs) are corrected through the EP method by assuming an evolution with redshift such that $L' = \frac{L}{(1+z)^{k}}$ and $T' = \frac{T}{(1+z)^{k}}$, where $L$ and $T$ are the measured quantities, $L'$ and $T'$ the corrected ones, and $k$ mimics the evolution. We here stress that the specific choice of the functional form, either a power-law or more complex functions, does not impact the results in more than 2 $\sigma$ \citep{2011ApJ...743..104S,Dainotti2013a,Dainotti2021ApJ...914L..40D,DainottiQSO}. 
The value of $k$ that removes the evolution is determined by applying Kendall's $\tau$ statistic, in which:
\begin{equation}
\label{tau}
    \tau =\frac{\sum_{i}{(\mathcal{R}_i-\mathcal{E}_i)}}{\sqrt{\sum_i{\mathcal{V}_i}}}.
\end{equation}
In this formula, $i$ runs on the sources that at $z_i$ have a luminosity greater than $L_{min, i}$, the lowest observable luminosity at that redshift, $z_i$; $L_{min, i}$ is obtained by assuming a flux limit of the observations. This value is chosen such that at least 90\% of the overall sample survives the selection and is compatible with the total
distribution according to the Kolmogorov-Smirnov test \citep{Dainotti2013a,Dainotti2015b,Dainotti2017A&A...600A..98D,Levine2022ApJ...925...15L,DainottiQSO,Dainotti2022MNRAS.514.1828D}. The rank $\mathcal{R}_i$ is the number of points in the associated set of the $i$-source, which is defined by the $j$-points with $z_j \leq z_i$ and $L_{z_j} \geq  L_{min, i}$. $\mathcal{E}_i = \frac{1}{2}(i+1)$ and $\mathcal{V}_i = \frac{1}{12}(i^{2}+1)$ are the expectation value and variance, respectively, when the evolution is removed. Hence, $\tau = 0$ corresponds to the absence of evolution and yields the value of $k$ that removes the correlation with the redshift. If $| \tau | > n$ the hypothesis of non-correlation is rejected at $n \sigma$ level, thus $|\tau| \leq 1$ provides us the 1 $\sigma$ uncertainty on $k$. Then, we use the obtained $k$ to compute $L'$ for the total sample. The same technique is also used for correcting the time variable for GRBs.
The $k$ values obtained for the original sample of 222 GRBs are provided in \citet{Dainotti2022MNRAS.tmp.2639D}, while here, for consistency, we derive for the first time the evolutionary parameters for the sample of 50 GRBs used in this work, as specified in the following section. For QSOs, we employ $k_{UV} = 4.36 \pm 0.08$ and $k_X = 3.36 \pm 0.07$ from \citet{DainottiQSO}. We use the corrected luminosities and times computed with these values of $k$ when accounting for a ``fixed" evolution. We here notice that in this procedure, $k$ is determined by assuming a specific cosmological model, required to obtain the luminosities $L$ from the measured fluxes. In \citet{Dainotti2022MNRAS.tmp.2639D} and \citet{DainottiQSO} the chosen model is a flat $\Lambda$CDM model with $\Omega_M =0.3$ and $H_0 = 70  \, \mathrm{km} \, \mathrm{s}^{-1} \, \mathrm{Mpc}^{-1}$. 
This originates from the well-known ``circularity problem": the assumptions of a specific cosmology impact the constraints obtained on the cosmological parameters since the fits are performed using the luminosities computed through this assumption.
This issue can be completely overcome by considering $k$ as a function of the cosmological parameters investigated, as shown in \citet{Dainotti2022MNRAS.tmp.2639D} (for GRBs) and \citet{DainottiQSO}, \citet{biasfreeQSO2022}, and \citet{DainottiGoldQSOApJ2023} (for QSOs). Indeed, in this innovative procedure, $k$ is computed over a grid of values of the cosmological parameters of interest (i.e. $\Omega_M$, $H_0$, $\Omega_k$, $w$), thus providing the functions $k(\Omega_M, H_0,\Omega_k)$ and $k(\Omega_M, H_0,w)$. These studies have proved that $k$ depends on $\Omega_M$, $\Omega_k$, and $w$, but not on $H_0$. This method overcomes the circularity problem since, when the MCMC inspects specific values of the cosmological parameters, it applies the correction with the $k$ corresponding to those values. In this work, we refer to this technique as ``varying" evolution. As in \citet{Dainotti2022MNRAS.tmp.2639D} and \citet{Bargiacchi2023arXiv230307076B}, we here compare the cosmological results from all possible approaches to the correction for the evolution: without correction, with ``fixed", and ``varying" evolution (see Table \ref{tab:bestfit}).  For sake of clarity, we here stress that the notations of ``fixed" and ``varying" evolution are only referred to the correction for redshift evolution of luminosities, and thus to astrophysical properties of the sources, and are not directly related to any possible evolution of dark energy in the cosmological models.

\subsubsection{The evolutionary parameters for the GRB Platinum sample}

\begin{figure}
\centering
\includegraphics[height=6.8cm,width=0.49\textwidth]{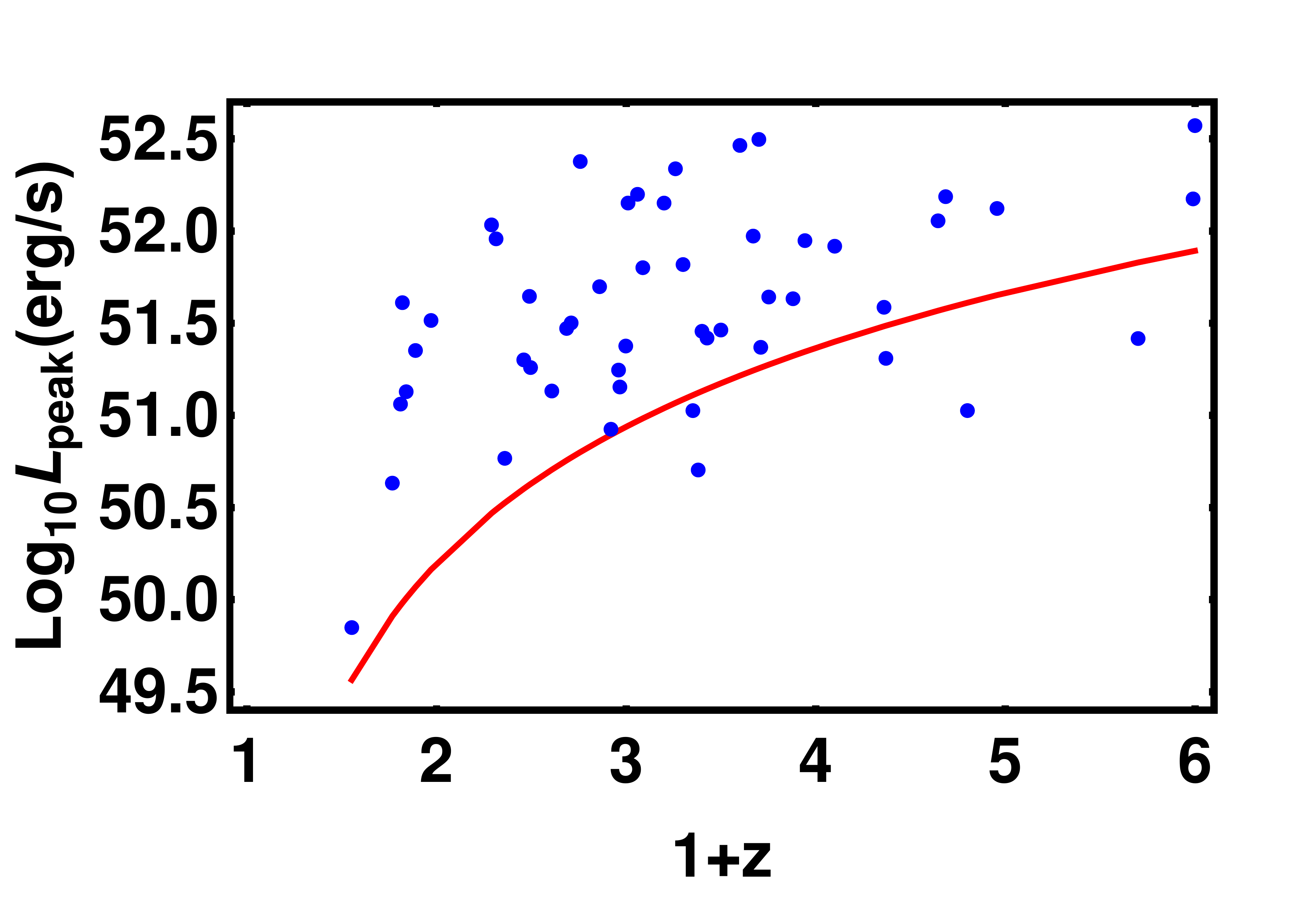}
\includegraphics[width=0.49\textwidth]{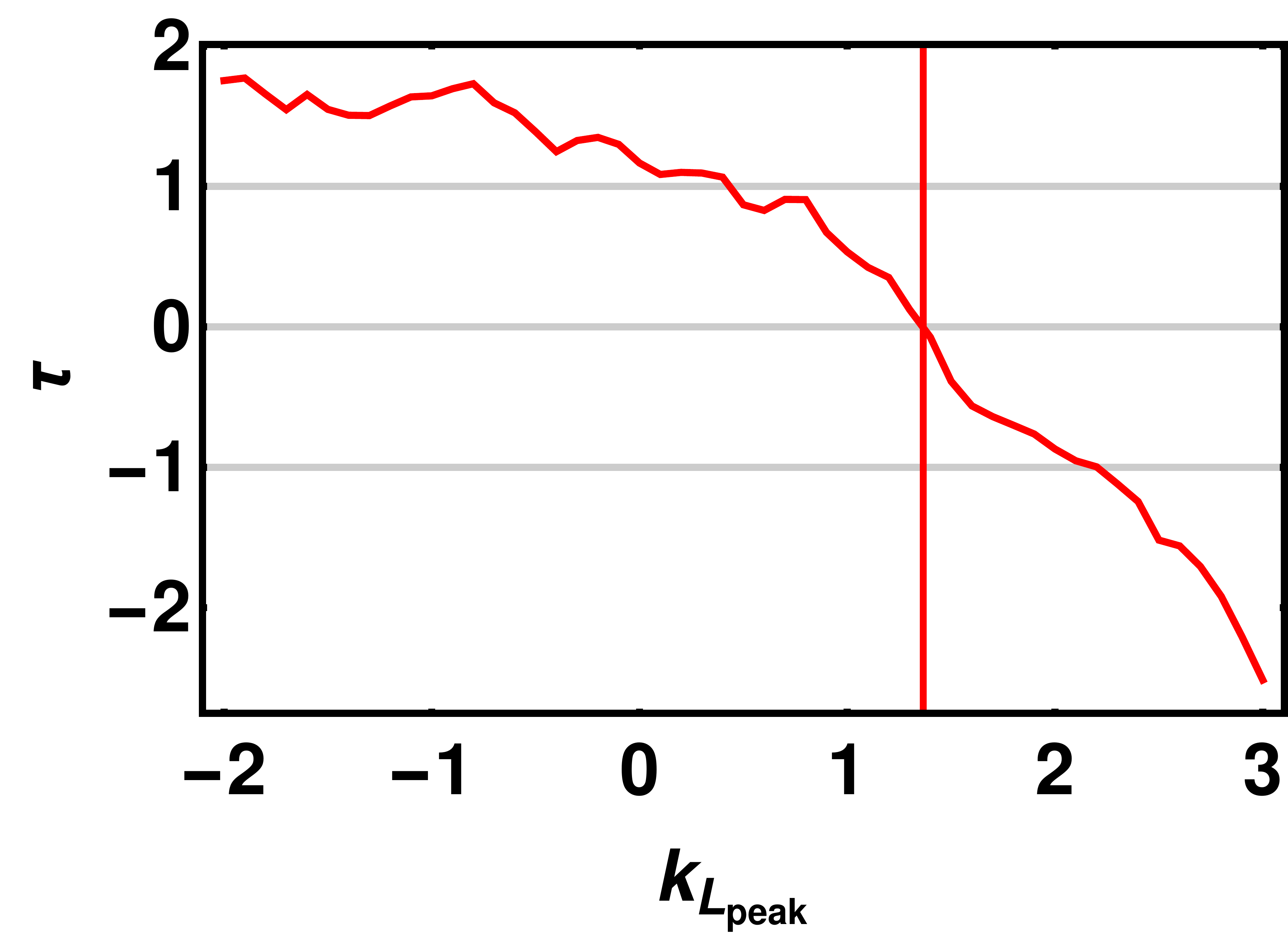}\\
\includegraphics[height=6.8cm,width=0.49\textwidth]{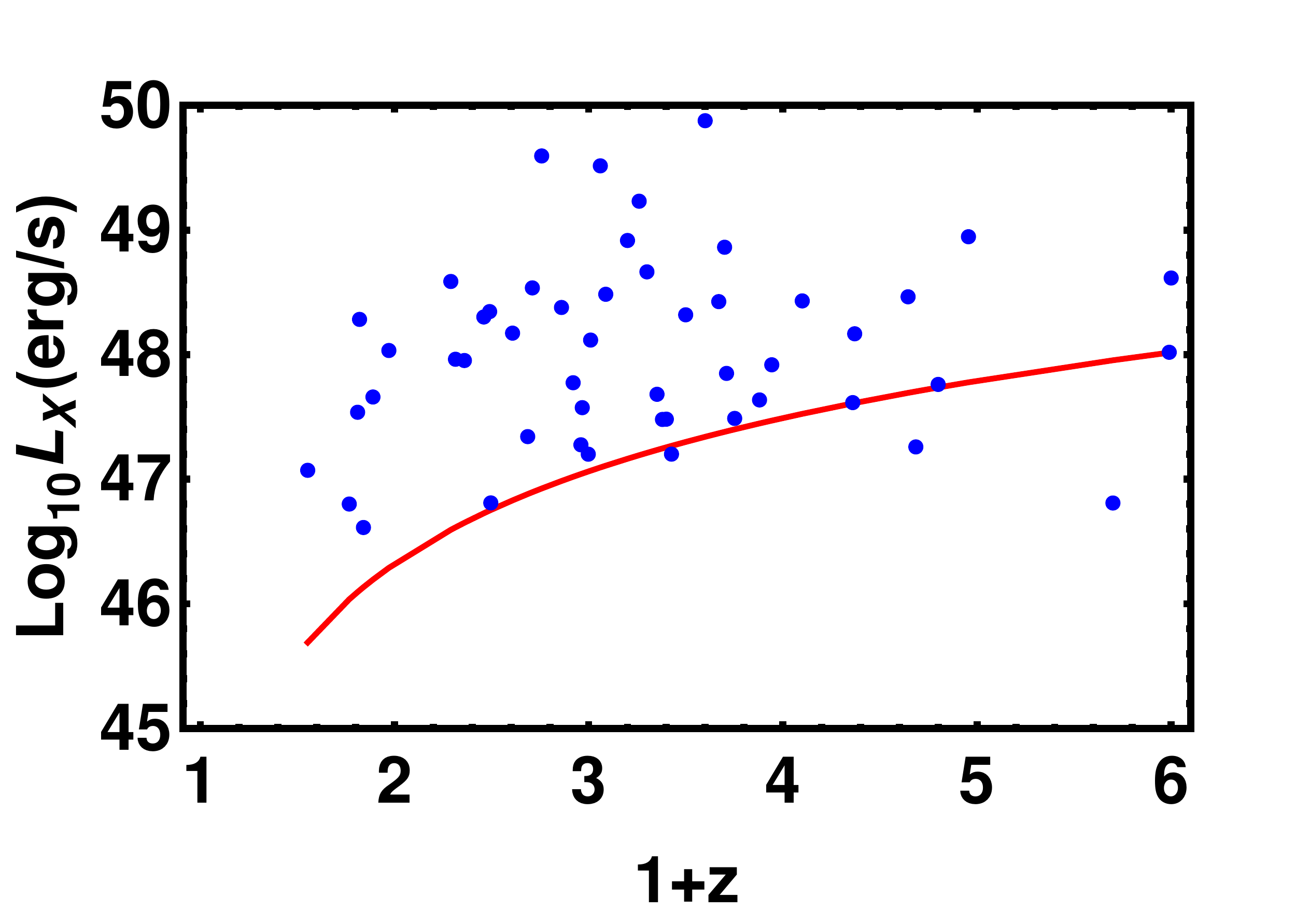}
\includegraphics[width=0.49\textwidth]{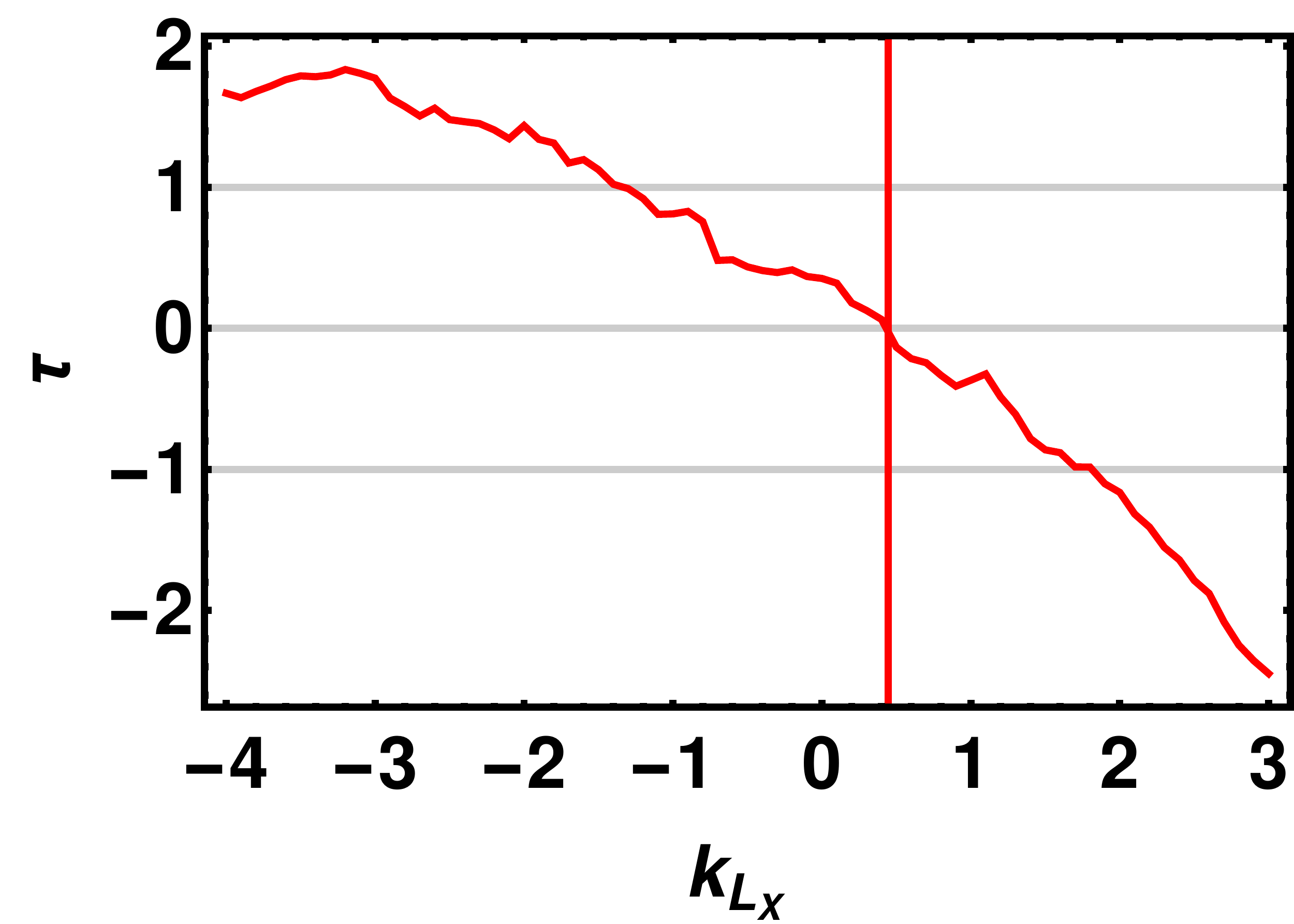}\\
\includegraphics[height=6.8cm,width=0.49\textwidth]{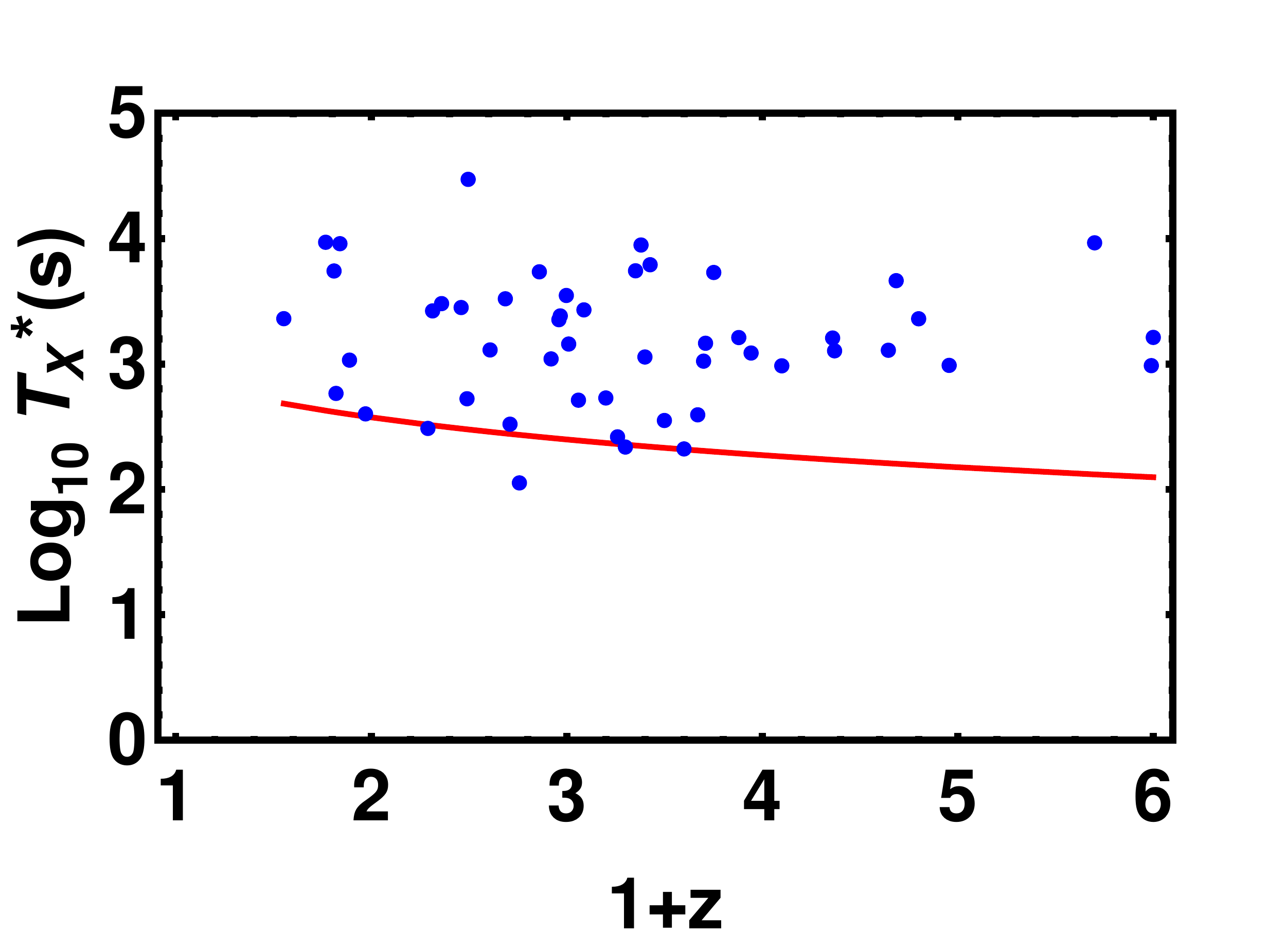}
\includegraphics[width=0.49\textwidth]{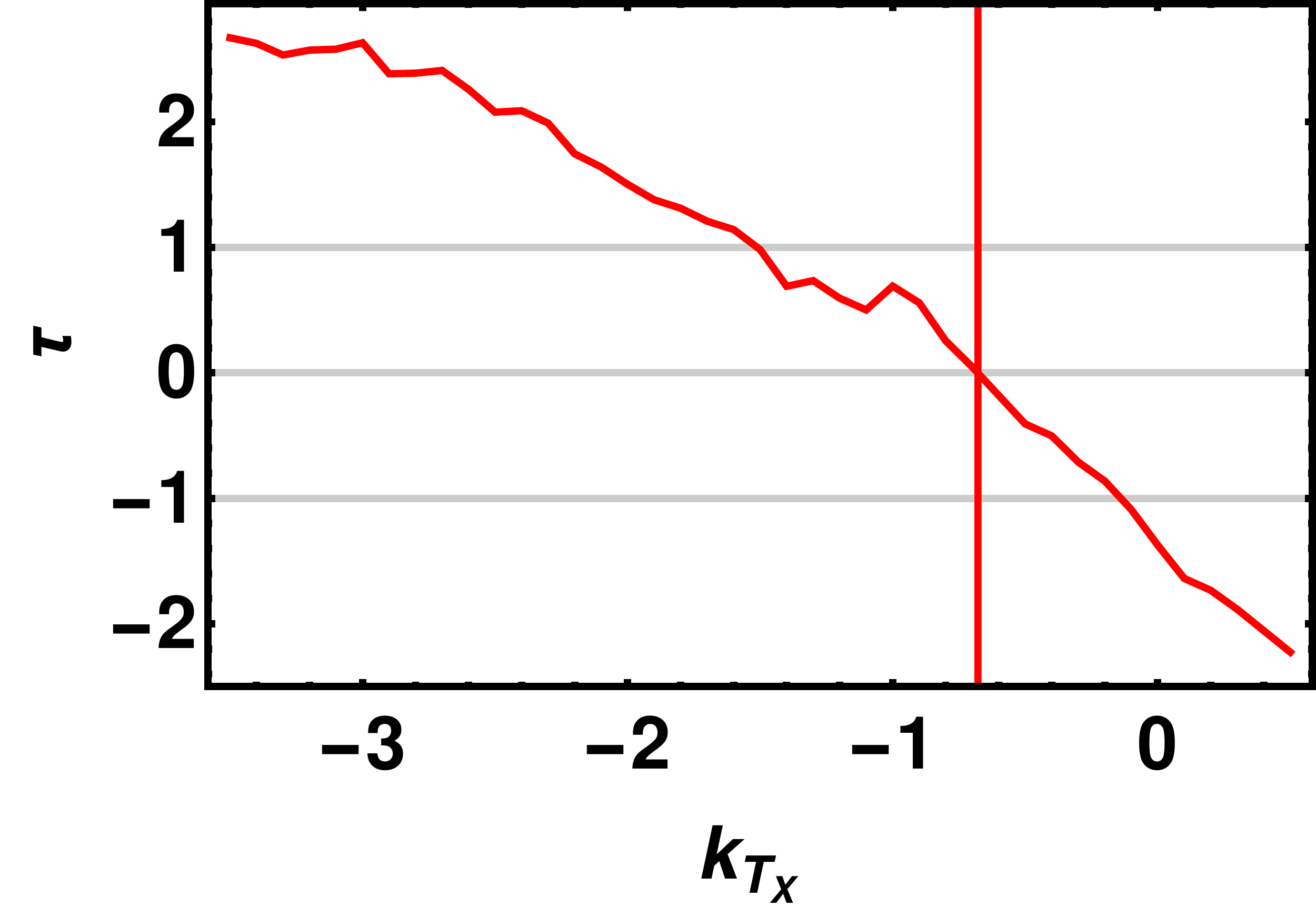}\\
\caption{
The application of the EP method to our Platinum sample for the parameters involved in the 3D fundamental relation.
The left panels show the distribution of the parameters in $\left(1+z\right)$ with limiting lines in red, while the right panels show the determination of the $\tau$ parameter.
The vertical red solid lines identify the value of $k$ corresponding to $\tau=0$, for which the redshift evolution is removed, and the gray horizontal lines provide $\tau=0$ and its 1 $\sigma$ uncertainty.
}

\label{fig:EP}
\end{figure}

Differently from \citet{Dainotti2022MNRAS.514.1828D} and \citet{Dainotti2022MNRAS.514.1828D}, here, for the first time, we have used the parameter of the evolution pertinent to this sample. We show in Fig. \ref{fig:EP} (left panels) the limiting luminosity for $L_{peak}$ and $L_{X}$, and the limiting time for $T_{X}$ along with their corresponding determination $\tau(k)$ (right panels).
In the previous analyses, the values of the evolution were produced considering the sample of 222 GRBs since we had demonstrated in \citet{Dainotti2022MNRAS.514.1828D}
that the Platinum is statistically similar to the full sample with the exception of the time. However, here we would like to be as much precise as possible and consider the exact evolution pertinent to these parameters both when we consider fixed and varying evolution. This new analysis leads to the following results for the evolutionary parameters of all quantities of interest when assuming a flat $\Lambda$CDM model with $\Omega_M=0.3$:  $k_{L_{peak}} = 1.37^{+0.83}_{-0.93}$, $k_{T_{X}} = - 0.68^{+0.54}_{-0.82}$, and $k_{L_{X}} = 0.44^{+1.37}_{-1.76}$. As expected, these values are compatible within 1 $\sigma$ with the ones obtained in \citet{Dainotti2022MNRAS.tmp.2639D} for the original sample of 222 GRBs, but in this case, we obtain larger uncertainties due to the reduction in the size of the sample. Additionally, also the functions $k(\Omega_M, H_0,\Omega_k)$ and $k(\Omega_M, H_0,w)$ above mentioned and used for the treatment of the varying evolution are coherently computed on the sample of 50 GRBs following the procedure already described.

\subsection{Fitting method and assumptions on free parameters}

All our cosmological fits are performed with the Kelly method. We apply this method in the most general case in which all free parameters of the investigated cosmological model are free to vary contemporaneously. More precisely, for the non-flat $\Lambda$CDM model, the free parameters are $\Omega_M$, $\Omega_k$, and $H_0$, while for the flat $w$CDM model $\Omega_M$, $w$, and $H_0$ vary together. To guarantee a complete exploration of the whole parameter space, we require very wide uniform priors on all the free parameters: $0 \leq \Omega_M \leq 1$, $50 \, \mathrm{km} \, \mathrm{s}^{-1} \, \mathrm{Mpc}^{-1} \leq H_0 \leq 100 \, \mathrm{km} \, \mathrm{s}^{-1} \, \mathrm{Mpc}^{-1}$,  $-2.5 \leq w \leq -0.34$, $-0.7 \leq \Omega_k \leq 0.2$, $0 < \nu_{\mathrm{SNe}} < 10$, $0 < \nu_{\mathrm{BAO}} <10$, $-2 < a < 0$, $0 < b < 2$, $0< sv < 2$, $0.1 < g_1 < 1$, $2< n_1 < 20$, and $0 < sv_1 < 2$. The limit $w=-0.34$ derives from the second Friedmann's equation, in which the current acceleration in the expansion of the Universe \citep{riess1998,perlmutter1999} can be ascribed to a dark energy contribution only if $ w(z)<-1/3$. 

We here also notice that in the non-flat $\Lambda$CDM model the region of the space parameter $(\Omega_{M}, \Omega_{k})$ that ensures the presence of an initial singularity (i.e. the Big-Bang) is defined by the following constraints \citep{1992ARA&A..30..499C}:
\begin{equation}
\label{nobigbang}
\Omega_{k} \geq
\left\{\begin{matrix}
 1 - \Omega_{M} - 4 \, \Omega_{M}\, \mathrm{cosh}^{3} \Bigg[\frac{1}{3} \, \mathrm{arccosh} \left( \frac{1- \Omega_{M}}{\Omega_{M}}\right)\Bigg] & \, \Omega_{M}\leq\frac{1}{2}, \\ \\
 1 - \Omega_{M} - 4 \, \Omega_{M}\, \mathrm{cos}^{3} \Bigg[\frac{1}{3} \, \mathrm{arccos} \left( \frac{1- \Omega_{M}}{\Omega_{M}}\right)\Bigg] & \, \Omega_{M}>\frac{1}{2}.
\end{matrix}\right.
\end{equation}
Moreover, in a flat $w$CDM model, the condition $w \geq (\Omega_{M}-1)^{-1}$ is the one that should be satisfied to guarantee $\frac{d\, H(z)}{d\, z}|_{z=0} \geq 0$, which verifies the null energy condition \citep{2000ppeu.conf...98V}.

Since the MCMC methods are based on a random exploration of the parameter's space, to guarantee the reliability of the obtained posterior distributions of the free parameters, we need to test the convergence of the algorithm. To this end, we have applied the Gelman-Rubin convergence diagnostic test, which uses the square root of the ratio between the variance intra-chain (within each chain) and the variance inter-chain (between one chain and another) ($R$) as a convergence probe: if $R$ is large, the former is greater than the latter and thus the chain has not yet converged. In our fitting procedure similarly to \citet{biasfreeQSO2022} and \citet{Bargiacchi2023arXiv230307076B}, we require $R-1 < 0.05$ for each free parameter, which is a more restrictive requirement compared to $R-1 < 0.1$ often employed in cosmological studies \citep{DainottiQSO}. This allows us to stop the MCMC run once the convergence has been reached for all free parameters.\\
\subsection{The joint likelihoods}
In our work, we perform all the fits with both the traditional $\cal L_G$ likelihoods and with the newly found likelihoods ($\cal L_N$) (see Table \ref{tab:bestfit}). In the case of $\cal L_G$, the likelihood for each probe separately is obtained from Eq. \eqref{gaussianlf}. Then, since we fit all the probes together, the individual likelihoods are multiplied (or summed if considered in logarithm) to obtain the joint likelihood function. As detailed in Section \ref{quantities}, we consider different quantities for the variables in Eq. \eqref{gaussianlf} according to the probe considered. Specifically, the likelihood ($\mathcal{L}$) for SNe Ia is (in units of natural logarithm ln):
\begin{equation} \label{lfsne}
\text{ln}(\mathcal{L})_{\mathrm{SNe Ia}} = -\frac{1}{2} \Bigg[\left(\boldsymbol{\mu_{obs, \mathrm{SNe \, Ia}}}-\boldsymbol{\mu_{th}}\right)^{T} \, \textit{C}^{-1} \, \left(\boldsymbol{\mu_{obs, \mathrm{SNe \, Ia}}}-\boldsymbol{\mu_{th}}\right)\Bigg]
\end{equation}
where $\boldsymbol{\mu_{obs, \mathrm{SNe \, Ia}}}$ is the distance modulus measured as expressed in Eq. \eqref{muobs}, $\textit{C}$ is the associated covariance matrix that includes both statistical and systematic uncertainties on the measured distance moduli provided by \textit{Pantheon} and \textit{Pantheon +} releases, and $\boldsymbol{\mu_{th}}$ is the distance modulus predicted by the cosmological model assumed, yet depending both on the free parameters of the model and the redshift as specified in Eq. \eqref{dmlcdm_corr}.
For GRBs instead $\mathcal{L}$ is defined as
\begin{equation} \label{lfgrb}
\text{ln}(\mathcal{L})_{\text{GRBs}} = -\frac{1}{2} \sum_{i=1}^{N} \left[ \frac{(\mu_{obs, \mathrm{GRBs},i} -\mu_{th,i})^{2}}{s^{2}_{i}} + \text{ln}(s^{2}_{i})\right]
\end{equation}
where $\mu_{obs, \mathrm{GRBs},i} $ is the one in Eq. \eqref{dmGRBs}, while $\mu_{th,i}$ is the one in Eq. \eqref{dmlcdm}. Moreover, $s^2_{i} = \sigma ^{2}_{\mu_{obs, \mathrm{GRBs}}} + sv^{2}$ which accounts for the statistical uncertainties on the measured quantities (i.e. $F_{a,X}$, $F_{peak}$, and $T^{*}_{X}$), but also the intrinsic dispersion $sv$ of the 3D fundamental plane relation, which is another free parameter of the fit. Similarly, in the case of QSOs, we define $\mathcal{L}$ as
\begin{equation} \label{lfqso}
\text{ln}(\mathcal{L})_{\text{QSOs}} = -\frac{1}{2} \sum_{i=1}^{N} \left[ \frac{(\mathrm{log_{10}}L_{X,i}-\mathrm{log_{10}}L_{X,th,i})^{2}}{\hat{s}^{2}_{i}} + \text{ln}(\hat{s}^{2}_{i})\right]
\end{equation}
where $\mathrm{log_{10}}L_{X,i}$ is given in Eq. \eqref{RL} and $\hat{s}^{2}_{i} = \sigma ^{2}_{\mathrm{log_{10}}L_{X,i}} + g_1^{2} \sigma ^{2}_{\mathrm{log_{10}}L_{UV,i}} + sv_1^{2}$ in which the statistical uncertainties on $\mathrm{log_{10}}L_{X}$ and $\mathrm{log_{10}}L_{UV}$ and the intrinsic dispersion $sv_1$ of the RL relation are taken into account.
In the case of BAO $\mathcal{L}$ reads as

\begin{equation} 
\label{lfbao2}
\text{ln}\left(\mathcal{L}\right)_{\mathrm{BAO}} = -\frac{1}{2} \left[\left(d_{z,obs}-d_{z,th}\right)^{T}\, \mathcal{C}^{-1}\, \left(d_{z,obs}-d_{z,th}\right)\right]
\end{equation}


with $\mathcal{C}$ the covariance matrix provided in \citet{2016JCAP...06..023S}.\\
When considering instead the $\cal L_N$ likelihoods in place of the $\cal L_G$ ones, we need to replace in Eqs. \eqref{lfsne}, \eqref{lfgrb}, \eqref{lfqso}, and \eqref{lfbao2} the Gaussian PDF with the proper PDF for each probe, namely with the logistic one (Eq. \eqref{logistic}) for \textit{Pantheon}
 SNe Ia and QSOs, the Student's T PDF (Eq. \eqref{student}) for \textit{Pantheon +} SNe Ia and BAO, while the Gaussian remains unchanged for GRBs.
 Once defined the likelihood for each source separately, we build the joint likelihood $\mathcal{L}_{\text{joint}}$ as the sum 
 $\text{ln}(\mathcal{L})_{\text{joint}} = \text{ln}(\mathcal{L})_{\text{SNe Ia}}+\text{ln}(\mathcal{L})_{\text{GRBs}}+\text{ln}(\mathcal{L})_{\text{QSOs}}+\text{ln}(\mathcal{L})_{\text{BAO}}$.\\
Indeed, we here would like to stress that in our case, the data for different probes (SNe Ia, GRBs, QSOs, and BAO) were collected independently, thus the joint density of all probes is simply the product of densities for individual probes. This means that the joint log likelihood ($\text{ln}(\mathcal{L})_{\text{joint}}$) is simply the sum of log-likelihoods from different probes. We do not consider different weights for different probes in the definition of the joint likelihood: as in the regular likelihood, all observations contribute equally to the joint likelihood. In practice, this means that the probes with the largest number of observations have the largest influence on the estimated parameters if the probes are able to constrain these parameters. As a matter of fact, it is worth noticing that QSOs, which are the data set with the largest number of sources, cannot constrain the cosmological parameters alone, as shown in \citet{biasfreeQSO2022}, but only if they are used jointly with SNe Ia. The same consideration also applies to the GRBs alone (see e.g. \citealt{Dainotti2022MNRAS.tmp.2639D}, \citealt{Dainotti2022MNRAS.514.1828D}, and \citealt{Dainotti2022PASJ}). Indeed, in \citet{Dainotti2022MNRAS.tmp.2639D}, it is shown that in order to obtain reliable cosmological parameters, 100 MCMC simulations must be performed and then averaged. This is the only way to compare the results of the single probes for QSOs and GRBs with the results obtained jointly. With these averages at our end, we can state that the estimates on the cosmological parameters obtained separately from the individual data sets of SNe Ia, GRBs with the averaged 100 MCMC runs, GRBs + SNe Ia, QSOs + SNe Ia, and BAO are consistent within $\sim 1 \sigma$ (see \citealt{biasfreeQSO2022}, \citealt{snelikelihood2022}, and \citealt{Dainotti2022MNRAS.tmp.2639D}).
However, since we obtain $H_0$ and $\Omega_M$ larger for QSOs this must be investigated further as we detailed in Sect. \ref{conclusions}.\\
We are aware that combining likelihoods based on different statistical models applied to data sets with different structures is not trivial and caution is needed when dealing with a ``composite likelihood" (i.e. a likelihood obtained as the product of sub-likelihoods), as pointed out for example in \citet{prabhu1988statistical}, \citet{varin2008composite}, \citet{larribe2011composite}, and \citet{lindsay2011issues}.
Nevertheless, these concerns are mainly related to situations in which the product of likelihoods is used as some approximation to unknown or some more complicated joint distributions. Instead, in our case, the joint likelihood corresponds to the true joint probability density for all probes due to the independence of observations for different data sources.\\
Additionally, concerning the construction of the joint likelihood, we do expect the presence of covariance between some of the free parameters fitted. More specifically, Eqs. \eqref{dmGRBs} and \eqref{RL} show the anti-correlation between the parameters $a$ and $b$ (since $c_2$ is fixed), and $g_1$ and $n_1$, while Eq. \eqref{nonflatdl} provides us with the covariance among the cosmological parameters of the model investigated. These correlations are clearly visible in Figs. \ref{fig: nonflat}, \ref{fig: nonflatnewlikelihoods}, \ref{fig: wCDM}, and \ref{fig: wCDMnewlikelihoods}. However, the covariance between $g_1$ and $n_1$ could be removed if we shift the variables $\mathrm{log_{10}}F_{UV}$ and $\mathrm{log_{10}}F_{X}$ by their mean values, thus normalizing them, as done in \citet{2020A&A...642A.150L} and \citet{2022MNRAS.515.1795B}. We performed the same analysis for GRBs, and the covariance can be removed when we subtract the average $L_{peak, X}$ and the $T_X$ from the best-fit values.

\section{Results}
\label{results}

We here present our main results for the two cosmological models studied and compare them with other works in the literature. In addition to the best-fit values and 1 $\sigma$ uncertainties of the cosmological free parameters, we present in Table \ref{tab:bestfit} also the z-scores $\zeta$ already adopted in previous cosmological studies \citep{Dainotti2022MNRAS.tmp.2639D,biasfreeQSO2022,snelikelihood2022} to draw a statistical interpretation of each value of $H_0$ obtained. Indeed, $\zeta$ is defined as $\zeta= |H_{0,i} - H_{0,our}|/ \sqrt{\sigma^2_{H_{0,i}} + \sigma^2_{H_{0,our}}}$ where $H_{0,our}$ is our $H_0$ value,  $\sigma_{H_{0,our}}$ its one $\sigma$ error, $H_{0,i}$ and $\sigma_{H_{0,i}}$ the reference value and its one $\sigma$ uncertainty, respectively. In this work, we employ this parameter to quantify the discrepancy between our results and three reference values in the literature: $H_0 = 67.4 \pm 0.5 \, \mathrm{km} \, \mathrm{s}^{-1} \, \mathrm{Mpc}^{-1}$, \citep{planck2018}, $H_0 = 73.04 \pm 1.04 \, \mathrm{km} \, \mathrm{s}^{-1} \, \mathrm{Mpc}^{-1}$ \citep{2022ApJ...934L...7R}, and $H_0 = 70.00 \pm 0.13 \, \mathrm{km} \, \mathrm{s}^{-1} \, \mathrm{Mpc}^{-1}$ provided by \citet{snelikelihood2022} within a flat $\Lambda$CDM model with $\Omega_M = 0.3$, as in \citet{scolnic2018}.
Moreover, the value of $H_0 = 70 \, \mathrm{km} \, \mathrm{s}^{-1} \, \mathrm{Mpc}^{-1}$ is also the one obtained in several works that combine SNe Ia from \textit{Pantheon} with other probes \citep[e.g.][]{2020JCAP...07..045D,2021MNRAS.504..300C,2022JHEAp..34...49A}.
For simplicity, we use the following notation to refer to the different fiducial values considered: $\zeta_{CMB}$, $\zeta_{P+}$, and $\zeta_P$ when comparing to the $H_0$ from the CMB, from \textit{Pantheon +}, and \textit{Pantheon}, respectively.

\subsection{Results of non-flat $\Lambda$CDM model with different likelihoods}
\label{resultsnonflat}
\begin{figure}
\centering
\gridline{
\fig{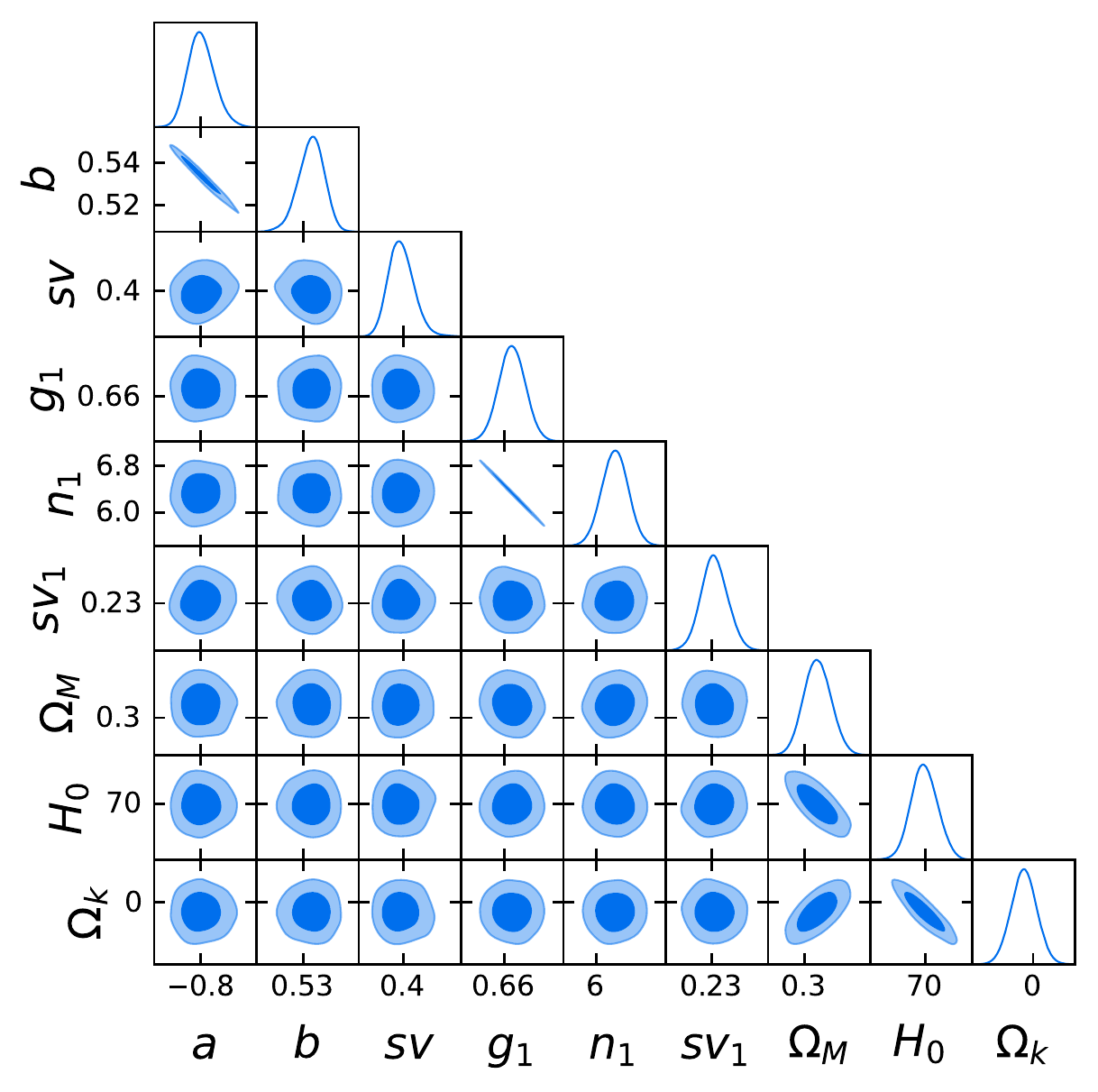}{0.364\textwidth}{(a) \textit{Pantheon} SNe Ia + GRBs + QSOs + BAO without correction for redshift evolution}
\fig{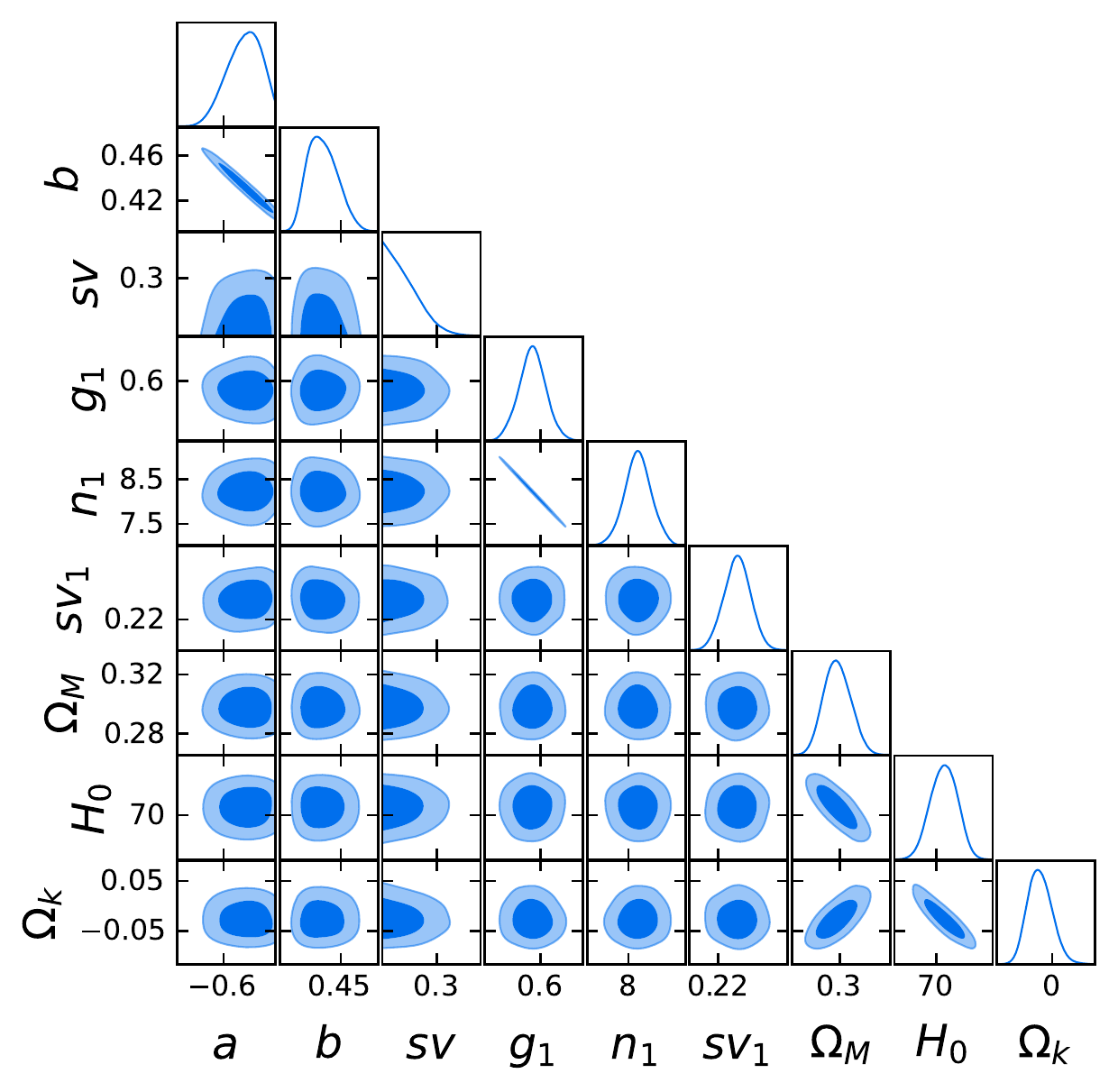}{0.364\textwidth}{ (b) \textit{Pantheon} SNe Ia + GRBs + QSOs + BAO with fixed correction for redshift evolution}}
\gridline{
\fig{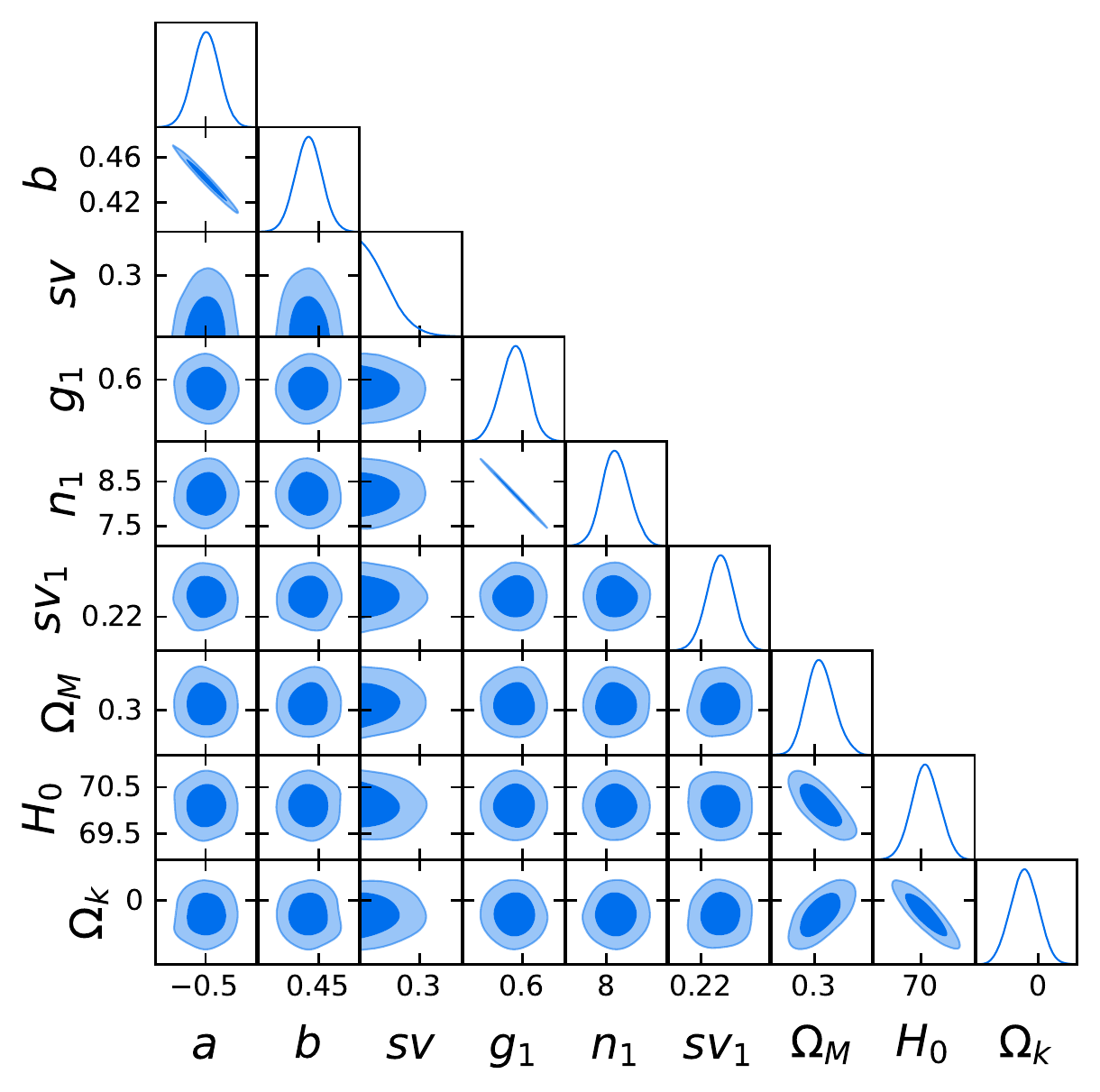}{0.364\textwidth}{ (c) \textit{Pantheon +} SNe Ia + GRBs + QSOs + BAO with varying correction for redshift evolution}\label{}
\fig{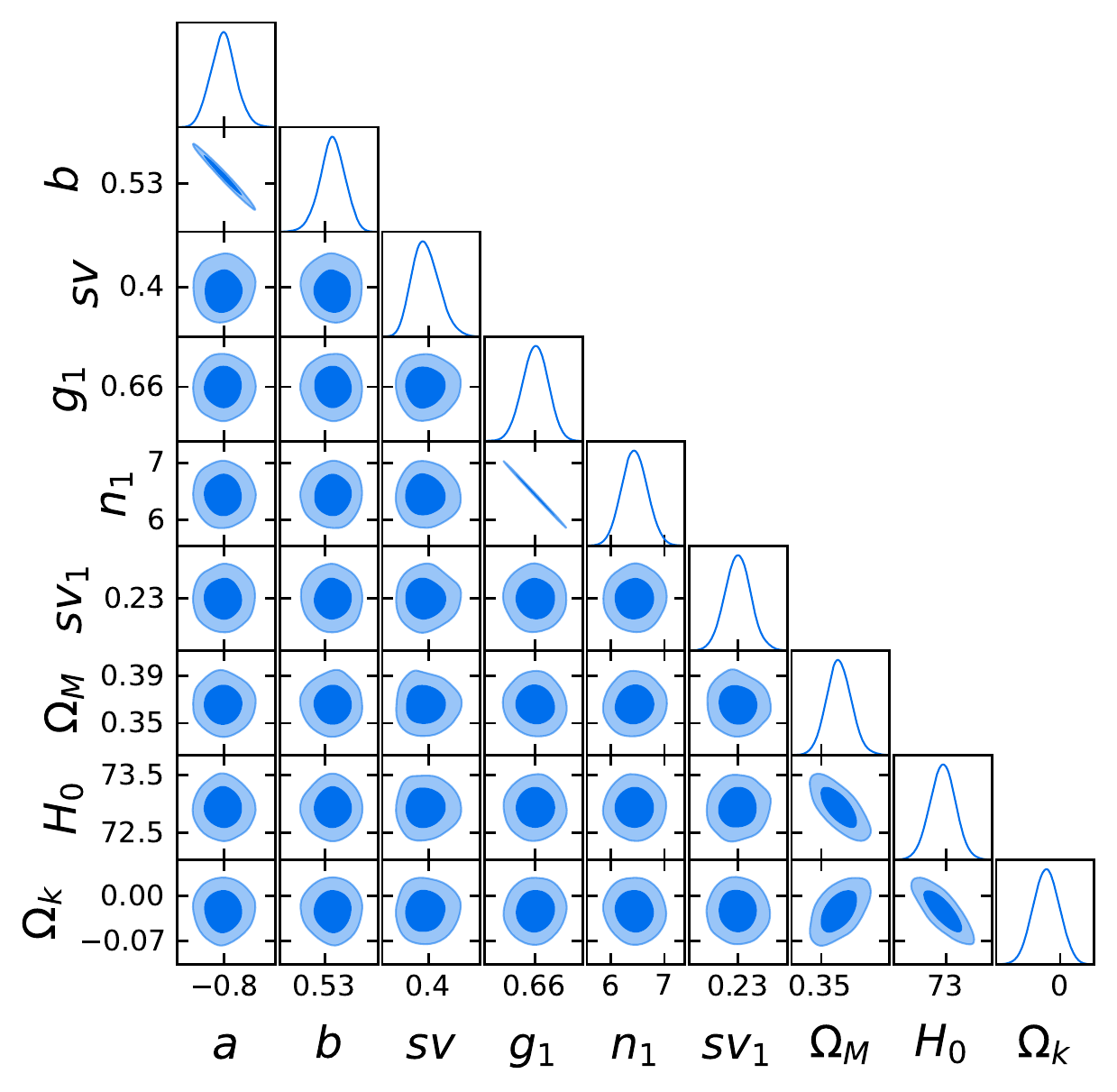}{0.364\textwidth}{ (d) \textit{Pantheon +} SNe Ia + GRBs + QSOs + BAO without correction for redshift evolution}\label{}}
\gridline{
\fig{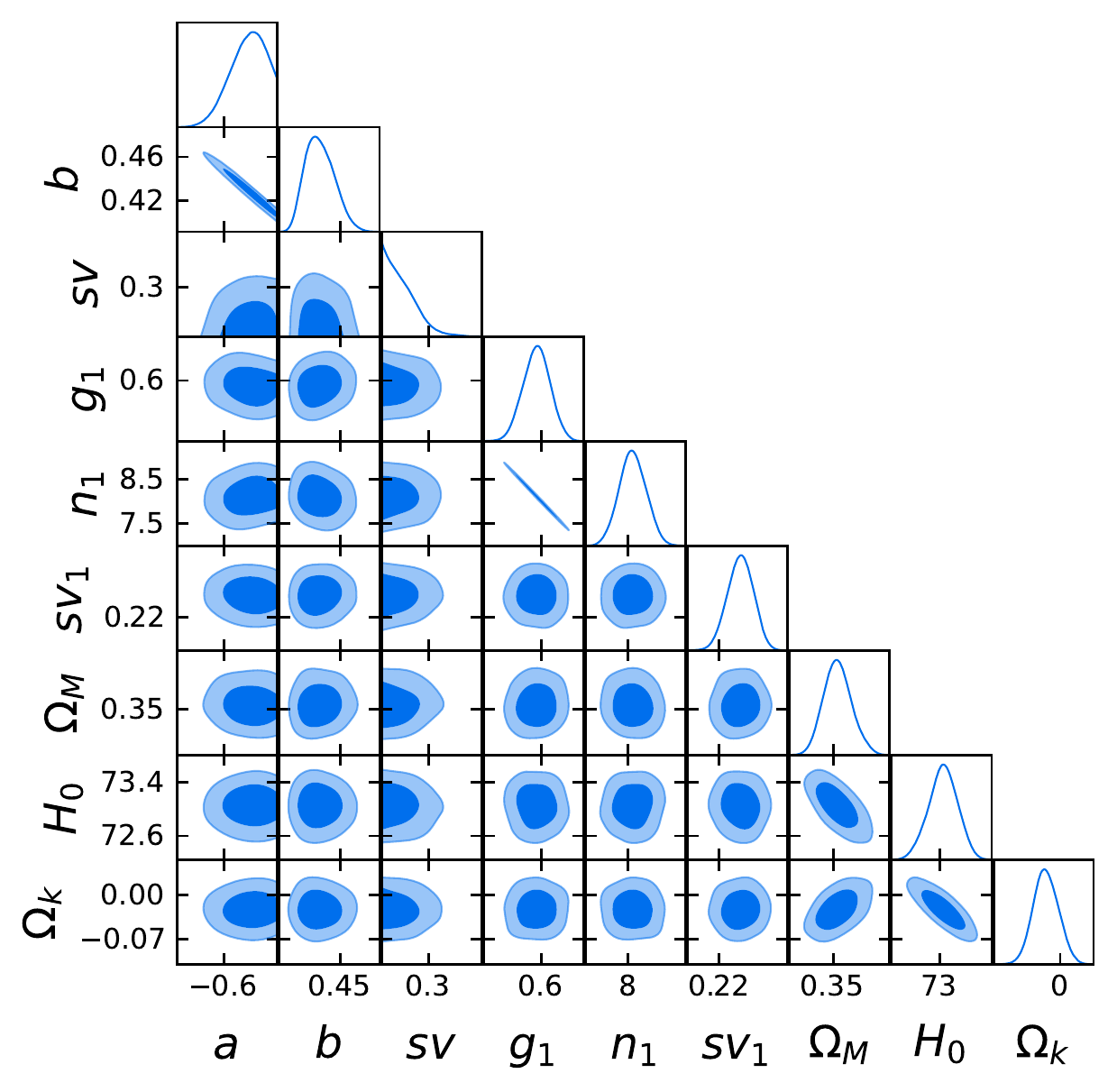}{0.364\textwidth}{  (e) \textit{Pantheon +} SNe Ia + GRBs + QSOs + BAO with fixed correction for redshift evolution}\label{}
\fig{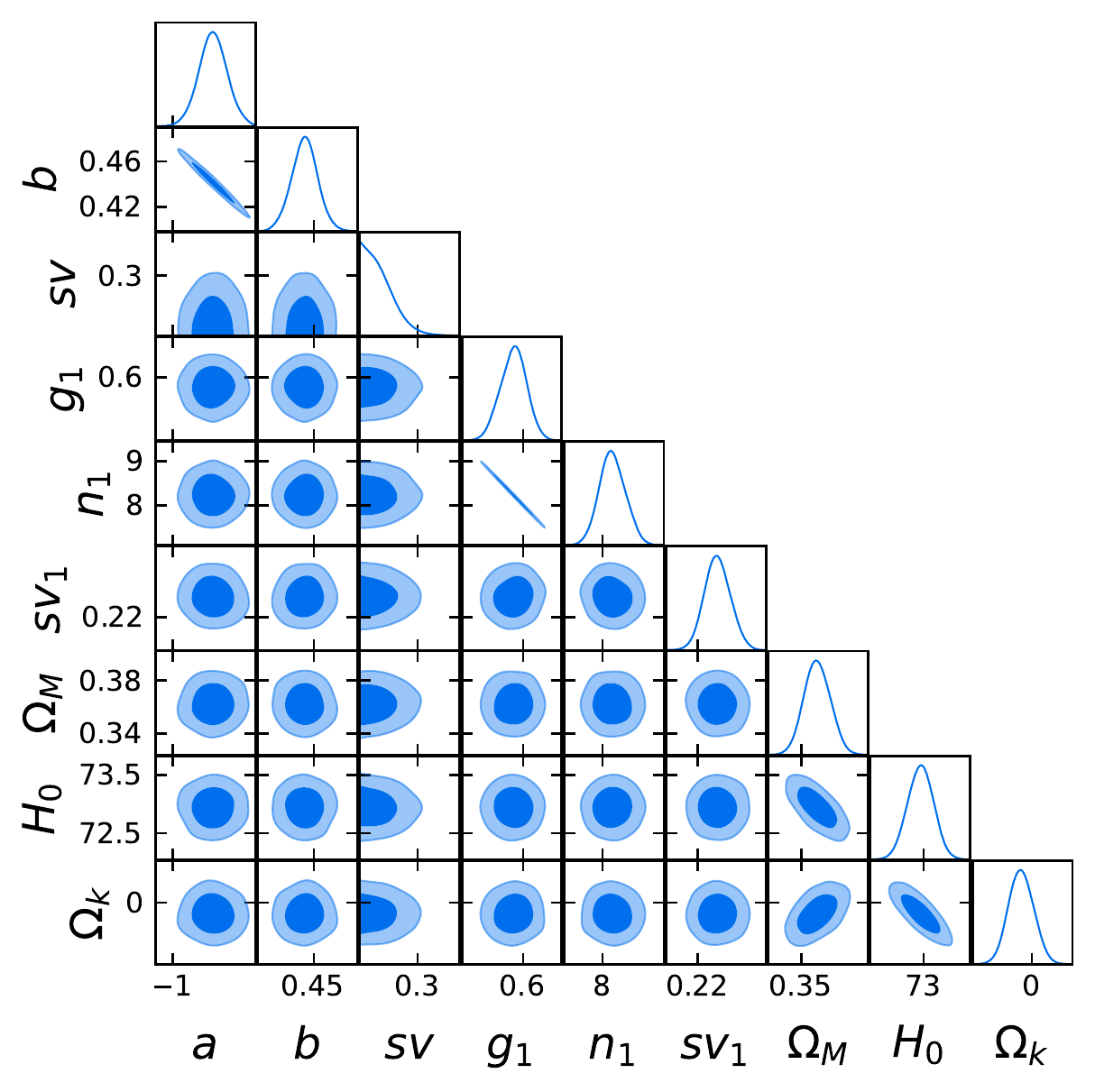}{0.364\textwidth}{ (f) \textit{Pantheon +} SNe Ia + GRBs + QSOs + BAO with varying correction for redshift evolution}\label{}}

\caption{Fits of the non-flat $\Lambda$CDM model.}
\label{fig: nonflat}
\end{figure}

Here we start the discussion considering the results obtained by applying the $\cal L_G$ likelihood to all investigated probes.
As shown in the upper part of Table \ref{tab:bestfit} and Fig. \ref{fig: nonflat}, the values obtained for $\Omega_k$ are always compatible within 1 $\sigma$ with $\Omega_k = 0$, independently on the SNe Ia sample considered and the treatment of the redshift evolution fixed at given $\Omega_M$ and $\Omega_k$ values employed. 
If we consider the evolution not fixed, but varying with a grid of values of $\Omega_M$ and $\Omega_k$ we can state that the results do not show substantial statistical differences compared to the cases with fixed evolution.
If we consider these scenarios, we could state that our results significantly favour a flat Universe, in agreement with \citet{planck2018} and some recent works \citep[e.g.][]{eboss2021,2021JCAP...11..060G}. Nevertheless, all our analyses show a common trend toward $\Omega_k <0$, suggesting a slight preference for a closed Universe, as claimed in \citet{Park:2017xbl}, \citet{2020NatAs...4..196D}, \citet{Handley:2019tkm}, and \citet{2021JCAP...10..008Y}. This hint at $\Omega_k <0$ is also in agreement with the analysis on a non-flat Universe reported in \citet{scolnic2018} with only SNe Ia, even if they have larger uncertainties compared to the ones obtained in this work (see their Table 8).

Concerning $\Omega_M$ and $H_0$, we recover the values assumed for the SNe Ia sample used (see Table \ref{tab:bestfit}). Specifically, when we consider \textit{Pantheon} SNe Ia, $\Omega_M$ and $H_0$ are compatible with 0.3 and 70 $\mathrm{km} \, \mathrm{s}^{-1} \, \mathrm{Mpc}^{-1}$, respectively, as imposed in \citet{scolnic2018}, while when using \textit{Pantheon +} SNe Ia, we obtain compatibility with $\Omega_M = 0.338 \pm 0.018$ and $H_0 = 73.04 \pm 1.04  \, \mathrm{km} \, \mathrm{s}^{-1} \, \mathrm{Mpc}^{-1}$, as in \citet{2022ApJ...938..110B} and \citet{2022ApJ...934L...7R}, respectively. This dependence of the results on the calibration choice for the SNe Ia samples was already mentioned in \citet{biasfreeQSO2022} and discussed in more detail in \citet{Bargiacchi2023arXiv230307076B}. This consideration also provides an interpretation of the $\zeta$ reported in Table \ref{tab:bestfit}. Indeed, since the two SNe Ia samples assume different calibrations, the $H_0$ values computed on the data set with \textit{Pantheon} SNe Ia show a discrepancy compared to the one from \textit{Pantheon +} ($\zeta_{P+} \sim 3$) and vice-versa the $H_0$ obtained by using \textit{Pantheon +} SNe Ia is discrepant with the one from \textit{Pantheon} ($\zeta_{P} \sim 11$). Both samples, with \textit{Pantheon } and \textit{Pantheon +}, manifest an $H_0$ incompatible with the one derived from the CMB, with $\zeta_{CMB} \sim 4$ when using \textit{Pantheon} SNe Ia and $\zeta_{CMB} \sim 10$ when considering \textit{Pantheon +} SNe Ia.

We also here compare our results with the ones of \citet{Dainotti2022MNRAS.tmp.2639D}. Compared to our work, in \citet{Dainotti2022MNRAS.tmp.2639D} the cases with no correction and fixed correction for redshift evolution are employed and the same samples of GRBs and BAOs are used combined with SNe Ia from \textit{Pantheon}, but only $\Omega_k$ is free to vary, while $\Omega_M$ and $H_0$ are fixed. The choice of leaving only $\Omega_k$ free to vary was dictated by the idea that testing only one free parameter, results could lead to incompatibility with a flat $\Lambda$CDM. Since here we test also the \textit{Pantheon +} sample, which enjoys a larger sample of SNe Ia, we then have allowed all parameters to be free to vary.
Besides these differences, the values of $\Omega_k$ provided in \citet{Dainotti2022MNRAS.tmp.2639D} (see their Table 6) and obtained in this work are compatible within less than 1 $\sigma$ for corresponding cases of redshift evolution, with the same hint at $\Omega_k <0$. We also stress that the differences between the two works in the uncertainties on $\Omega_k$ are driven by the fact that here we marginalize also over the other cosmological parameters $\Omega_M$ and $H_0$, thus obtaining larger uncertainties.

\begin{figure}
\centering
\gridline{
\fig{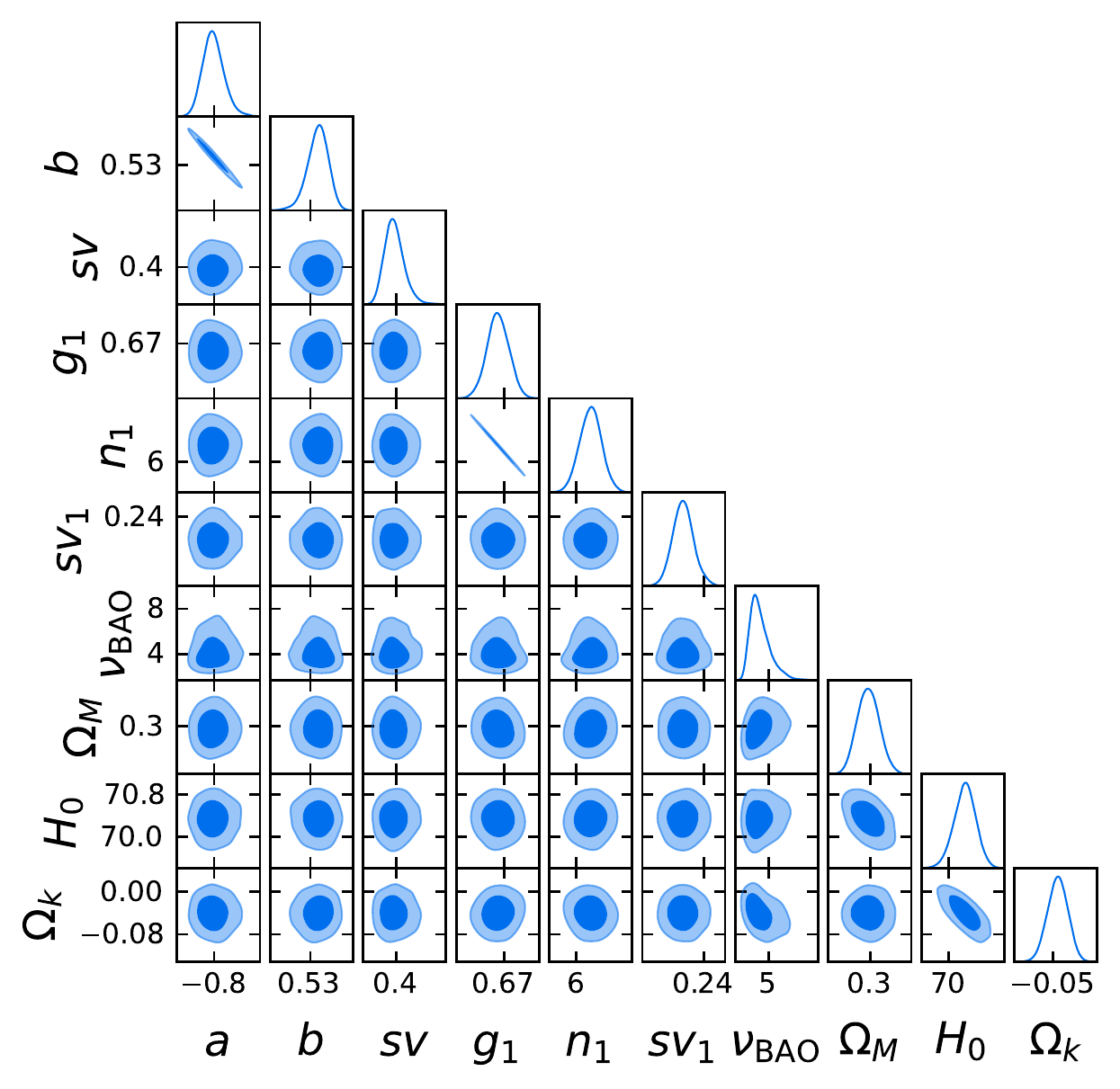}{0.364\textwidth}
{ (a) \textit{Pantheon} SNe Ia + GRBs + QSOs + BAO without correction for redshift evolution}
\fig{PantheonfixedEvo_OmH0Ok_NewLikelihoods.pdf}{0.364\textwidth}
{ (b) \textit{Pantheon} SNe Ia + GRBs + QSOs + BAO with fixed correction for redshift evolution}}
\gridline{
\fig{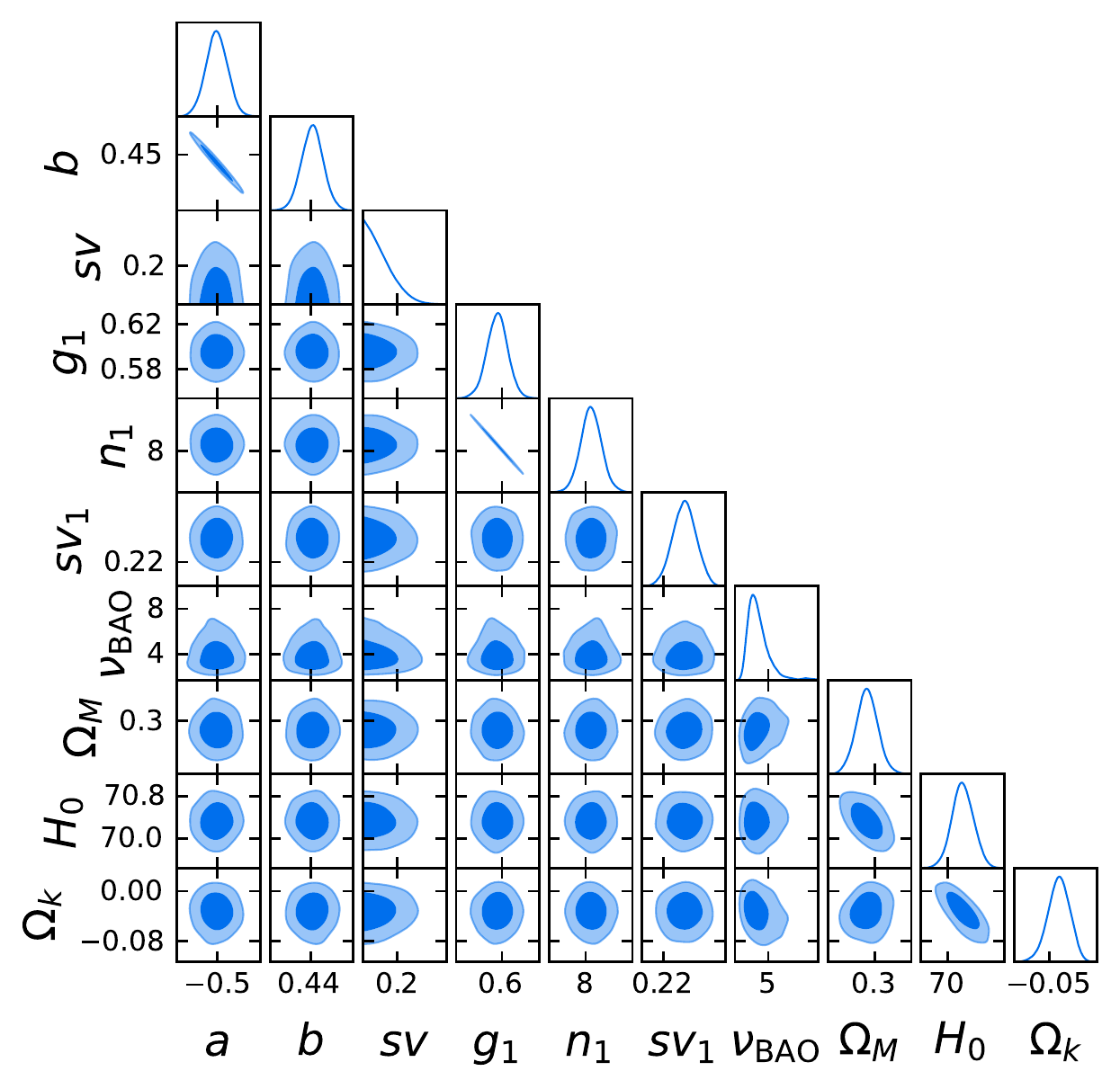}{0.364\textwidth}
{(c) \textit{Pantheon} SNe Ia + GRBs + QSOs + BAO with varying correction for redshift evolution}
\fig{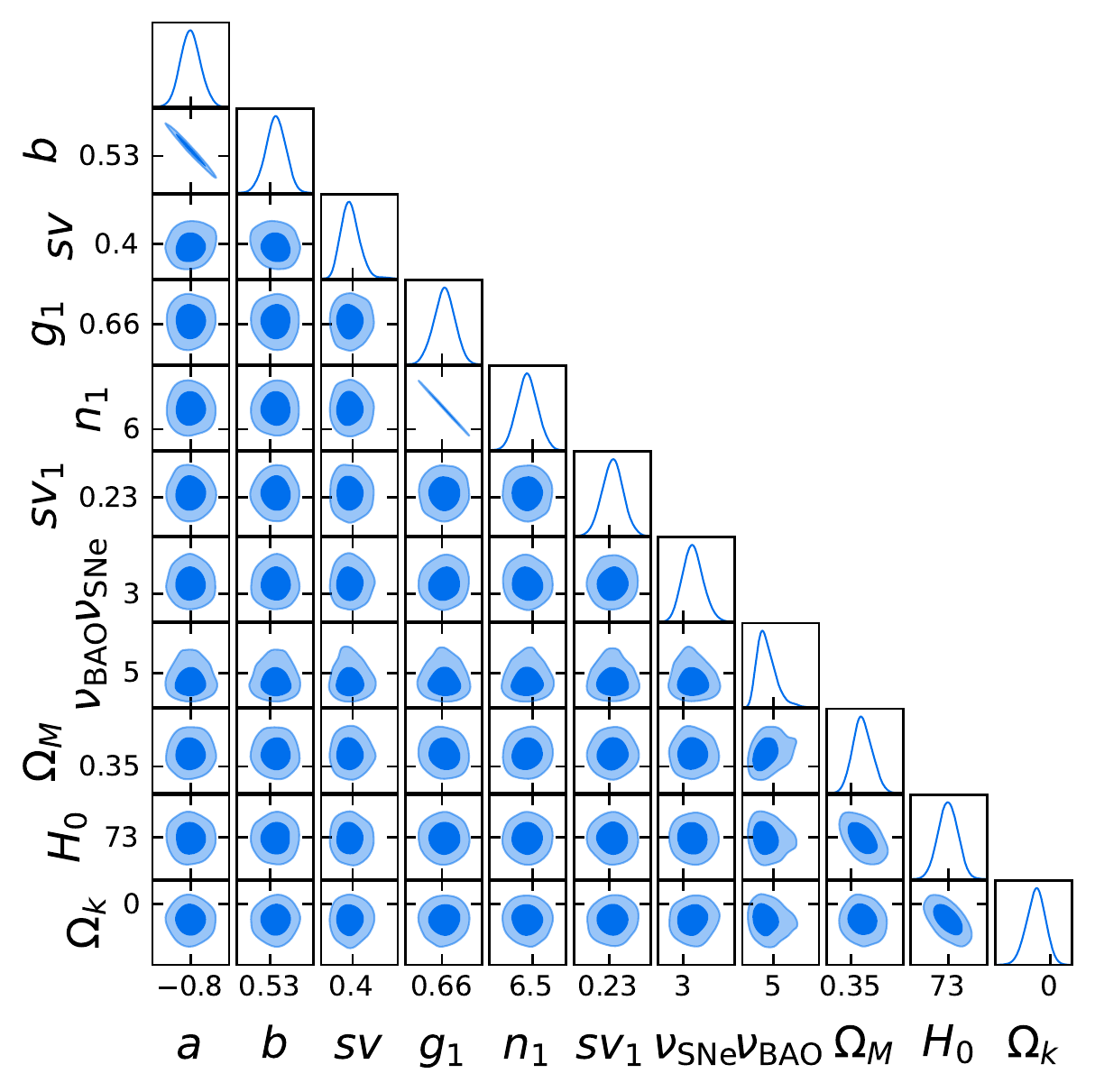}{0.364\textwidth}
{ (d) \textit{Pantheon +} SNe Ia + GRBs + QSOs + BAO without correction for redshift evolution}}
\gridline{
\fig{PantheonPlusfixedEvo_OmH0Ok_NewLikelihoods.pdf}{0.364\textwidth}
{(e) \textit{Pantheon +} SNe Ia + GRBs + QSOs + BAO with fixed correction for redshift evolution}
\fig{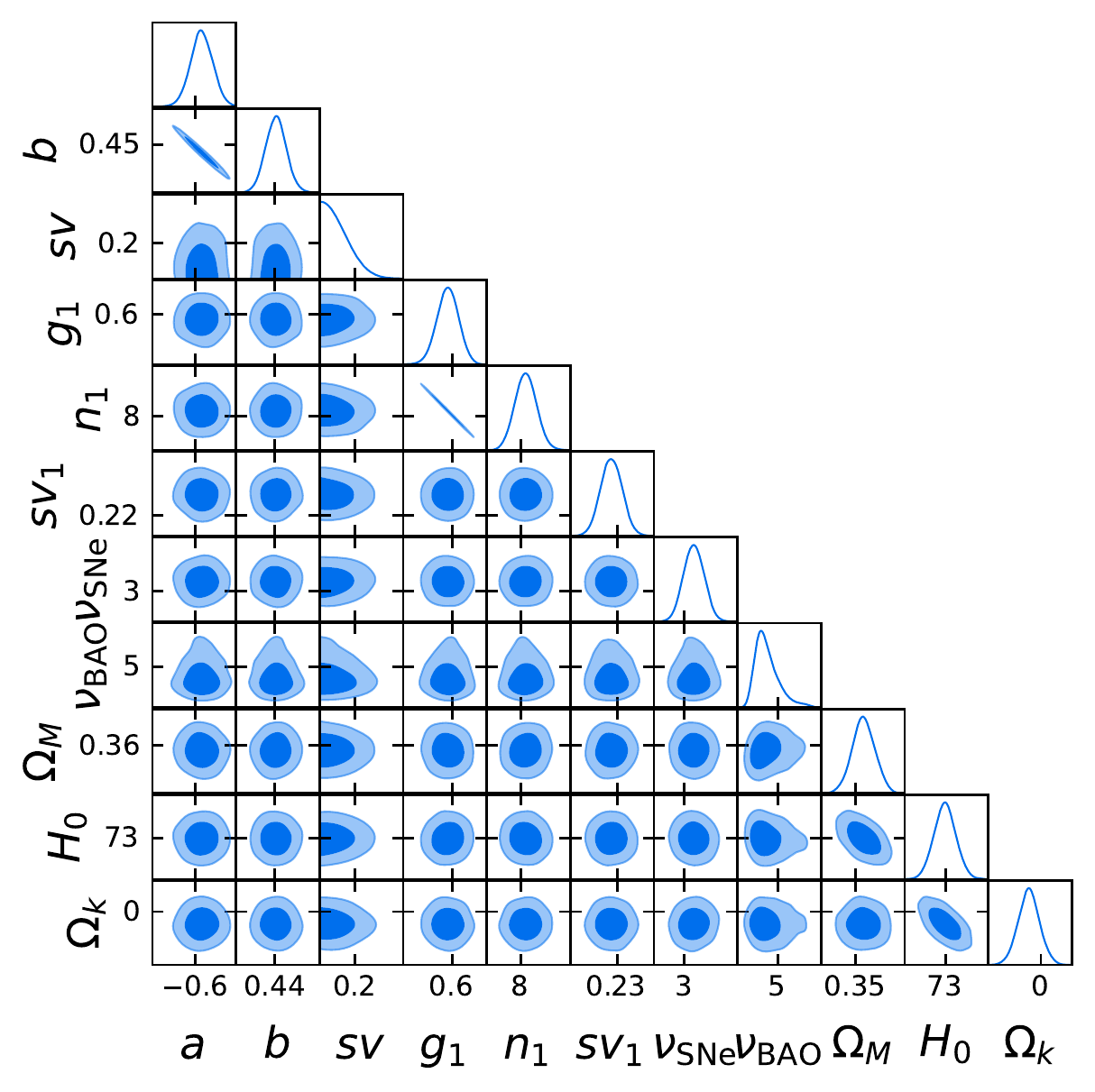}{0.364\textwidth}
{(f) \textit{Pantheon +} SNe Ia + GRBs + QSOs + BAO with varying correction for redshift evolution}}
\caption{Fits of the non-flat $\Lambda$CDM model with $\cal L_N$ likelihoods.}
\label{fig: nonflatnewlikelihoods}
\end{figure}

Moving the discussion to the results obtained by applying the $\cal L_N$ likelihoods, shown in the lower part of Table \ref{tab:bestfit} and Fig. \ref{fig: nonflatnewlikelihoods}, we can state that these results confirm the ones from the $\cal L_G$ since the best-fit values of the cosmological parameters are consistent in the two cases. Besides this, the use of the new $\cal L_N$ likelihoods remarkably reduces the uncertainties on $H_0$, $\Omega_M$, and $\Omega_k$, as proved by the computation of the percentage difference provided in Table \ref{tab:comparison}. This value ($\Delta_{\%}$) is derived as $\Delta_{\%} = (\Delta_{\mathrm{comparing}} - \Delta_{\mathrm{reference}}) / \Delta_{\mathrm{reference}}$, where $\Delta_{\mathrm{comparing}}$ is the uncertainty on the parameter considered obtained with the $\cal L_N$ likelihoods and $\Delta_{\mathrm{reference}}$ the one obtained with the $\cal L_G$. Specifically, the uncertainties on $\Omega_M$ are reduced up to $27 \%$, the ones on $H_0$ up to $30 \%$, and the ones on $\Omega_k$ up to $32 \%$, when considering both \textit{Pantheon} and \textit{Pantheon +} samples.
All the results above described are valid independently of the treatment of the correction for redshift evolution applied.

\subsection{Results of flat $w$CDM model  with different likelihoods}

\begin{figure}
\centering
\gridline{
\fig{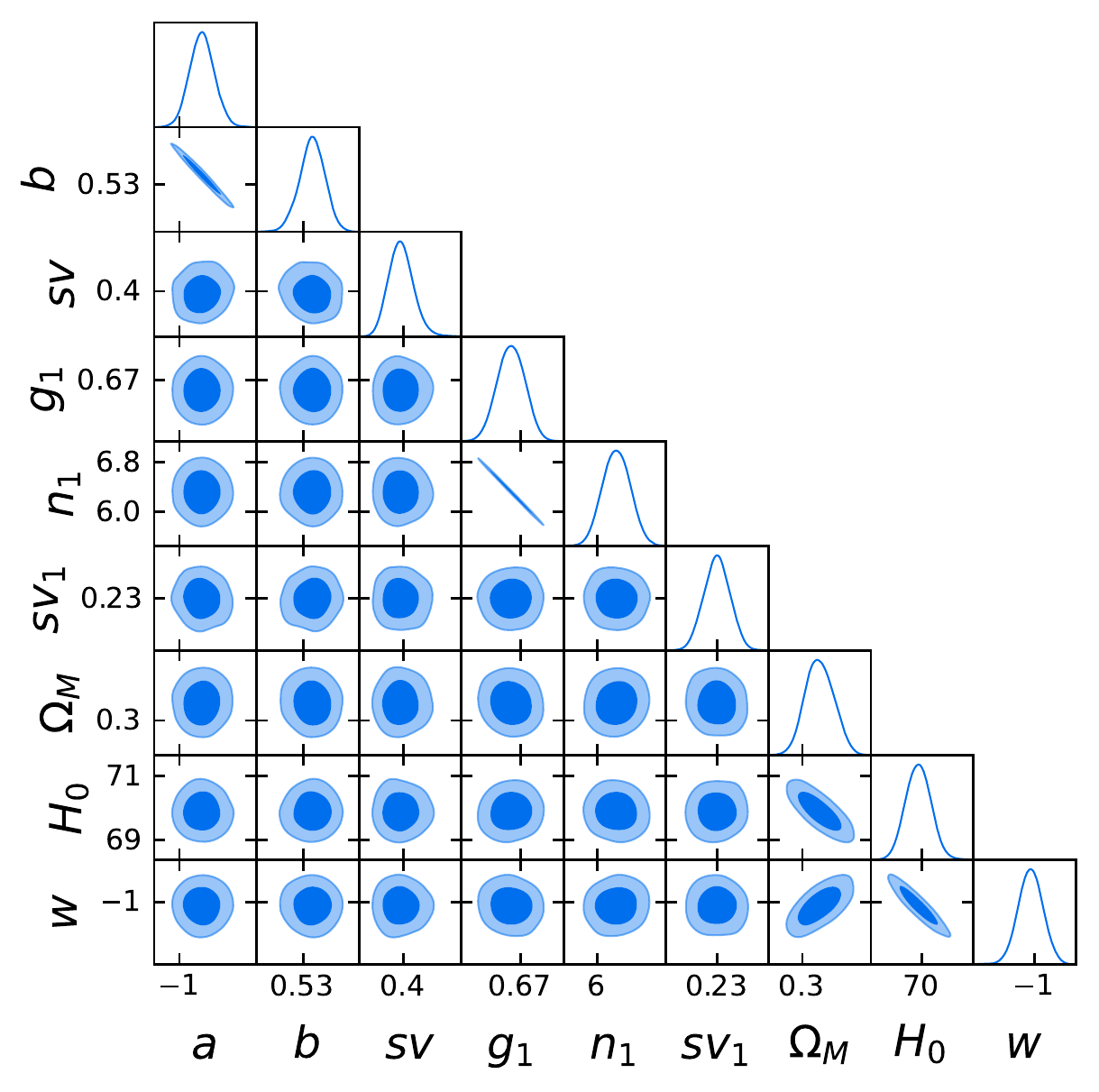}{0.364\textwidth}{(a) \textit{Pantheon} SNe Ia + GRBs + QSOs + BAO without correction for redshift evolution}
\fig{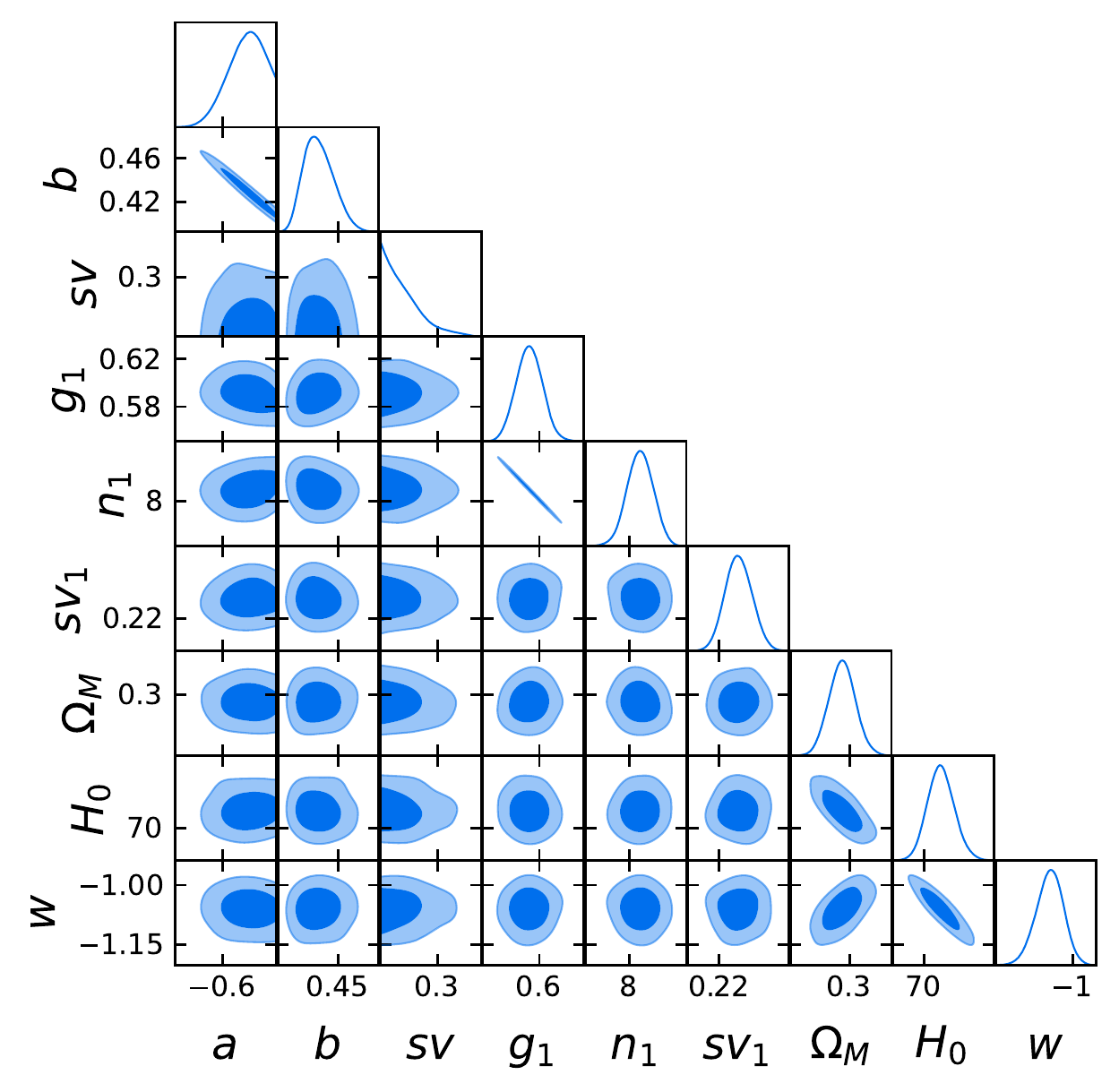}{0.364\textwidth}{(b) \textit{Pantheon} SNe Ia + GRBs + QSOs + BAO with fixed correction for redshift evolution}}
\gridline{
\fig{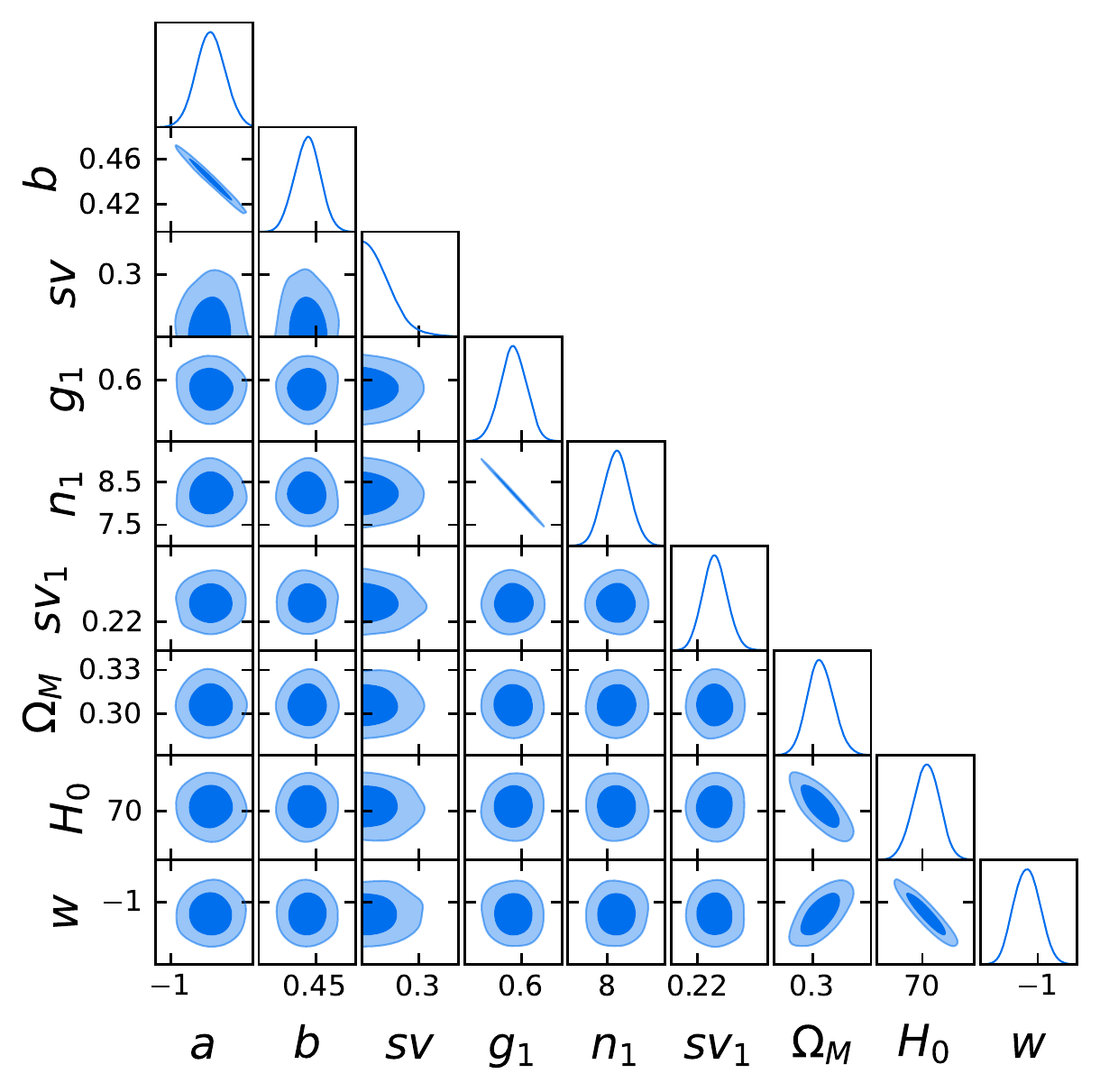}{0.364\textwidth}{(c) \textit{Pantheon} SNe Ia + GRBs + QSOs + BAO with varying correction for redshift evolution}
\fig{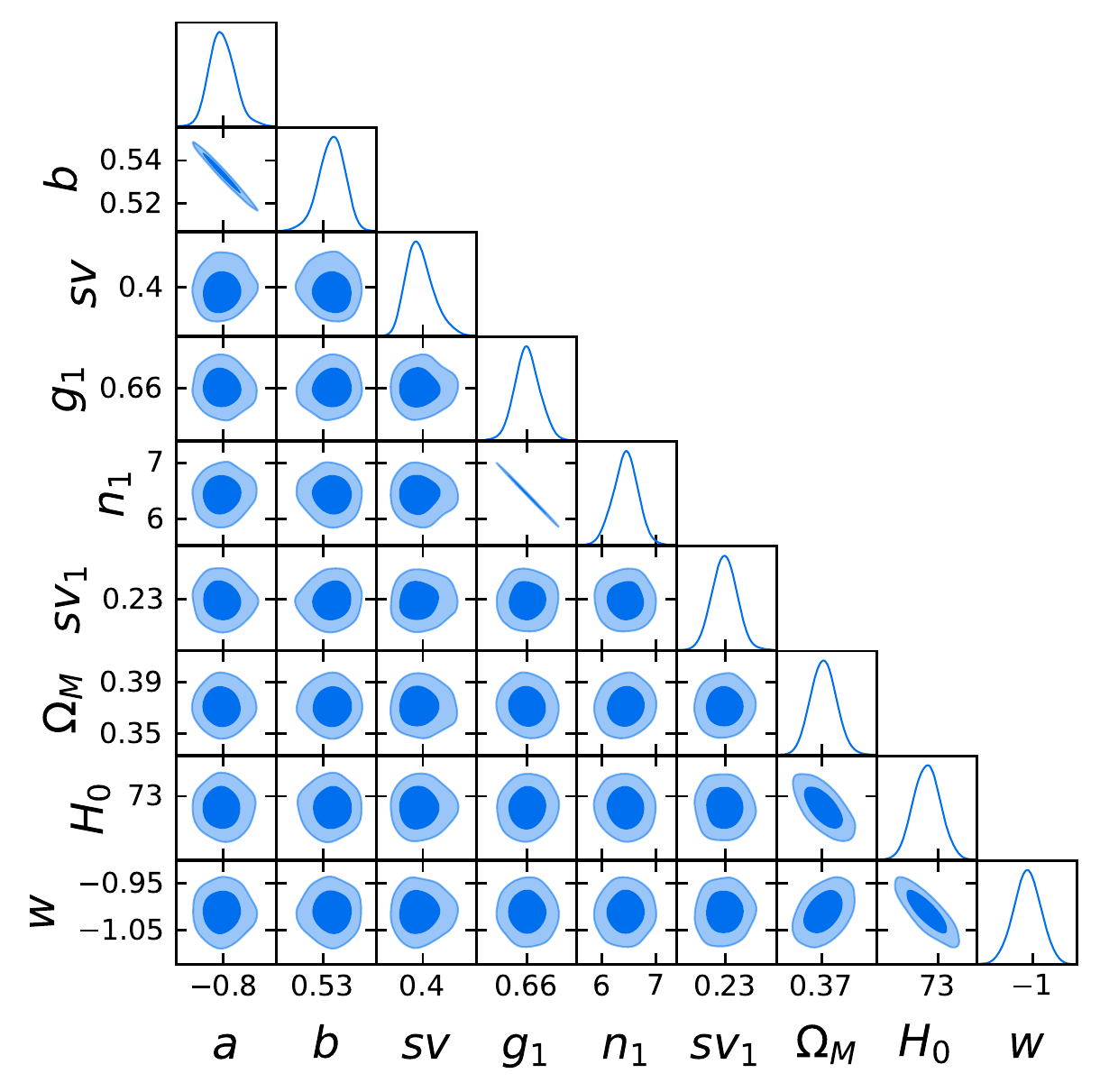}{0.364\textwidth}{(d) \textit{Pantheon +} SNe Ia + GRBs + QSOs + BAO without correction for redshift evolution}}
\gridline{
\fig{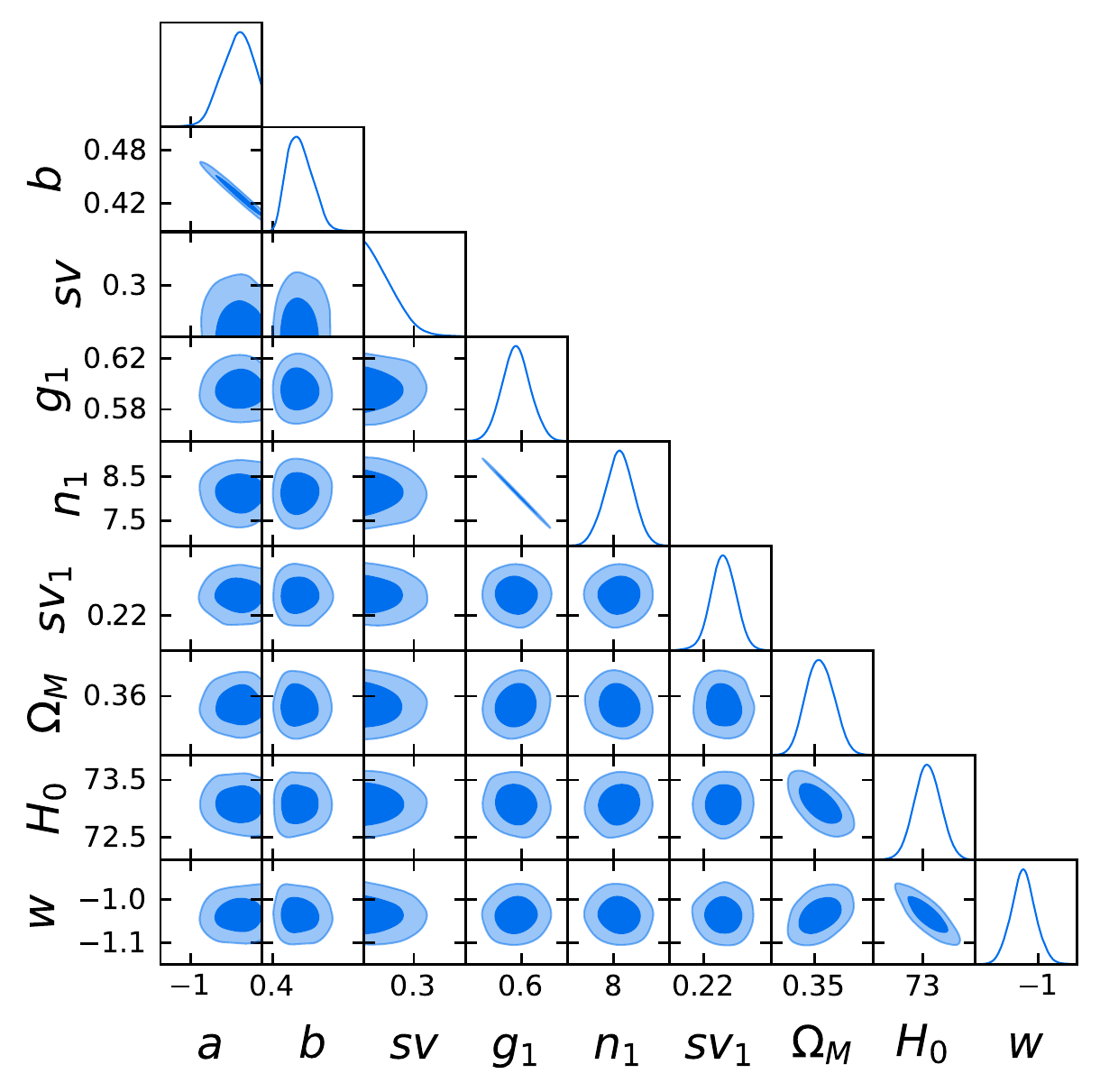}{0.364\textwidth}{(e) \textit{Pantheon +} SNe Ia + GRBs + QSOs + BAO with fixed correction for redshift evolution}
\fig{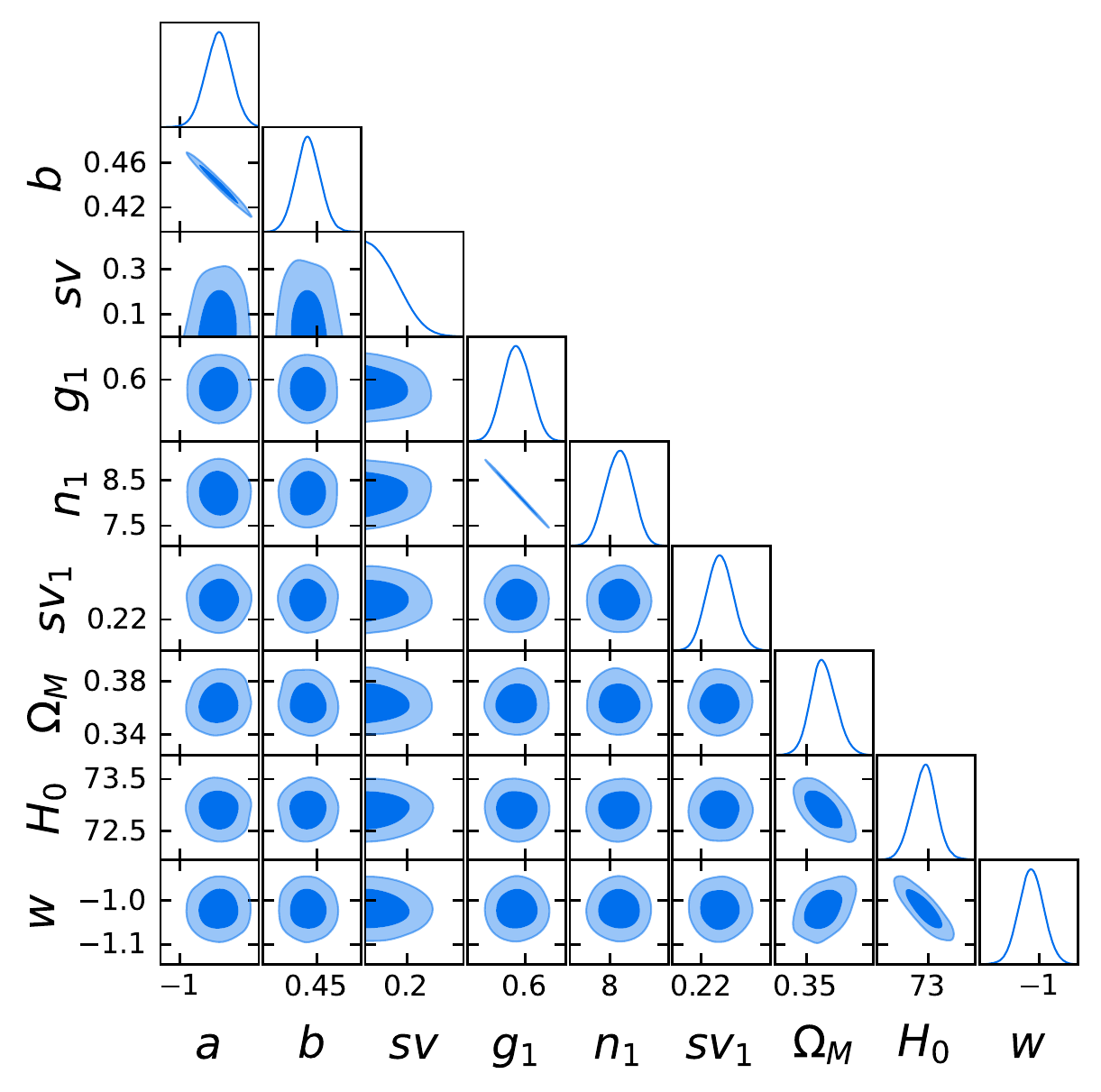}{0.364\textwidth}{(f) \textit{Pantheon +} SNe Ia + GRBs + QSOs + BAO with varying correction for redshift evolution}}
\caption{Fits of the flat $w$CDM model.}
\label{fig: wCDM}
\end{figure}

Starting from the case of $\cal L_G$ likelihoods, the upper part of Table \ref{tab:bestfit} and Fig. \ref{fig: wCDM} show the results obtained from the flat $w$CDM model. All the values of $w$ are consistent with $w=-1$, reaching the maximum discrepancy of 1.7 $\sigma$ in the case with \textit{Pantheon} SNe Ia and fixed correction for evolution. 
This corroborates the standard flat $\Lambda$CDM model, even if with a visible hint at values $w<-1$ pointing towards a phantom dark energy scenario. 
Concerning the values of $\Omega_M$, $H_0$, and $\zeta$, the same considerations drawn for the non-flat $\Lambda$CDM model are still valid since the determination of these parameters is driven mainly by the SNe Ia samples and their corresponding calibration.
As for the non-flat $\Lambda$CDM model, the results just described are not significantly affected by the different cases studied for the correction for redshift evolution.

Our results on $\Omega_M$ and $w$ are compatible within less than 1 $\sigma$ with the ones provided for only SNe Ia in \citet{scolnic2018} (see their Table 8), in which $w$ again points towards $w<-1$ even if with large uncertainties (i.e. $w = -1.090 \pm 0.220$). Additionally, \citet{Dainotti2022MNRAS.tmp.2639D}, for which similarities and differences with our work have already been described in Section \ref{resultsnonflat}, have also investigated a flat $w$CDM model. Thus, we can compare the corresponding results considering that in \citet{Dainotti2022MNRAS.tmp.2639D} $\Omega_M$ and $H_0$ are fixed and $w$ is the only free parameters, which leads to smaller uncertainties on $w$ compared to ours. Nonetheless, the values of $w$ obtained in  \citet{Dainotti2022MNRAS.tmp.2639D} are consistent within 1 $\sigma$ with our values and show the same hint at a phantom dark energy regime.

\begin{figure}
\centering
\gridline{
\fig{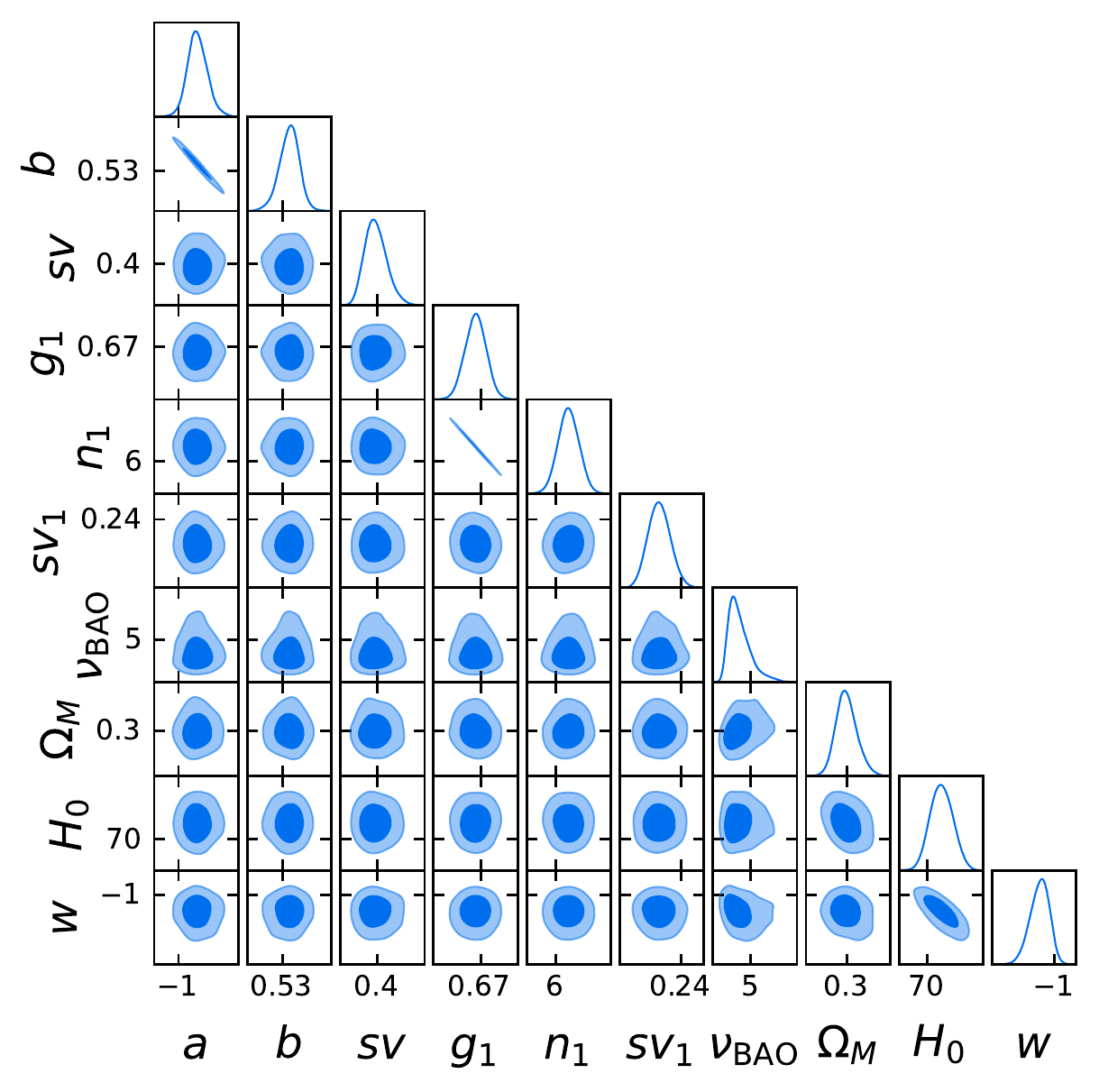}{0.364\textwidth}{(a) \textit{Pantheon} SNe Ia + GRBs + QSOs + BAO without correction for redshift evolution}
\fig{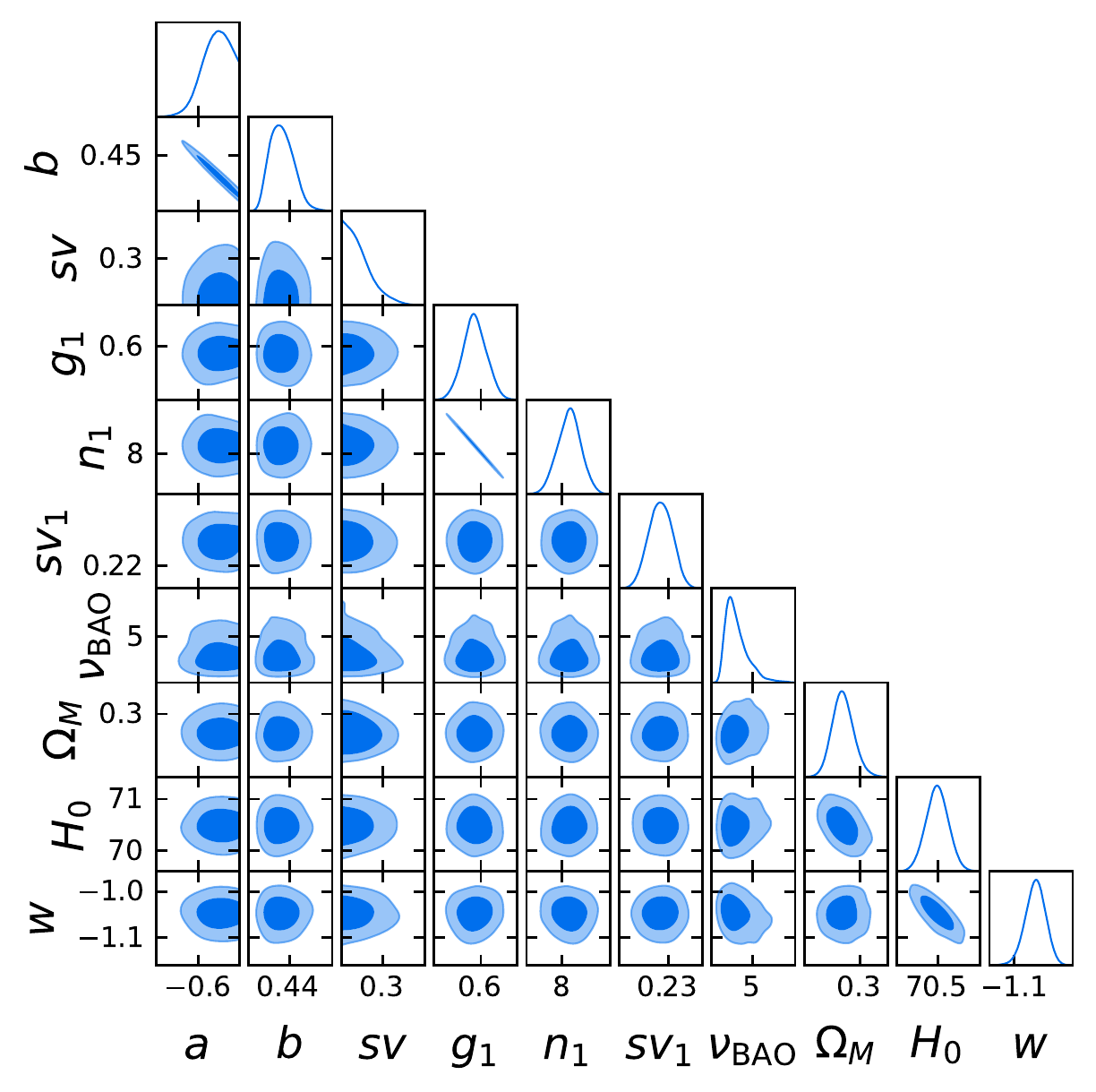}{0.364\textwidth}{(b) \textit{Pantheon} SNe Ia + GRBs + QSOs + BAO with fixed correction for redshift evolution}}
\gridline{
\fig{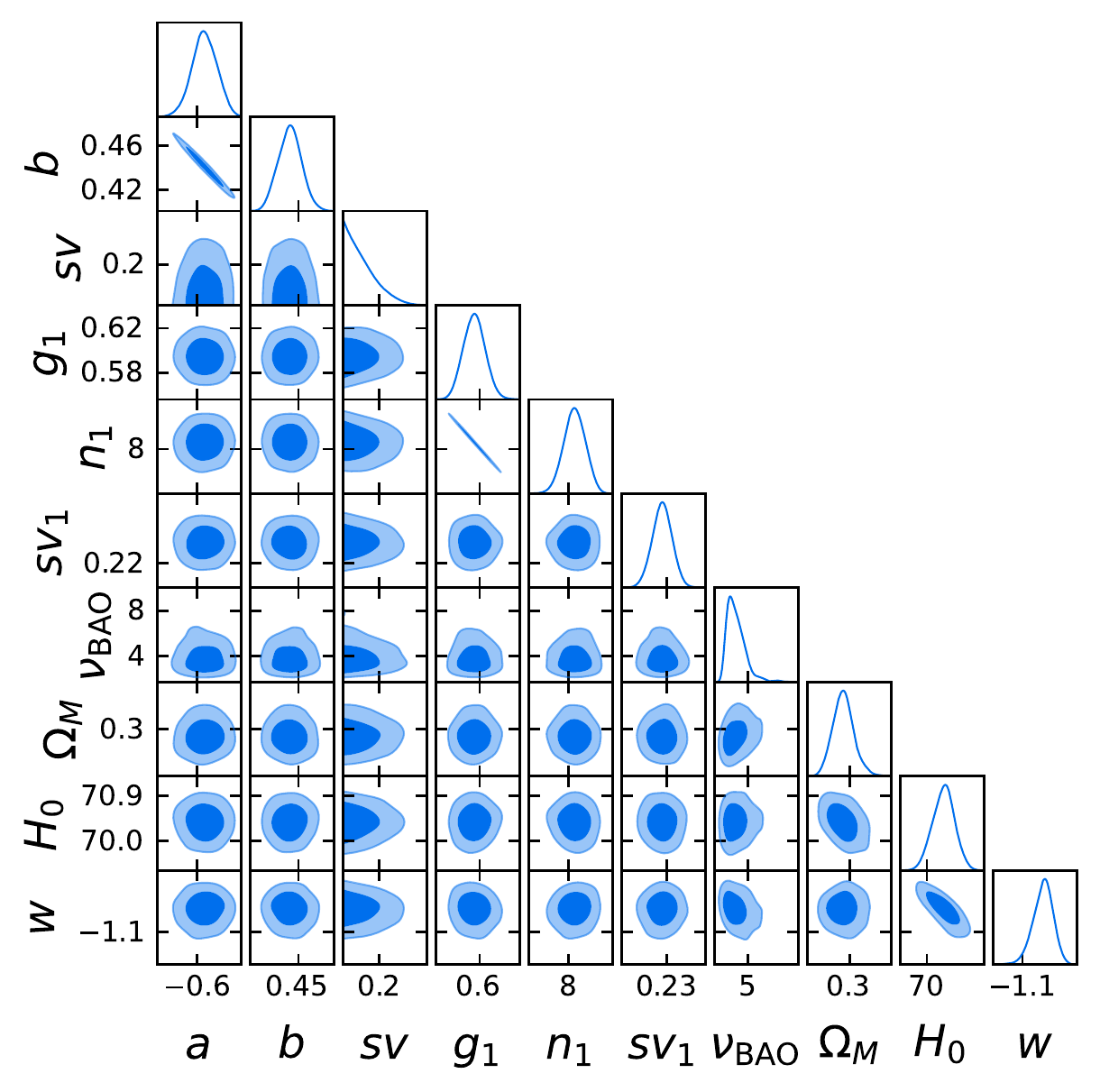}{0.364\textwidth}{(c) \textit{Pantheon} SNe Ia + GRBs + QSOs + BAO with varying correction for redshift evolution}
\fig{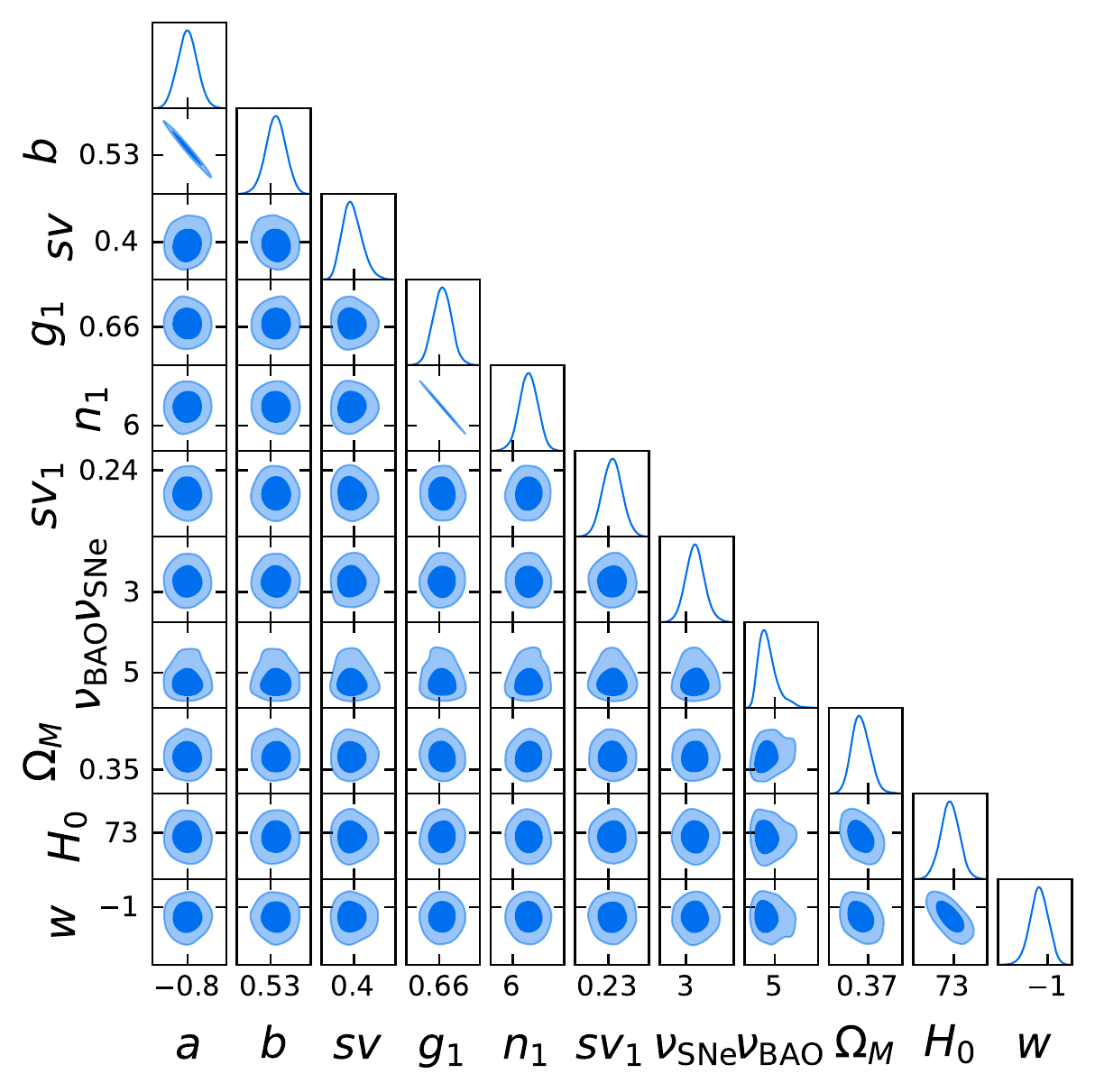}{0.364\textwidth}{(d) \textit{Pantheon +} SNe Ia + GRBs + QSOs + BAO without correction for redshift evolution}}
\gridline{
\fig{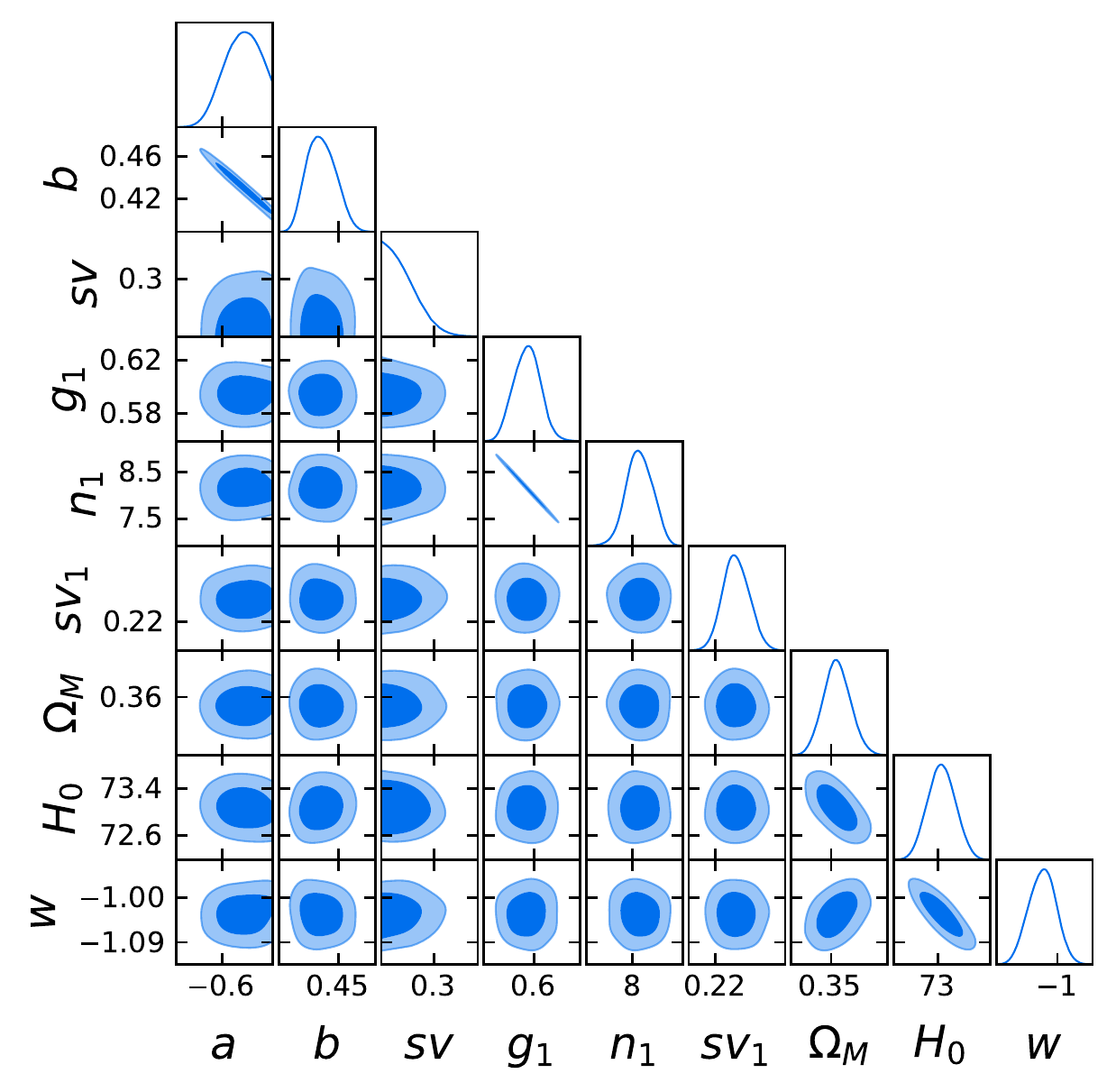}{0.364\textwidth}{(e) \textit{Pantheon +} SNe Ia + GRBs + QSOs + BAO with fixed correction for redshift evolution}
\fig{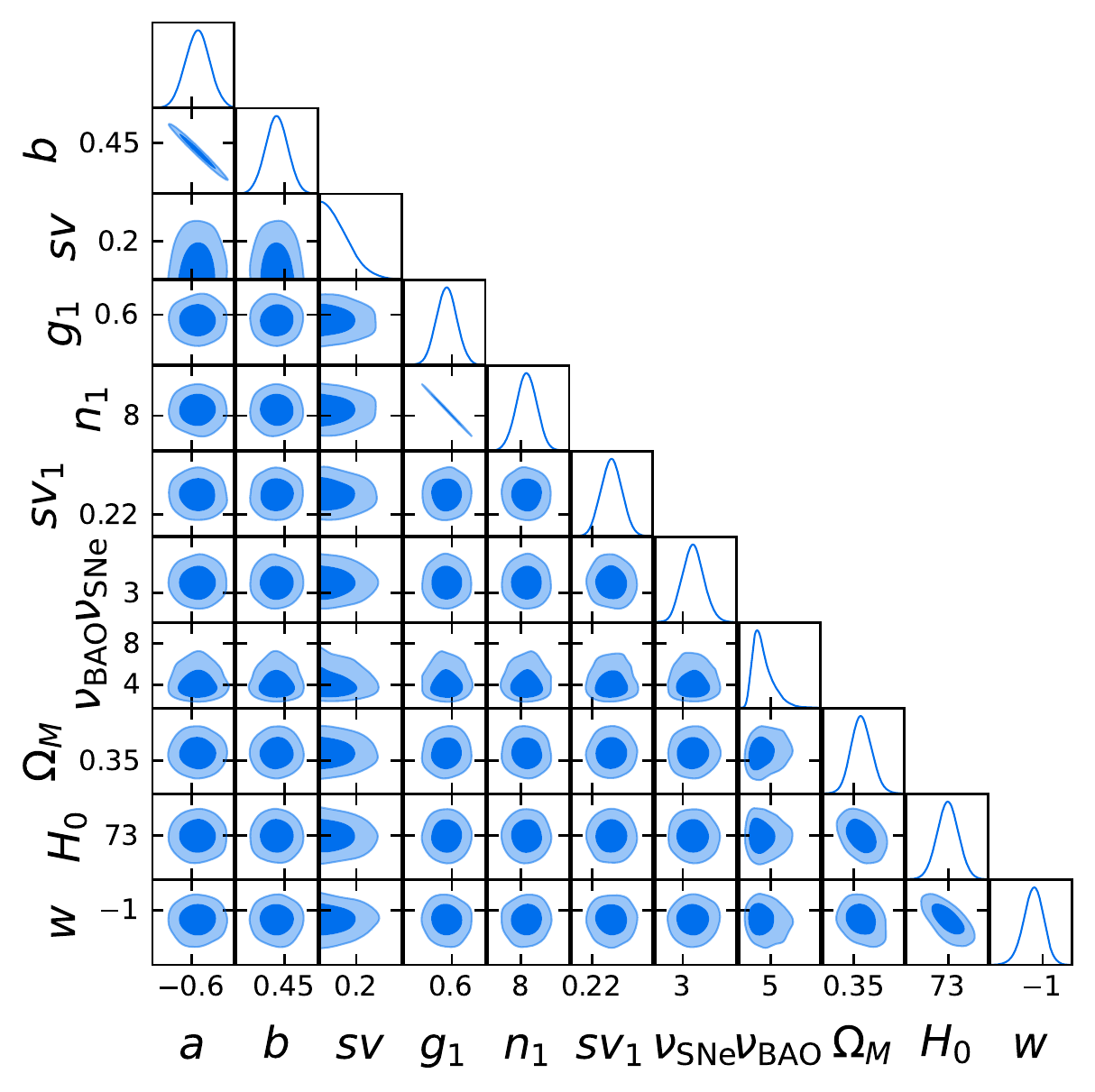}{0.364\textwidth}{(f) \textit{Pantheon +} SNe Ia + GRBs + QSOs + BAO with varying correction for redshift evolution}}
\caption{Fits of the flat $w$CDM model with $\cal L_N$ likelihoods.}
\label{fig: wCDMnewlikelihoods}
\end{figure}

We now compare these results obtained with the $\cal L_G$ likelihoods with the ones obtained when we consider the new $\cal L_N$ likelihoods, which are shown in the lower part of Table \ref{tab:bestfit} and Fig. \ref{fig: wCDMnewlikelihoods}. As for the non-flat $\Lambda$CDM model, the best-fit values of the cosmological parameters are still not impacted by the change in the likelihood, but we reach a significantly higher precision on these parameters with the new $\cal L_N$ likelihoods. Indeed, as shown in the lower part of Table \ref{tab:comparison}, the uncertainty on $H_0$ is reduced up to the $35 \%$, the one on $\Omega_M$ up to $22 \%$, and the one on $w$ up to $31 \%$, when considering both SNe Ia samples.

\begin{table*}
\caption{Best-fit values and 1 $\sigma$ uncertainty of the cosmological free parameters of the non-flat $\Lambda$CDM and flat $w$CDM model for GRBs+QSOs+BAO together with \textit{Pantheon} or \textit{Pantheon +} SNe Ia, as specified in the left column, for no evolution, fixed and varying evolutionary cases studied with the $\cal L_G$ likelihood (upper part of the Table) and $\cal L_N$ one (lower part of the Table).}
\begin{centering}
\begingroup
\setlength{\tabcolsep}{0.5\tabcolsep}
\scriptsize
\begin{tabular}{ccccccccccccc}
\hline
\multicolumn{1}{c}{$\cal L_G$ likelihoods:}&\multicolumn{5}{c}{Non-flat $\Lambda$CDM }&\multicolumn{5}{c}{flat $w$CDM}\tabularnewline
\hline
\hline
GRBs+QSOs+BAO+\textit{Pantheon} & $H_0$ & $\Omega_M$ & $\Omega_k$ & $\zeta_{CMB}$ & $\zeta_{P}$ & $\zeta_{P+}$ & $H_0$ & $\Omega_M$ & $w$ & $\zeta_{CMB}$ & $\zeta_{P}$ & $\zeta_{P+}$
\tabularnewline
\hline
\hline
No Evolution & $69.98 \pm 0.32$ & $0.310 \pm 0.010$ & $-0.018 \pm 0.025$ & 4.3 & 0.06 & 2.8 & $69.90 \pm 0.40$ & $0.312 \pm 0.010$  & $-1.012 \pm 0.038$  & 3.9 & 0.2 & 2.8 \tabularnewline
\hline
Fixed Evolution & $70.20 \pm 0.33$ & $0.297 \pm 0.009$ & $-0.027 \pm 0.025$ & 4.7 & 0.6 & 2.6 & $70.45 \pm 0.37$ & $0.295 \pm 0.009$  & $-1.058 \pm 0.035$  & 4.9  & 1.1 & 2.3\tabularnewline
\hline
Varying Evolution & $70.10 \pm 0.30$ & $0.304 \pm 0.010$ & $-0.024 \pm 0.024$ & 4.6 & 0.3 & 2.7 & $70.12 \pm 0.38$ & $0.306 \pm 0.010$ & $-1.031 \pm 0.036$ & 4.3 & 0.3 & 2.6\tabularnewline
\hline
\hline
GRBs+QSOs+BAO+\textit{Pantheon +} & $H_0$ & $\Omega_M$ & $\Omega_k$ & $\zeta_{CMB}$ & $\zeta_{P}$ & $\zeta_{P+}$ & $H_0$ & $\Omega_M$ & $w$ & $\zeta_{CMB}$ & $\zeta_{P}$ & $\zeta_{P+}$
\tabularnewline
\hline
\hline
No Evolution & $72.94 \pm 0.23$ & $0.366 \pm 0.011$ & $-0.023 \pm 0.021$ & 10.1 & 11.1 & 0.1 & $72.80 \pm 0.24$ & $0.371 \pm 0.010$ & $-1.011 \pm 0.030$ & 9.7 & 10.3 & 0.2 \tabularnewline
\hline
Fixed Evolution & $73.02 \pm 0.23$ & $0.354 \pm 0.010$ & $-0.021 \pm 0.021$ & 10.2 & 11.4 & 0 & $73.07 \pm 0.25$  & $0.354 \pm 0.010$  & $-1.035 \pm 0.029$ & 10.1 & 10.9 & 0.03 \tabularnewline
\hline
Varying Evolution & $72.94 \pm 0.24$ & $0.362 \pm 0.011$ & $-0.021 \pm 0.022$ & 10 & 10.8 & 0.09 & $72.91 \pm 0.25$ &  $0.364 \pm 0.010$ & $-1.020 \pm 0.030$ & 9.9& 10.3 & 0.1\tabularnewline
\hline
\hline
\multicolumn{1}{c}{The $\cal L_N$ likelihoods:}&\multicolumn{5}{c}{Non-flat $\Lambda$CDM}&\multicolumn{5}{c}{flat $w$CDM}\tabularnewline
\hline
\hline
GRBs+QSOs+BAO+\textit{Pantheon} & $H_0$ & $\Omega_M$ & $\Omega_k$ & $\zeta_{CMB}$ & $\zeta_{P}$ & $\zeta_{P+}$ & $H_0$ & $\Omega_M$ & $w$ & $\zeta_{CMB}$ & $\zeta_{P}$ & $\zeta_{P+}$
\tabularnewline
\hline
\hline
No Evolution & $70.34 \pm 0.23$ & $0.299 \pm 0.008$ & $-0.040\pm 0.022$ & 5.3 & 1.3 & 2.5 & $70.31 \pm 0.24$ & $0.300 \pm 0.008$  & $-1.043 \pm 0.028$  & 5.2 & 1.1 & 2.6 \tabularnewline
\hline
Fixed Evolution & $70.37 \pm 0.22$ & $0.287 \pm 0.007$ & $-0.027 \pm 0.017$ & 5.4 & 1.4 & 2.5 & $70.47 \pm 0.24$ & $0.289 \pm 0.007$  & $-1.046 \pm 0.024$  & 5.5  & 1.7 & 2.4\tabularnewline
\hline
Varying Evolution & $70.33 \pm 0.23$ & $0.294 \pm 0.008$ & $-0.033 \pm 0.020$  & 5.3 & 1.2 & 2.5 & $70.37 \pm 0.25$ & $0.295 \pm 0.008$ & $-1.046 \pm 0.027$ & 5.3 & 1.3   & 2.5\tabularnewline
\hline
\hline
GRBs+QSOs+BAO+\textit{Pantheon +} & $H_0$ & $\Omega_M$ & $\Omega_k$ & $\zeta_{CMB}$ & $\zeta_{P}$ & $\zeta_{P+}$ & $H_0$ & $\Omega_M$ & $w$ & $\zeta_{CMB}$ & $\zeta_{P}$ & $\zeta_{P+}$
\tabularnewline
\hline
\hline
No Evolution & $72.99 \pm 0.17$ & $0.361 \pm 0.009$ & $-0.026 \pm 0.017$ & 10.6 & 14 & 0.05 & $72.93 \pm 0.18$ & $0.362 \pm 0.010$ & $-1.025 \pm 0.026$ & 10.4 & 13 & 0.1 \tabularnewline
\hline
Fixed Evolution & $73.03 \pm 0.17$ & $0.347 \pm 0.008$ & $-0.011 \pm 0.018$ & 10.7 & 14 & 0.01 & $73.06 \pm 0.19$  &  $0.348 \pm 0.009$ & $-1.019 \pm 0.024$ & 10.6 & 13.3 & 0.02  \tabularnewline
\hline
Varying Evolution & $72.99 \pm 0.17$ & $0.356 \pm 0.009$ & $-0.019 \pm 0.017$ & 10.6 & 14 & 0.05 & $72.99 \pm 0.18$ & $0.357 \pm 0.009$ & $-1.023 \pm 0.025$ & 10.5 & 13.5 & 0.05\tabularnewline
\hline
\end{tabular}
\endgroup
\tablecomments{
$\zeta_{CMB}$, $\zeta_{P}$, and $\zeta_{P+}$ are the $\zeta$ computed as described in Section \ref{results}. $H_0$ is in units of $\mathrm{km} \, \mathrm{s}^{-1} \, \mathrm{Mpc}^{-1}$.}
\label{tab:bestfit}
\par\end{centering}
\end{table*}

\begin{table*}
\caption{Percentage difference of the uncertainties on the best-fit values of cosmological parameters obtained when using the $\cal L_N$ instead of the $\cal L_G$ likelihood.}
\begin{centering}
\begin{tabular}{ccccccc}
\hline
\multicolumn{1}{c}{}&\multicolumn{3}{c}{Non-flat $\Lambda$CDM }&\multicolumn{3}{c}{flat $w$CDM}\tabularnewline
\hline
\hline
GRBs+QSOs+BAO+\textit{Pantheon} & $\Delta_{\%}(H_0)$ & $\Delta_{\%}(\Omega_M)$ & $\Delta_{\%}(\Omega_k)$ & $\Delta_{\%}(H_0)$ & $\Delta_{\%}(\Omega_M)$ & $\Delta_{\%}(w)$
\tabularnewline
\hline
\hline
No Evolution & -0.28 & -0.20 & -0.16 & -0.34 & -0.20 & -0.26 \tabularnewline
\hline
Fixed Evolution &  -0.30 & -0.22 & -0.32 & -0.35 & -0.22 & -0.31 \tabularnewline
\hline
Varying Evolution & -0.23  & -0.20 & -0.17 & -0.34 & -0.20 &  -0.25 \tabularnewline
\hline
\hline
GRBs+QSOs+BAO+\textit{Pantheon +} & $\Delta_{\%}(H_0)$ & $\Delta_{\%}(\Omega_M)$ & $\Delta_{\%}(\Omega_k)$ & $\Delta_{\%}(H_0)$ & $\Delta_{\%}(\Omega_M)$ & $\Delta_{\%}(w)$
\tabularnewline
\hline
\hline
No Evolution & -0.26 & -0.27 & -0.19 & -0.25 & 0 & -0.13\tabularnewline
\hline
Fixed Evolution & -0.26 & -0.20 & -0.14 & -0.24 & -0.10 & -0.17  \tabularnewline
\hline
Varying Evolution & -0.29 & -0.18 & -0.23 & -0.28 & -0.10 & -0.17 \tabularnewline
\hline
\end{tabular}
\tablecomments{
$\Delta_{\%}$ is computed as defined in Section \ref{resultsnonflat}.}
\label{tab:comparison}
\par\end{centering}
\end{table*}

\section{The Goodness of the fits of the results}
\label{sec5}
\begin{figure}
\centering
\gridline{
 \fig{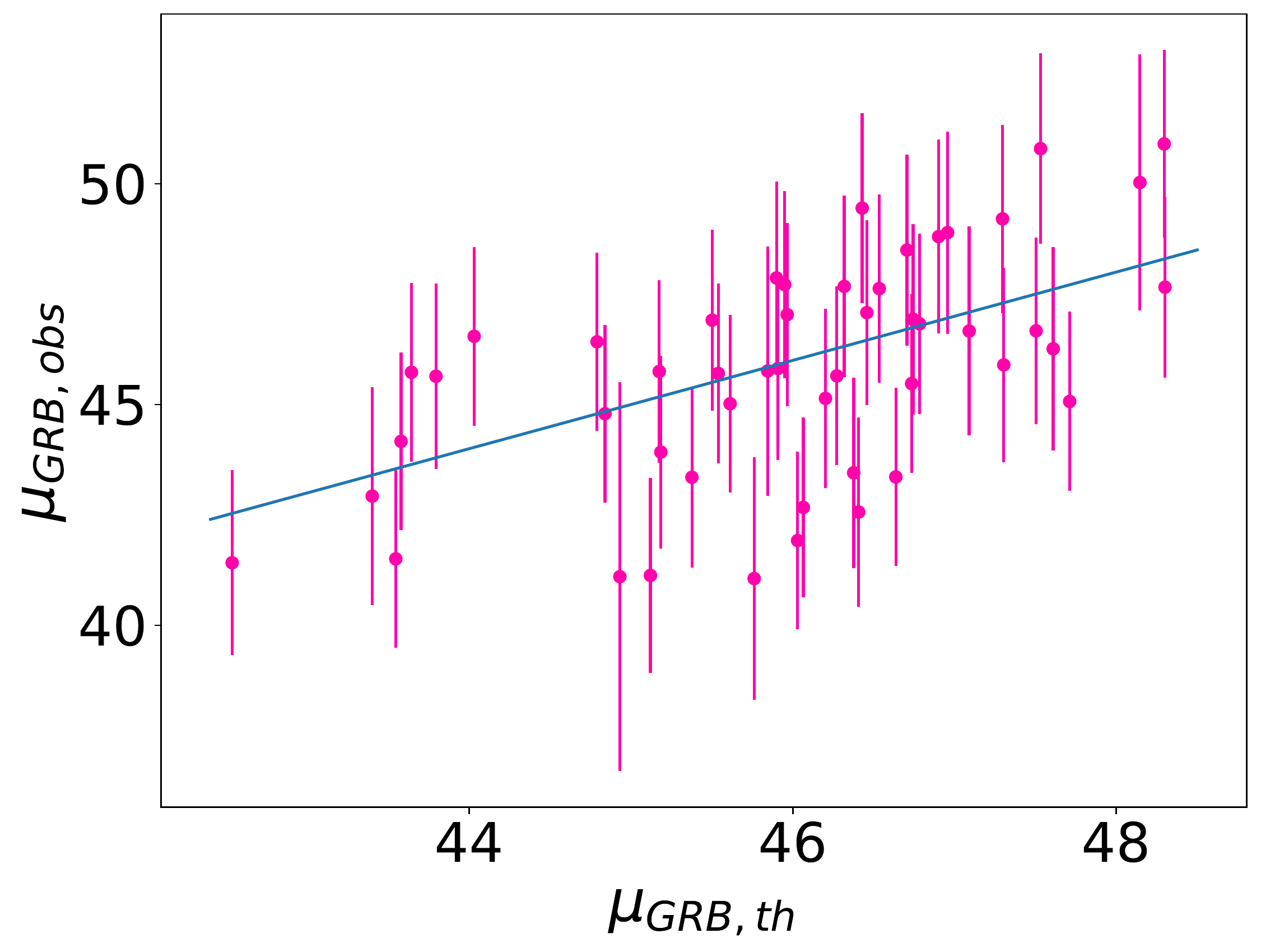}{0.34\textwidth}{(a) $\mu_{GRB,obs}$ vs $\mu_{GRB,th}$ with $\cal L_G$ likelihood.}\label{a1}
\fig{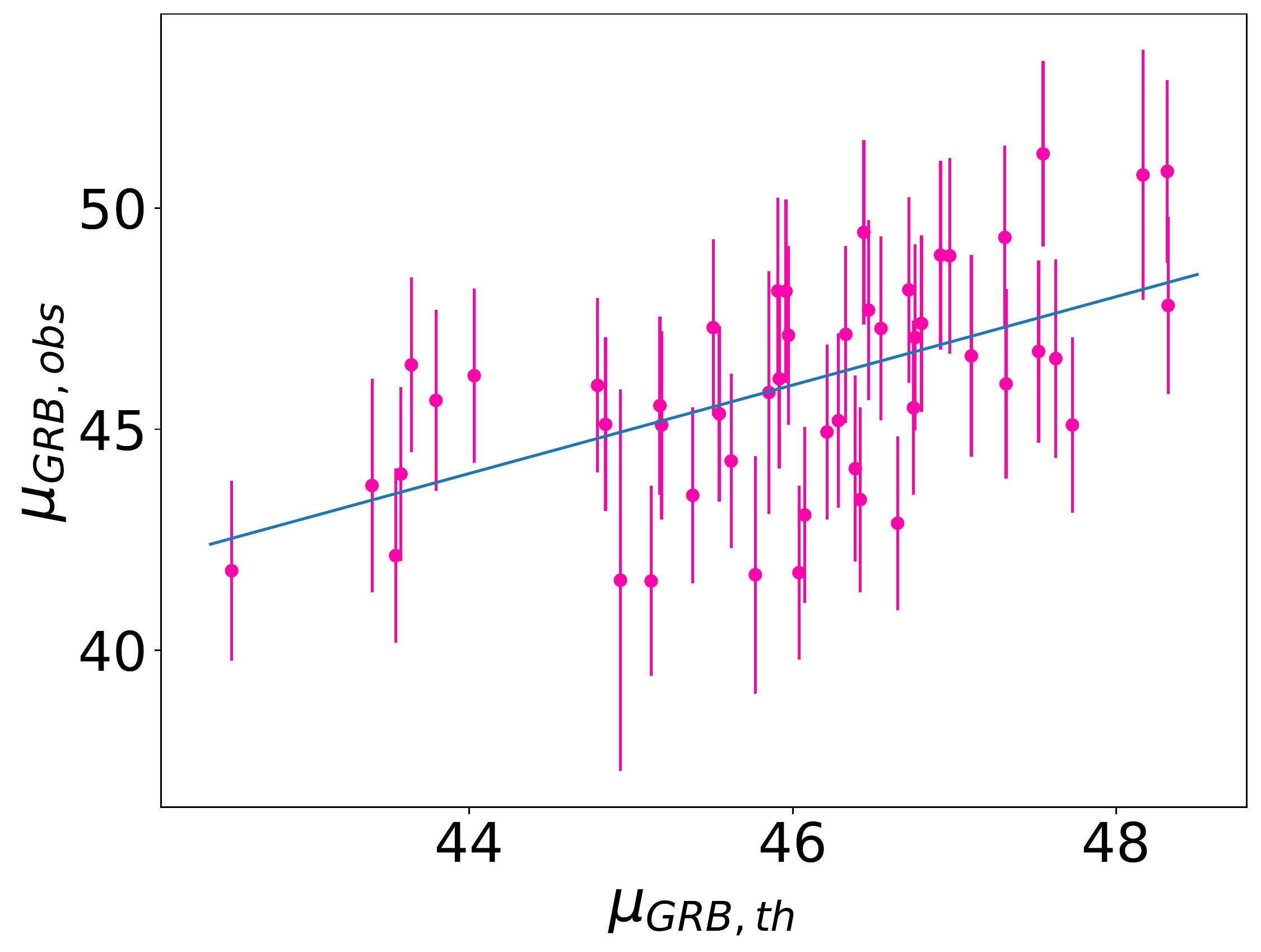}{0.34\textwidth}{(b) $\mu_{GRB,obs}$ vs $\mu_{GRB,th}$ with $\cal L_N$ likelihood .}\label{a2}}
\gridline{
\fig{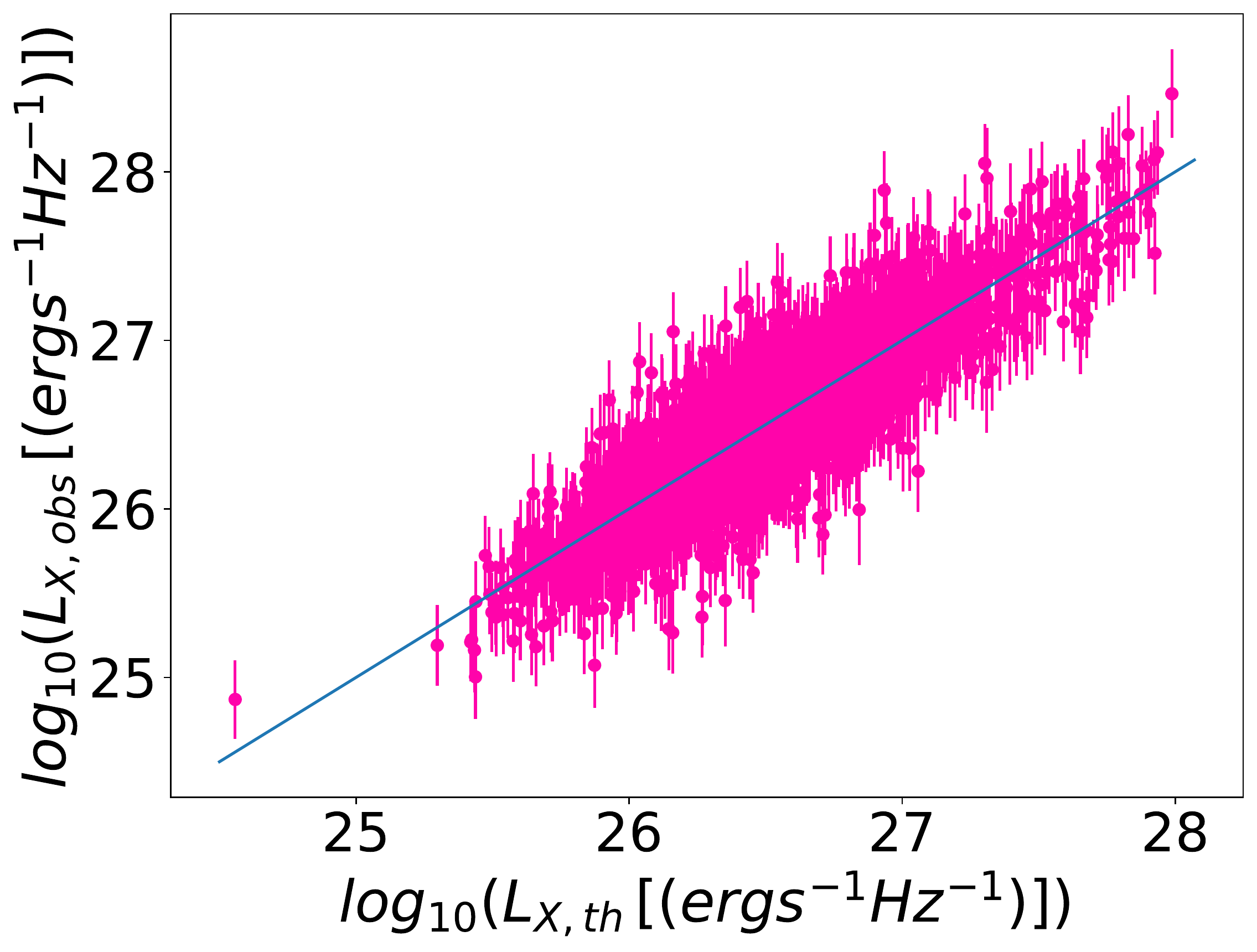}{0.34\textwidth}{(c) $\log L_{X,th}$ vs $\log L_{X,th}$ for QSOs with $\cal L_G$ likelihood.}\label{a3}
\fig{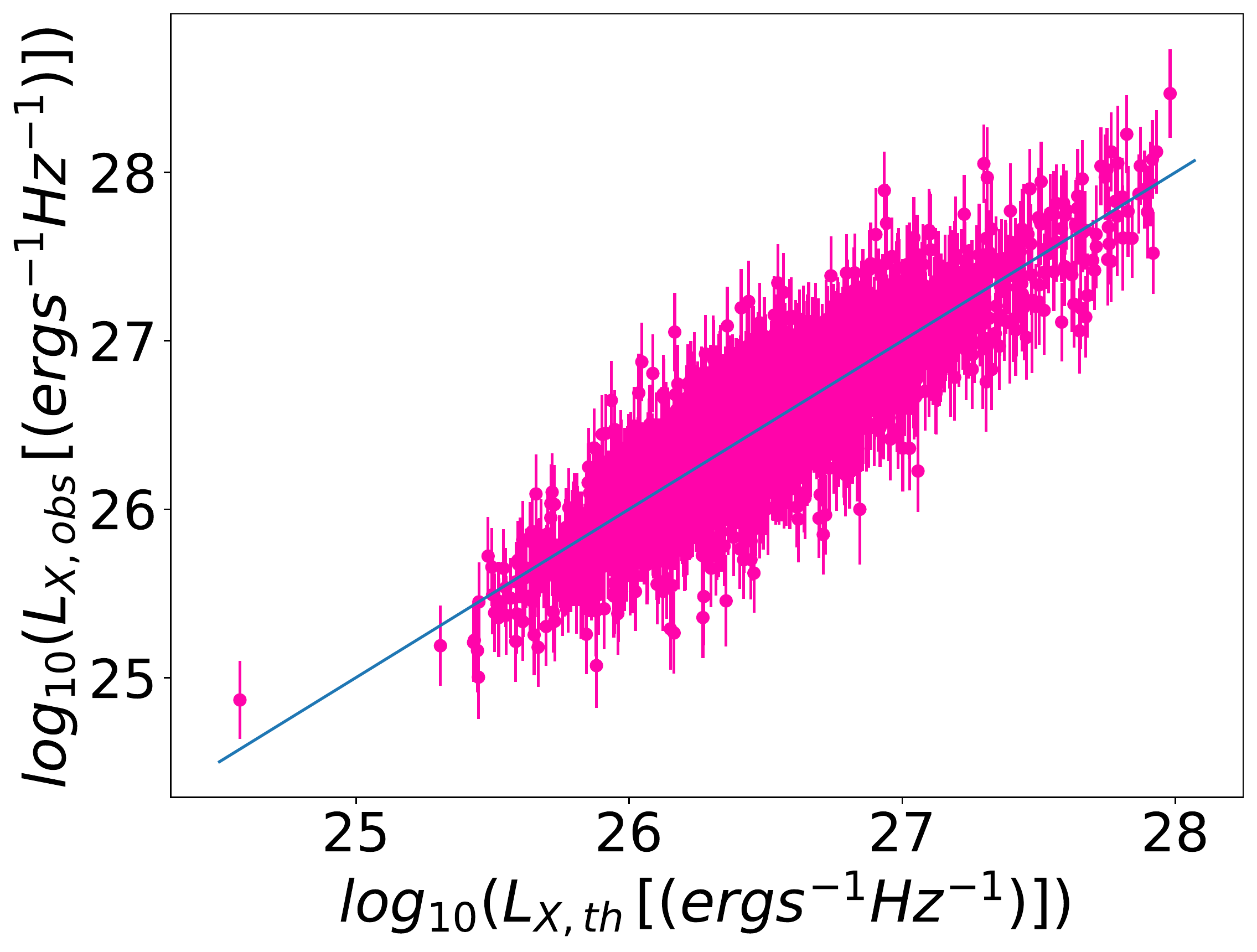}{0.34\textwidth}{(d) $\log L_{X,th}$ vs $\log L_{X,th}$ for QSOs with $\cal L_N$ likelihood.}\label{a4}}
\gridline{
\fig{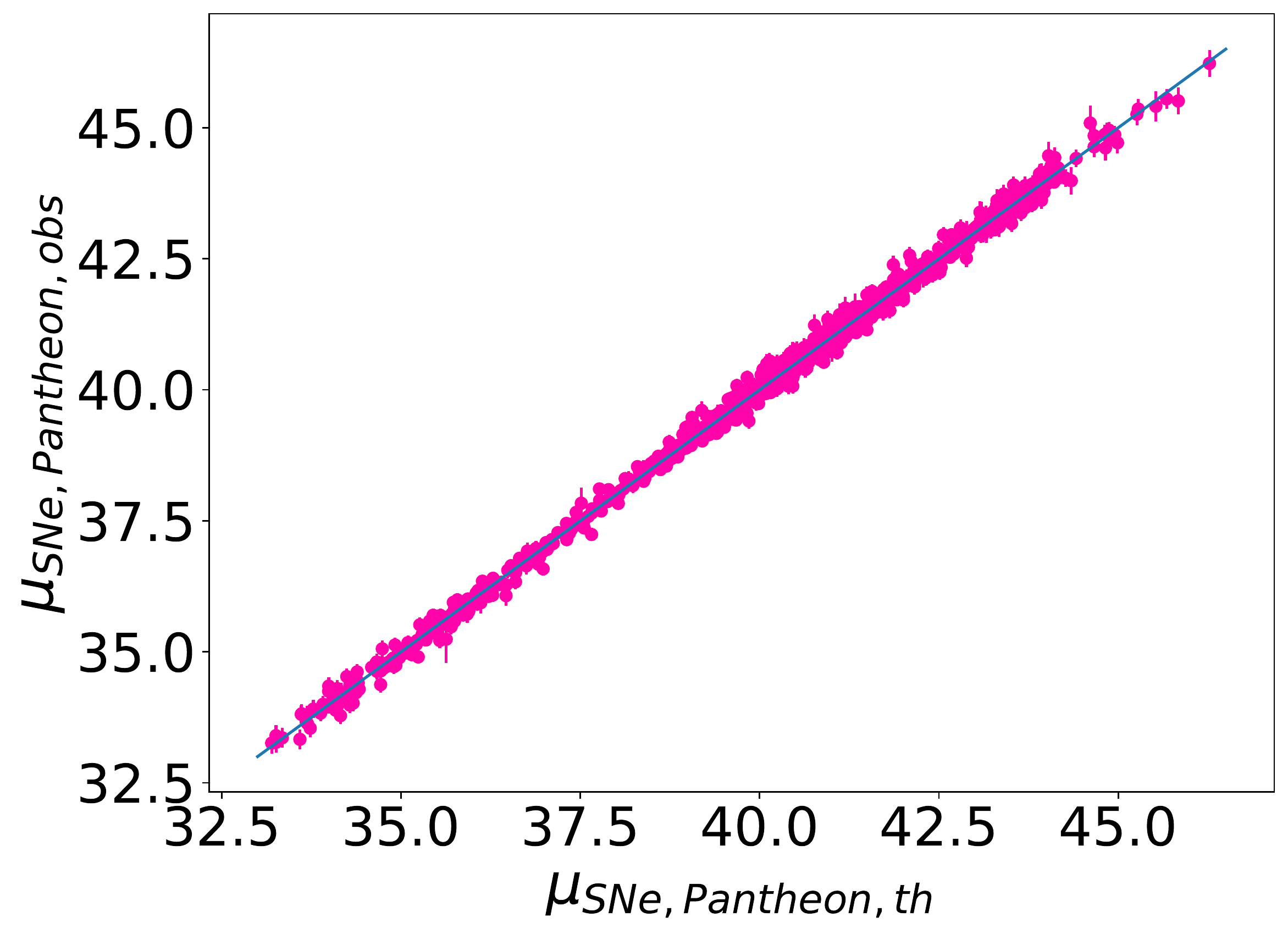}{0.34\textwidth}{(e) $\mu_{SNe Ia, obs}$ vs $\mu_{SNe Ia, th}$ with $\cal L_G$ likelihood.}\label{a6}
\fig{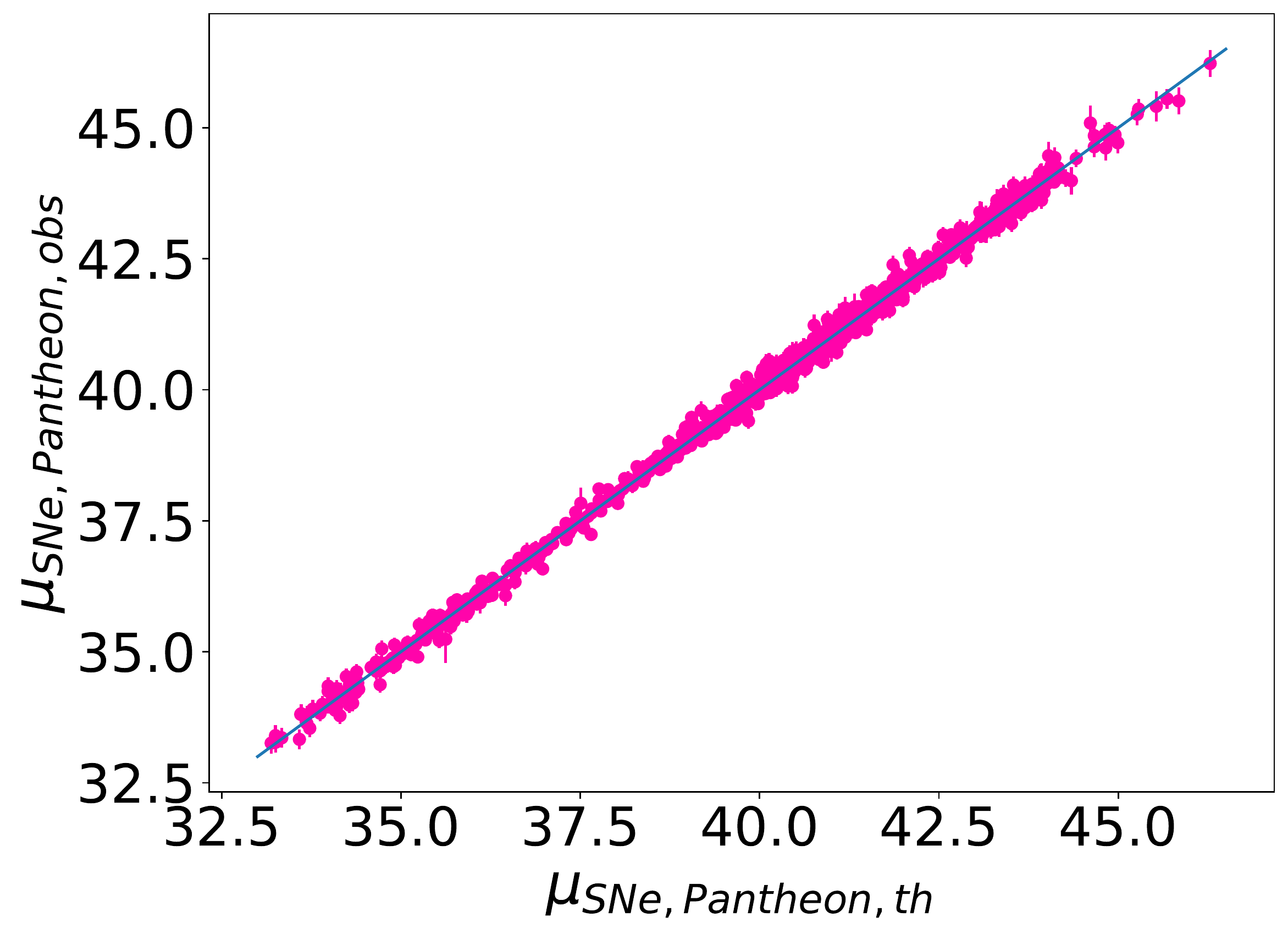}{0.34\textwidth}{(f) $\mu_{SNe Ia, obs}$ vs $\mu_{SNe Ia, th}$ with $\cal L_N$ likelihood.}\label{a5}}
\gridline{
\fig{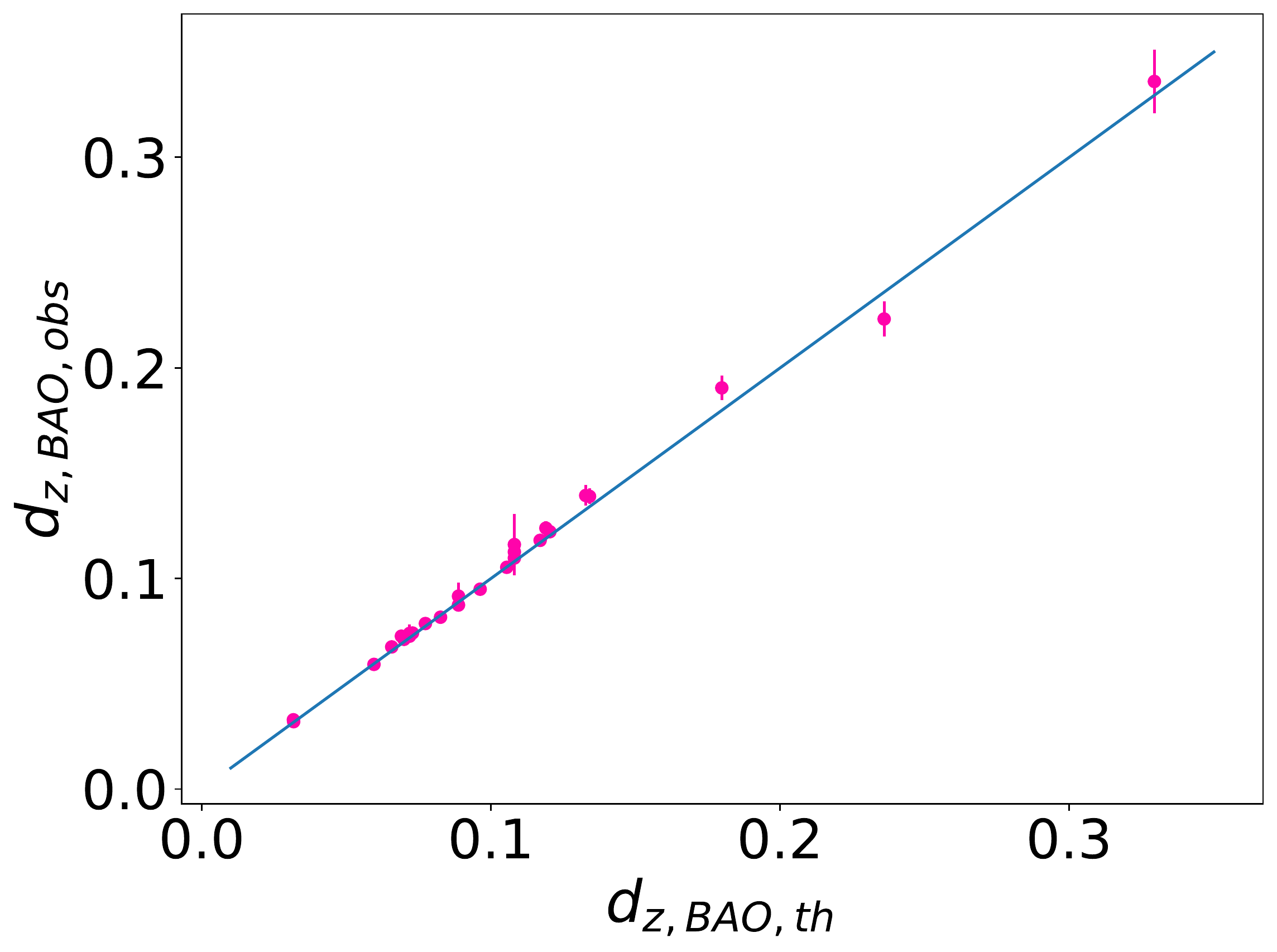}{0.34\textwidth}{(g) $d_{z,BAO,obs}$ vs $d_{z,BAO,th}$ with $\cal L_G$ likelihood.}\label{a8}
\fig{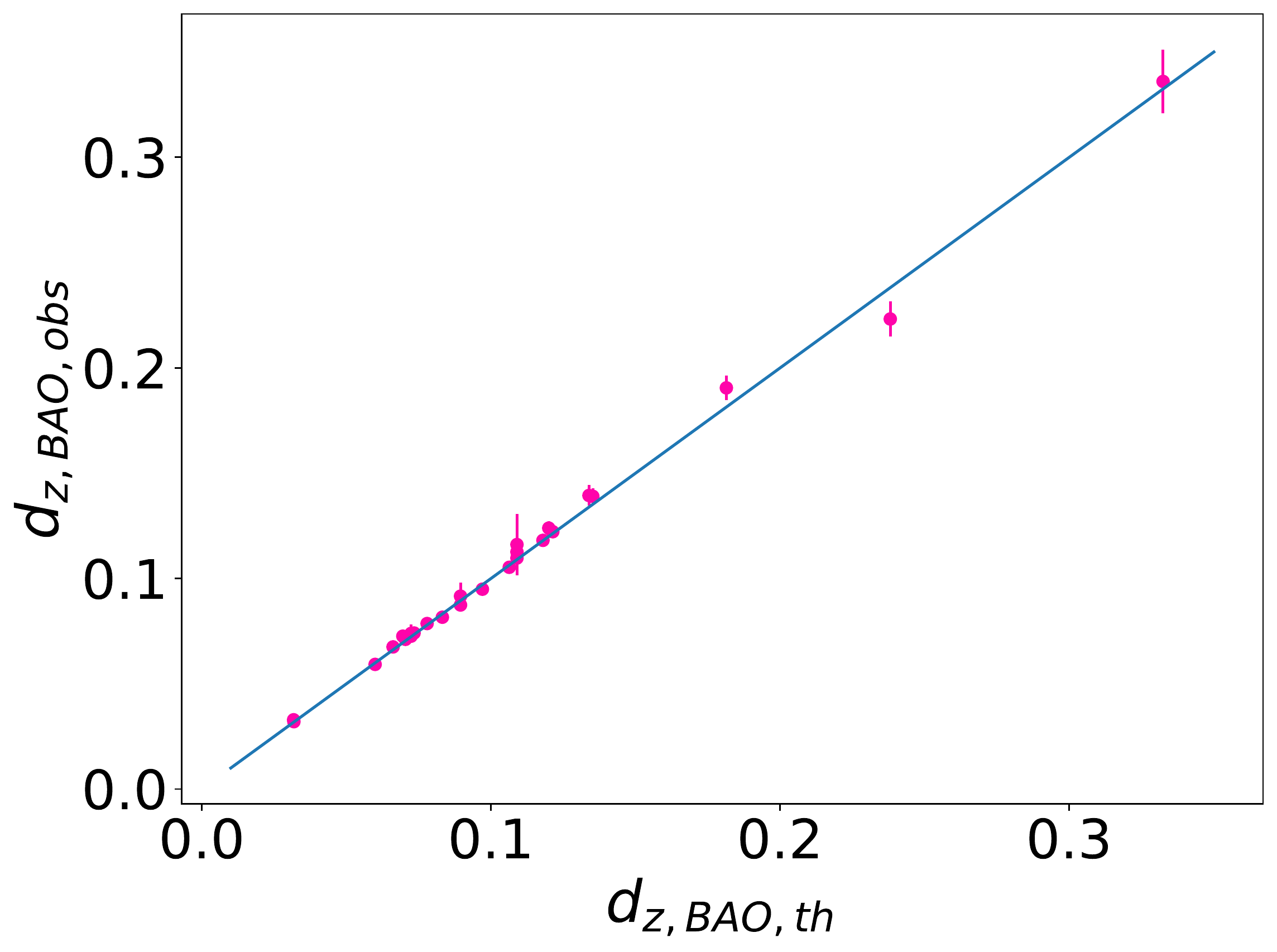}{0.34\textwidth}{(h) $d_{z,BAO,obs}$ vs $d_{z,BAO,th}$ with $\cal L_N$ likelihood.}\label{a7}}
\caption{The observed vs the predicted distance moduli, $\mu_{SNe Ia}$ for SNe Ia and $\mu_{GRBs}$ for GRBs, luminosities, $\mathrm{log_{10}}(L_{X})$ for QSOs, and distances $d_{z}$ for BAO, computed with the best-fit values obtained with the likelihoods suitable for all probes for the case without correction for evolution and the non flat $\Lambda$  CDM model.}
\label{scatterplots}
\end{figure}

\begin{figure}
\gridline{
\fig{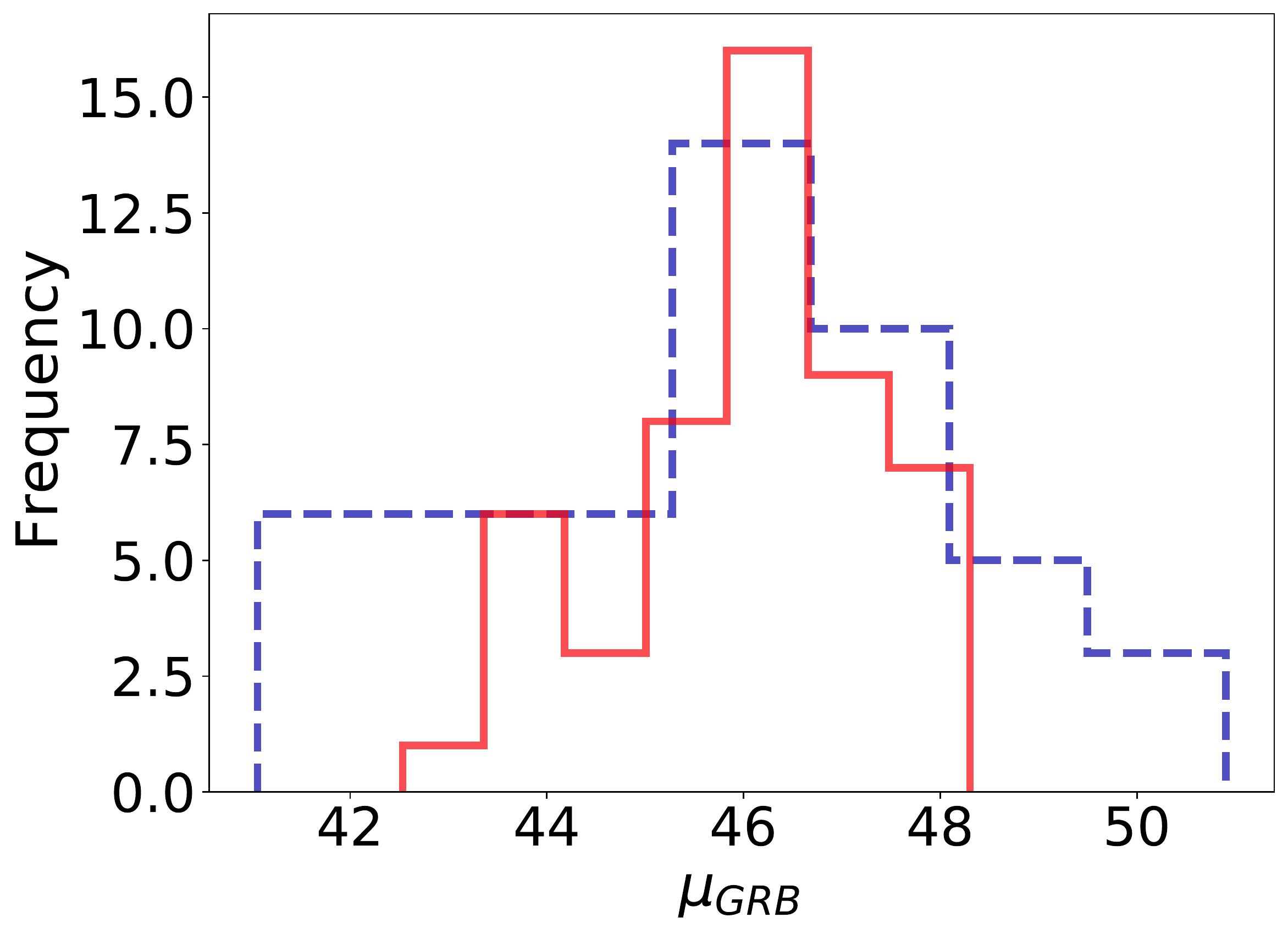}{0.34\textwidth}{(a) $\mu_{GRB,obs}$ vs $\mu_{GRB,th}$ with $\cal L_G$ likelihood.}\label{a2}
\fig{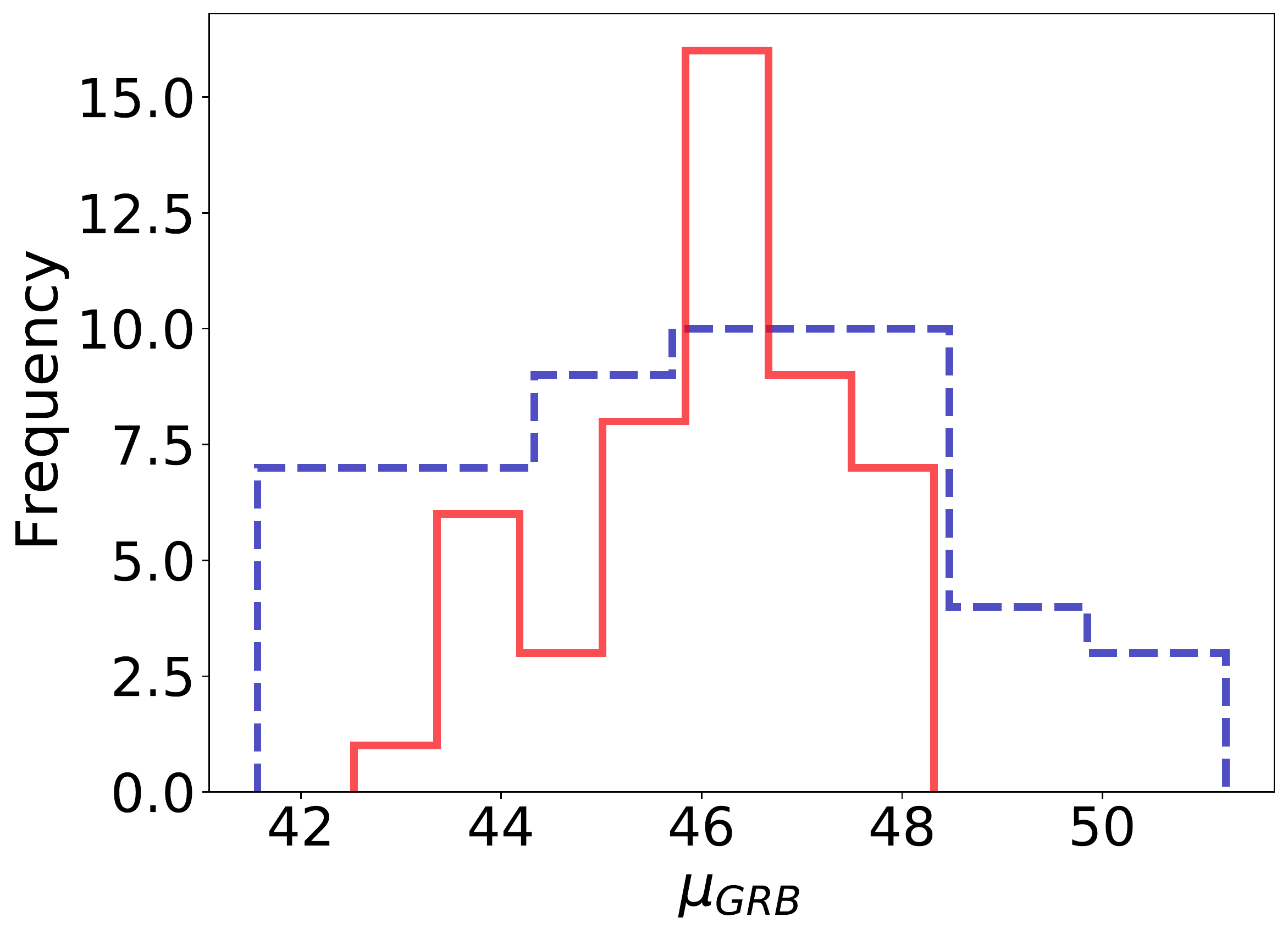}{0.34\textwidth}{(b) $\mu_{GRB,obs}$ vs $\mu_{GRB,th}$ with $\cal L_N$ likelihood .}\label{a1}}
\gridline{
\fig{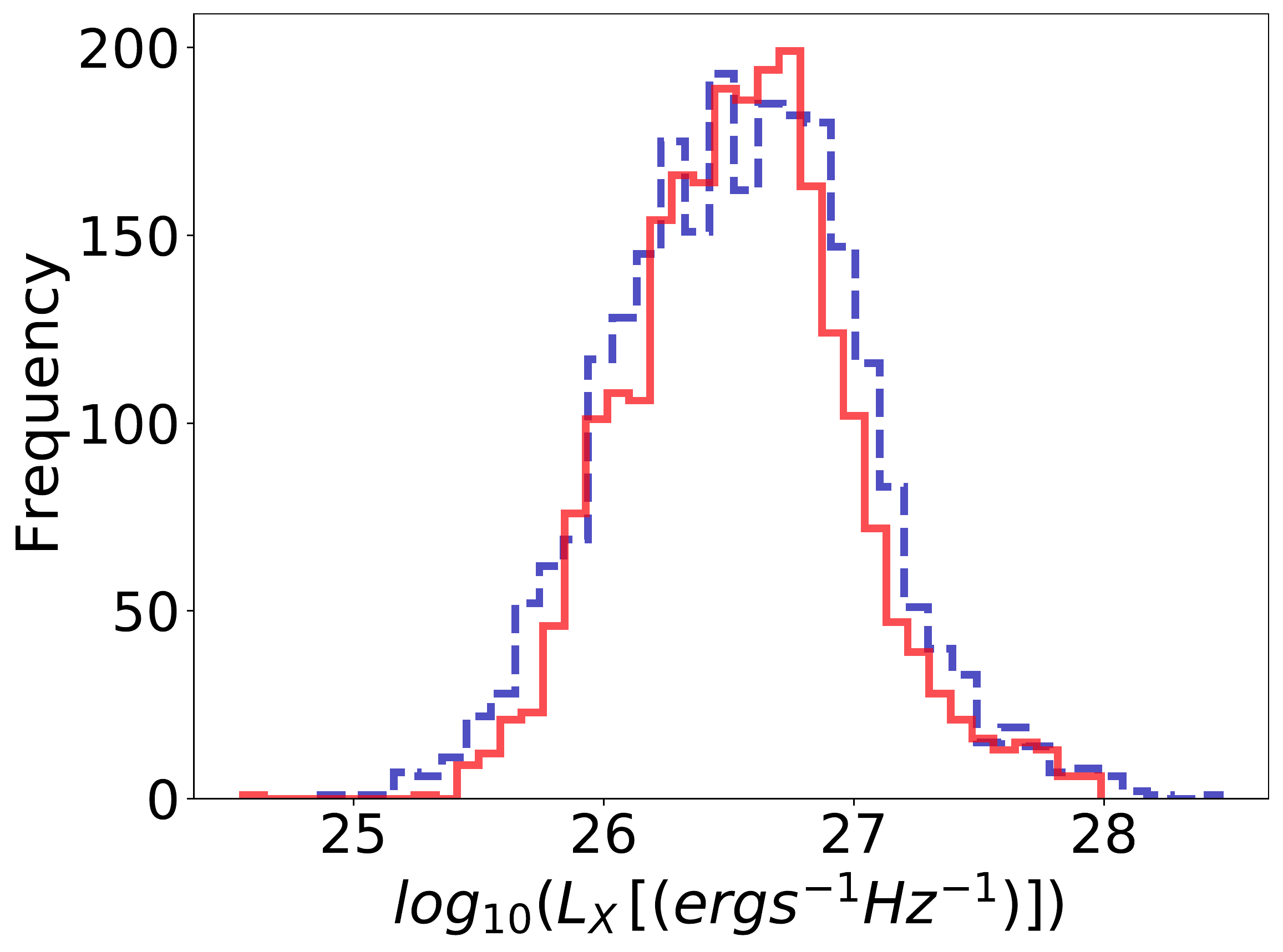}{0.34\textwidth}{(c) $\log L_{X,th}$ vs $\log L_{X,th}$ for QSOs  with $\cal L_G$ likelihood.}\label{a4}
\fig{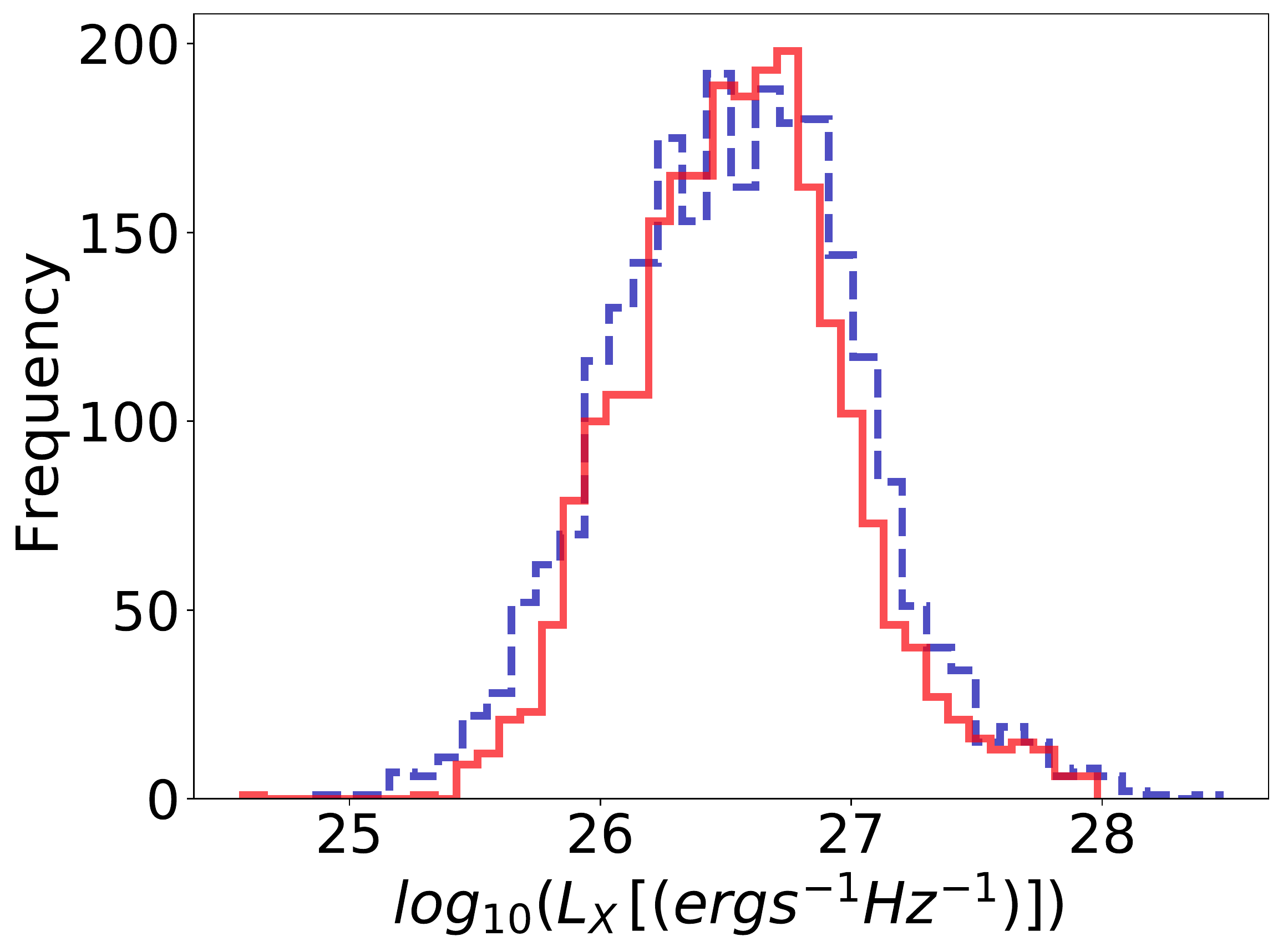}{0.34\textwidth}{(d) $\log L_{X,th}$ vs $\log L_{X,th}$ for QSOs  with $\cal L_N$ likelihood.}\label{a3}}
\gridline{
\fig{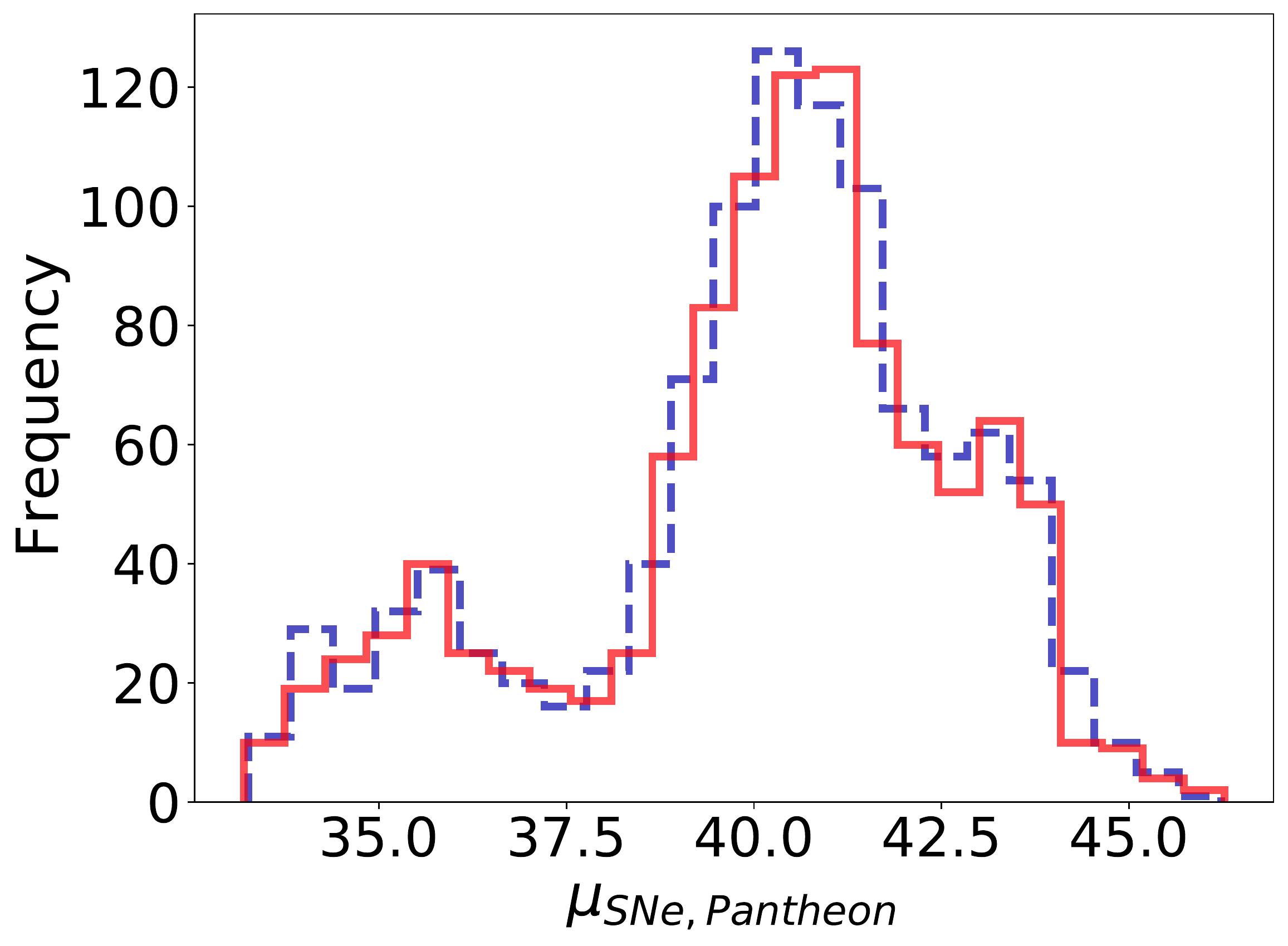}{0.34\textwidth}{(e) $\mu_{SNe Ia, obs}$ vs $\mu_{SNe Ia, th}$ with $\cal L_G$ likelihood.}\label{a6}
\fig{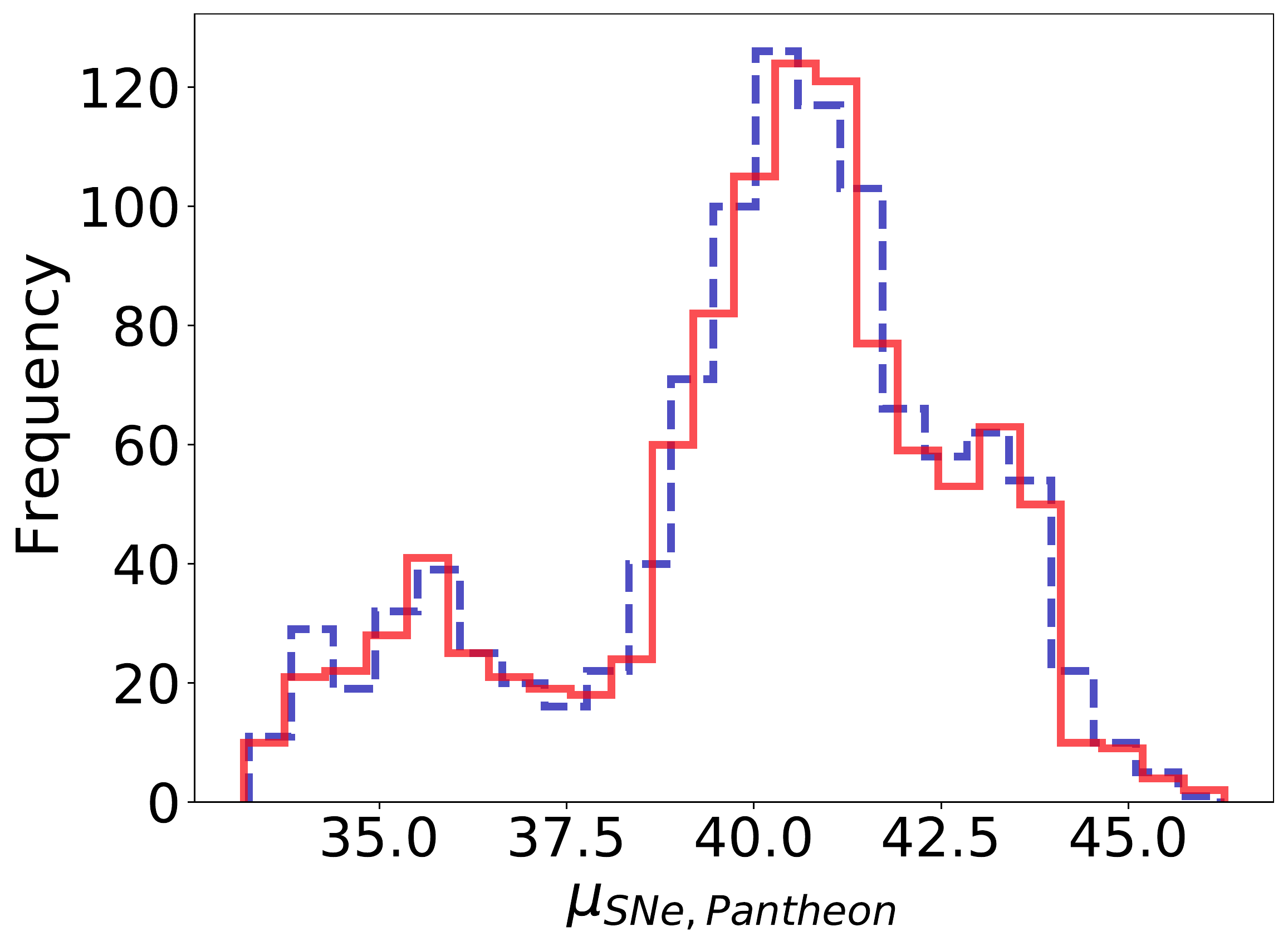}{0.34\textwidth}{(f) $\mu_{SNe Ia, obs}$ vs $\mu_{SNe Ia, th}$ with $\cal L_N$ likelihood.}\label{a5}}
\gridline{
\fig{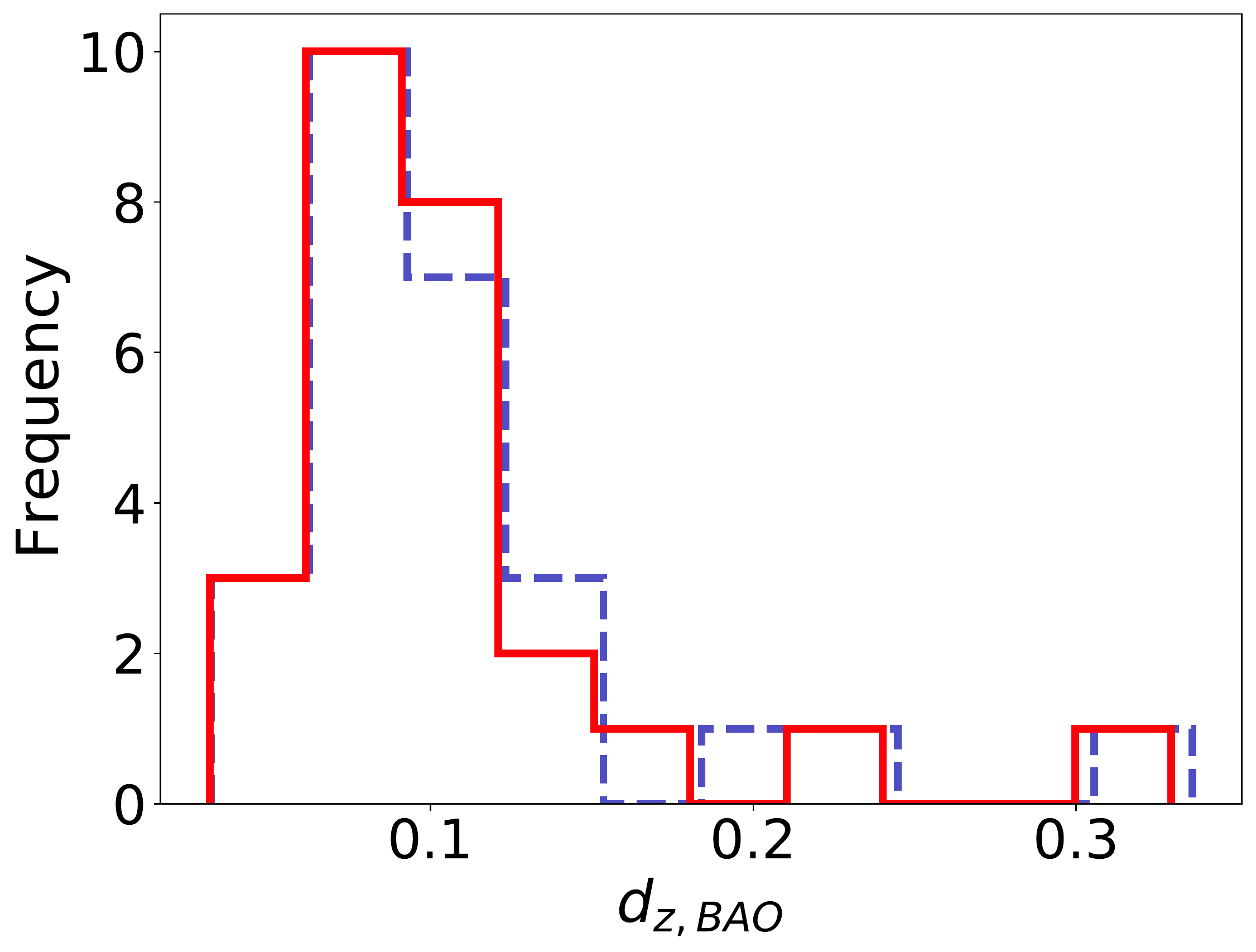}{0.34\textwidth}{(g) $d_{z,BAO,obs}$ vs $d_{z,BAO,th}$ with $\cal L_G$ likelihood.}\label{a8}
\fig{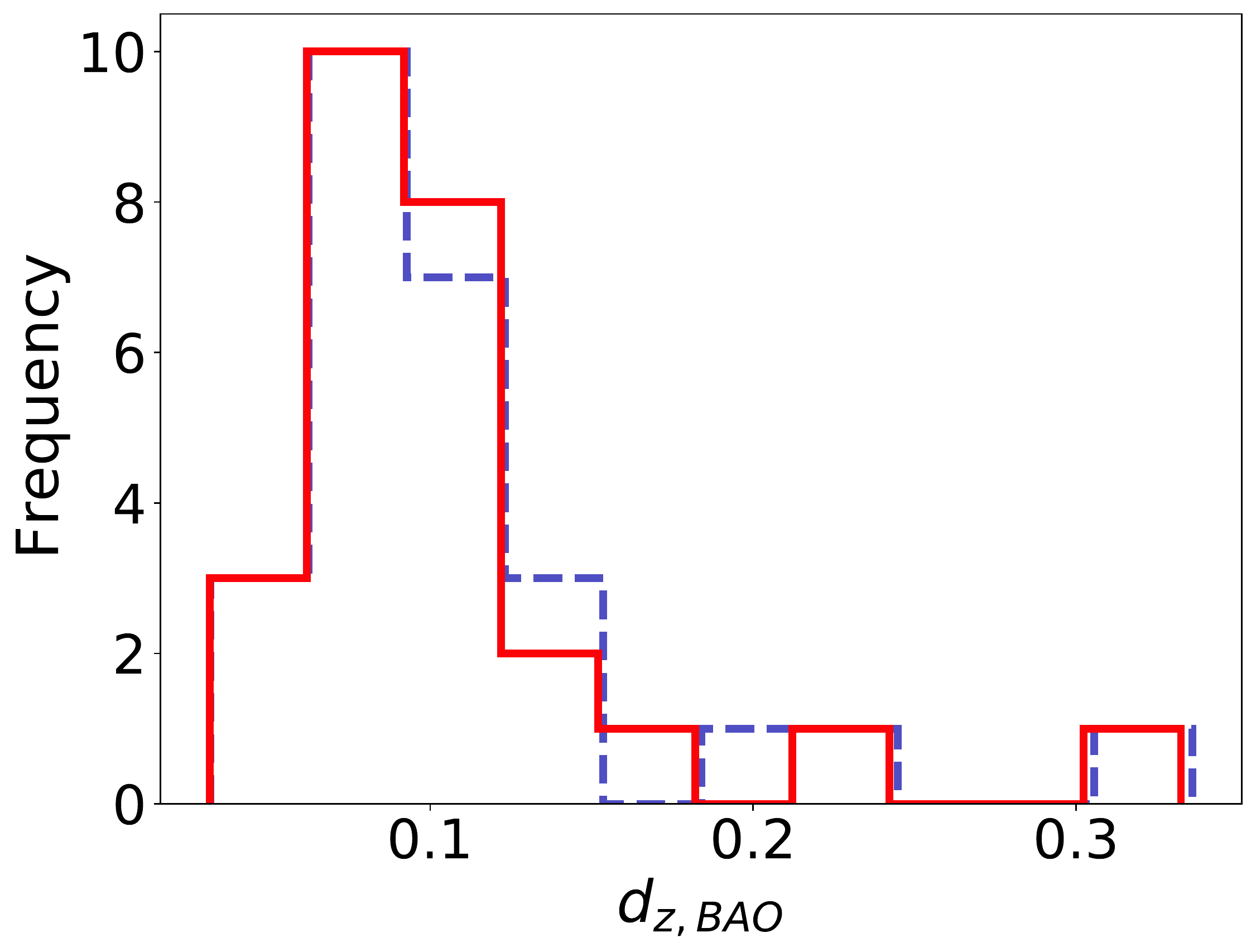}{0.34\textwidth}{(h) $d_{z,BAO,obs}$ vs $d_{z,BAO,th}$ with $\cal L_N$ likelihood.}\label{a7}}
\caption{The observed vs the predicted distance moduli, $\mu_{SNe Ia}$ for SNe Ia and $\mu_{GRBs}$ for GRBs, luminosities, $\mathrm{log_{10}}(L_{X})$ for QSOs, and distances $d_{z}$ for BAO, computed with the best-fit values obtained with the likelihoods suitable for all probes for the case without correction for evolution and the non-flat $\Lambda$  CDM model. The observed quantities are marked with a red continuous line, while the predicted ones are with a blue dashed line. }
\label{histogramsobth}
\end{figure}

We here further investigate quantitatively the goodness of our fits. To this end, we have employed both the Anderson-Darling and Cramer Von Mises tests to compare the measured and modelled values of the cosmological parameters in each configuration studied in this work. The results of these tests are summarized in Table \ref{tab:test}. In all cases, SNe Ia (from both \textit{Pantheon} and \textit{Pantheon +} samples) and BAO pass both the statistical tests at a significance level $> 25 \%$ and with p-value $>99\%$. For the GRBs, the model is well-specified in all cases at the level of acceptance of the 5\% as commonly used, with the exception of the following cases: $\cal L_G$ likelihood with varying evolution and both \textit{Pantheon} and \textit{Pantheon +} SNe Ia sample. This may be due to the fact that the correction for evolution brings larger uncertainties on the cosmological parameters and thus could lead to slightly different values of the cosmological parameters, larger for $\Omega_M$ and $H_0$, as shown in Table 3 of \citet{Dainotti2022MNRAS.tmp.2639D}. Concerning QSOs, they never pass the Anderson-Darling and Cramer Von Mises tests. However, the fact that measured and fitted quantities do not come from the same distribution for GRBs (in some cases) and QSOs is not surprising. 
To further investigate the applicability and limitations of these tests and to have a visual inspection of the test results, we have plotted in Figs. \ref{scatterplots} and \ref{histogramsobth} the scatter plots and histograms of the observed and predicted distance moduli of the GRBs, SNe Ia, the observed and predicted angular distance of the BAOs, and the observed and predicted luminosities for QSOs.
Indeed, if we compare the distributions of measured and modelled quantities for these probes, we can notice that the observed distribution (in blue) presents a larger variance than the theoretical distribution (in red). This variance originated from the larger uncertainties and the intrinsic dispersion of the measurements. This discrepancy between the two distributions causes the rejection of the null hypothesis for the statistical tests. On the other hand, the same discrepancy is not present in the case of SNe Ia and BAO, as shown in Figs. \ref{scatterplots} and \ref{histogramsobth}, since their measurements are more precise and SNe Ia have a negligible dispersion compared to GRBs and QSOs. Thus, for SNe Ia, BAO and in most cases for GRBs the two distributions of observed and fitted quantities are compatible with each other and hence the Anderson-Darling and Cramer Von Mises tests are passed.

\section{The cosmological analysis with QSOs alone}\label{sec6}
\begin{figure}
\centering
\gridline{
\fig{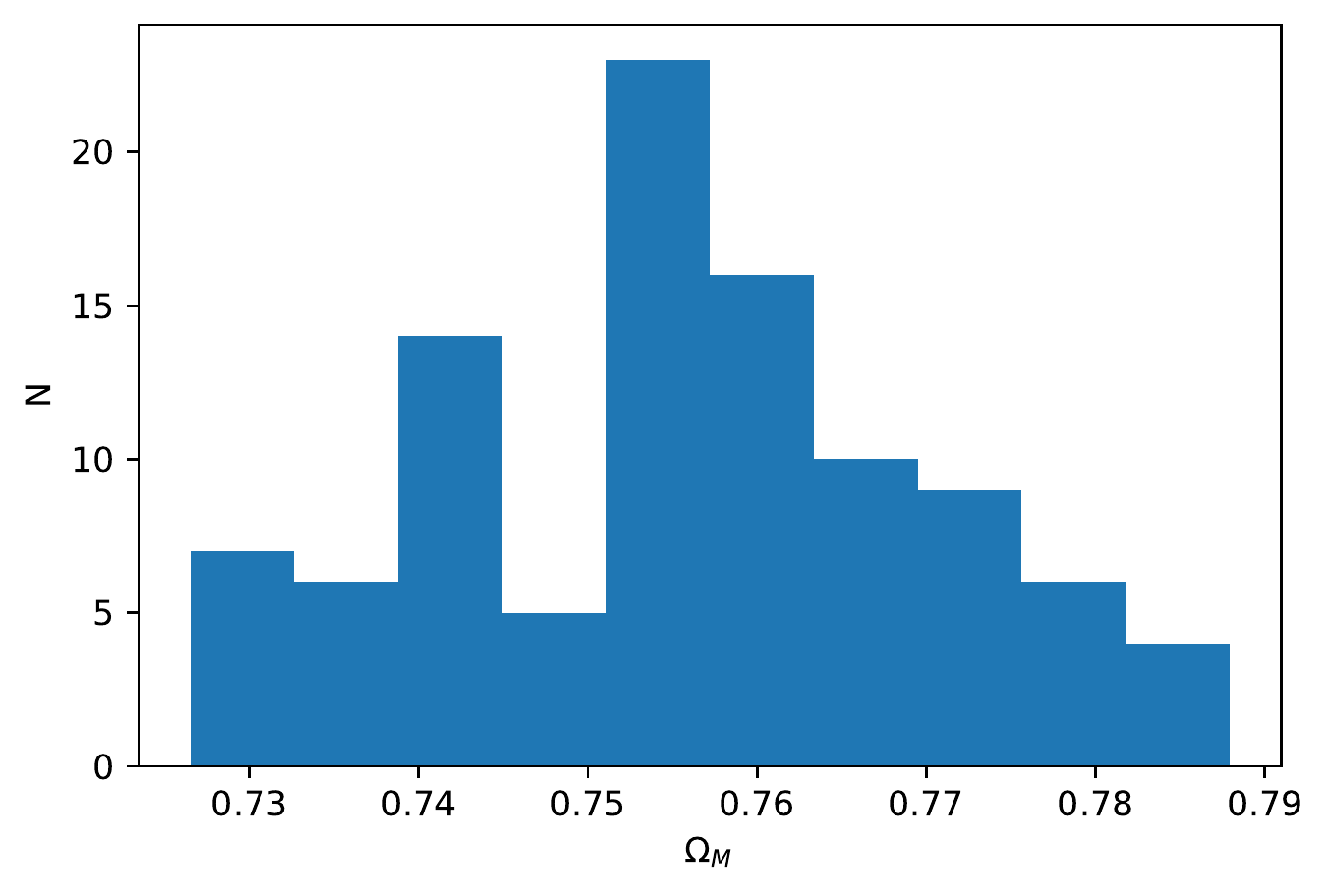}{0.3\textwidth}{(a) $\Omega_M$ without correction for redshift evolution}
\fig{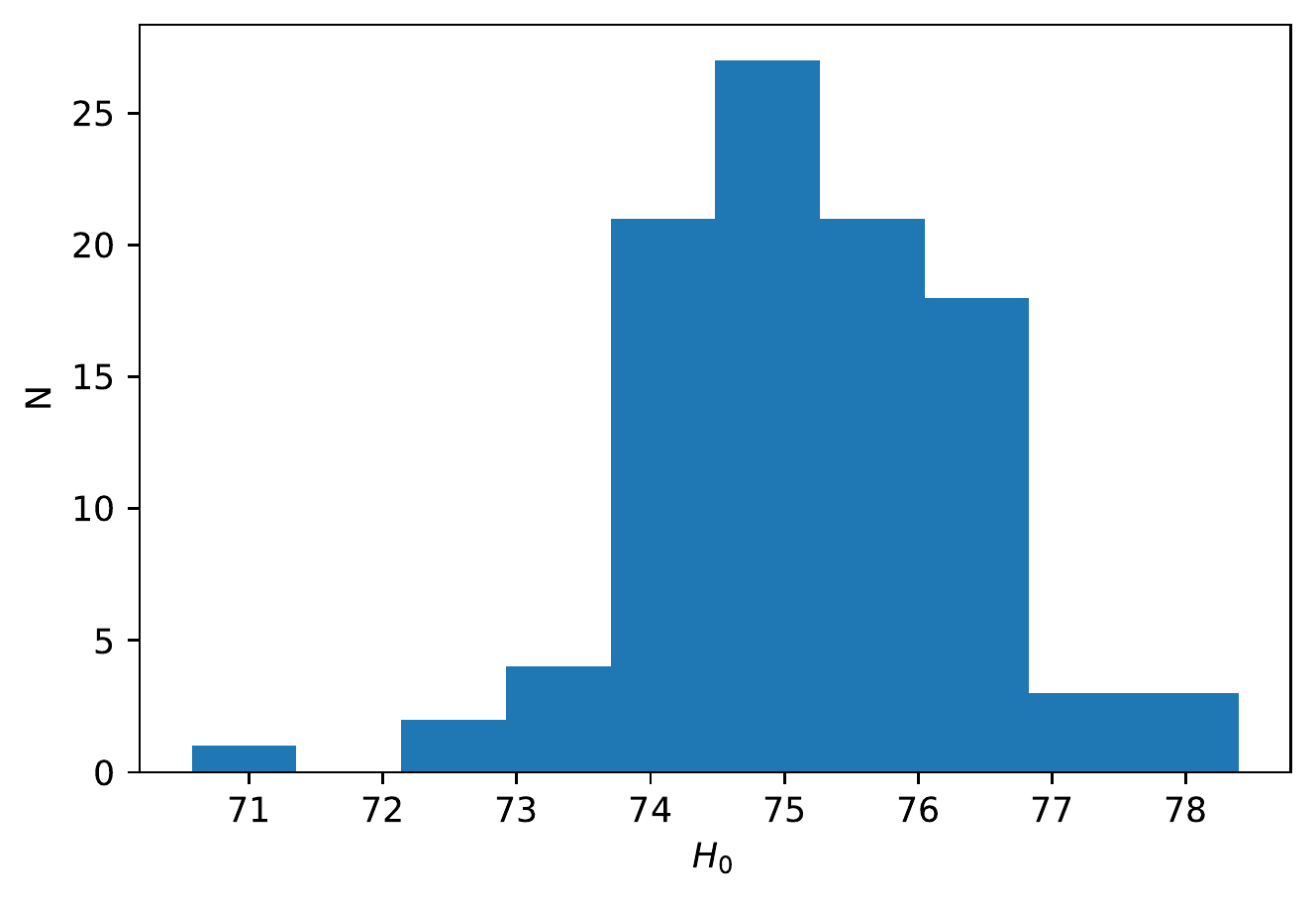}{0.3\textwidth}{(b) $H_0$ without correction for redshift evolution}
\fig{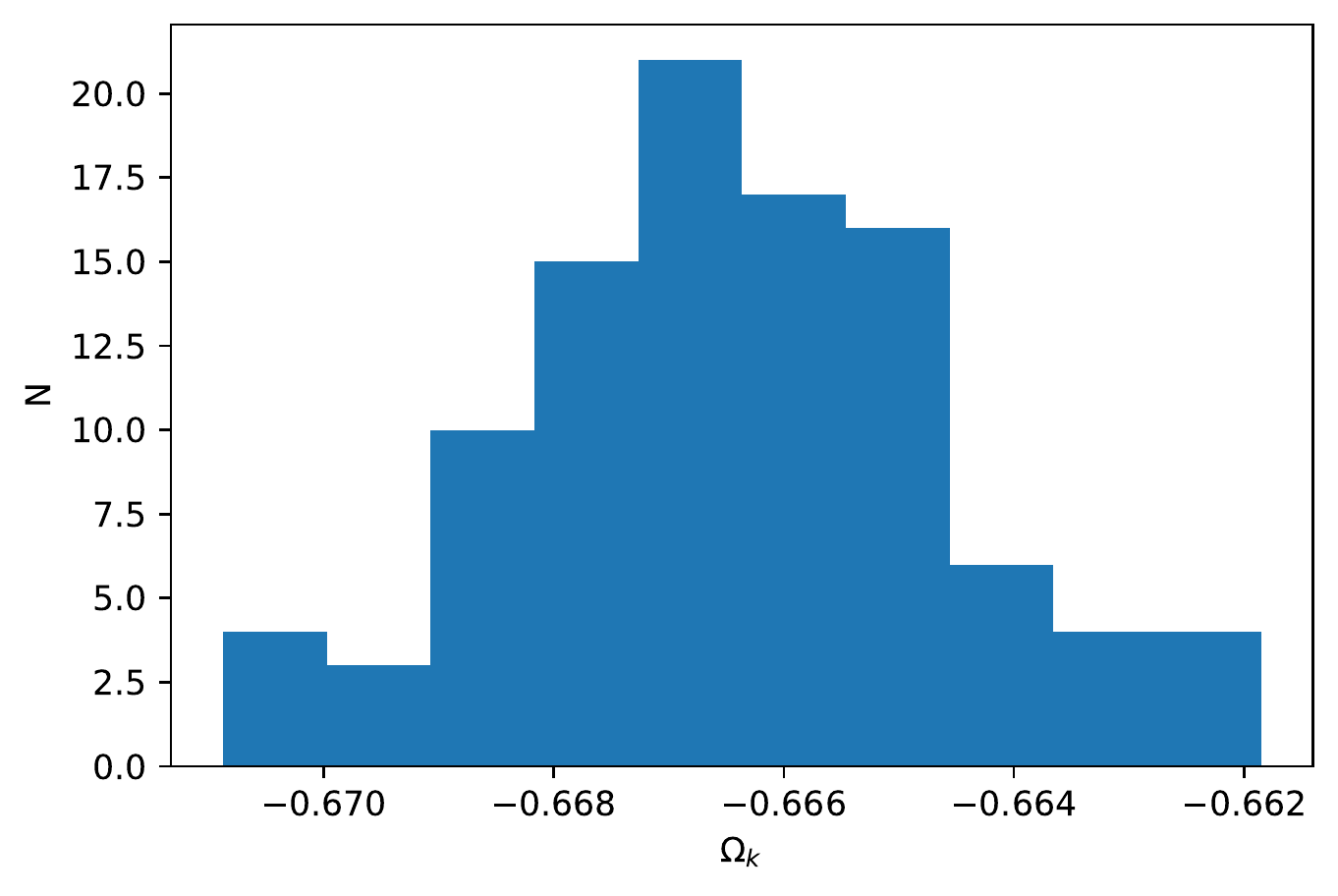}{0.3\textwidth}{(c) $\Omega_k$ without correction for redshift evolution}}
\gridline{
\fig{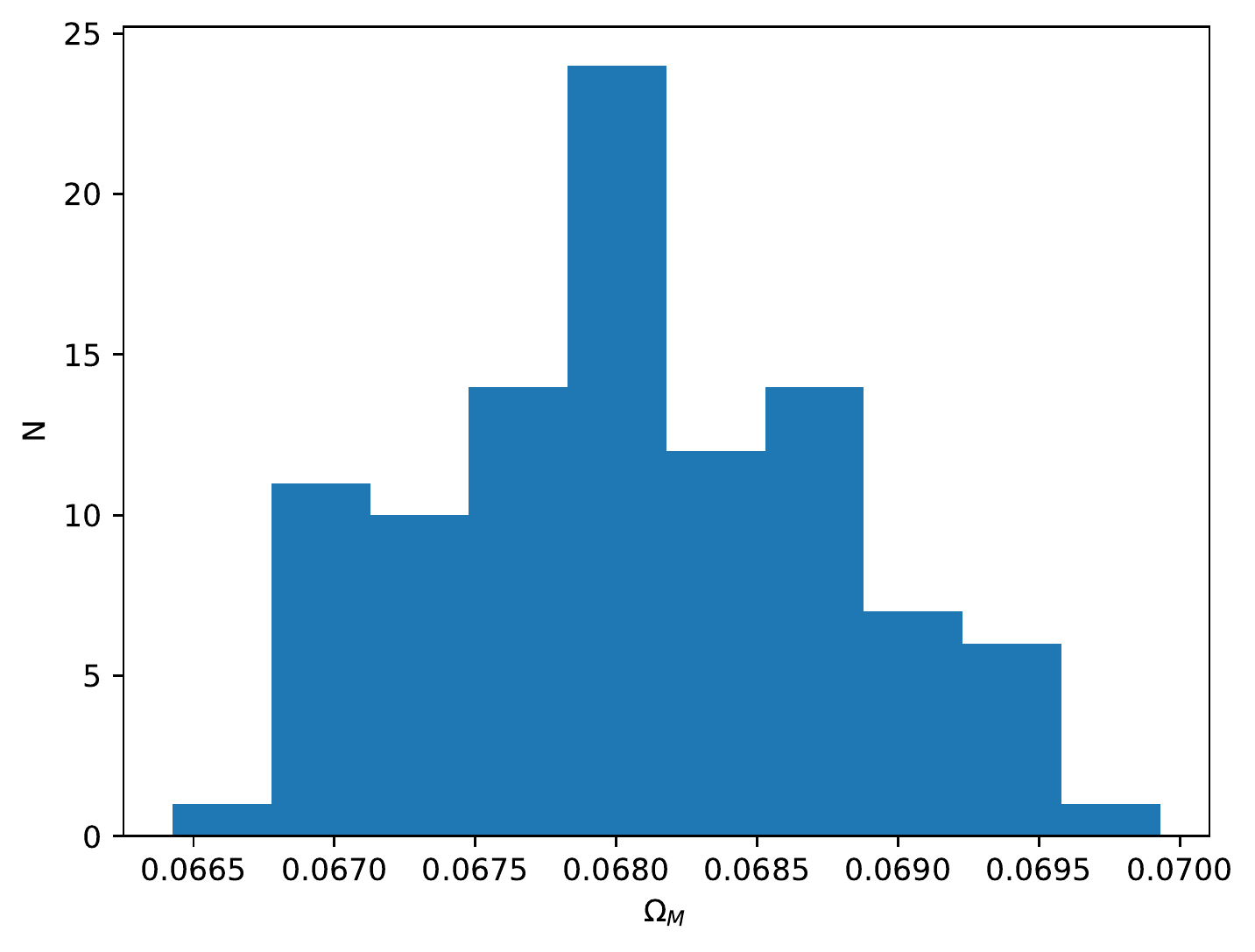}{0.3\textwidth}{(d) $\Omega_M$ with fixed correction for redshift evolution}
\fig{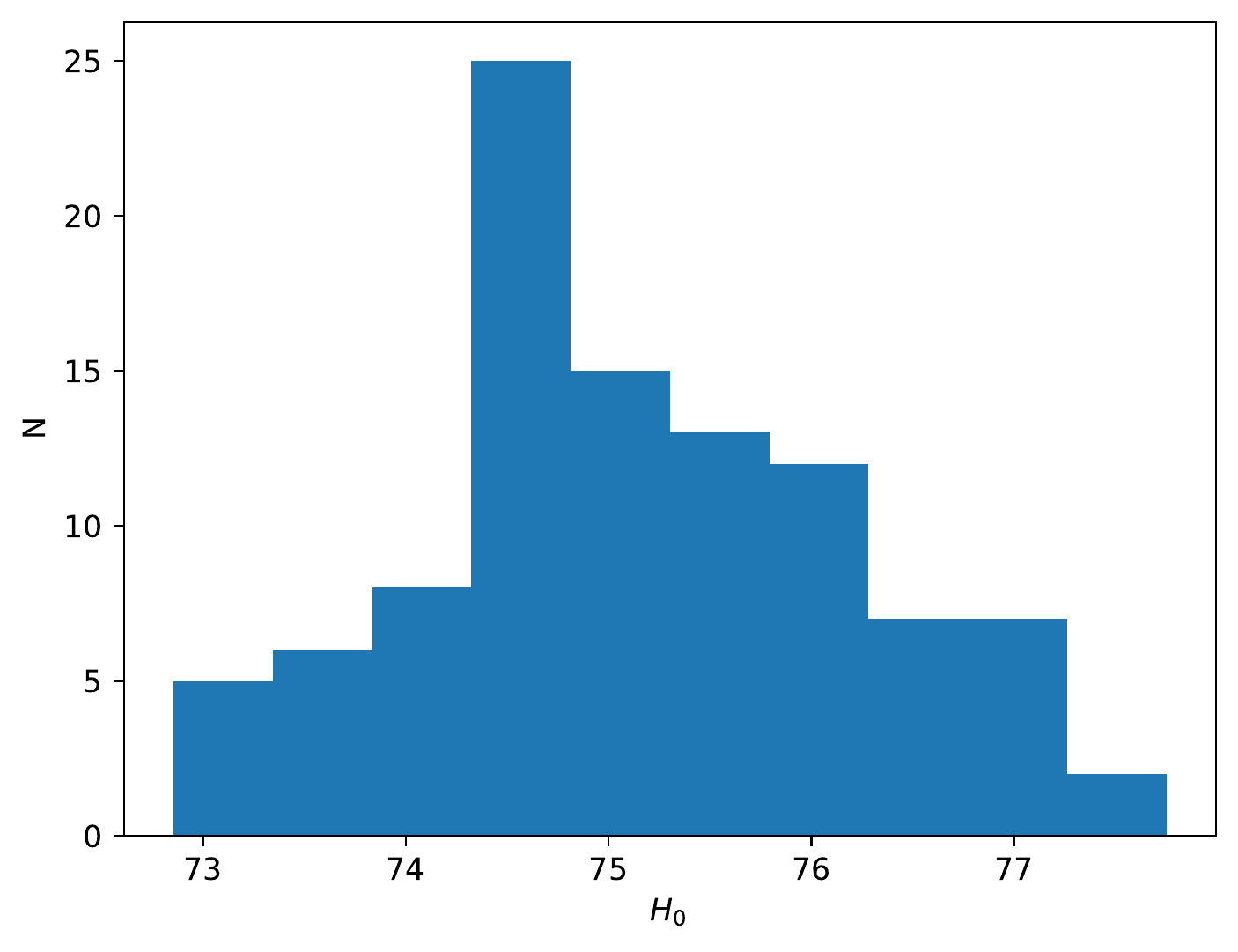}{0.3\textwidth}{(e) $H_0$ with fixed correction for redshift evolution}\label{}
\fig{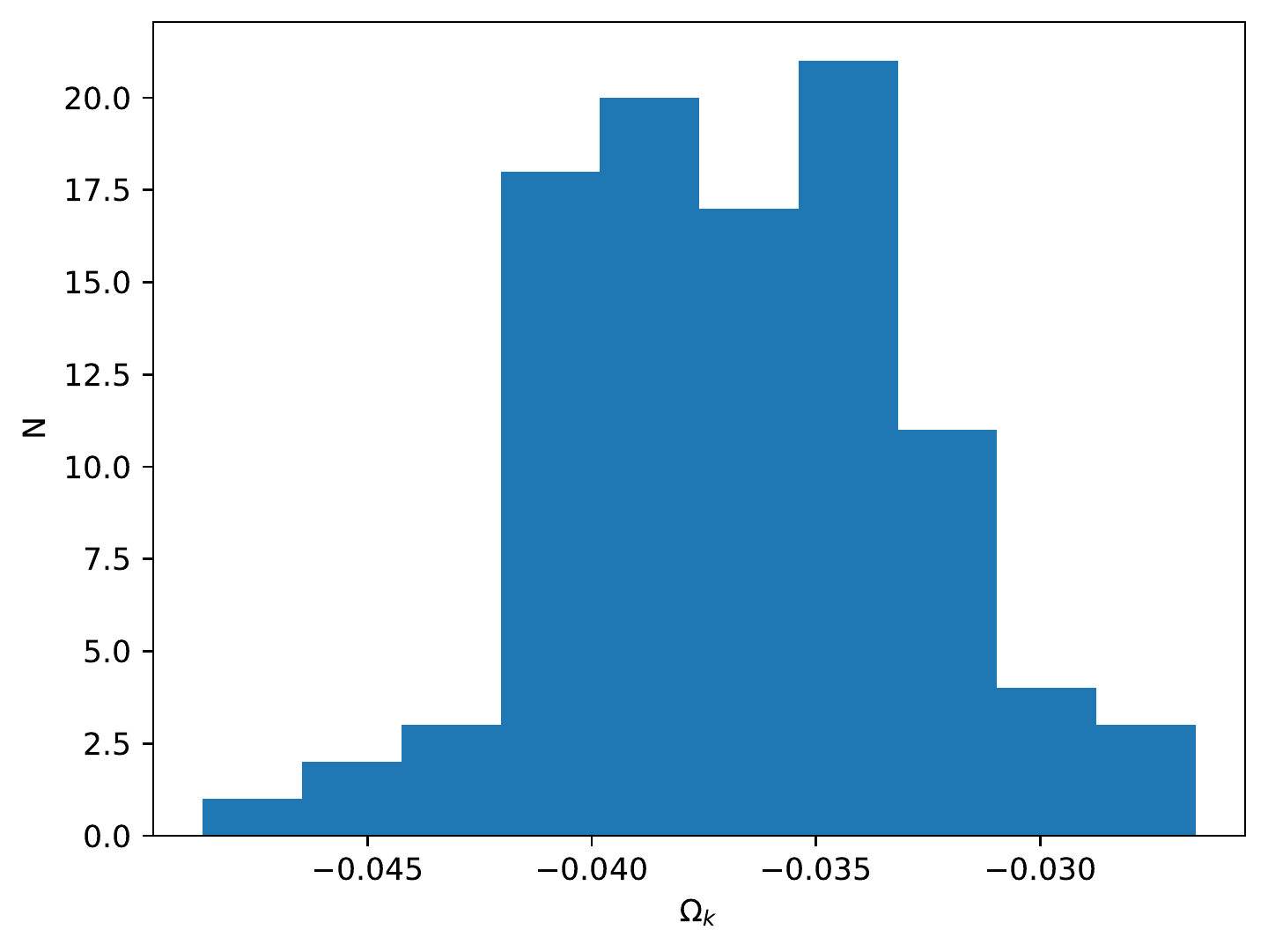}{0.3\textwidth}{(f) $\Omega_k$ with fixed correction for redshift evolution}}
\gridline{
\fig{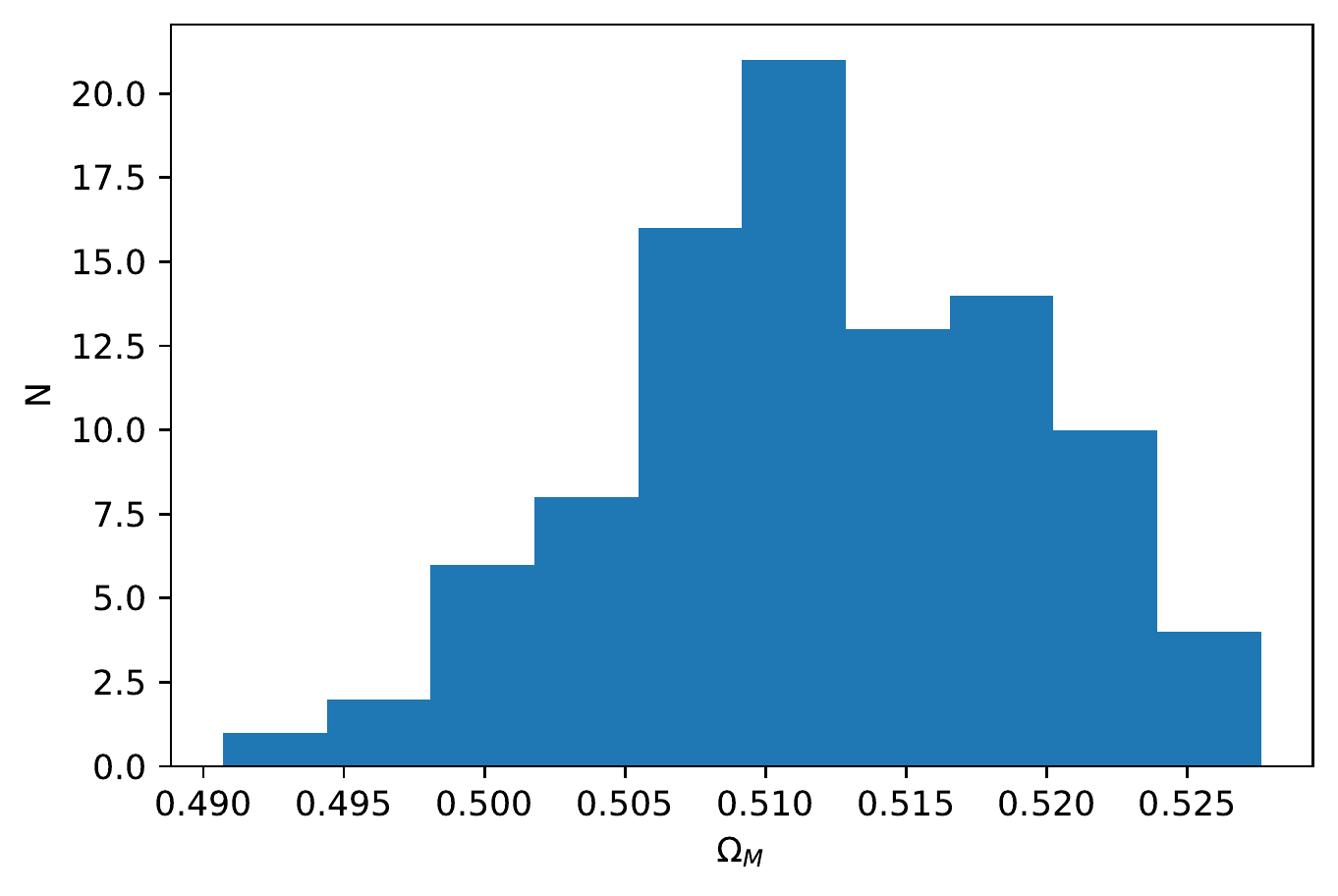}{0.3\textwidth}{(g) $\Omega_M$ with varying correction for redshift evolution}
\fig{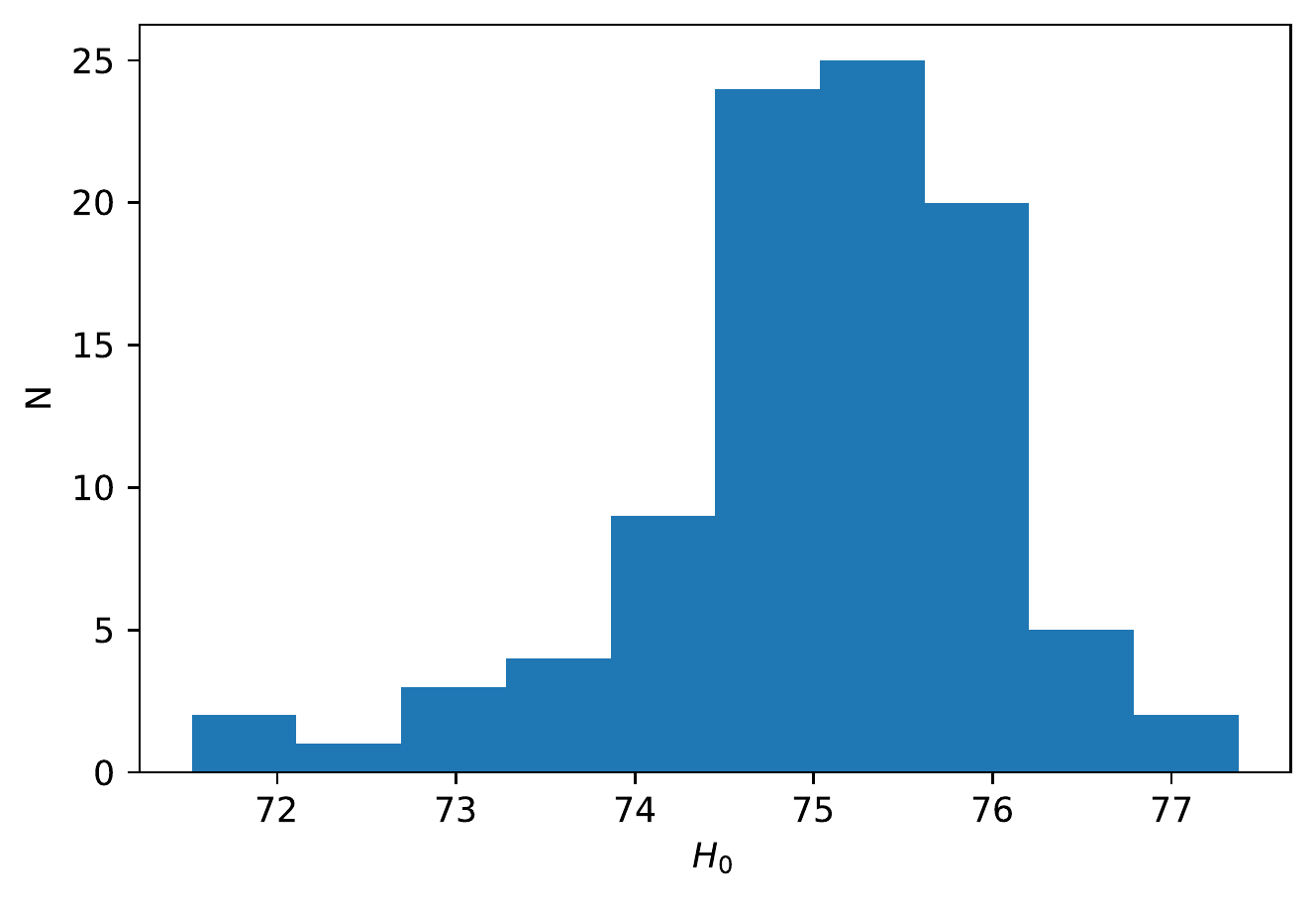}{0.3\textwidth}{(h) $H_0$ with varying correction for redshift evolution}
\fig{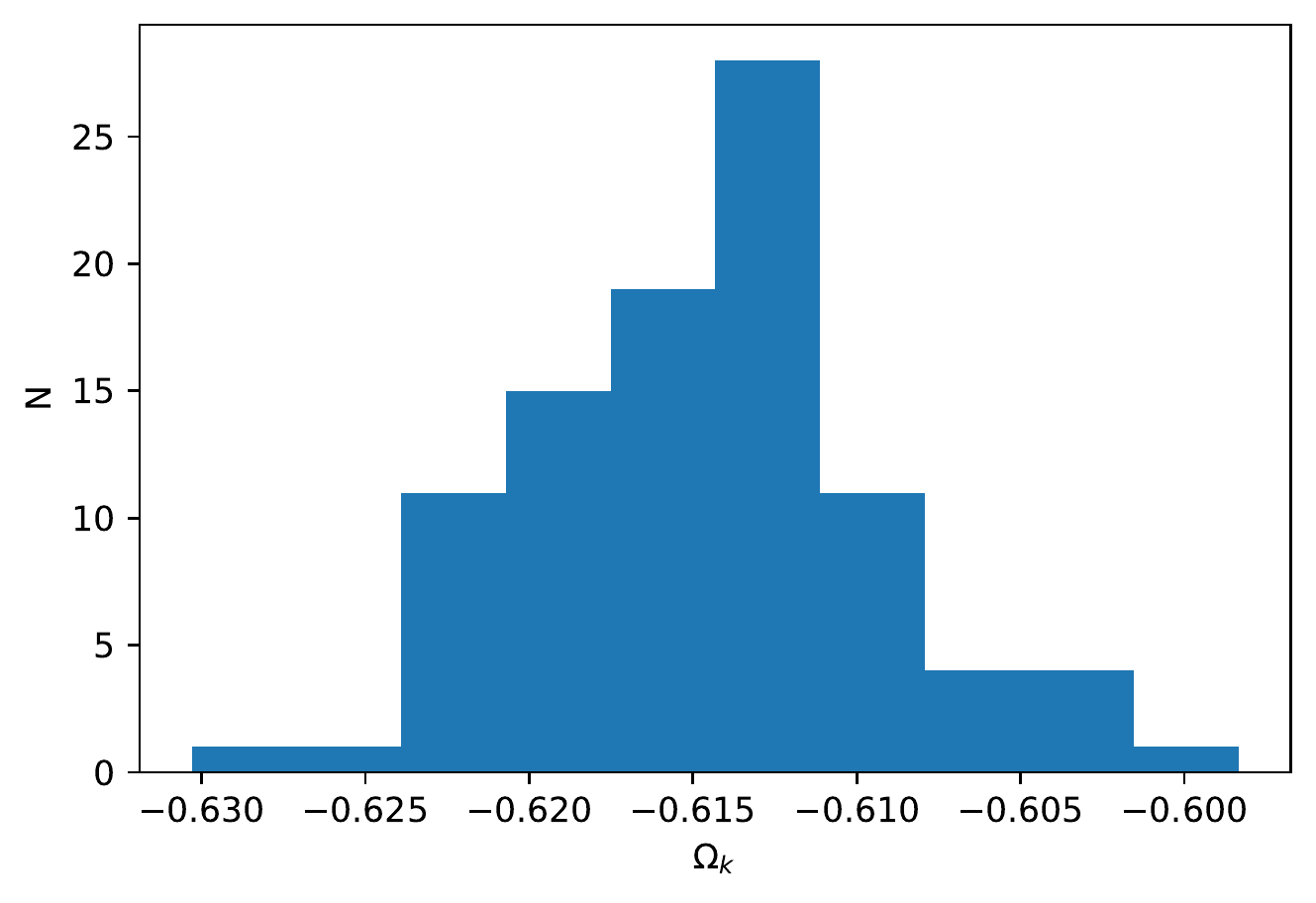}{0.3\textwidth}{(i) $\Omega_k$ with varying correction for redshift evolution}}
\caption{Histograms from 100 loops on QSO sample for no evolution, fixed evolution, and varying evolution and $\cal L_G$ likelihoods in the non-flat $\Lambda$CDM model.}
\label{fig: nonflatGaussloops}
\end{figure}

\begin{figure}
\centering
\gridline{
\fig{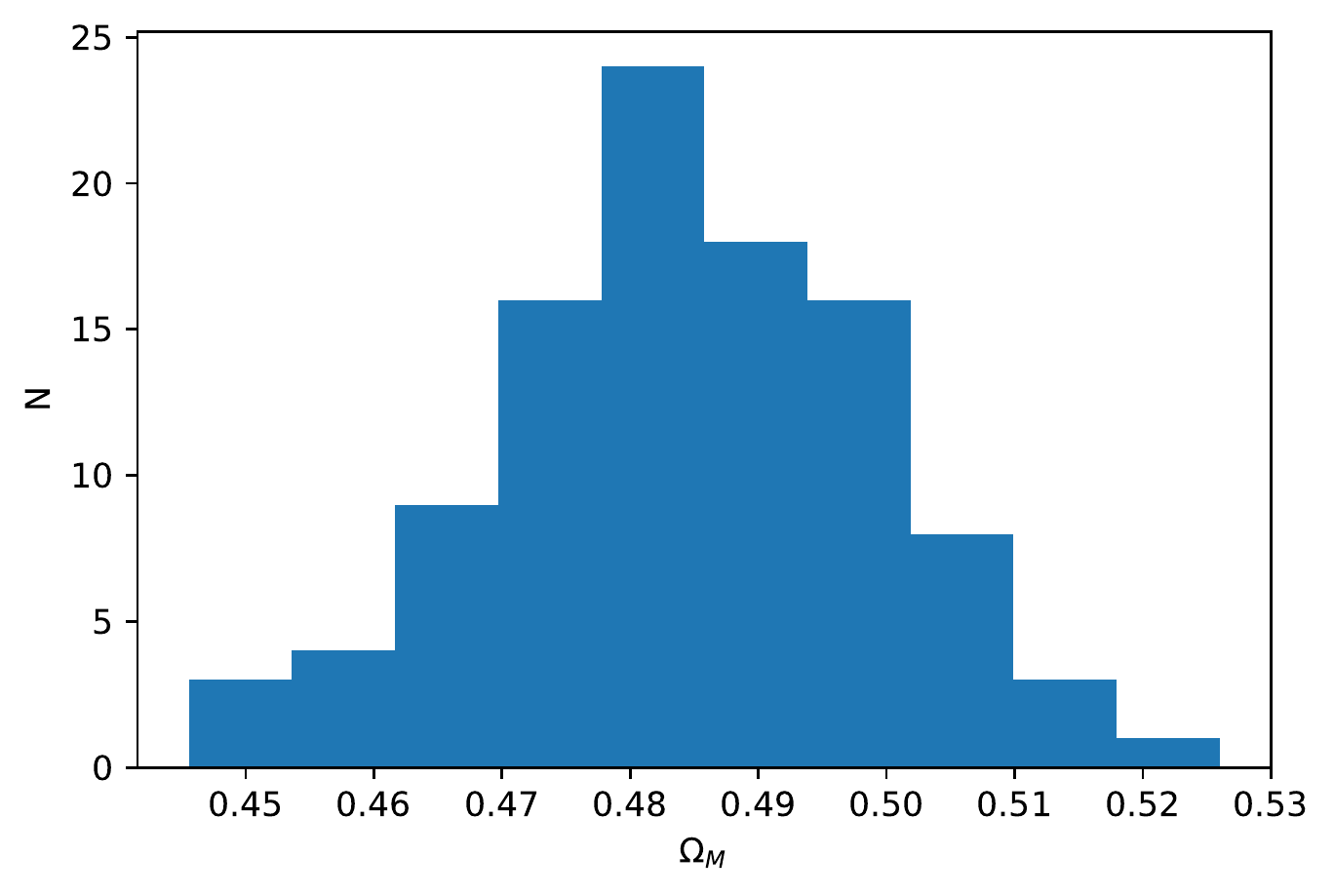}{0.3\textwidth}{(a) $\Omega_M$ without correction for redshift evolution}
\fig{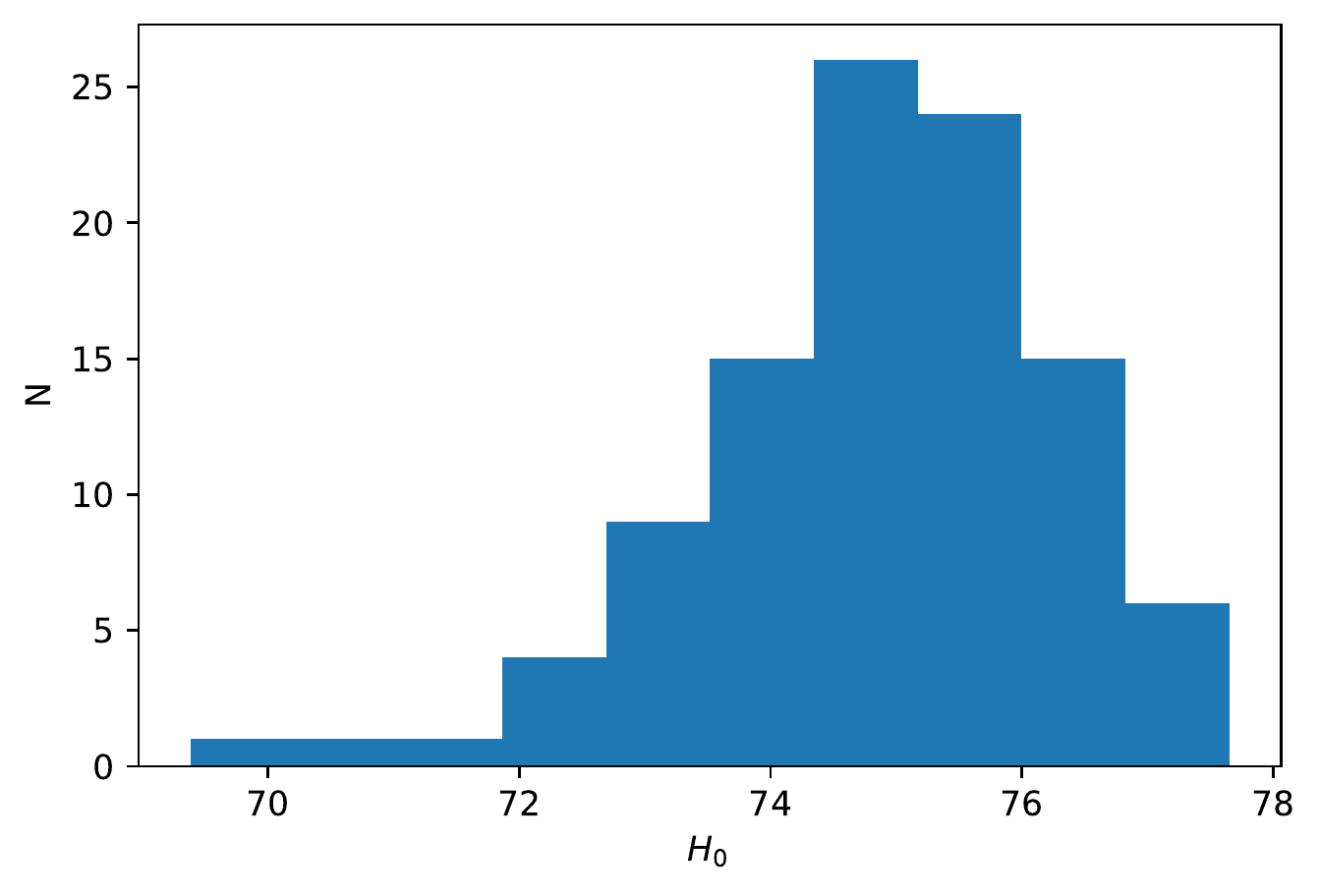}{0.3\textwidth}{(b) $H_0$ without correction for redshift evolution}\label{}
\fig{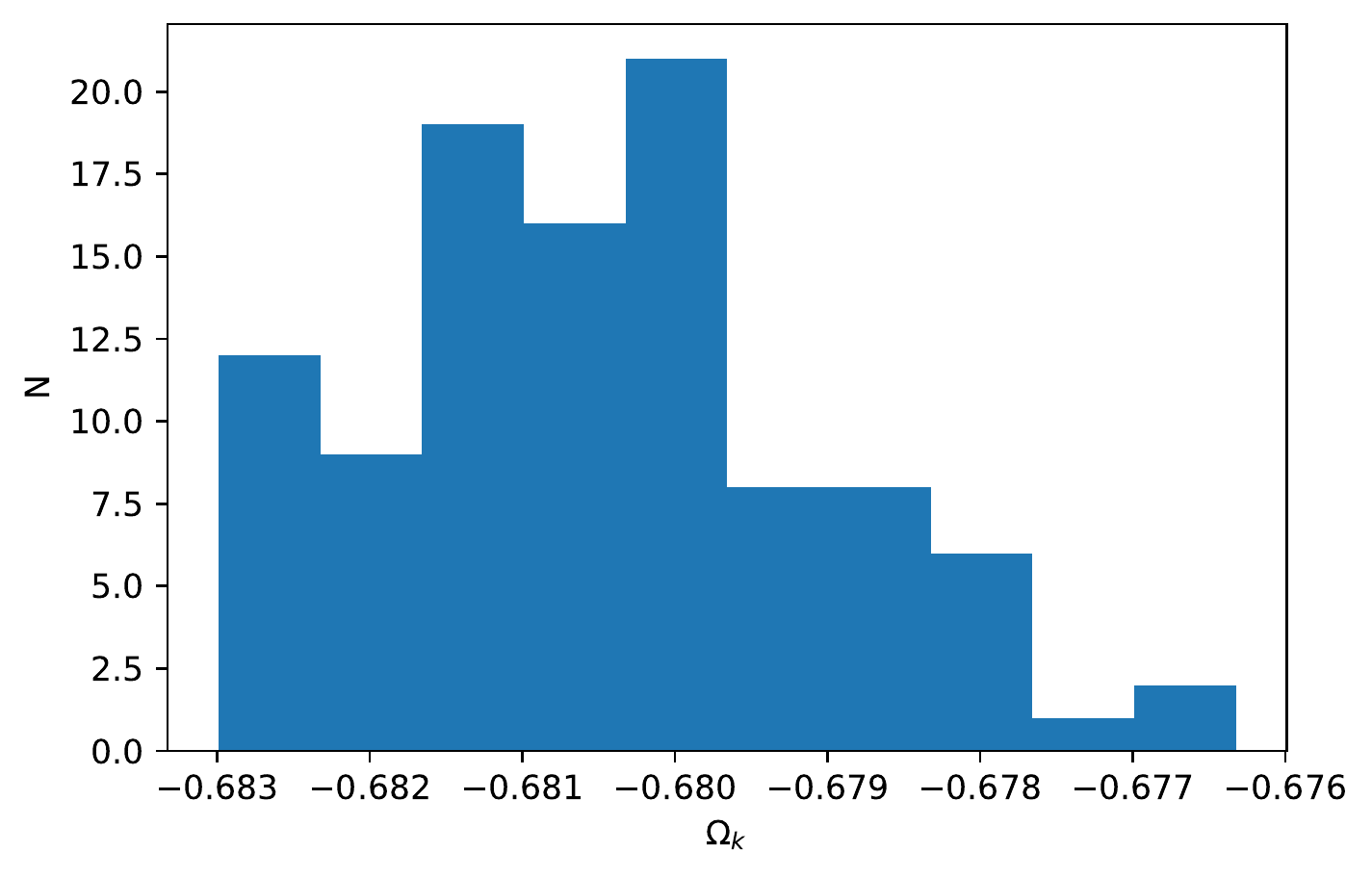}{0.3\textwidth}{(c) $\Omega_k$ without correction for redshift evolution}\label{}}
\gridline{
\fig{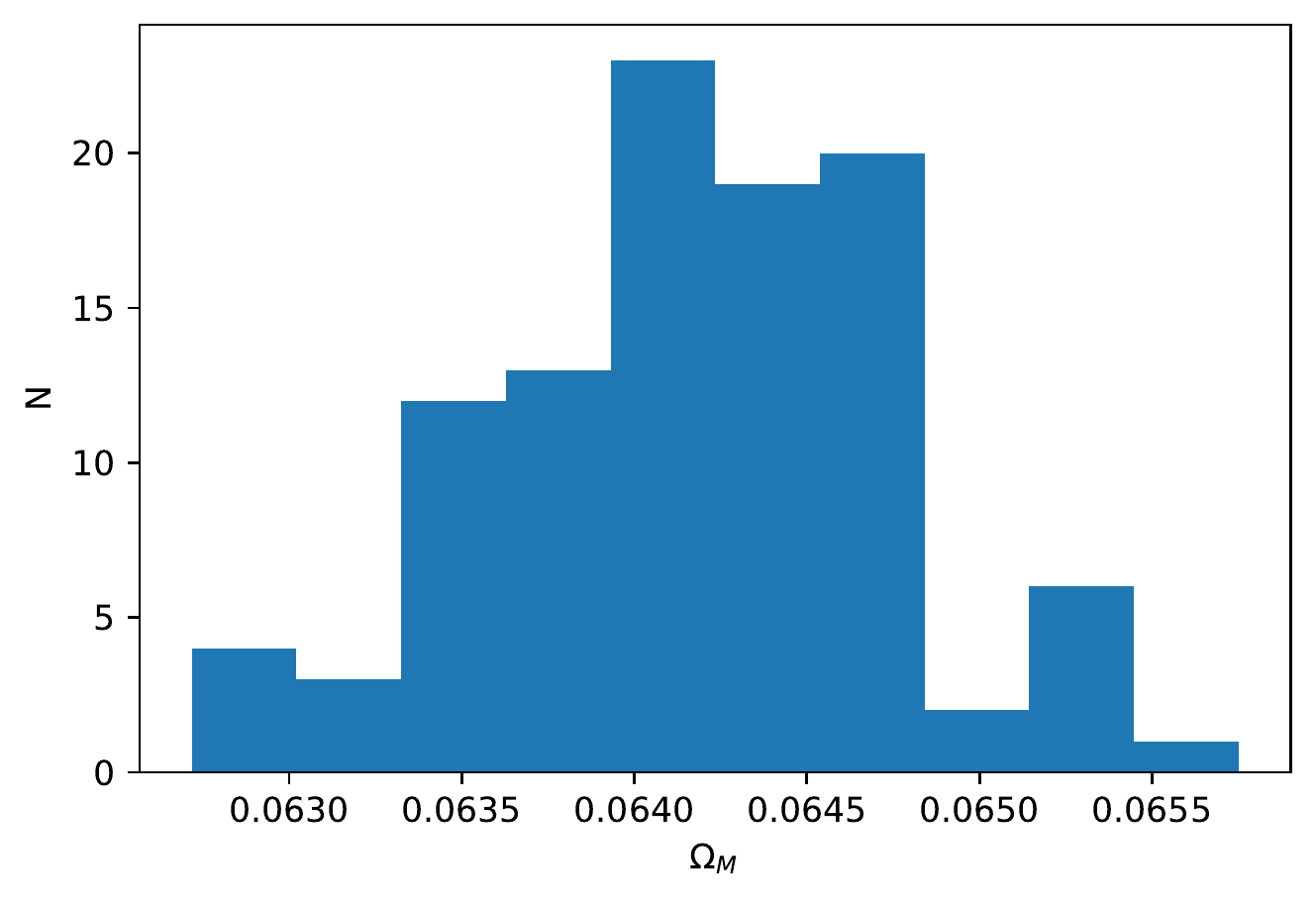}{0.3\textwidth}{(d) $\Omega_M$ with fixed correction for redshift evolution}\label{}
\fig{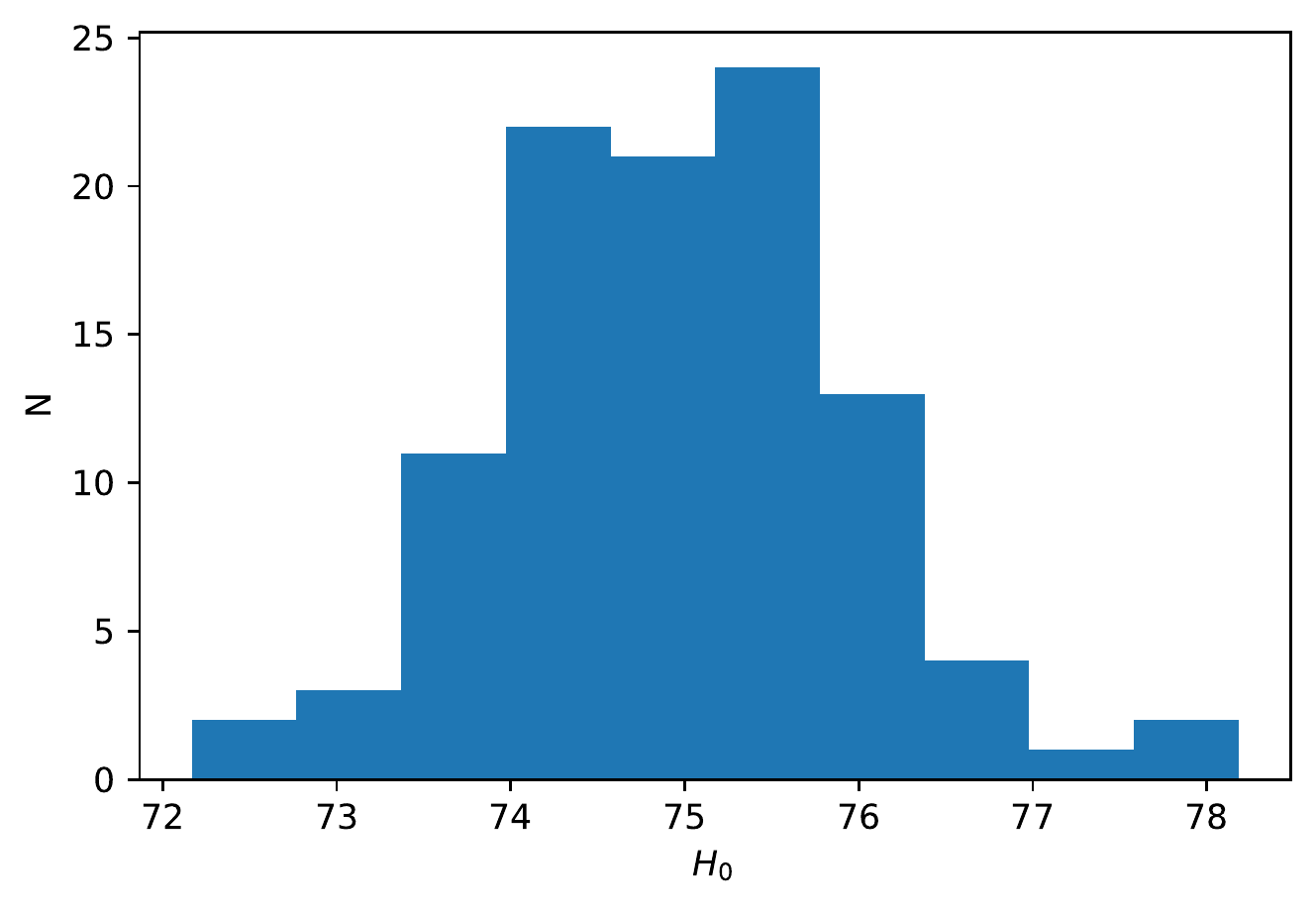}{0.3\textwidth}{(e) $H_0$ with fixed correction for redshift evolution}\label{}
\fig{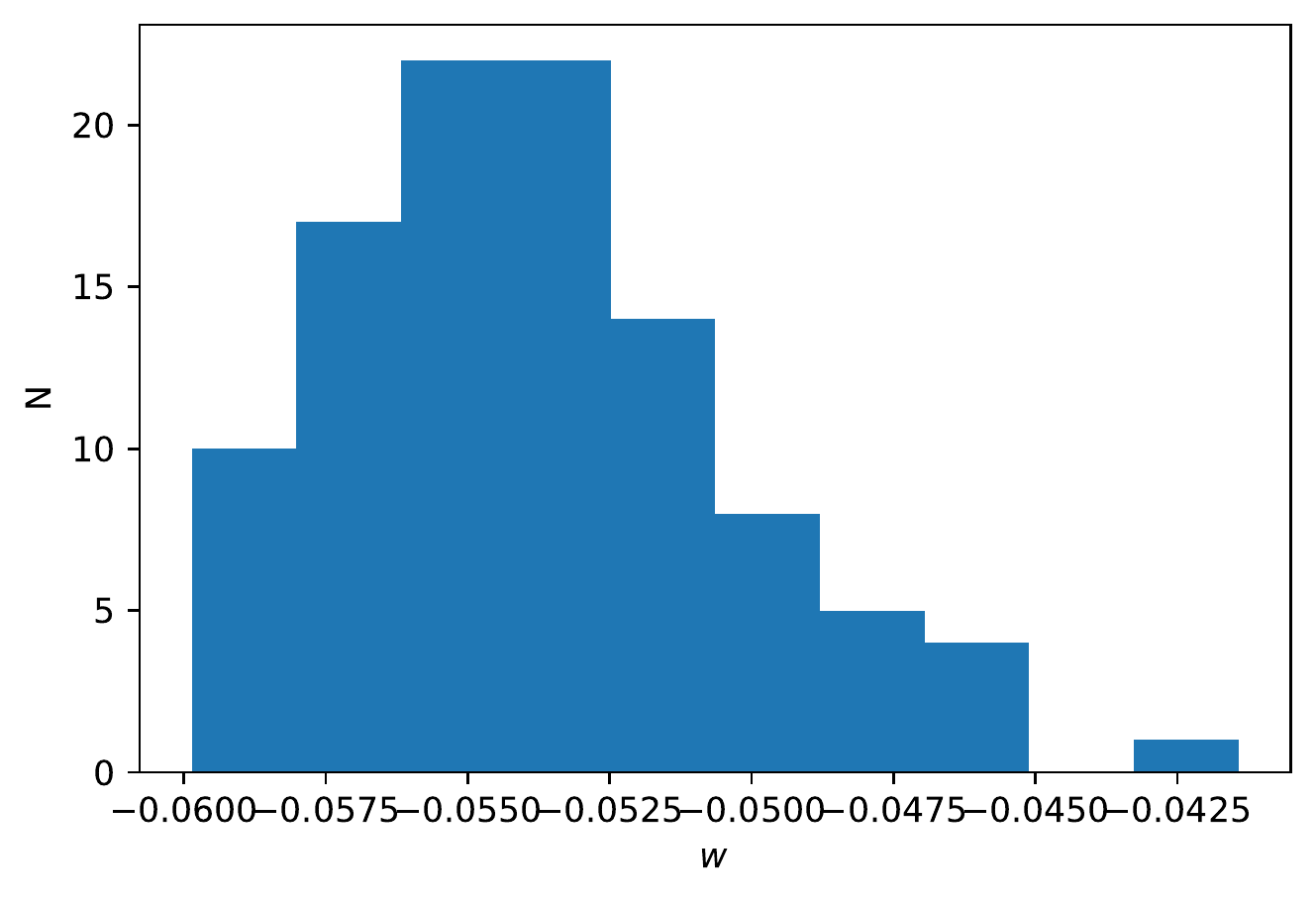}{0.3\textwidth}{(f) $\Omega_k$ with fixed correction for redshift evolution}\label{}}
\gridline{
\fig{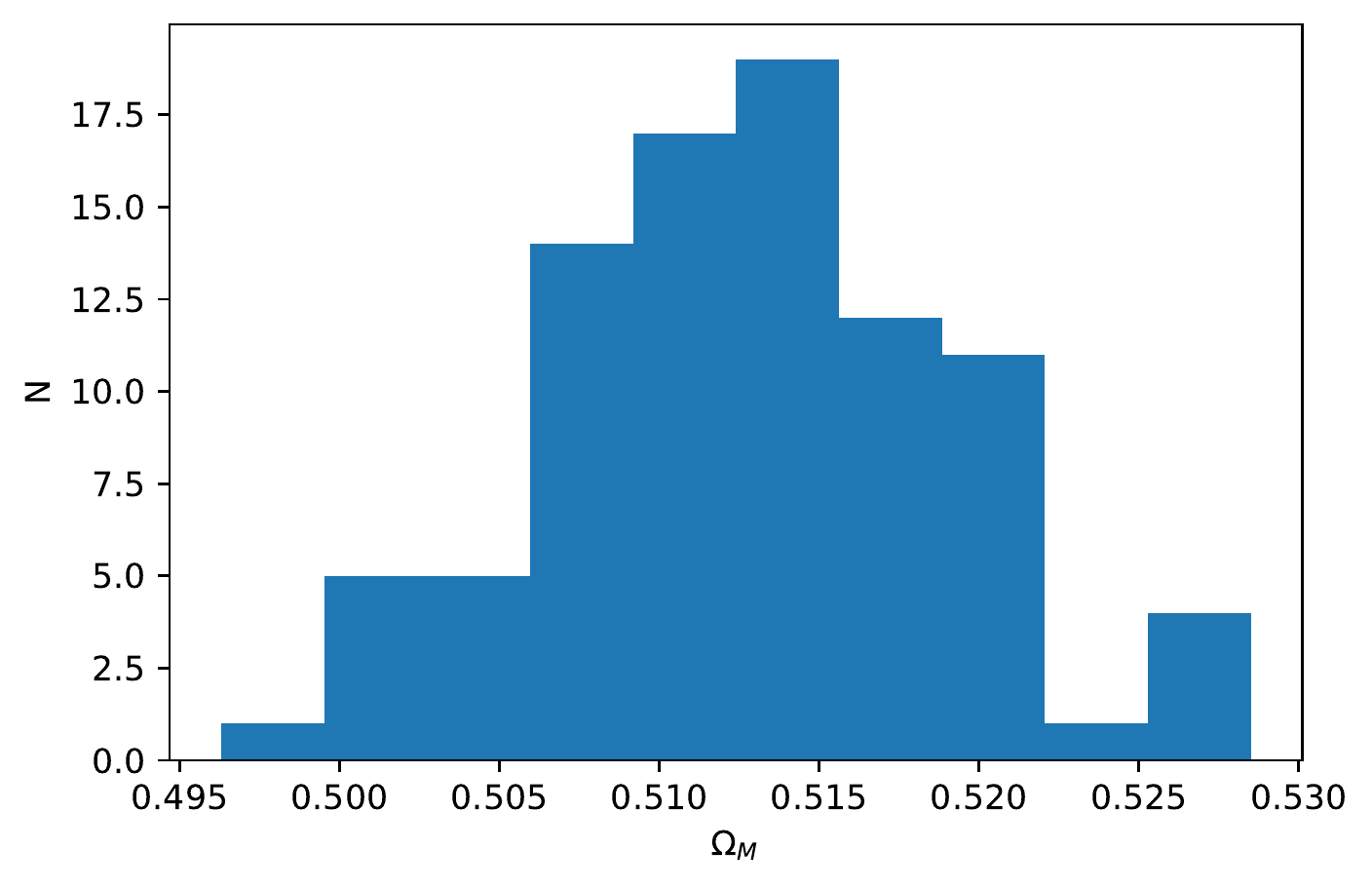}{0.3\textwidth}{(g) $\Omega_M$ with varying correction for redshift evolution}\label{}
\fig{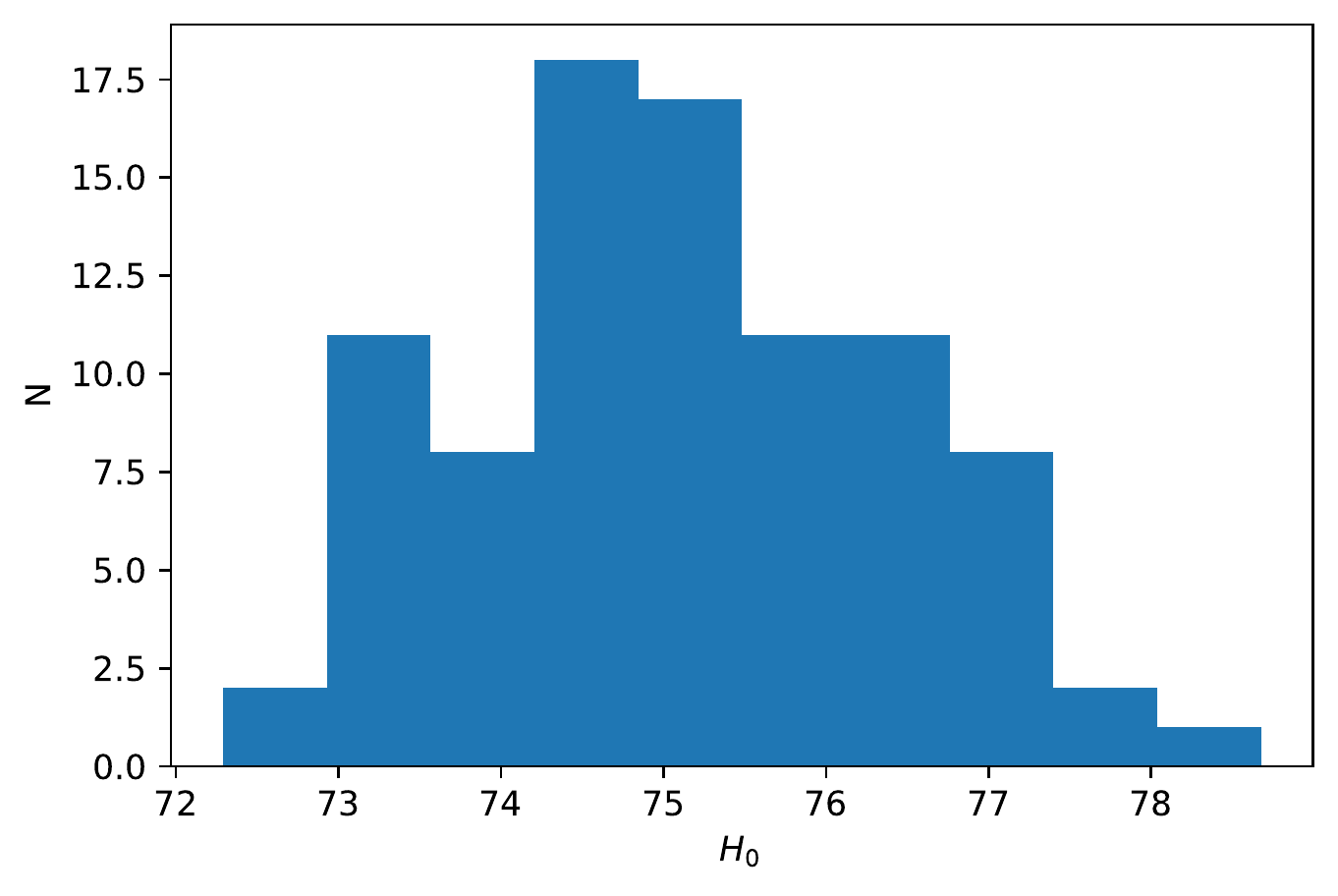}{0.3\textwidth}{(i) $H_0$ with varying correction for redshift evolution}\label{}
\fig{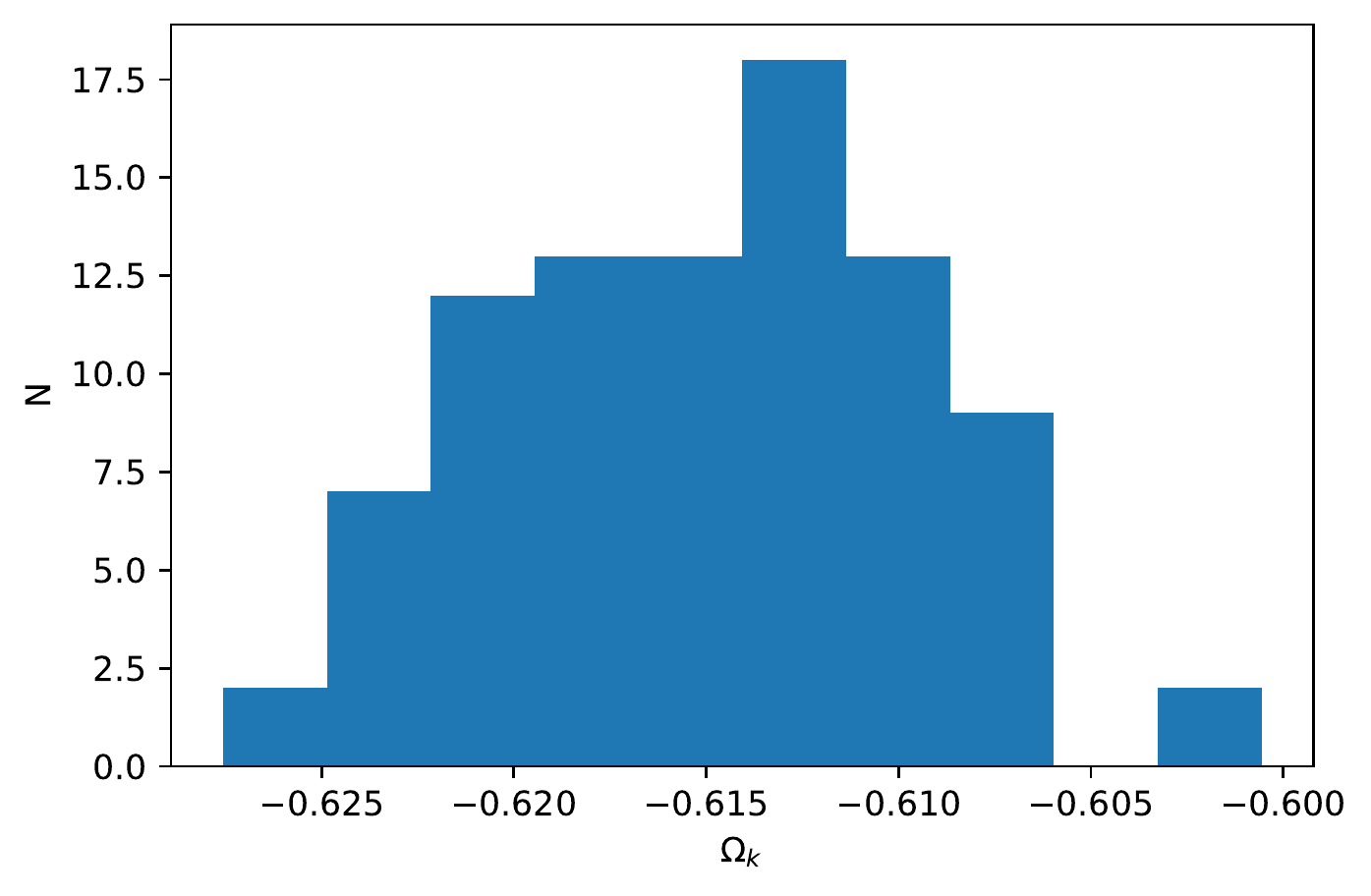}{0.3\textwidth}{(h) $\Omega_k$ with varying correction for redshift evolution}\label{}}
\caption{Histograms from 100 loops on QSO sample for no evolution, fixed evolution,and varying evolution and new $\cal L_N$ likelihoods in the non-flat $\Lambda$CDM model.}
\label{fig: nonflatNewloops}
\end{figure}

\begin{figure}
\centering
\gridline{
\fig{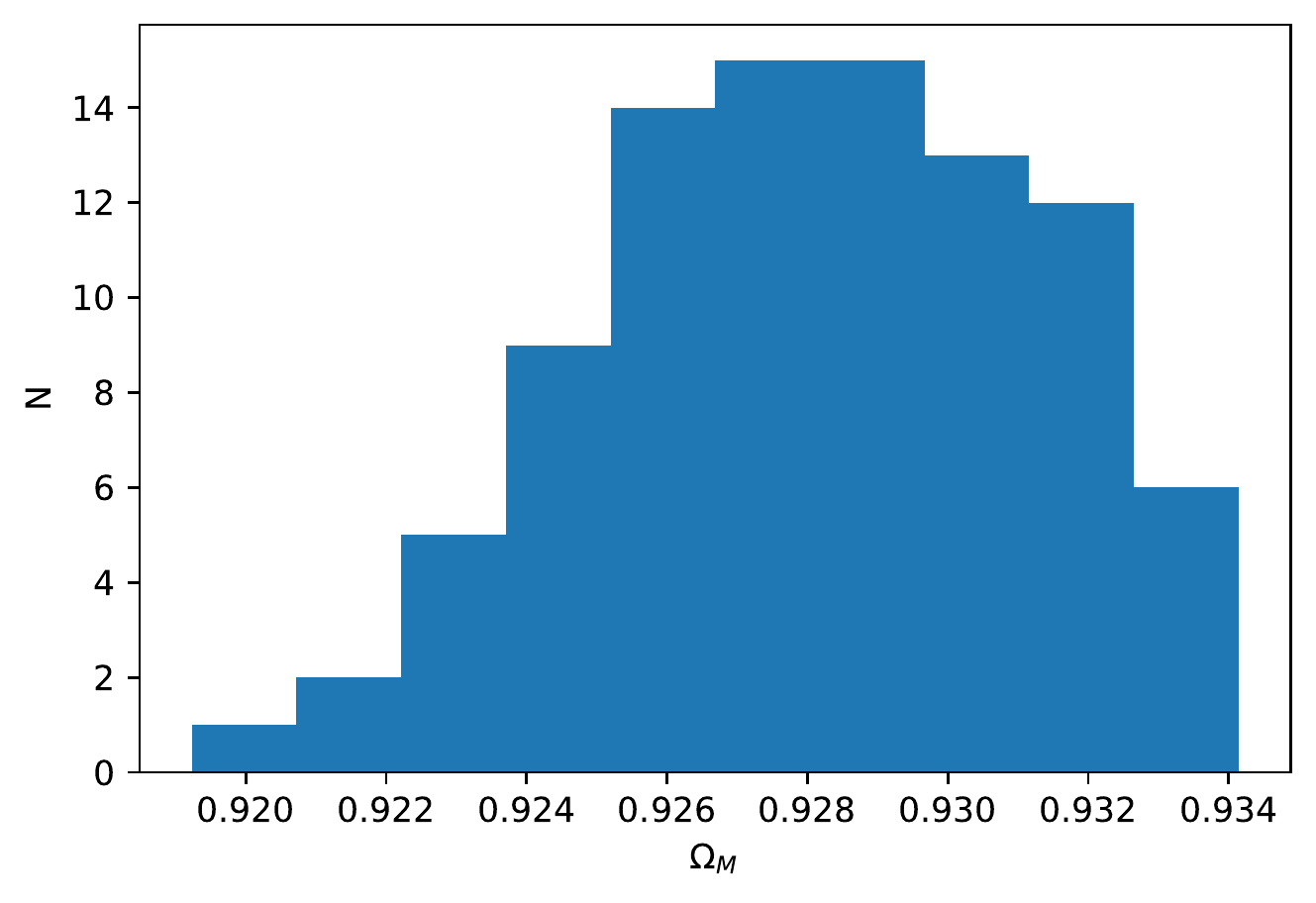}{0.3\textwidth}{(a) $\Omega_M$ without correction for redshift evolution}\label{}
\fig{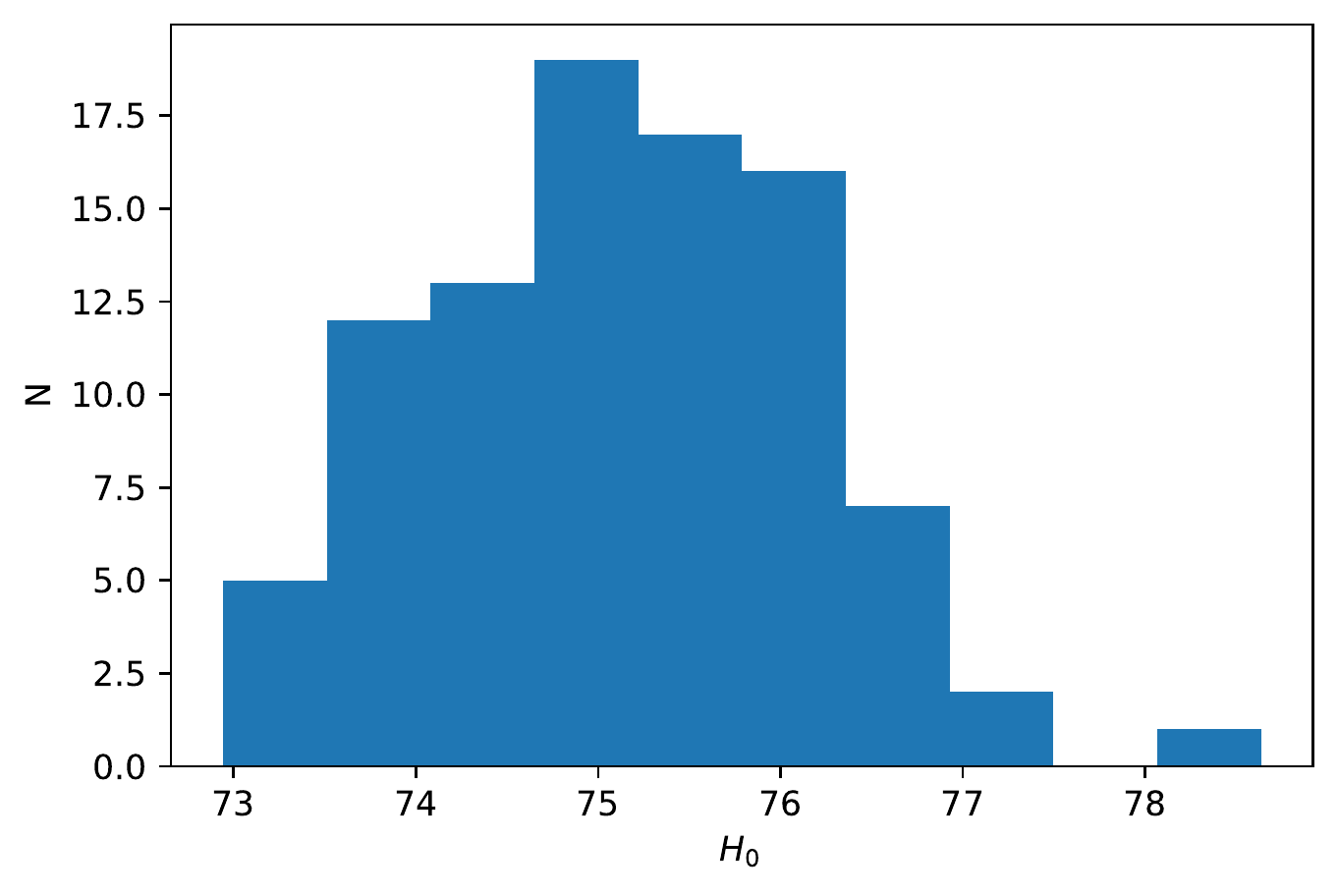}{0.3\textwidth}{(b) $H_0$ without correction for redshift evolution}\label{}
\fig{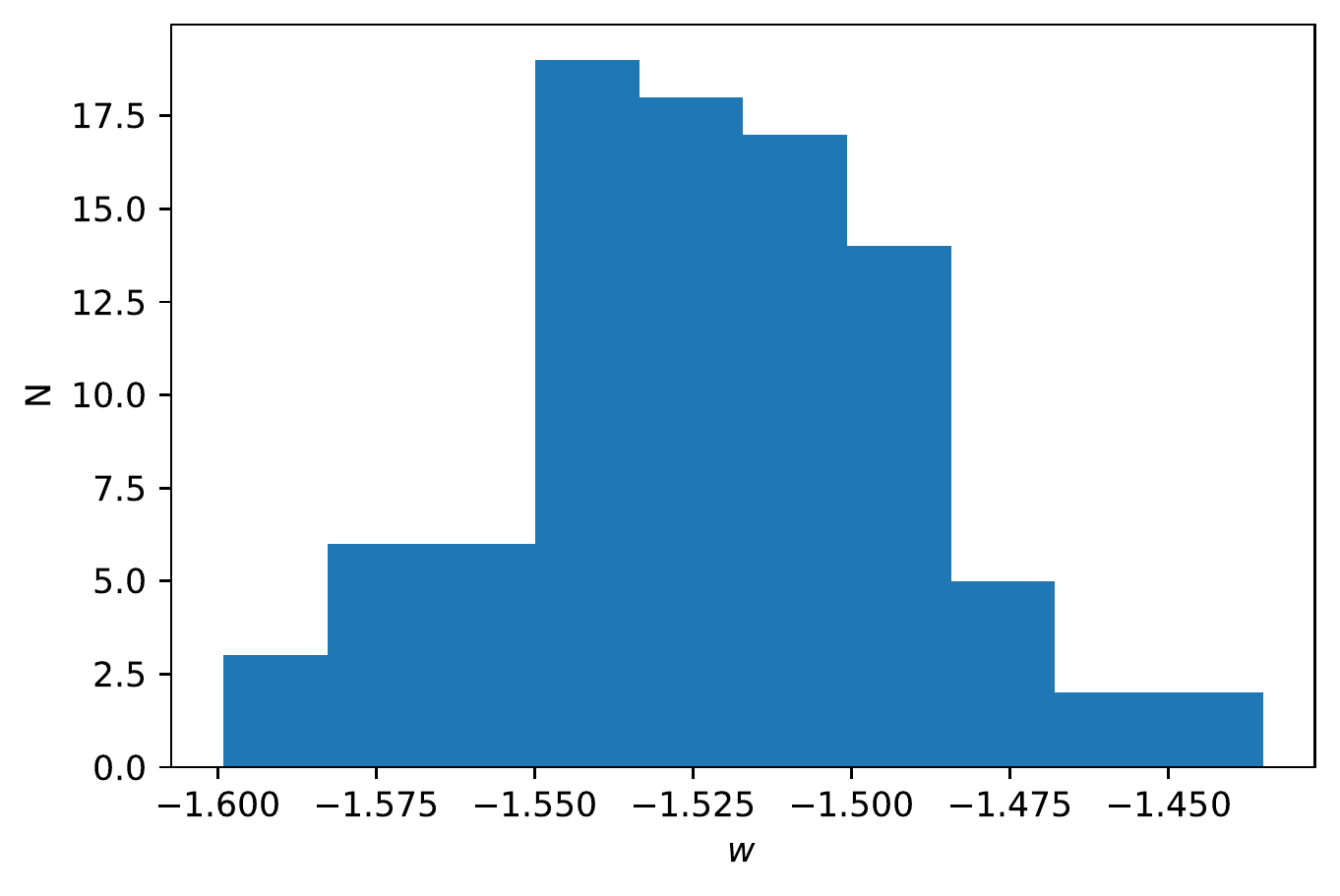}{0.3\textwidth}{(c) $w$ without correction for redshift \\evolution}\label{}}
\gridline{
\fig{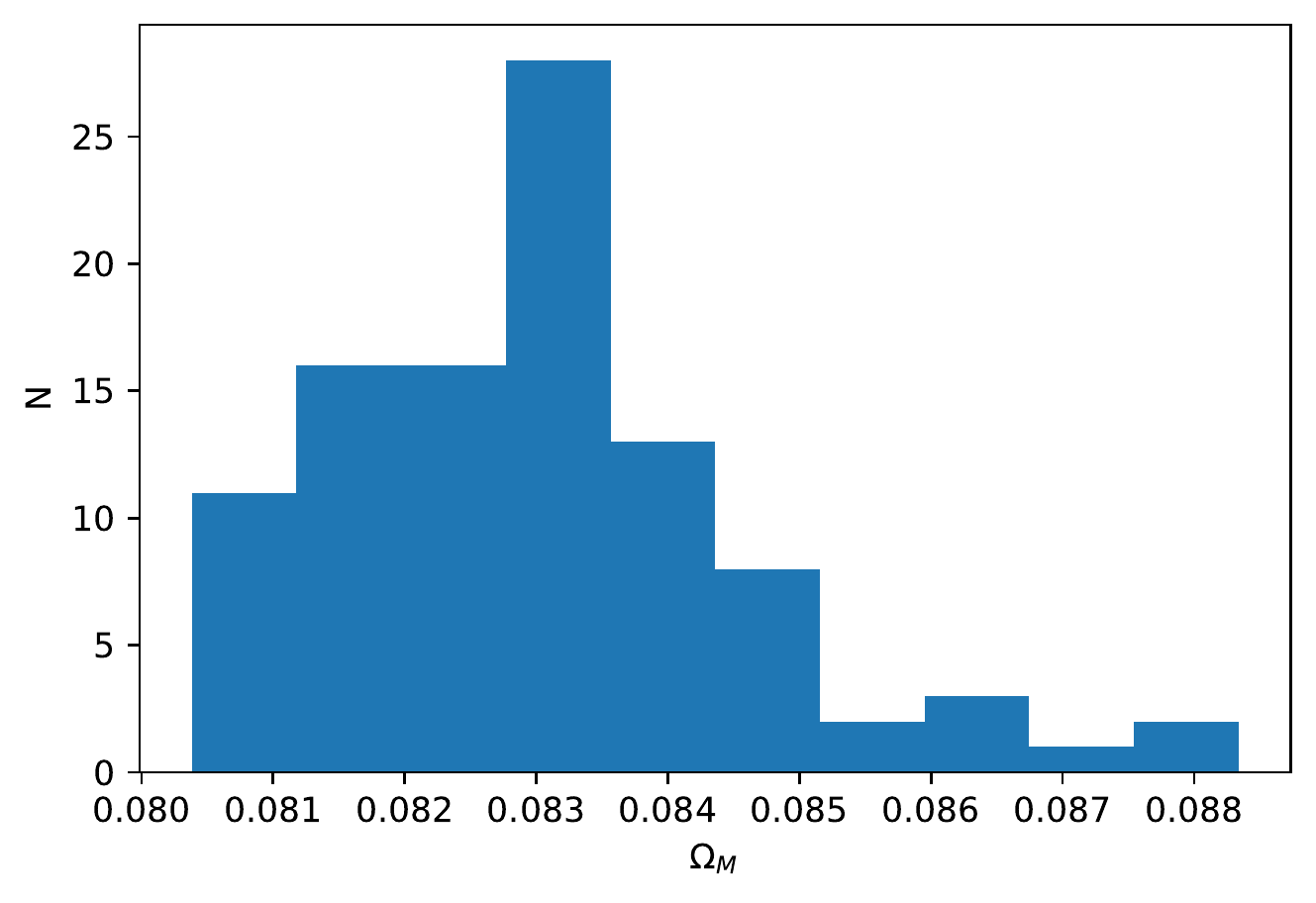}{0.3\textwidth}{(d) $\Omega_M$ with fixed correction for redshift evolution}\label{}
\fig{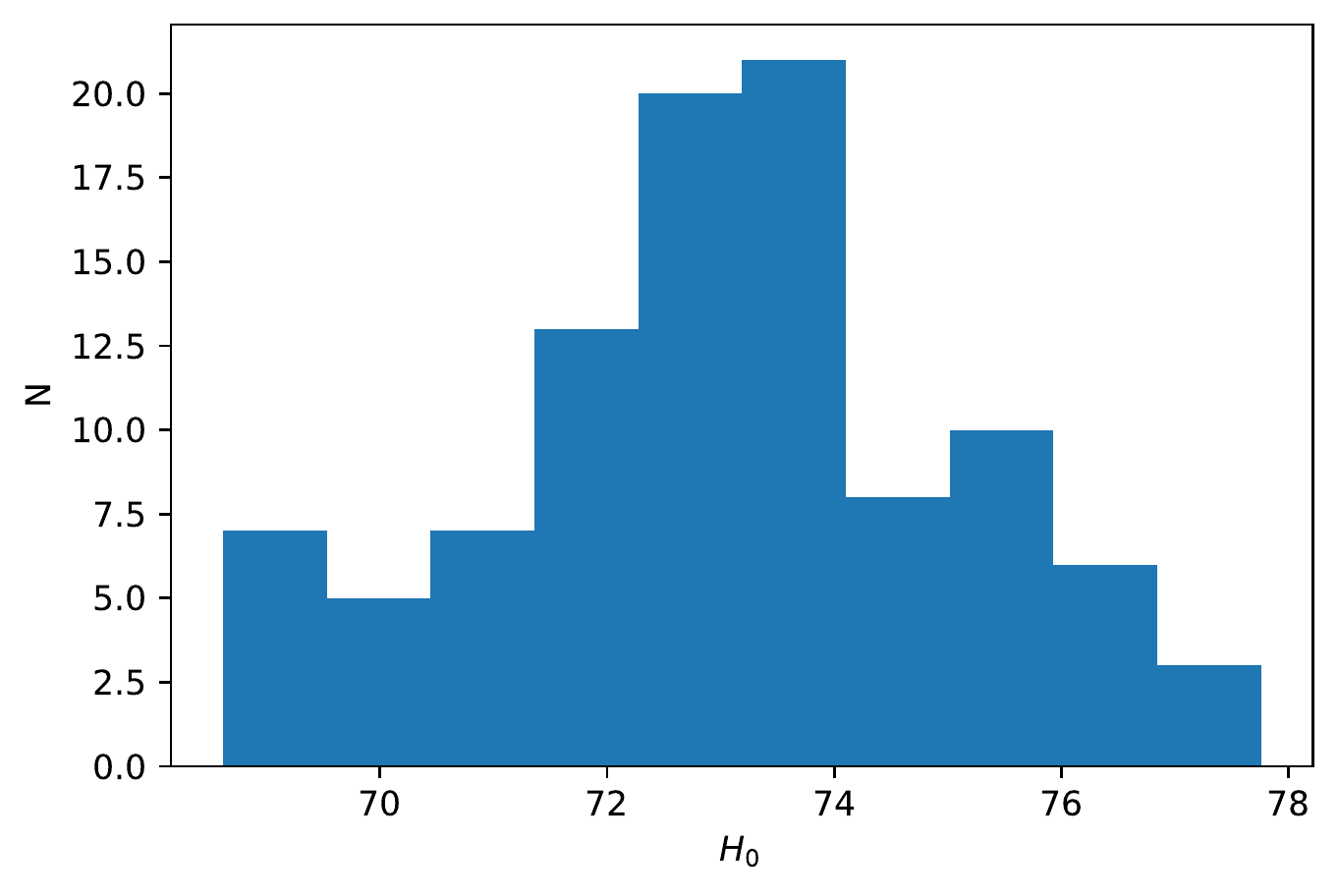}{0.3\textwidth}{(e) $H_0$ with fixed correction for redshift evolution}\label{}
\fig{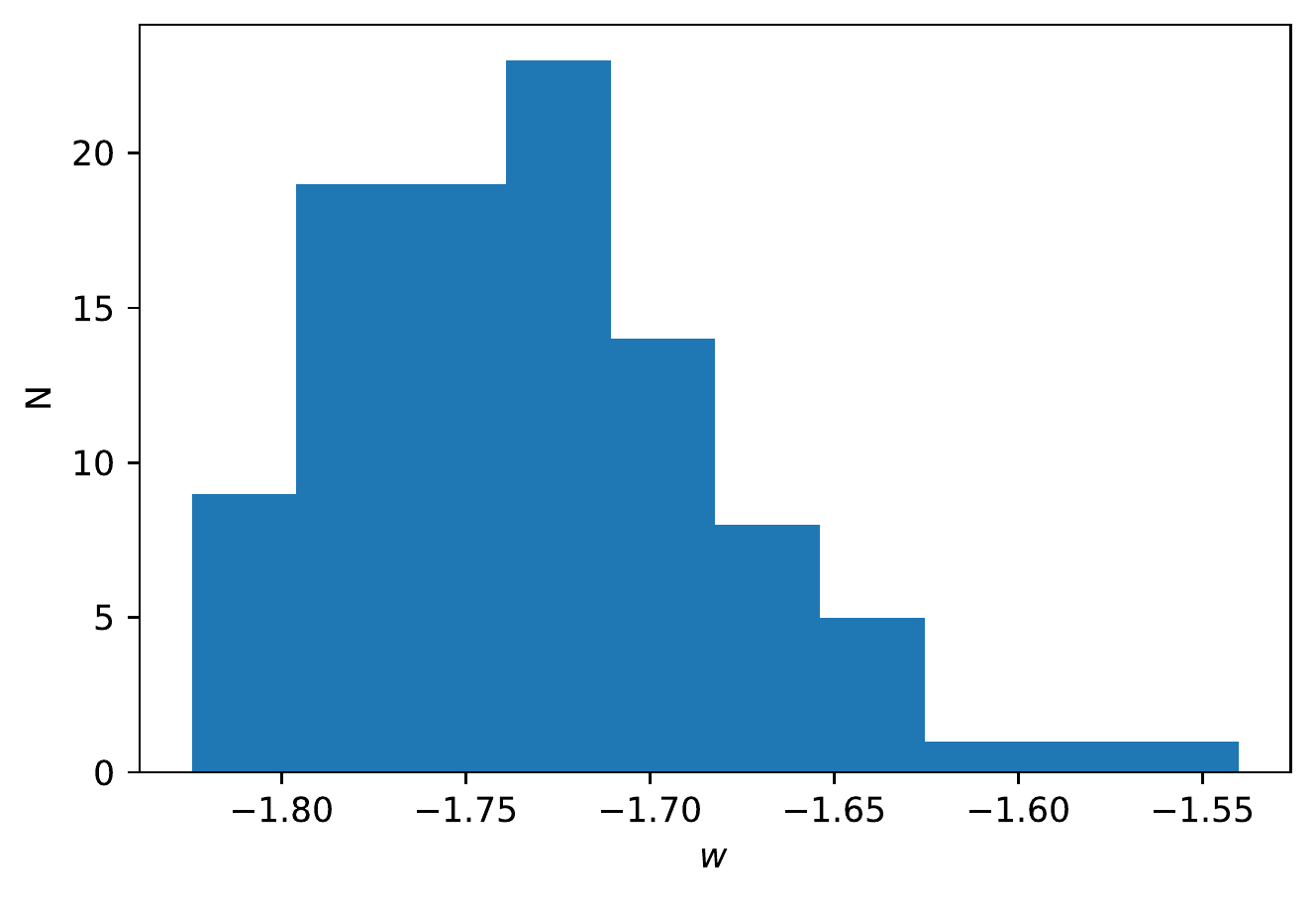}{0.3\textwidth}{(f) $w$ with fixed correction for redshift evolution}\label{}}
\gridline{
\fig{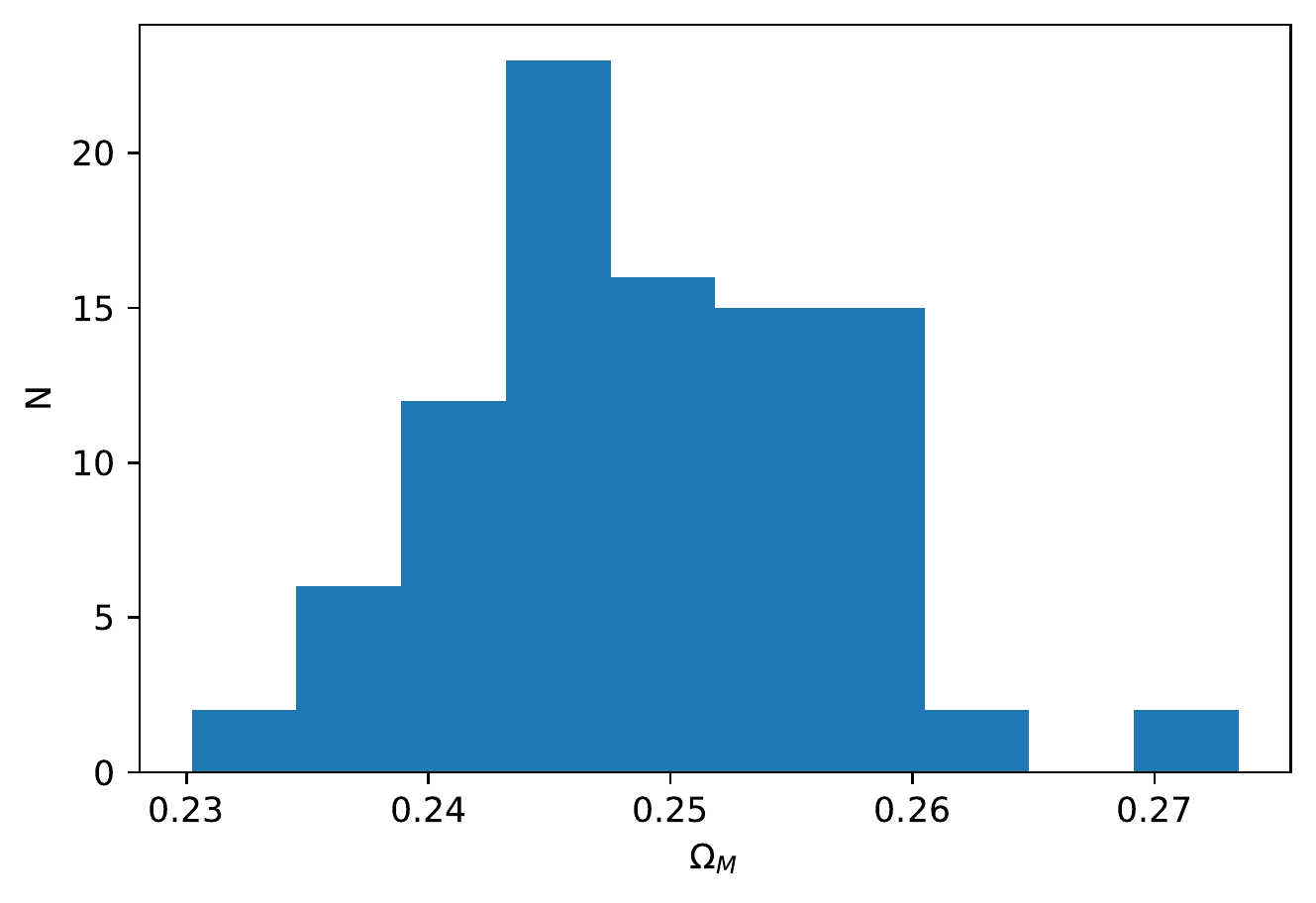}{0.3\textwidth}{(g) $\Omega_M$ with varying correction for redshift evolution}\label{}
\fig{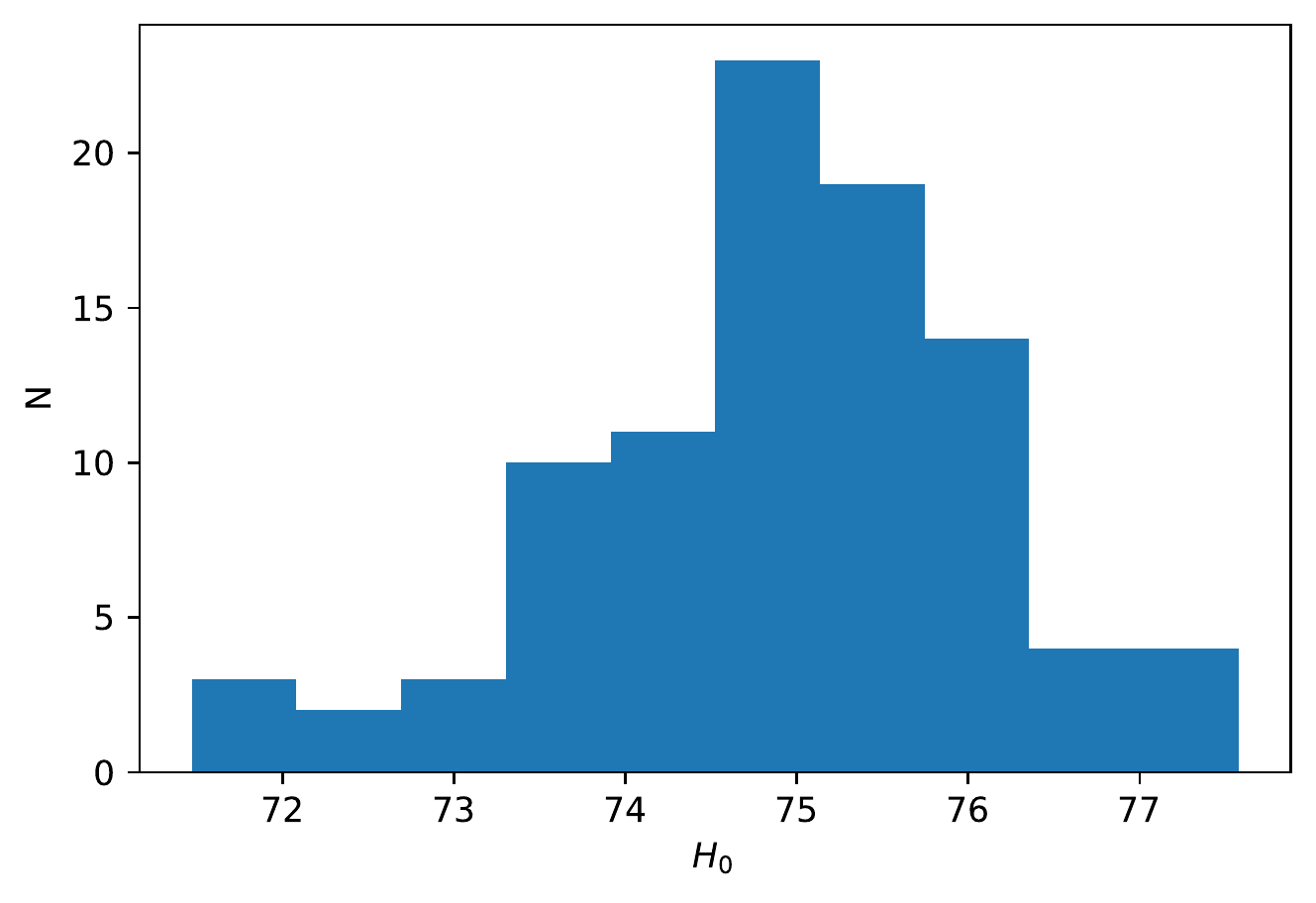}{0.3\textwidth}{(h) $H_0$ with varying correction for redshift evolution}\label{}
\fig{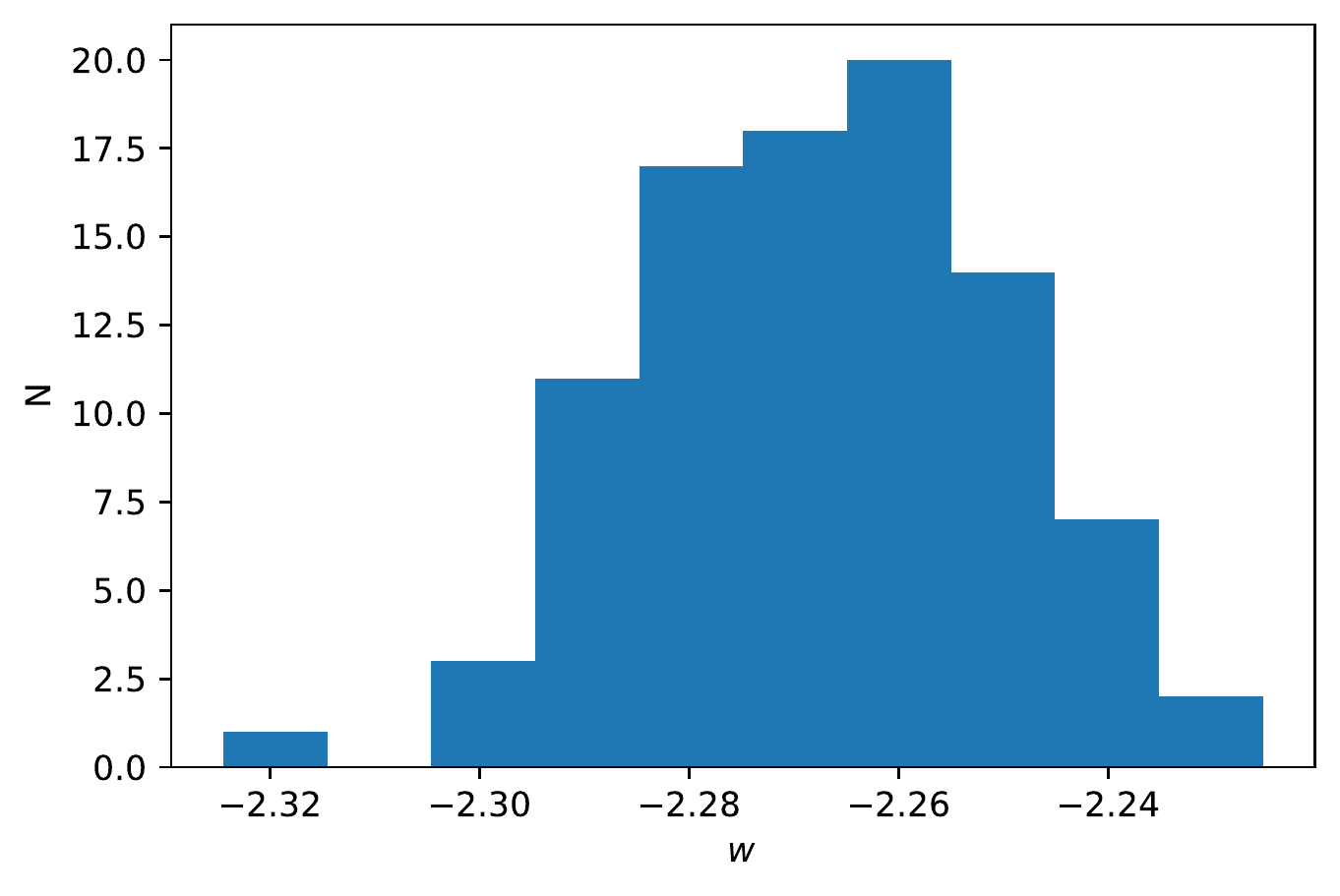}{0.3\textwidth}{(i) $w$ with varying correction for redshift evolution}\label{}}
\caption{Histograms from 100 loops on QSO sample for no evolution, fixed evolution,and varying evolution and $\cal L_G$ likelihoods in the flat $w$CDM model.}
\label{fig: wcdmGaussloops}
\end{figure}

\begin{figure}
\centering
\gridline{
\fig{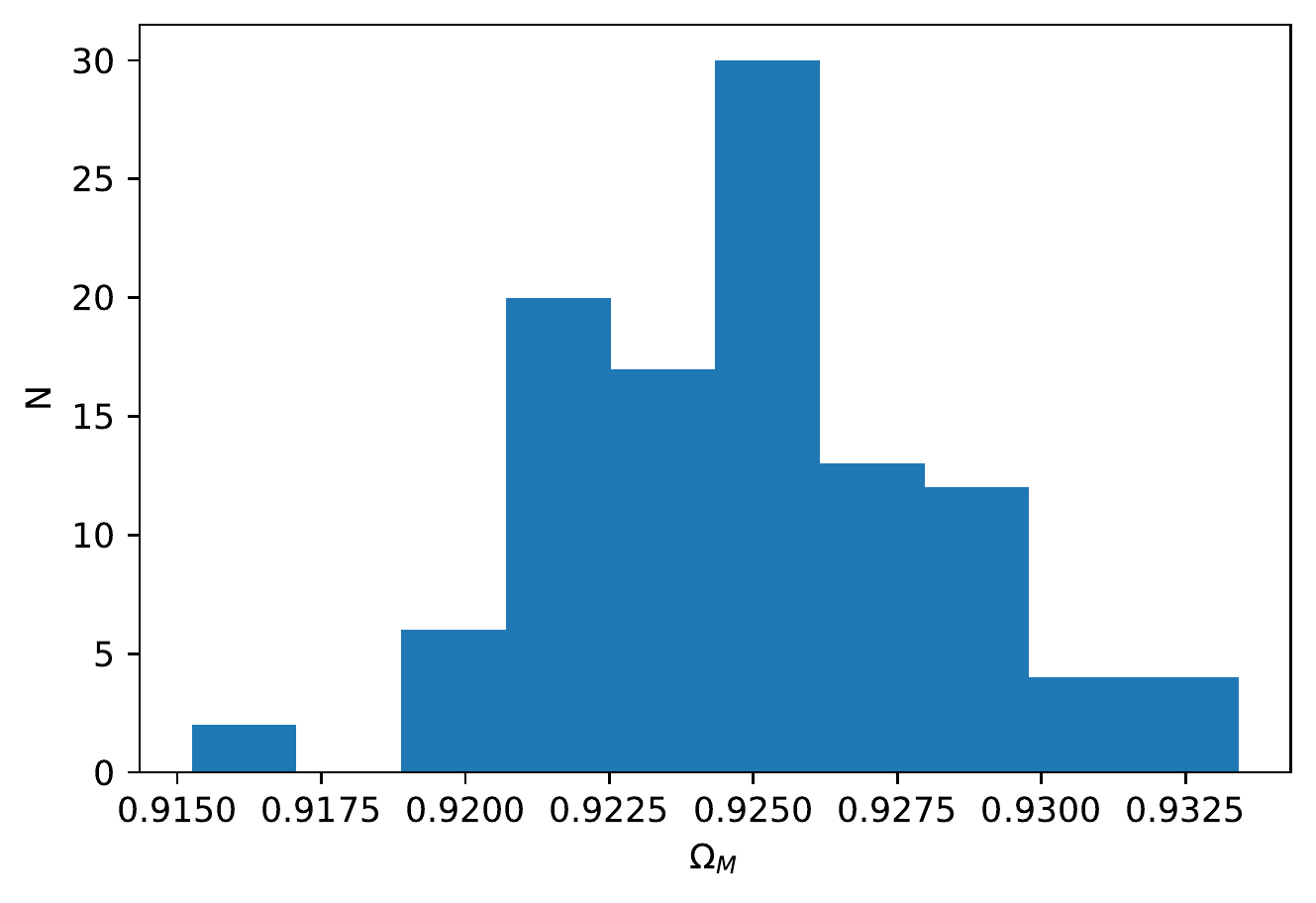}{0.3\textwidth}{(a) $\Omega_M$ without correction for redshift evolution}\label{}
\fig{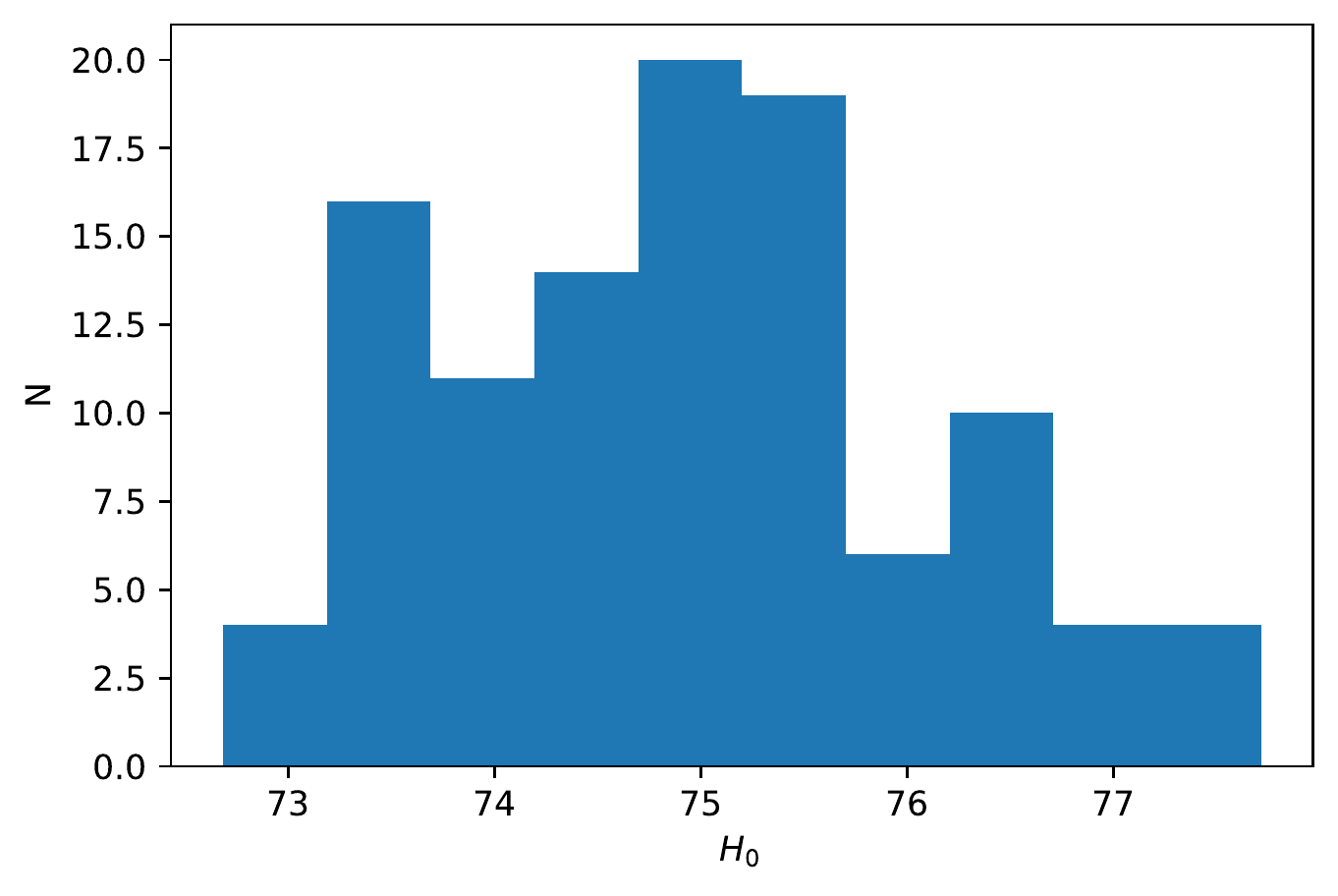}{0.3\textwidth}{(b) $H_0$ without correction for redshift evolution}\label{}
\fig{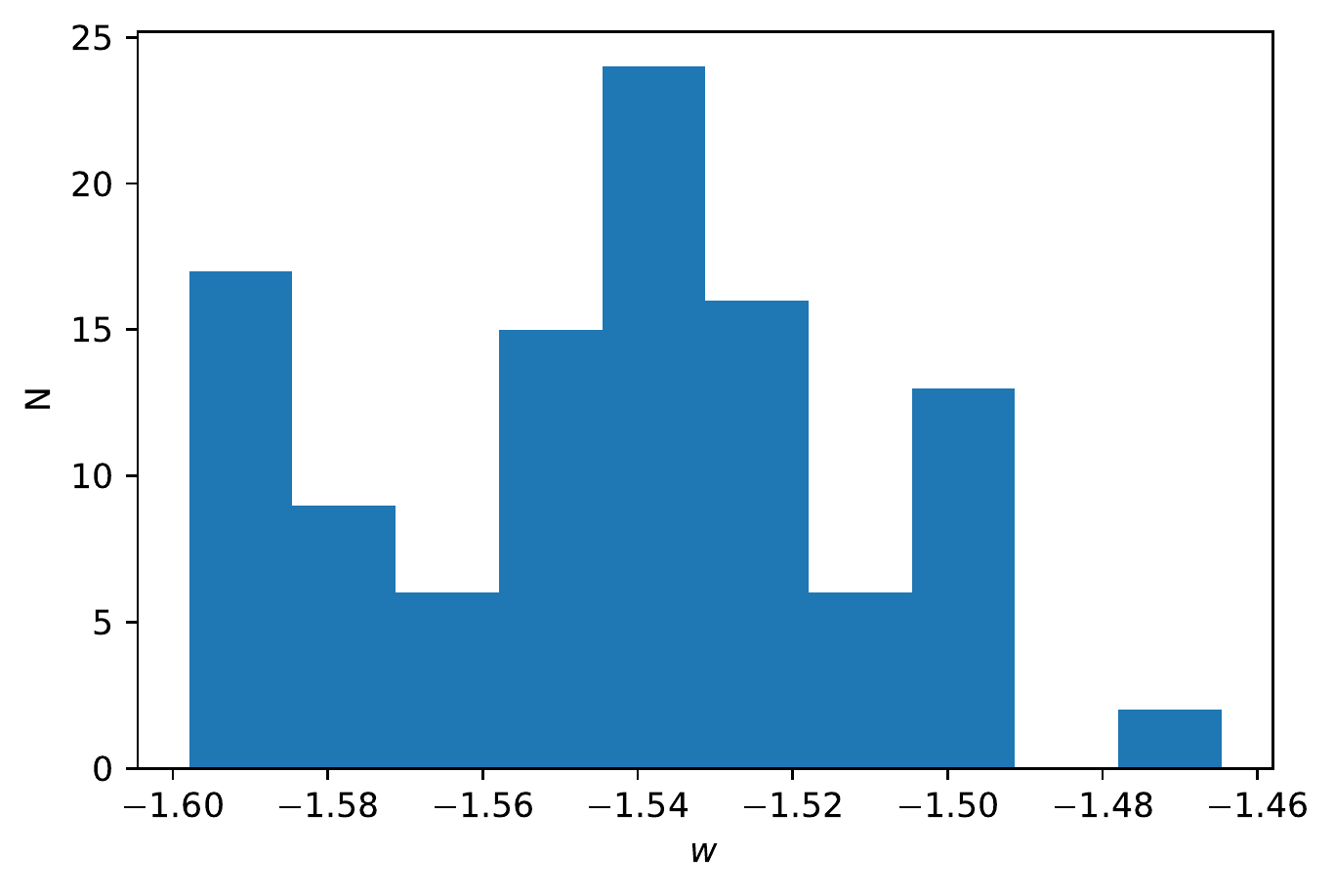}{0.3\textwidth}{(c) $w$ without correction for redshift evolution}\label{}}
\gridline{
\fig{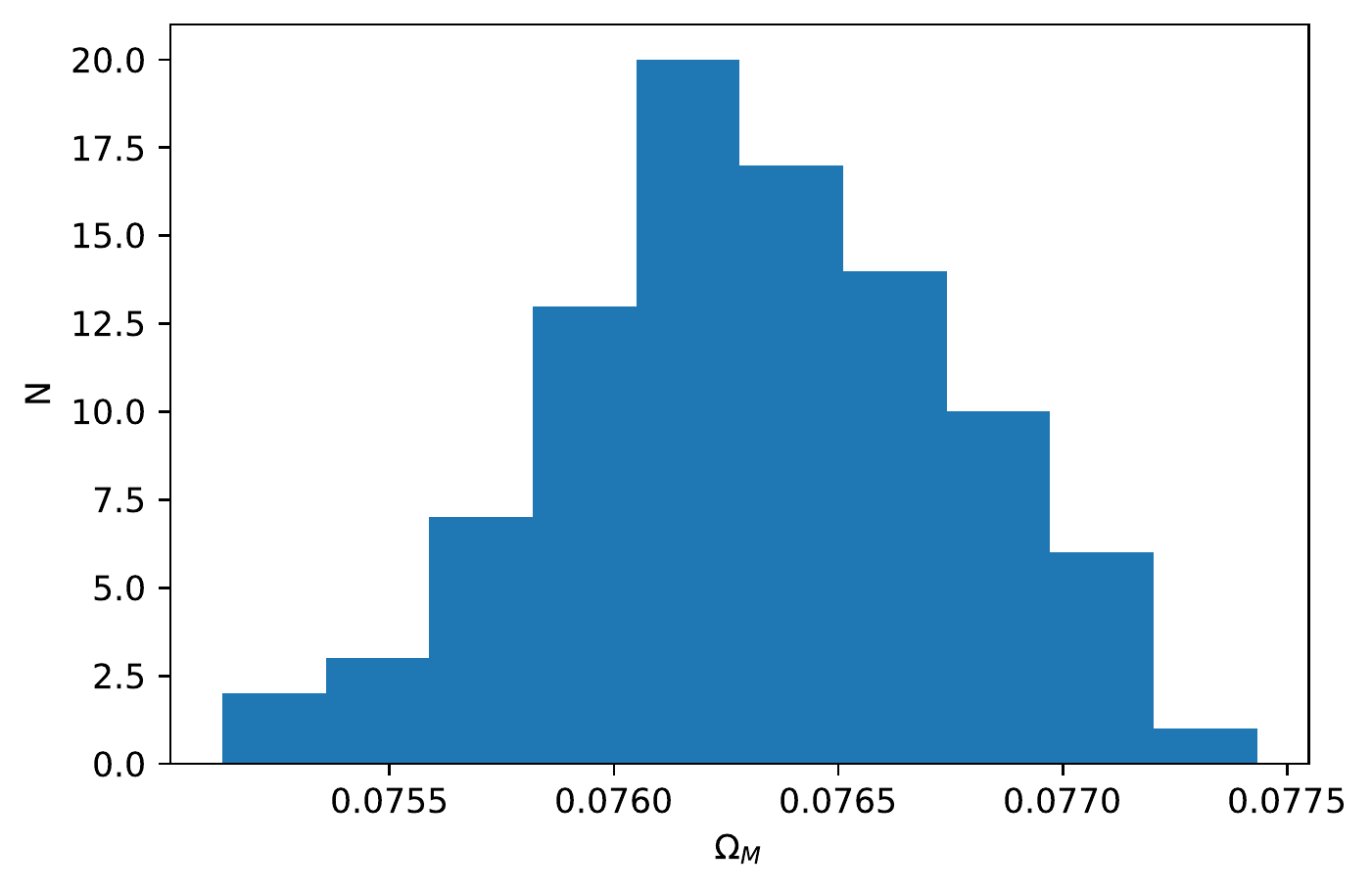}{0.3\textwidth}{(d) $\Omega_M$ with fixed correction for redshift evolution}\label{}
\fig{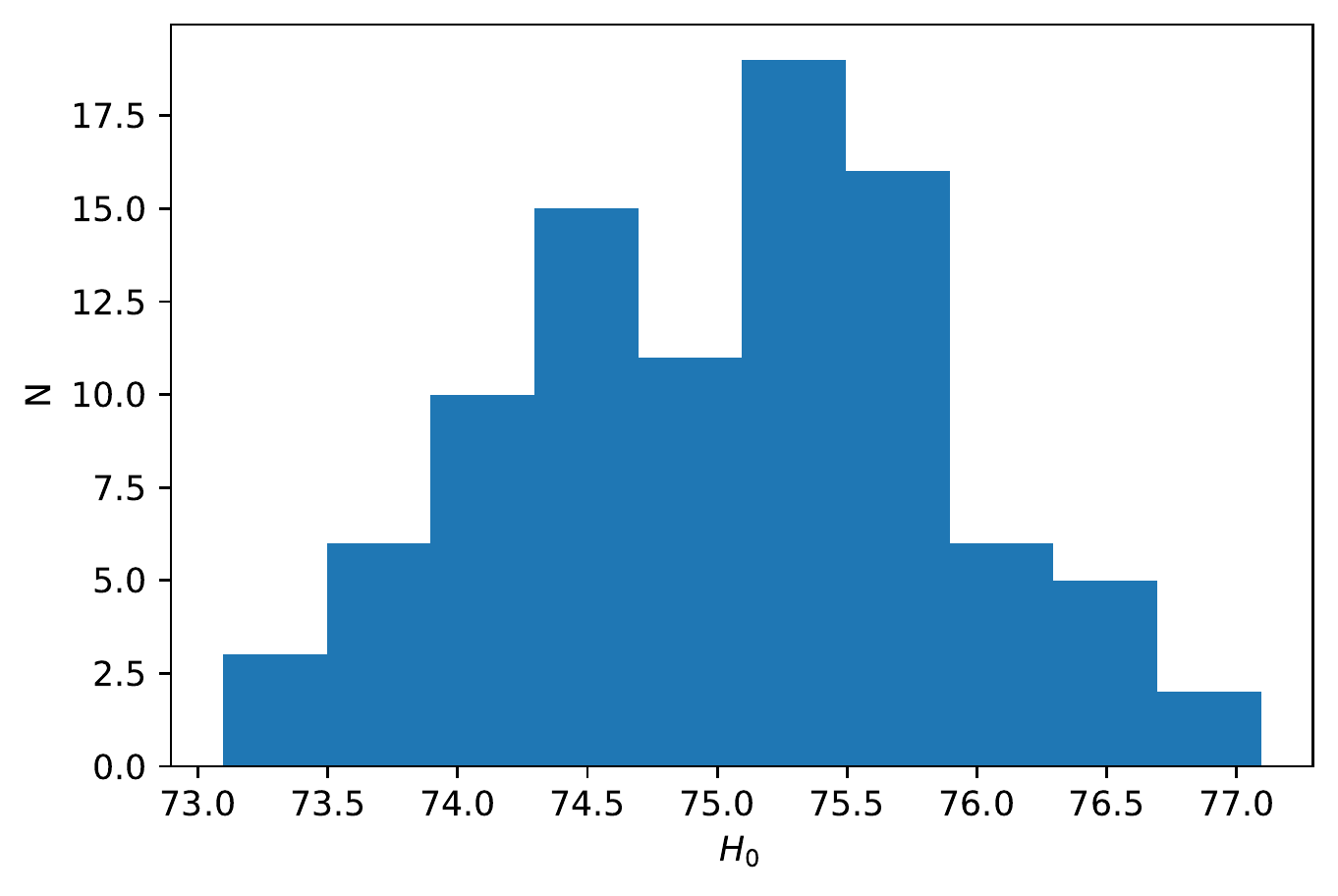}{0.3\textwidth}{(e) $H_0$ with fixed correction for redshift evolution}\label{}
\fig{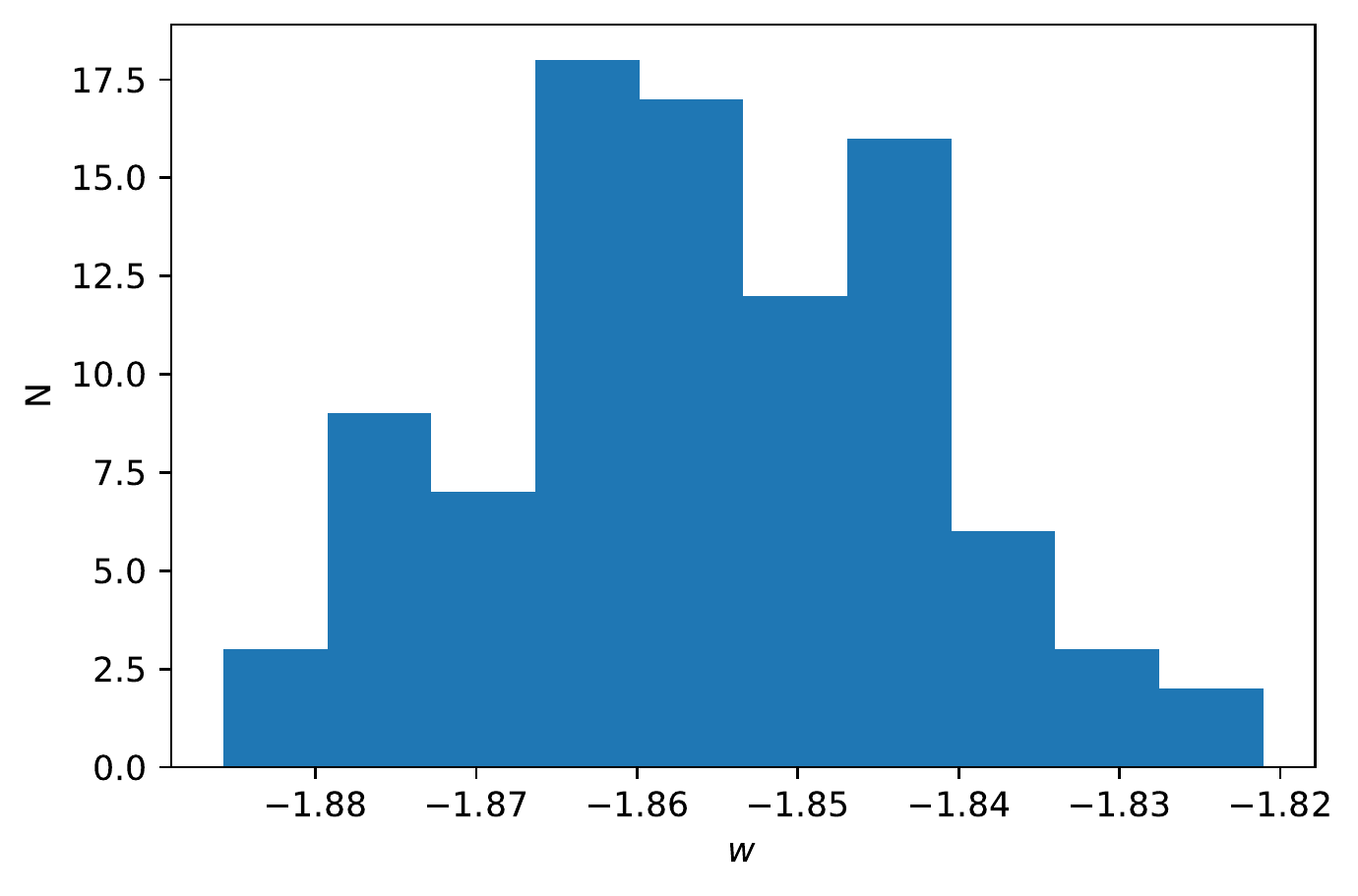}{0.3\textwidth}{(f) $w$ with fixed correction for redshift evolution}\label{}}
\gridline{
\fig{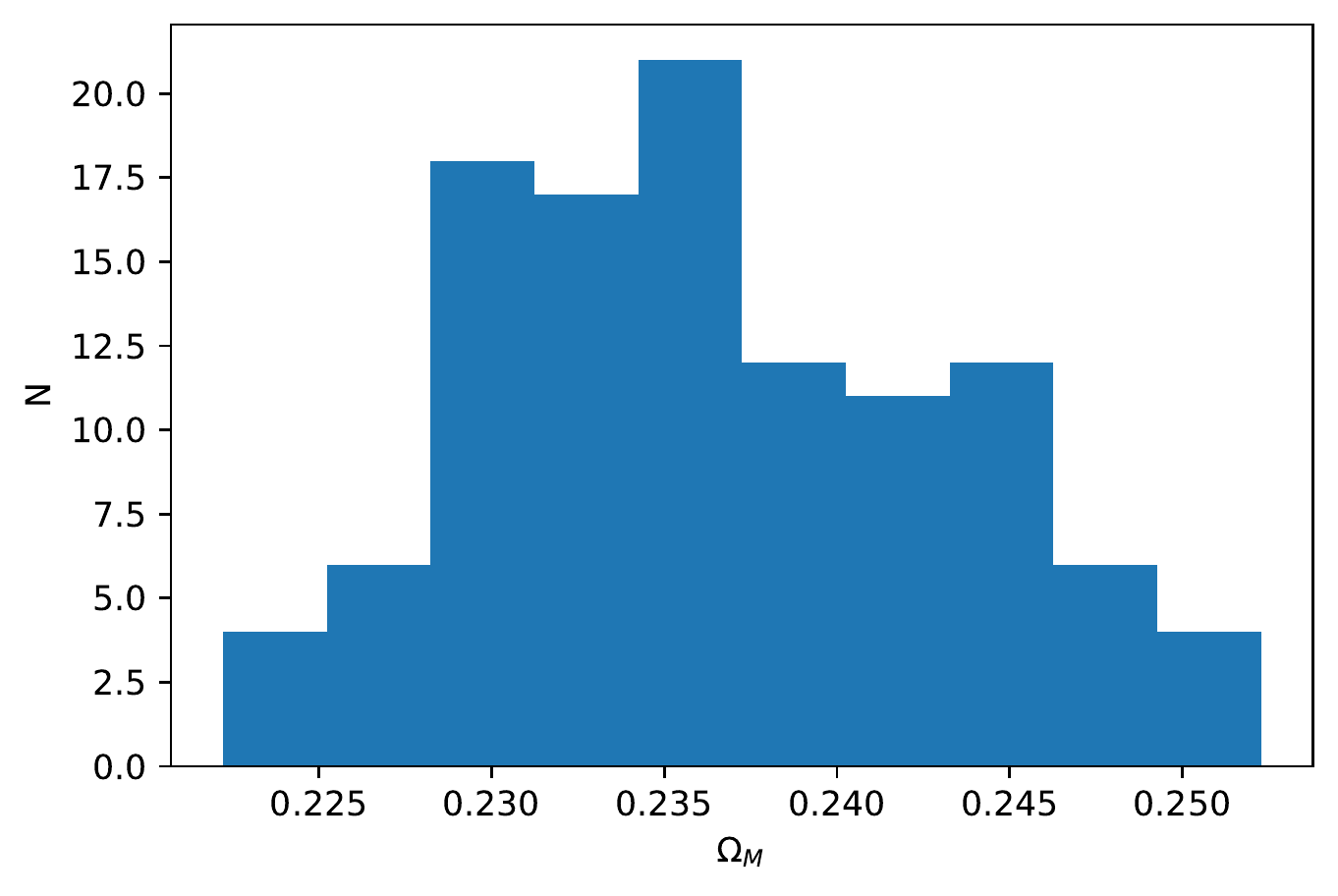}{0.3\textwidth}{(g) $\Omega_M$ with varying correction for redshift evolution}\label{}
\fig{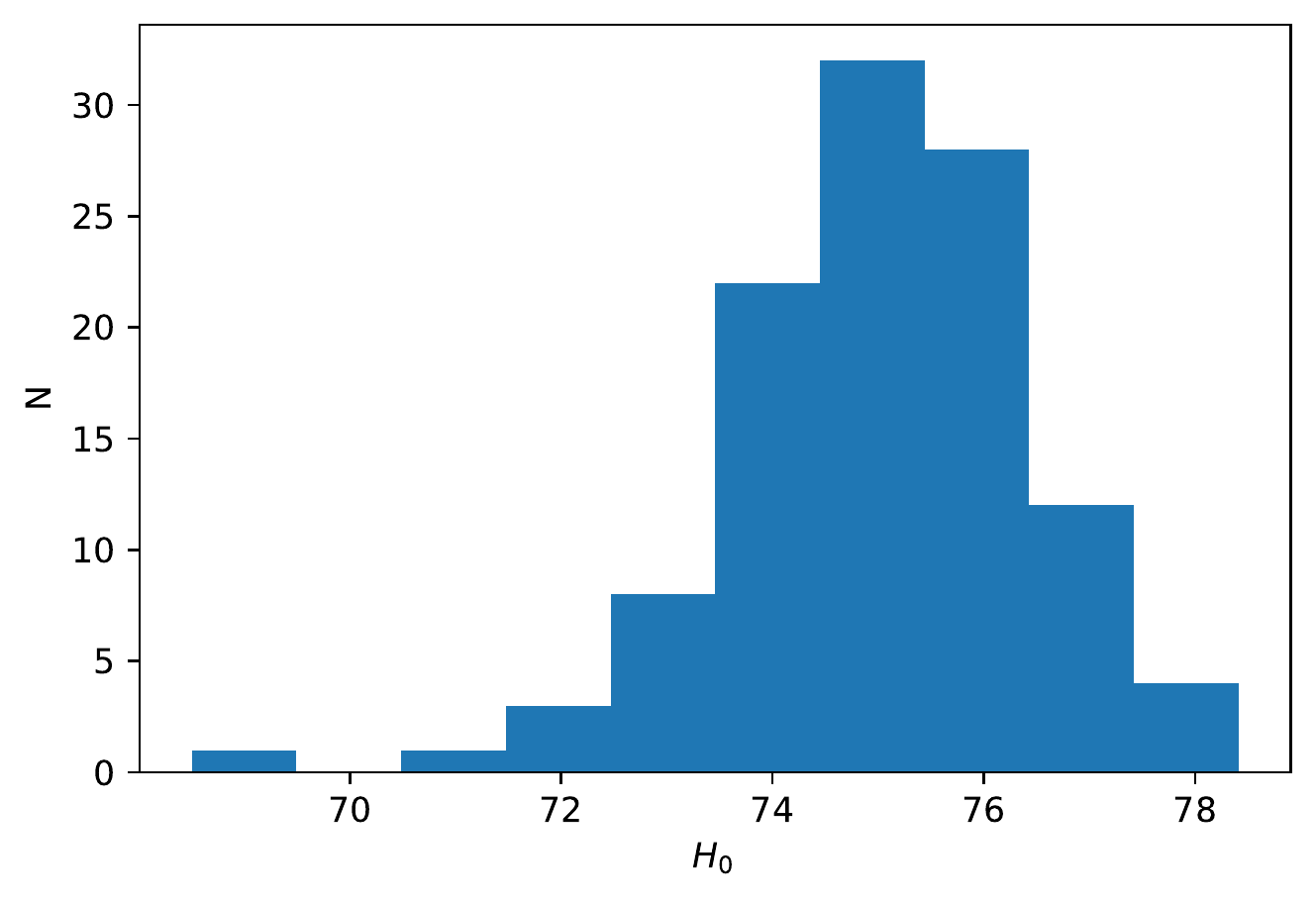}{0.3\textwidth}{(h) $H_0$ with varying correction for redshift evolution}\label{}
\fig{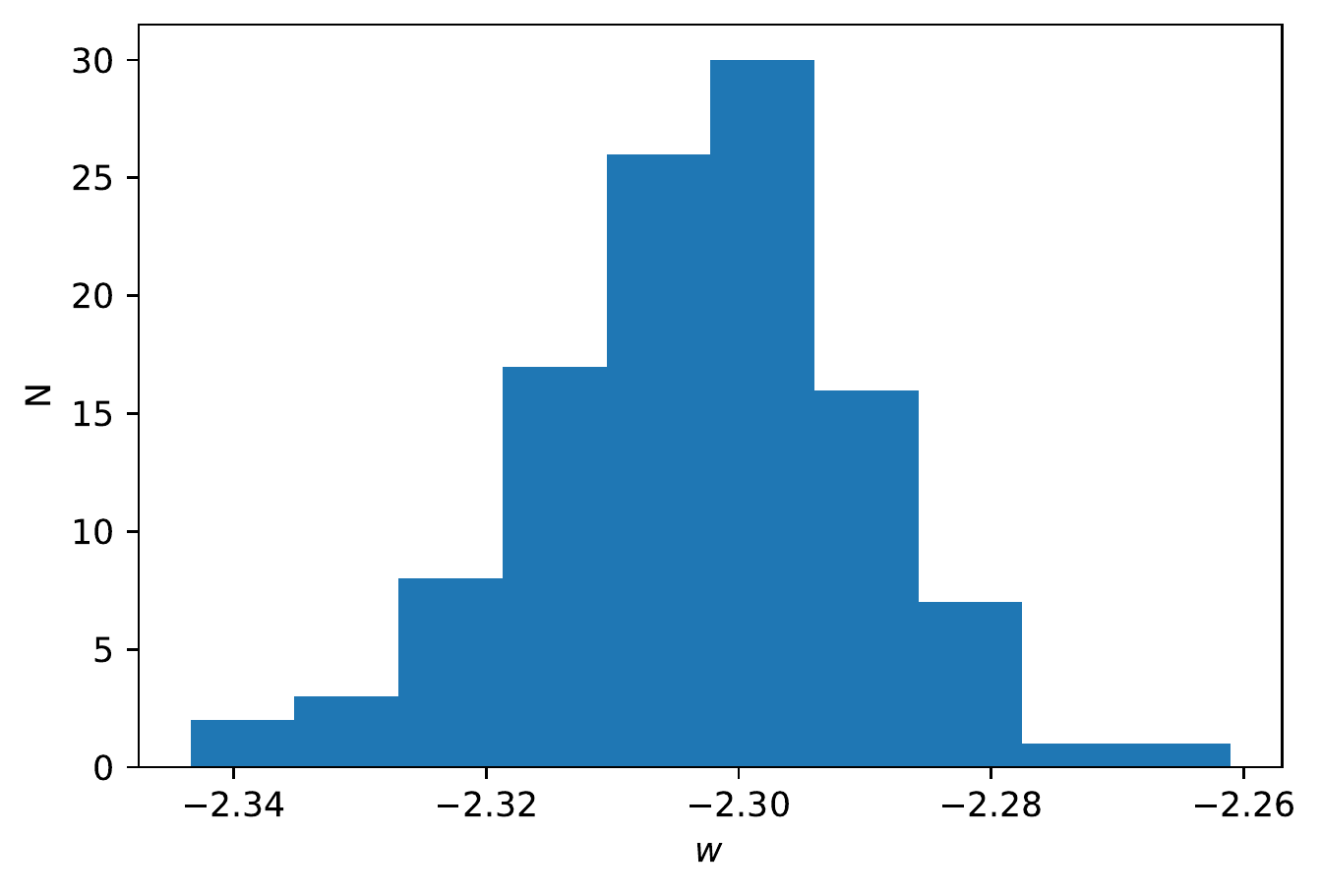}{0.3\textwidth}{(i) $w$ with varying correction for redshift evolution}\label{}}
\caption{Histograms from 100 loops on QSO sample for no evolution, fixed evolution, and varying evolution and new $\cal L_N$ likelihoods in the flat $w$CDM model.}
\label{fig: wcdmNewloops}
\end{figure}

Similarly to GRBs, see \citet{Dainotti2022MNRAS.tmp.2639D}, for QSOs alone if we do not calibrate with SNe Ia the constraints on the cosmological parameters are so large that they are not informative if we consider a single run with the MCMC. If however, we consider, similarly to our analysis for GRBs, a set of 100 runs, we are able to pinpoint where the most probable values are centered. Thus, for clarity, we here present the histograms of results in Figs. \ref{fig: nonflatGaussloops}, \ref{fig: nonflatNewloops}, \ref{fig: wcdmGaussloops}, and \ref{fig: wcdmNewloops} with the distributions of the cosmological parameters. We show the non-flat $\Lambda$CDM model with the $\cal L_G$ likelihoods in Fig. \ref{fig: nonflatGaussloops} and with logistic distribution in Fig. \ref{fig: nonflatNewloops}. We show the $w$CDM model with the $\cal L_G$ likelihoods in Fig. \ref{fig: wcdmGaussloops} and the logistic distribution in Fig. \ref{fig: wcdmNewloops}. In all these figures, the upper panels show the cases without correction for the redshift evolution, the middle panels are the cases with fixed correction for evolution and the lower panels are the cases with varying evolution. From these distributions, regardless of the cosmological model and of the likelihoods, the central values of $H_0$ are shifted at higher values compared to the ones from SNe Ia (an average of 74.5-75 $\mathrm{km  \,s^{-1}  \, Mpc^{-1}}$ for QSOs vs 73 $\mathrm{km  \,s^{-1}  \, Mpc^{-1}}$ for SNe Ia). We should however note that only in the case of the wCDM model with $\cal L_G$ likelihood and fixed evolution the central value of the distribution is 73, thus compatible with the SNe Ia. It is not clear the reason for this behaviour, but its investigation goes far beyond the scope of the current paper. Regarding the $\Omega_M$ parameter, the values are shifted at higher values as seen in \citet{Colgain2022arXiv220310558C,Colgain2022arXiv220611447C}, and \citet{biasfreeQSO2022}. The only exception is for the $w$CDM model with varying evolution with $\cal L_G$ likelihood, for which $\Omega_M=0.25 \pm 0.11$, and with $\cal L_N$ one, for which $\Omega_M=0.24 \pm 0.10$. These values are compatible within 1 $\sigma$ with the values of $\Omega_M$ obtained by SNe Ia. It appears that for the cases of fixed evolution, the trend of $\Omega_M$ points towards a Universe filled only with baryonic matter and where the contribution of Dark Matter is negligible. Indeed, these results are compatible with the values of the observed baryonic matter. Regarding the parameters inherent to the curvature of the Universe, $\Omega_k$ values are negative indicating a closed Universe both for the cases of no evolution and for the cases of varying evolution. For the cases with fixed evolution, the results are compatible with a zero curvature. We here stress that when a fixed evolution is considered, the parameters of the evolution are obtained by fixing a given cosmological model, which in this case is a flat $\Lambda$CDM model with $\Omega_k=0$ and $w=-1$. Thus, we should bear in mind that the results in relation to the curvature for the fixed evolution are driven by fixing the cosmological parameters, and we acknowledge, as we have discussed in \citet{Dainotti2022MNRAS.514.1828D, Dainotti2022MNRAS.tmp.2639D}, that the most reliable approach is the one that takes into account the varying evolution. Regarding the values of the equation of the state parameter $w$, we note that these are compatible with -1 for the case of no evolution and fixed evolution, but they are largely incompatible with -1 in the case of varying evolution pointing towards phantom dark energy models, as already discussed in \citet{biasfreeQSO2022}.  

\begin{table*}
\caption{Results from the Anderson-Darling (first value in parenthesis) and Cramer von Mises (second value in parenthesis) goodness-of-fit tests for all configurations investigated in this work.}
\begin{centering}
\begingroup
\setlength{\tabcolsep}{0.5\tabcolsep}
\scriptsize
\begin{tabular}{ccccccccc}
\hline
\multicolumn{1}{c}{$\cal L_G$ likelihoods:}&\multicolumn{4}{c}{Non-flat $\Lambda$CDM }&\multicolumn{4}{c}{flat $w$CDM}\tabularnewline
\hline
\hline
GRBs+QSOs+BAO+\textit{Pantheon} & GRBs & BAO & SNe Ia & QSOs & GRBs & BAO & SNe Ia & QSOs
\tabularnewline
\hline
\hline
No Evolution & $(>2.5\%, 5\%)$ &  $(>25\%,99\%)$ & $(>25\%,99\%)$  & $(<0.1\%,0.7\%)$  & $(>5\%,9\%)$ & $(>25\%,99\%)$ & $(>25\%,99\%)$ & $(<0.1\%,0.4\%)$ \tabularnewline
\hline
Fixed Evolution & $(>2.5\%,5\%)$ & $(>25\%,99\%)$ & $(>25\%,99\%)$ & $(<0.1\%,10^{-5}\%)$ & $(>2.5\%,6\%)$ & $(>25\%,99\%)$ & $(>25\%,99\%)$ & $(<0.1\&, 10^{-7}\%)$\tabularnewline
\hline
Varying Evolution & $(>2.5\%,2.5\%)$ & $(>25\%, 99\%)$ & $(>25\%, 99\%)$ & $(<0.1\%,10^{-4}\%)$ & $(>2.5\%,5\%)$ & $(>25\%,99\%)$ & $(>25\%,99\%)$ & $(<0.1\%, 10^{-7}\%)$\tabularnewline
\hline
\hline
GRBs+QSOs+BAO+\textit{Pantheon +} & GRBs & BAO & SNe Ia & QSOs & GRBs & BAO & SNe Ia & QSOs
\tabularnewline
\hline
\hline
No Evolution & $(>5\%, 11\%)$ & $(>25\%, 99\%)$ & $(>25\%, 99\%)$ & $(>0.1\%,1\%)$ & $(>5\%,16\%)$ & $(>25\%, 99\%)$ & $(>25\%, 99\%)$ & $(<0.1\%,0.3\%)$ \tabularnewline
\hline
Fixed Evolution & $(>5\%,16\%)$ & $(>25\%,99\%)$ & $(>25\%,99\%)$ & $(<0.1\%, 10^{-5}\%)$ & $(>2.5\%,5\%)$ & $(>25\%,99\%)$ & $(>25\%,99\%)$ & $(<0.1\%,10^{-5}\%)$ \tabularnewline
\hline
Varying Evolution & $(>2.5\%,3\%)$ & $(>25\%,99\%)$ & $(>25\%,99\%)$ & $(<0.1\%,10^{-5}\%)$ & $(>2.5\%,5\%)$ & $(>25\%,99\%)$ & $(>25\%,99\%)$ & $(<0.1\%, 10^{-6}\%)$ \tabularnewline
\hline
\hline
\multicolumn{1}{c}{The $\cal L_N$ likelihoods:}&\multicolumn{4}{c}{Non-flat $\Lambda$CDM}&\multicolumn{4}{c}{flat $w$CDM}\tabularnewline
\hline
\hline
GRBs+QSOs+BAO+\textit{Pantheon} & GRBs & BAO & SNe Ia & QSOs & GRBs & BAO & SNe Ia & QSOs
\tabularnewline
\hline
\hline
No Evolution & $(>5\%, 13\%)$ & $(>25\%,99\%)$ & $(>25\%,99\%)$ & $(<0.1\%,0.2\%)$ & $(>2.5\%,7\%)$ & $(>25\%,99\%)$ & $(>25\%,99\%)$ & $(<0.1\%,0.5\%)$  \tabularnewline
\hline
Fixed Evolution & $(>2.5\%,8\%)$ & $(>25\%,99\%)$ & $(>25\%,99\%)$ & $(<0.1\%, 10^{-6}\%)$ & $(>2.5\%,5\%)$ & $(>25\%,99\%)$ & $(>25\%,99\%)$ & $(<0.1\%, 10^{-5}\%)$ \tabularnewline
\hline
Varying Evolution & $(>10\%,17\%)$ & $(>25\%,99\%)$ & $(>25\%,99\%)$ & $(<0.1\%,10^{-4}\%)$ & $(>25\%,31\%)$ & $(>25\%,99\%)$ & $(>25\%,99\%)$ & $(<0.1\%,10^{-4}\%)$ \tabularnewline
\hline
\hline
GRBs+QSOs+BAO+\textit{Pantheon +} & GRBs & BAO & SNe Ia & QSOs & GRBs & BAO & SNe Ia & QSOs
\tabularnewline
\hline
\hline
No Evolution & $(>2.5\%,7\%)$ & $(>25\%,99\%)$ & $(>25\%,99\%)$ & $(<0.1\%,0.6\%)$ & $(>5\%,7\%)$ & $(>25\%,99\%)$ & $(>25\%,99\%)$ & $(<0.1\%,0.2\%)$  \tabularnewline
\hline
Fixed Evolution & $(>2.5\%,5\%)$  & $(>25\%,99\%)$ & $(>25\%,99\%)$ & $(<0.1\%,10^{-4}\%)$ & $(>5\%,8\%)$ & $(>25\%,99\%)$ & $(>25\%,99\%)$ & $(<0.1\%,10^{-4}\%)$  \tabularnewline
\hline
Varying Evolution & $(>10\%.18\%)$ &  $(>25\%,99\%)$ & $(>25\%,99\%)$  & $(<0.1\%,10^{-5}\%)$ & $(>25\%,23\%)$ & $(>25\%,99\%)$ & $(>25\%,99\%)$ & $(<0.1\%,10^{-4}\%)$\tabularnewline
\hline
\end{tabular}
\endgroup
\label{tab:test}
\par\end{centering}
\end{table*}

\begin{table*}
\caption{Mean values of cosmological parameters and their 1 $\sigma$ uncertainties for all cosmological configurations obtained from the 100 loops on the QSO sample.}
\begin{centering}
\begin{tabular}{ccccccc}
\hline
\multicolumn{1}{c}{$\cal L_G$ likelihoods:}&\multicolumn{3}{c}{Non-flat $\Lambda$CDM }&\multicolumn{3}{c}{flat $w$CDM}\tabularnewline
\hline
\hline
 & $<H_0>$ & $<\Omega_M>$ & $<\Omega_k>$ & $<H_0>$ & $<\Omega_M>$ & $<w>$
\tabularnewline
\hline
\hline
No Evolution & $75.2 \pm 14.4$ &  $0.76 \pm 0.21 $& $-0.67 \pm 0.04$  & $75.1 \pm 14.4$  & $0.93 \pm 0.07$ & $-1.52 \pm 0.63$ \tabularnewline
\hline
Fixed Evolution & $75.1 \pm 14.3$ & $0.07 \pm 0.02$ & $-0.04 \pm 0.08$ & $73.0 \pm 14.4$ & $0.08 \pm 0.03$ & $-1.73 \pm 0.65$ \tabularnewline
\hline
Varying Evolution & $75.0 \pm 14.4$ & $0.51 \pm 0.14$ & $-0.61 \pm 0.09$ & $74.9 \pm 14.3$ & $0.25 \pm 0.11$ & $-2.27 \pm 0.26$ \tabularnewline
\hline
\hline
\multicolumn{1}{c}{The $\cal L_N$ likelihoods:}&\multicolumn{3}{c}{}&\multicolumn{3}{c}{}\tabularnewline
\hline
\hline
No Evolution & $74.8 \pm 14.3$ & $0.48 \pm 0.23$ & $-0.68 \pm 0.02$  & $75.0 \pm 14.3$ & $0.92 \pm 0.07$ & $-1.54 \pm 0.63$  \tabularnewline
\hline
Fixed Evolution & $75.0 \pm 14.3$ & $0.06 \pm 0.01$ & $-0.05 \pm 0.08$ & $75.0 \pm 14.4$ & $0.08 \pm 0.04$ & $-1.86 \pm 0.66$  \tabularnewline
\hline
Varying Evolution & $75.2 \pm 14.3$ & $0.51 \pm 0.14$ & $-0.62 \pm 0.09$ & $73.0 \pm 14.3$ & $0.24 \pm 0.10$ & $-2.30 \pm 0.23$ \tabularnewline 
\hline
\end{tabular}
\label{tab:loops}
\par\end{centering}
\end{table*}

\section{Discussion}\label{sec7}
Looking at Table \ref{tab:bestfit} and Table \ref{tab:loops} we see that the contribution of SNe Ia in terms of the results of the cosmological parameters dominates compared to the contribution of QSOs because otherwise, we would have had values of the cosmological parameters averaged among the various probes. Each probe is weighted in the same way, so each observation brings the weight according to the number of sources and their scatter with the scatter weighting more compared to the number of sources. Thus, the sources that are less reliable with a larger scatter, weigh less than the ones that are more precise. This is clearly visible in the likelihoods in Eqs. \eqref{lfsne}, \eqref{lfgrb}, \eqref{lfqso}, and \eqref{lfbao2} and where the larger the denominator, the smaller is the overall contribution to the likelihood. 
However, since we do obtain different results in terms of the cosmological parameters between SNe Ia and QSOs, we would need to think more carefully about joining the likelihoods together, since the approach of joining all probes could be not the most appropriate method. Our approach of combining likelihoods is based on the assumption that the cosmological parameters are expected to be the same regardless of the probes and this is the reason why we have used combined likelihoods.
Indeed, for GRBs this is not an issue since in a previous paper \citep{Dainotti2022MNRAS.tmp.2639D} the cosmological parameters obtained as an average of 100 with MCMC are compatible with the one obtained by SNe Ia.
To assess the inclusion of QSOs in the joint likelihood further, for future approaches, we would need to find a way to reduce as much as possible the scatter of QSOs at the same level as the SNe Ia. In such a case once we achieve this result we will be able to clarify if the probes can be added together. Although this is a topic of high interest, it goes far beyond the scope of the current investigation.

\section{Summary and Conclusions}
\label{conclusions}
In this work, we have compared the traditional approach of a $\cal L_G$ likelihood with the use of the best-fit likelihood for the considered samples of QSOs, BAO, GRBs, and SNe Ia to test if alternative cosmologies like the open or closed Universe, or the extension of the flat $\Lambda$CDM model, the flat $w$CDM, show a statistically significant deviation from the standard $\Lambda$CDM model.
To this end, we have used the methodology provided in \citet{Dainotti2022MNRAS.tmp.2639D}, \citet{biasfreeQSO2022}, \citet{Bargiacchi2023arXiv230307076B} for the computation of the evolution in the redshift of the physical variables of QSOs and GRBs, but computing the evolutionary parameters for GRBs on the sample of 50 sources used in this work.

We notice that, when applying the varying evolution, the $\zeta_{CMB}$ (see Table \ref{tab:bestfit}) computed between the Planck measurement and our results of combined probes is slightly smaller for the cases with fixed evolution for both Pantheon and Pantheon+ samples than the cases with varying evolution (with a reduction between $1\%$ and $12\%$) and slightly larger for the cases without evolution (with an increase between $1 \%$ and $9\%$). This can be naturally ascribed to the fact that the fixed evolution in general leads to slightly larger uncertainties on $H_0$ (between 3 and 5 $\%$) compared to the ones of no-evolutionary cases, but slightly smaller (between 3 and 4 $\%$) than the varying evolution. This depends on the uncertainties on the evolutionary parameters for both GRBs and QSOs (i.e $\Delta k_{Lpeak} = 0.88$, $\Delta k_{L_{a, X}} = 1.6$, $\Delta k_{LUV} = 0.08$, $\Delta k_{L_X} = 0.07$, where with the symbol $\Delta$ we indicate the uncertainties) which propagates in the computation of luminosities and thus increase the uncertainties on the fitted cosmological parameters compared to the case in which we do not include any correction for the redshift evolution. 

When comparing the results of the $\cal L_G$ likelihoods commonly used and the ones of the $\cal L_N$ ones, we note that the values obtained for the cosmological parameters, in both cosmological models investigated and all evolutionary treatments applied, are completely consistent. Besides this, there is a significant decrease in the uncertainties on all parameters. This reduction ranges from $10\%$ on $\Omega_M$ in the case of fixed and varying evolution in the flat $w$CDM model with \textit{Pantheon +} data to the highest decrease of $35\%$ for $H_0$ in the flat $w$CDM model with fixed evolution and the \textit{Pantheon} sample. The only case in which there is no difference between the $\cal L_G$ likelihoods and the $\cal L_N$ ones is the case of $\Omega_M$ in the flat $w$CDM model without correction for evolution and with \textit{Pantheon +} SNe Ia. Detailed percentage decreases are shown in Table \ref{tab:comparison}.

Overall, our results show compatibility with a flat $\Lambda$CDM model with hints pointing toward a closed Universe as already pointed out by \citet{2020NatAs...4..196D} and toward a $w<-1$ in a phantom dark energy scenario.
In addition, we have investigated the contribution of QSOs alone to the results of the cosmological parameters obtaining the average of 100 MCMC runs to compare these averaged values to the ones of SNe Ia. We have found that the values of both $H_0$ and $\Omega_M$ are larger compared to the ones of SNe Ia. Thus, this work poses also a caveat on the use of the joint likelihood functions in relation to QSOs, although the approach is to assume that all probes should in principle, give the same cosmological parameters with a reduced scatter. In an epoch of cosmological tensions, it is important to re-think carefully the approaches of both the joint likelihoods and the definition of the likelihood themselves ($\cal L_G$ likelihood versus $\cal L_N$).
In conclusion, when fitting cosmological models, and especially alternatives and extensions to the flat $\Lambda$CDM model, the use of the best-fit likelihood function for each probe considered is important not only from a statistical point of view, but also to obtain reliable and intrinsic results and reduced uncertainties on cosmological parameters with the current data available, as also pointed out in \citet{snelikelihood2022}. This paper focuses on the likelihoods used in the literature so far, and to what extent this can influence results when different likelihoods. Our results show the importance of taking into account these issues in future analysis when also considering new samples, for which the condition of Gaussianity should be checked.
Indeed, in the epoch of precision cosmology in which the community strives to carefully reduce systematic biases, building new observing facilities for pushing further the observability limits and the precision on the observables, a reduction in the uncertainties of cosmological of 35\% with the current sample is a step forward in this direction.

\section{Acknowledgments}
The SNe Ia data are supplied by the GitHub repositories \url{https://github.com/dscolnic/Pantheon} and \url{https://github.com/PantheonPlusSH0ES}, for \textit{Pantheon} and \textit{Pantheon +} respectively.
GB acknowledges the Istituto Nazionale di Fisica Nucleare (INFN), Naples, and Scuola Superiore Meridionale for supporting her visit to NAOJ. GB is grateful to be hosted by the Division of Science. MGD acknowledges the Division of Science and NAOJ.
This paper is based upon work from COST Action CA21136 {\it Addressing
observational tensions in cosmology with systematics and fundamental
physics} (CosmoVerse) supported by COST (European Cooperation in Science and
Technology). We are particularly grateful to B. De Simone for the help on the cosmological computations.
Numerical computations were in part carried out on Small Parallel Computers at the Center for Computational Astrophysics, National Astronomical Observatory of Japan.
We thank Beta Lusso and Risaliti Guido for the discussion on the non-Gaussianity of the QSO's likelihood.

\bibliography{bibliografia}{}

\begin{thebibliography}{}
\expandafter\ifx\csname natexlab\endcsname\relax\def\natexlab#1{#1}\fi
\providecommand{\url}[1]{\href{#1}{#1}}
\providecommand{\dodoi}[1]{doi:~\href{http://doi.org/#1}{\nolinkurl{#1}}}
\providecommand{\doeprint}[1]{\href{http://ascl.net/#1}{\nolinkurl{http://ascl.net/#1}}}
\providecommand{\doarXiv}[1]{\href{https://arxiv.org/abs/#1}{\nolinkurl{https://arxiv.org/abs/#1}}}

\bibitem[{{Abdalla} {et~al.}(2022){Abdalla}, {Abell{\'a}n}, {Aboubrahim},
  {Agnello}, {Akarsu}, {Akrami}, {Alestas}, {Aloni}, {Amendola}, {Anchordoqui},
  {Anderson}, {Arendse}, {Asgari}, {Ballardini}, {Barger}, {Basilakos},
  {Batista}, {Battistelli}, {Battye}, {Benetti}, {Benisty}, {Berlin}, {de
  Bernardis}, {Berti}, {Bidenko}, {Birrer}, {Blakeslee}, {Boddy}, {Bom},
  {Bonilla}, {Borghi}, {Bouchet}, {Braglia}, {Buchert}, {Buckley-Geer},
  {Calabrese}, {Caldwell}, {Camarena}, {Capozziello}, {Casertano}, {Chen},
  {Chluba}, {Chen}, {Chen}, {Chudaykin}, {Cicoli}, {Copi}, {Courbin},
  {Cyr-Racine}, {Czerny}, {Dainotti}, {D'Amico}, {Davis}, {de Cruz P{\'e}rez},
  {de Haro}, {Delabrouille}, {Denton}, {Dhawan}, {Dienes}, {Di Valentino},
  {Du}, {Eckert}, {Escamilla-Rivera}, {Fert{\'e}}, {Finelli}, {Fosalba},
  {Freedman}, {Frusciante}, {Gazta{\~n}aga}, {Giar{\`e}}, {Giusarma},
  {G{\'o}mez-Valent}, {Handley}, {Harrison}, {Hart}, {Hazra}, {Heavens},
  {Heinesen}, {Hildebrandt}, {Hill}, {Hogg}, {Holz}, {Hooper}, {Hosseininejad},
  {Huterer}, {Ishak}, {Ivanov}, {Jaffe}, {Jang}, {Jedamzik}, {Jimenez},
  {Joseph}, {Joudaki}, {Kamionkowski}, {Karwal}, {Kazantzidis}, {Keeley},
  {Klasen}, {Komatsu}, {Koopmans}, {Kumar}, {Lamagna}, {Lazkoz}, {Lee},
  {Lesgourgues}, {Levi Said}, {Lewis}, {L'Huillier}, {Lucca}, {Maartens},
  {Macri}, {Marfatia}, {Marra}, {Martins}, {Masi}, {Matarrese}, {Mazumdar},
  {Melchiorri}, {Mena}, {Mersini-Houghton}, {Mertens}, {Milakovi{\'c}},
  {Minami}, {Miranda}, {Moreno-Pulido}, {Moresco}, {Mota}, {Mottola}, {Mozzon},
  {Muir}, {Mukherjee}, {Mukherjee}, {Naselsky}, {Nath}, {Nesseris},
  {Niedermann}, {Notari}, {Nunes}, {{\'O} Colg{\'a}in}, {Owens},
  {{\"O}z{\"u}lker}, {Pace}, {Paliathanasis}, {Palmese}, {Pan}, {Paoletti},
  {Perez Bergliaffa}, {Perivolaropoulos}, {Pesce}, {Pettorino}, {Philcox},
  {Pogosian}, {Poulin}, {Poulot}, {Raveri}, {Reid}, {Renzi}, {Riess}, {Sabla},
  {Salucci}, {Salzano}, {Saridakis}, {Sathyaprakash}, {Schmaltz},
  {Sch{\"o}neberg}, {Scolnic}, {Sen}, {Sehgal}, {Shafieloo}, {Sheikh-Jabbari},
  {Silk}, {Silvestri}, {Skara}, {Sloth}, {Soares-Santos}, {Sol{\`a} Peracaula},
  {Songsheng}, {Soriano}, {Staicova}, {Starkman}, {Szapudi}, {Teixeira},
  {Thomas}, {Treu}, {Trott}, {van de Bruck}, {Vazquez}, {Verde}, {Visinelli},
  {Wang}, {Wang}, {Wang}, {Watkins}, {Watson}, {Webb}, {Weiner}, {Weltman},
  {Witte}, {Wojtak}, {Yadav}, {Yang}, {Zhao}, \&
  {Zumalac{\'a}rregui}}]{2022JHEAp..34...49A}
{Abdalla}, E., {Abell{\'a}n}, G.~F., {Aboubrahim}, A., {et~al.} 2022, Journal
  of High Energy Astrophysics, 34, 49, \dodoi{10.1016/j.jheap.2022.04.002}

\bibitem[{{Alam} {et~al.}(2021){Alam}, {Aubert}, {Avila}, {Balland},
  {Bautista}, {Bershady}, {Bizyaev}, {Blanton}, {Bolton}, {Bovy}, {Brinkmann},
  {Brownstein}, {Burtin}, {Chabanier}, {Chapman}, {Choi}, {Chuang}, {Comparat},
  {Cousinou}, {Cuceu}, {Dawson}, {de la Torre}, {de Mattia}, {Agathe}, {des
  Bourboux}, {Escoffier}, {Etourneau}, {Farr}, {Font-Ribera}, {Frinchaboy},
  {Fromenteau}, {Gil-Mar{\'\i}n}, {Le Goff}, {Gonzalez-Morales},
  {Gonzalez-Perez}, {Grabowski}, {Guy}, {Hawken}, {Hou}, {Kong}, {Parker},
  {Klaene}, {Kneib}, {Lin}, {Long}, {Lyke}, {de la Macorra}, {Martini},
  {Masters}, {Mohammad}, {Moon}, {Mueller}, {Mu{\~n}oz-Guti{\'e}rrez}, {Myers},
  {Nadathur}, {Neveux}, {Newman}, {Noterdaeme}, {Oravetz}, {Oravetz},
  {Palanque-Delabrouille}, {Pan}, {Paviot}, {Percival}, {P{\'e}rez-R{\`a}fols},
  {Petitjean}, {Pieri}, {Prakash}, {Raichoor}, {Ravoux}, {Rezaie}, {Rich},
  {Ross}, {Rossi}, {Ruggeri}, {Ruhlmann-Kleider}, {S{\'a}nchez}, {S{\'a}nchez},
  {S{\'a}nchez-Gallego}, {Sayres}, {Schneider}, {Seo}, {Shafieloo}, {Slosar},
  {Smith}, {Stermer}, {Tamone}, {Tinker}, {Tojeiro}, {Vargas-Maga{\~n}a},
  {Variu}, {Wang}, {Weaver}, {Weijmans}, {Y{\`e}che}, {Zarrouk}, {Zhao},
  {Zhao}, \& {Zheng}}]{eboss2021}
{Alam}, S., {Aubert}, M., {Avila}, S., {et~al.} 2021, \prd, 103, 083533,
  \dodoi{10.1103/PhysRevD.103.083533}

\bibitem[{{Aubourg} {et~al.}(2015){Aubourg}, {Bailey}, {Bautista}, {Beutler},
  {Bhardwaj}, {Bizyaev}, {Blanton}, {Blomqvist}, {Bolton}, {Bovy},
  {Brewington}, {Brinkmann}, {Brownstein}, {Burden}, {Busca}, {Carithers},
  {Chuang}, {Comparat}, {Croft}, {Cuesta}, {Dawson}, {Delubac}, {Eisenstein},
  {Font-Ribera}, {Ge}, {Le Goff}, {Gontcho}, {Gott}, {Gunn}, {Guo}, {Guy},
  {Hamilton}, {Ho}, {Honscheid}, {Howlett}, {Kirkby}, {Kitaura}, {Kneib},
  {Lee}, {Long}, {Lupton}, {Maga{\~n}a}, {Malanushenko}, {Malanushenko},
  {Manera}, {Maraston}, {Margala}, {McBride}, {Miralda-Escud{\'e}}, {Myers},
  {Nichol}, {Noterdaeme}, {Nuza}, {Olmstead}, {Oravetz}, {P{\^a}ris},
  {Padmanabhan}, {Palanque-Delabrouille}, {Pan}, {Pellejero-Ibanez},
  {Percival}, {Petitjean}, {Pieri}, {Prada}, {Reid}, {Rich}, {Roe}, {Ross},
  {Ross}, {Rossi}, {Rubi{\~n}o-Mart{\'\i}n}, {S{\'a}nchez}, {Samushia},
  {G{\'e}nova-Santos}, {Sc{\'o}ccola}, {Schlegel}, {Schneider}, {Seo},
  {Sheldon}, {Simmons}, {Skibba}, {Slosar}, {Strauss}, {Thomas}, {Tinker},
  {Tojeiro}, {Vazquez}, {Viel}, {Wake}, {Weaver}, {Weinberg}, {Wood-Vasey},
  {Y{\`e}che}, {Zehavi}, {Zhao}, \& {BOSS Collaboration}}]{2015PhRvD..92l3516A}
{Aubourg}, {\'E}., {Bailey}, S., {Bautista}, J.~E., {et~al.} 2015, \prd, 92,
  123516, \dodoi{10.1103/PhysRevD.92.123516}

\bibitem[{{Avni} \& {Tananbaum}(1986)}]{1986ApJ...305...83A}
{Avni}, Y., \& {Tananbaum}, H. 1986, \apj, 305, 83, \dodoi{10.1086/164230}

\bibitem[{{Ba{\~n}ados} {et~al.}(2018){Ba{\~n}ados}, {Venemans},
  {Mazzucchelli}, {Farina}, {Walter}, {Wang}, {Decarli}, {Stern}, {Fan},
  {Davies}, {Hennawi}, {Simcoe}, {Turner}, {Rix}, {Yang}, {Kelson}, {Rudie}, \&
  {Winters}}]{banados2018}
{Ba{\~n}ados}, E., {Venemans}, B.~P., {Mazzucchelli}, C., {et~al.} 2018, \nat,
  553, 473, \dodoi{10.1038/nature25180}

\bibitem[{{Bargiacchi} {et~al.}(2022){Bargiacchi}, {Benetti}, {Capozziello},
  {Lusso}, {Risaliti}, \& {Signorini}}]{2022MNRAS.515.1795B}
{Bargiacchi}, G., {Benetti}, M., {Capozziello}, S., {et~al.} 2022, \mnras, 515,
  1795, \dodoi{10.1093/mnras/stac1941}

\bibitem[{{Bargiacchi} {et~al.}(2023){Bargiacchi}, {Dainotti}, {Nagataki}, \&
  {Capozziello}}]{Bargiacchi2023arXiv230307076B}
{Bargiacchi}, G., {Dainotti}, M.~G., {Nagataki}, S., \& {Capozziello}, S. 2023,
  \mnras, \dodoi{10.1093/mnras/stad763}

\bibitem[{{Bernardini} {et~al.}(2012){Bernardini}, {Margutti}, {Zaninoni}, \&
  {Chincarini}}]{2012MNRAS.425.1199B}
{Bernardini}, M.~G., {Margutti}, R., {Zaninoni}, E., \& {Chincarini}, G. 2012,
  \mnras, 425, 1199, \dodoi{10.1111/j.1365-2966.2012.21487.x}

\bibitem[{{Bisogni} {et~al.}(2021){Bisogni}, {Lusso}, {Civano}, {Nardini},
  {Risaliti}, {Elvis}, \& {Fabbiano}}]{2021A&A...655A.109B}
{Bisogni}, S., {Lusso}, E., {Civano}, F., {et~al.} 2021, \aap, 655, A109,
  \dodoi{10.1051/0004-6361/202140852}

\bibitem[{{Bloom} {et~al.}(2001){Bloom}, {Frail}, \&
  {Sari}}]{2001AJ....121.2879B}
{Bloom}, J.~S., {Frail}, D.~A., \& {Sari}, R. 2001, \aj, 121, 2879,
  \dodoi{10.1086/321093}

\bibitem[{{Brout} {et~al.}(2022){Brout}, {Scolnic}, {Popovic}, {Riess}, {Carr},
  {Zuntz}, {Kessler}, {Davis}, {Hinton}, {Jones}, {Kenworthy}, {Peterson},
  {Said}, {Taylor}, {Ali}, {Armstrong}, {Charvu}, {Dwomoh}, {Meldorf},
  {Palmese}, {Qu}, {Rose}, {Sanchez}, {Stubbs}, {Vincenzi}, {Wood}, {Brown},
  {Chen}, {Chambers}, {Coulter}, {Dai}, {Dimitriadis}, {Filippenko}, {Foley},
  {Jha}, {Kelsey}, {Kirshner}, {M{\"o}ller}, {Muir}, {Nadathur}, {Pan}, {Rest},
  {Rojas-Bravo}, {Sako}, {Siebert}, {Smith}, {Stahl}, \&
  {Wiseman}}]{2022ApJ...938..110B}
{Brout}, D., {Scolnic}, D., {Popovic}, B., {et~al.} 2022, \apj, 938, 110,
  \dodoi{10.3847/1538-4357/ac8e04}

\bibitem[{{Camarena} \& {Marra}(2020)}]{2020PhRvR...2a3028C}
{Camarena}, D., \& {Marra}, V. 2020, Physical Review Research, 2, 013028,
  \dodoi{10.1103/PhysRevResearch.2.013028}

\bibitem[{{Cao} {et~al.}(2022{\natexlab{a}}){Cao}, {Dainotti}, \&
  {Ratra}}]{Cao2022arXiv220105245C}
{Cao}, S., {Dainotti}, M., \& {Ratra}, B. 2022{\natexlab{a}}, arXiv e-prints,
  arXiv:2201.05245.
\newblock \doarXiv{2201.05245}

\bibitem[{{Cao} {et~al.}(2021{\natexlab{a}}){Cao}, {Khadka}, \&
  {Ratra}}]{Cao2021arXiv211014840C}
{Cao}, S., {Khadka}, N., \& {Ratra}, B. 2021{\natexlab{a}}, arXiv e-prints,
  arXiv:2110.14840.
\newblock \doarXiv{2110.14840}

\bibitem[{{Cao} {et~al.}(2022{\natexlab{b}}){Cao}, {Khadka}, \&
  {Ratra}}]{Cao2022MNRAS.510.2928C}
---. 2022{\natexlab{b}}, \mnras, 510, 2928, \dodoi{10.1093/mnras/stab3559}

\bibitem[{{Cao} \& {Ratra}(2023)}]{Cao2023}
{Cao}, S., \& {Ratra}, B. 2023, arXiv e-prints, arXiv:2302.14203,
  \dodoi{10.48550/arXiv.2302.14203}

\bibitem[{{Cao} {et~al.}(2021{\natexlab{b}}){Cao}, {Ryan}, \&
  {Ratra}}]{2021MNRAS.504..300C}
{Cao}, S., {Ryan}, J., \& {Ratra}, B. 2021{\natexlab{b}}, \mnras, 504, 300,
  \dodoi{10.1093/mnras/stab942}

\bibitem[{{Cardone} {et~al.}(2009{\natexlab{a}}){Cardone}, {Capozziello}, \&
  {Dainotti}}]{Cardone2009MNRAS.400..775C}
{Cardone}, V.~F., {Capozziello}, S., \& {Dainotti}, M.~G. 2009{\natexlab{a}},
  \mnras, 400, 775, \dodoi{10.1111/j.1365-2966.2009.15456.x}

\bibitem[{{Cardone} {et~al.}(2009{\natexlab{b}}){Cardone}, {Capozziello}, \&
  {Dainotti}}]{cardone09}
---. 2009{\natexlab{b}}, \mnras, 400, 775,
  \dodoi{10.1111/j.1365-2966.2009.15456.x}

\bibitem[{{Cardone} {et~al.}(2010){Cardone}, {Dainotti}, {Capozziello}, \&
  {Willingale}}]{cardone10}
{Cardone}, V.~F., {Dainotti}, M.~G., {Capozziello}, S., \& {Willingale}, R.
  2010, \mnras, 408, 1181, \dodoi{10.1111/j.1365-2966.2010.17197.x}

\bibitem[{{Carroll}(2001)}]{2001LRR.....4....1C}
{Carroll}, S.~M. 2001, Living Reviews in Relativity, 4, 1,
  \dodoi{10.12942/lrr-2001-1}

\bibitem[{{Carroll} {et~al.}(1992){Carroll}, {Press}, \&
  {Turner}}]{1992ARA&A..30..499C}
{Carroll}, S.~M., {Press}, W.~H., \& {Turner}, E.~L. 1992, \araa, 30, 499,
  \dodoi{10.1146/annurev.aa.30.090192.002435}

\bibitem[{{Colg{\'a}in} {et~al.}(2022{\natexlab{a}}){Colg{\'a}in},
  {Sheikh-Jabbari}, {Solomon}, {Bargiacchi}, {Capozziello}, {Dainotti}, \&
  {Stojkovic}}]{Colgain2022arXiv220310558C}
{Colg{\'a}in}, E.~{\'O}., {Sheikh-Jabbari}, M.~M., {Solomon}, R., {et~al.}
  2022{\natexlab{a}}, arXiv e-prints, arXiv:2203.10558.
\newblock \doarXiv{2203.10558}

\bibitem[{{Colg{\'a}in} {et~al.}(2022{\natexlab{b}}){Colg{\'a}in},
  {Sheikh-Jabbari}, {Solomon}, {Dainotti}, \&
  {Stojkovic}}]{Colgain2022arXiv220611447C}
{Colg{\'a}in}, E.~{\'O}., {Sheikh-Jabbari}, M.~M., {Solomon}, R., {Dainotti},
  M.~G., \& {Stojkovic}, D. 2022{\natexlab{b}}, arXiv e-prints,
  arXiv:2206.11447.
\newblock \doarXiv{2206.11447}

\bibitem[{{Cuceu} {et~al.}(2019){Cuceu}, {Farr}, {Lemos}, \&
  {Font-Ribera}}]{2019JCAP...10..044C}
{Cuceu}, A., {Farr}, J., {Lemos}, P., \& {Font-Ribera}, A. 2019, \jcap, 2019,
  044, \dodoi{10.1088/1475-7516/2019/10/044}

\bibitem[{{Dainotti} {et~al.}(2020{\natexlab{a}}){Dainotti}, {Lenart},
  {Sarracino}, {Nagataki}, {Capozziello}, \& {Fraija}}]{Dainotti2020a}
{Dainotti}, M., {Lenart}, A., {Sarracino}, G., {et~al.} 2020{\natexlab{a}},
  \apj, 904, 19, \dodoi{doi:10.3847/1538-4357/abbe8a}

\bibitem[{{Dainotti} {et~al.}(2021{\natexlab{a}}){Dainotti}, {Levine},
  {Fraija}, \& {Chandra}}]{Dainotti2021Galax...9...95D}
{Dainotti}, M., {Levine}, D., {Fraija}, N., \& {Chandra}, P.
  2021{\natexlab{a}}, Galaxies, 9, 95, \dodoi{10.3390/galaxies9040095}

\bibitem[{{Dainotti} {et~al.}(2015){Dainotti}, {Petrosian}, {Willingale},
  {O'Brien}, {Ostrowski}, \& {Nagataki}}]{Dainotti2015b}
{Dainotti}, M., {Petrosian}, V., {Willingale}, R., {et~al.} 2015, \mnras, 451,
  3898, \dodoi{10.1093/mnras/stv1229}

\bibitem[{{Dainotti} \& {Amati}(2018)}]{Dainotti2018PASP..130e1001D}
{Dainotti}, M.~G., \& {Amati}, L. 2018, \pasp, 130, 051001,
  \dodoi{10.1088/1538-3873/aaa8d7}

\bibitem[{Dainotti {et~al.}(2023)Dainotti, Bargiacchi, Lenart, Nagataki, \&
  Capozziello}]{DainottiGoldQSOApJ2023}
Dainotti, M.~G., Bargiacchi, G., Lenart, A., Nagataki, S., \& Capozziello, S.
  2023, \apj

\bibitem[{{Dainotti} {et~al.}(2022{\natexlab{a}}){Dainotti}, {Bargiacchi},
  {Lenart}, {Capozziello}, {{\'O} Colg{\'a}in}, {Solomon}, {Stojkovic}, \&
  {Sheikh-Jabbari}}]{DainottiQSO}
{Dainotti}, M.~G., {Bargiacchi}, G., {Lenart}, A.~{\L}., {et~al.}
  2022{\natexlab{a}}, \apj, 931, 106, \dodoi{10.3847/1538-4357/ac6593}

\bibitem[{{Dainotti} {et~al.}(2022{\natexlab{b}}){Dainotti}, {Bargiacchi},
  {Nagataki}, {Bogdan}, \& {Capozziello}}]{snelikelihood2022}
{Dainotti}, M.~G., {Bargiacchi}, G., {Nagataki}, S., {Bogdan}, M., \&
  {Capozziello}, S. 2022{\natexlab{b}}, ApJL submitted

\bibitem[{{Dainotti} {et~al.}(2008){Dainotti}, {Cardone}, \&
  {Capozziello}}]{Dainotti2008}
{Dainotti}, M.~G., {Cardone}, V.~F., \& {Capozziello}, S. 2008, \mnras, 391,
  L79, \dodoi{10.1111/j.1745-3933.2008.00560.x}

\bibitem[{{Dainotti} {et~al.}(2013{\natexlab{a}}){Dainotti}, {Cardone},
  {Piedipalumbo}, \& {Capozziello}}]{Dainotti2013a}
{Dainotti}, M.~G., {Cardone}, V.~F., {Piedipalumbo}, E., \& {Capozziello}, S.
  2013{\natexlab{a}}, \mnras, 436, 82, \dodoi{10.1093/mnras/stt1516}

\bibitem[{{Dainotti} {et~al.}(2021{\natexlab{b}}){Dainotti}, {De Simone},
  {Schiavone}, {Montani}, {Rinaldi}, \&
  {Lambiase}}]{Dainotti2021ApJ...912..150D}
{Dainotti}, M.~G., {De Simone}, B., {Schiavone}, T., {et~al.}
  2021{\natexlab{b}}, \apj, 912, 150, \dodoi{10.3847/1538-4357/abeb73}

\bibitem[{{Dainotti} {et~al.}(2022{\natexlab{c}}){Dainotti}, {De Simone},
  {Schiavone}, {Montani}, {Rinaldi}, {Lambiase}, {Bogdan}, \&
  {Ugale}}]{Dainotti2022Galax..10...24D}
---. 2022{\natexlab{c}}, Galaxies, 10, 24, \dodoi{doi:10.3390/galaxies10010024}

\bibitem[{{Dainotti} \& {Del Vecchio}(2017)}]{Dainotti2017NewAR..77...23D}
{Dainotti}, M.~G., \& {Del Vecchio}, R. 2017, \nar, 77, 23,
  \dodoi{10.1016/j.newar.2017.04.001}

\bibitem[{{Dainotti} {et~al.}(2018){Dainotti}, {Del Vecchio}, \&
  {Tarnopolski}}]{Dainotti2018AdAst2018E...1D}
{Dainotti}, M.~G., {Del Vecchio}, R., \& {Tarnopolski}, M. 2018, Advances in
  Astronomy, 2018, 4969503, \dodoi{10.1155/2018/4969503}

\bibitem[{{Dainotti} {et~al.}(2011){Dainotti}, {Fabrizio Cardone},
  {Capozziello}, {Ostrowski}, \& {Willingale}}]{dainotti11a}
{Dainotti}, M.~G., {Fabrizio Cardone}, V., {Capozziello}, S., {Ostrowski}, M.,
  \& {Willingale}, R. 2011, \apj, 730, 135, \dodoi{10.1088/0004-637X/730/2/135}

\bibitem[{{Dainotti} {et~al.}(2017{\natexlab{a}}){Dainotti}, {Hernandez},
  {Postnikov}, {Nagataki}, {O'brien}, {Willingale}, \&
  {Striegel}}]{Dainotti2017ApJ...848...88D}
{Dainotti}, M.~G., {Hernandez}, X., {Postnikov}, S., {et~al.}
  2017{\natexlab{a}}, \apj, 848, 88, \dodoi{10.3847/1538-4357/aa8a6b}

\bibitem[{{Dainotti} {et~al.}(2022{\natexlab{d}}){Dainotti}, {Lenart},
  {Chraya}, {Sarracino}, {Nagataki}, {Fraija}, {Capozziello}, \&
  {Bogdan}}]{Dainotti2022MNRAS.tmp.2639D}
{Dainotti}, M.~G., {Lenart}, A.~L., {Chraya}, A., {et~al.} 2022{\natexlab{d}},
  \mnras, \dodoi{10.1093/mnras/stac2752}

\bibitem[{{Dainotti} {et~al.}(2020{\natexlab{b}}){Dainotti}, {Lenart},
  {Sarracino}, {Nagataki}, {Capozziello}, \&
  {Fraija}}]{Dainotti2020ApJ...904...97D}
{Dainotti}, M.~G., {Lenart}, A.~{\L}., {Sarracino}, G., {et~al.}
  2020{\natexlab{b}}, \apj, 904, 97, \dodoi{10.3847/1538-4357/abbe8a}

\bibitem[{{Dainotti} {et~al.}(2022{\natexlab{e}}){Dainotti}, {Levine},
  {Fraija}, {Warren}, \& {Sourav}}]{Dainotticlosureoptical2022ApJ...940..169D}
{Dainotti}, M.~G., {Levine}, D., {Fraija}, N., {Warren}, D., \& {Sourav}, S.
  2022{\natexlab{e}}, \apj, 940, 169, \dodoi{10.3847/1538-4357/ac9b11}

\bibitem[{{Dainotti} {et~al.}(2017{\natexlab{b}}){Dainotti}, {Nagataki},
  {Maeda}, {Postnikov}, \& {Pian}}]{Dainotti2017A&A...600A..98D}
{Dainotti}, M.~G., {Nagataki}, S., {Maeda}, K., {Postnikov}, S., \& {Pian}, E.
  2017{\natexlab{b}}, \aap, 600, A98, \dodoi{10.1051/0004-6361/201628384}

\bibitem[{{Dainotti} {et~al.}(2022{\natexlab{f}}){Dainotti}, {Nielson},
  {Sarracino}, {Rinaldi}, {Nagataki}, {Capozziello}, {Gnedin}, \&
  {Bargiacchi}}]{Dainotti2022MNRAS.514.1828D}
{Dainotti}, M.~G., {Nielson}, V., {Sarracino}, G., {et~al.} 2022{\natexlab{f}},
  \mnras, 514, 1828, \dodoi{10.1093/mnras/stac1141}

\bibitem[{{Dainotti} {et~al.}(2021{\natexlab{c}}){Dainotti}, {Petrosian}, \&
  {Bowden}}]{Dainotti2021ApJ...914L..40D}
{Dainotti}, M.~G., {Petrosian}, V., \& {Bowden}, L. 2021{\natexlab{c}}, \apjl,
  914, L40, \dodoi{10.3847/2041-8213/abf5e4}

\bibitem[{{Dainotti} {et~al.}(2013{\natexlab{b}}){Dainotti}, {Petrosian},
  {Singal}, \& {Ostrowski}}]{Dainotti2013b}
{Dainotti}, M.~G., {Petrosian}, V., {Singal}, J., \& {Ostrowski}, M.
  2013{\natexlab{b}}, \apj, 774, 157, \dodoi{10.1088/0004-637X/774/2/157}

\bibitem[{{Dainotti} {et~al.}(2016){Dainotti}, {Postnikov}, {Hernandez}, \&
  {Ostrowski}}]{Dainotti2016ApJ...825L..20D}
{Dainotti}, M.~G., {Postnikov}, S., {Hernandez}, X., \& {Ostrowski}, M. 2016,
  \apjl, 825, L20, \dodoi{10.3847/2041-8205/825/2/L20}

\bibitem[{{Dainotti} {et~al.}(2022{\natexlab{g}}){Dainotti}, {Sarracino}, \&
  {Capozziello}}]{Dainotti2022PASJ}
{Dainotti}, M.~G., {Sarracino}, G., \& {Capozziello}, S. 2022{\natexlab{g}},
  \pasj, \dodoi{10.1093/pasj/psac057}

\bibitem[{{Dainotti} {et~al.}(2010){Dainotti}, {Willingale}, {Capozziello},
  {Fabrizio Cardone}, \& {Ostrowski}}]{Dainotti2010ApJ...722L.215D}
{Dainotti}, M.~G., {Willingale}, R., {Capozziello}, S., {Fabrizio Cardone}, V.,
  \& {Ostrowski}, M. 2010, \apjl, 722, L215,
  \dodoi{10.1088/2041-8205/722/2/L215}

\bibitem[{{Dainotti} {et~al.}(2020{\natexlab{c}}){Dainotti}, {Livermore},
  {Kann}, {Li}, {Oates}, {Yi}, {Zhang}, {Gendre}, {Cenko}, \&
  {Fraija}}]{Dainotti2020b}
{Dainotti}, M.~G., {Livermore}, S., {Kann}, D.~A., {et~al.} 2020{\natexlab{c}},
  \apjl, 905, L26, \dodoi{10.3847/2041-8213/abcda9}

\bibitem[{{Dainotti} {et~al.}(2021{\natexlab{d}}){Dainotti}, {Omodei},
  {Srinivasaragavan}, {Vianello}, {Willingale}, {O'Brien}, {Nagataki},
  {Petrosian}, {Nuygen}, {Hernandez}, {Axelsson}, {Bissaldi}, \&
  {Longo}}]{Dainotti2021ApJS..255...13D}
{Dainotti}, M.~G., {Omodei}, N., {Srinivasaragavan}, G.~P., {et~al.}
  2021{\natexlab{d}}, \apjs, 255, 13, \dodoi{10.3847/1538-4365/abfe17}

\bibitem[{{Dainotti} {et~al.}(2022{\natexlab{h}}){Dainotti}, {Young}, {Li},
  {Levine}, {Kalinowski}, {Kann}, {Tran}, {Zambrano-Tapia}, {Zambrano-Tapia},
  {Cenko}, {Fuentes}, {S{\'a}nchez-V{\'a}zquez}, {Oates}, {Fraija}, {Becerra},
  {Watson}, {Butler}, {Gonz{\'a}lez}, {Kutyrev}, {Lee}, {Prochaska},
  {Ramirez-Ruiz}, {Richer}, \& {Zola}}]{Dainotti2022ApJS..261...25D}
{Dainotti}, M.~G., {Young}, S., {Li}, L., {et~al.} 2022{\natexlab{h}}, \apjs,
  261, 25, \dodoi{10.3847/1538-4365/ac7c64}

\bibitem[{{Dall'Osso} {et~al.}(2011){Dall'Osso}, {Stratta}, {Guetta}, {Covino},
  {De Cesare}, \& {Stella}}]{2011A&A...526A.121D}
{Dall'Osso}, S., {Stratta}, G., {Guetta}, D., {et~al.} 2011, \aap, 526, A121,
  \dodoi{10.1051/0004-6361/201014168}

\bibitem[{{Di Valentino} {et~al.}(2020{\natexlab{a}}){Di Valentino},
  {Gariazzo}, {Mena}, \& {Vagnozzi}}]{2020JCAP...07..045D}
{Di Valentino}, E., {Gariazzo}, S., {Mena}, O., \& {Vagnozzi}, S.
  2020{\natexlab{a}}, \jcap, 2020, 045, \dodoi{10.1088/1475-7516/2020/07/045}

\bibitem[{{Di Valentino} {et~al.}(2020{\natexlab{b}}){Di Valentino},
  {Melchiorri}, \& {Silk}}]{2020NatAs...4..196D}
{Di Valentino}, E., {Melchiorri}, A., \& {Silk}, J. 2020{\natexlab{b}}, Nature
  Astronomy, 4, 196, \dodoi{10.1038/s41550-019-0906-9}

\bibitem[{{Djorgovski} \& {Davis}(1987)}]{1987ApJ...313...59D}
{Djorgovski}, S., \& {Davis}, M. 1987, \apj, 313, 59, \dodoi{10.1086/164948}

\bibitem[{{Efron} \& {Petrosian}(1992)}]{1992ApJ...399..345E}
{Efron}, B., \& {Petrosian}, V. 1992, \apj, 399, 345, \dodoi{10.1086/171931}

\bibitem[{{Eisenstein} {et~al.}(2005){Eisenstein}, {Zehavi}, {Hogg},
  {Scoccimarro}, {Blanton}, {Nichol}, {Scranton}, {Seo}, {Tegmark}, {Zheng},
  {Anderson}, {Annis}, {Bahcall}, {Brinkmann}, {Burles}, {Castander},
  {Connolly}, {Csabai}, {Doi}, {Fukugita}, {Frieman}, {Glazebrook}, {Gunn},
  {Hendry}, {Hennessy}, {Ivezi{\'c}}, {Kent}, {Knapp}, {Lin}, {Loh}, {Lupton},
  {Margon}, {McKay}, {Meiksin}, {Munn}, {Pope}, {Richmond}, {Schlegel},
  {Schneider}, {Shimasaku}, {Stoughton}, {Strauss}, {SubbaRao}, {Szalay},
  {Szapudi}, {Tucker}, {Yanny}, \& {York}}]{2005ApJ...633..560E}
{Eisenstein}, D.~J., {Zehavi}, I., {Hogg}, D.~W., {et~al.} 2005, \apj, 633,
  560, \dodoi{10.1086/466512}

\bibitem[{{Evans} {et~al.}(2009){Evans}, {Beardmore}, {Page}, {Osborne},
  {O'Brien}, {Willingale}, {Starling}, {Burrows}, {Godet}, {Vetere}, {Racusin},
  {Goad}, {Wiersema}, {Angelini}, {Capalbi}, {Chincarini}, {Gehrels}, {Kennea},
  {Margutti}, {Morris}, {Mountford}, {Pagani}, {Perri}, {Romano}, \&
  {Tanvir}}]{Evans2009}
{Evans}, P.~A., {Beardmore}, A.~P., {Page}, K.~L., {et~al.} 2009, Monthly
  Notices of the Royal Astronomical Society, 397, 1177,
  \dodoi{10.1111/j.1365-2966.2009.14913.x}

\bibitem[{{Freedman}(2021)}]{2021ApJ...919...16F}
{Freedman}, W.~L. 2021, \apj, 919, 16, \dodoi{10.3847/1538-4357/ac0e95}

\bibitem[{{G{\'o}mez-Valent} \& {Amendola}(2018)}]{2018JCAP...04..051G}
{G{\'o}mez-Valent}, A., \& {Amendola}, L. 2018, \jcap, 2018, 051,
  \dodoi{10.1088/1475-7516/2018/04/051}

\bibitem[{{Gonzalez} {et~al.}(2021){Gonzalez}, {Benetti}, {von Marttens}, \&
  {Alcaniz}}]{2021JCAP...11..060G}
{Gonzalez}, J.~E., {Benetti}, M., {von Marttens}, R., \& {Alcaniz}, J. 2021,
  \jcap, 2021, 060, \dodoi{10.1088/1475-7516/2021/11/060}

\bibitem[{Handley(2021)}]{Handley:2019tkm}
Handley, W. 2021, Phys. Rev. D, 103, L041301,
  \dodoi{10.1103/PhysRevD.103.L041301}

\bibitem[{Hinshaw {et~al.}(2013)Hinshaw, Larson, Komatsu, Spergel, Bennett,
  Dunkley, Nolta, Halpern, Hill, Odegard, \& et~al.}]{Hinshaw_2013}
Hinshaw, G., Larson, D., Komatsu, E., {et~al.} 2013, The Astrophysical Journal
  Supplement Series, 208, 19, \dodoi{10.1088/0067-0049/208/2/19}

\bibitem[{{Horowitz} \& {Teukolsky}(1999)}]{1999RvMPS..71..180H}
{Horowitz}, G.~T., \& {Teukolsky}, S.~A. 1999, Reviews of Modern Physics
  Supplement, 71, S180, \dodoi{10.1103/RevModPhys.71.S180}

\bibitem[{{Just} {et~al.}(2007){Just}, {Brandt}, {Shemmer}, {Steffen},
  {Schneider}, {Chartas}, \& {Garmire}}]{just07}
{Just}, D.~W., {Brandt}, W.~N., {Shemmer}, O., {et~al.} 2007, \apj, 665, 1004,
  \dodoi{10.1086/519990}

\bibitem[{{Kelly}(2007)}]{Kelly2007}
{Kelly}, B.~C. 2007, \apj, 665, 1489, \dodoi{10.1086/519947}

\bibitem[{{Kenworthy} {et~al.}(2019){Kenworthy}, {Scolnic}, \&
  {Riess}}]{2019ApJ...875..145K}
{Kenworthy}, W.~D., {Scolnic}, D., \& {Riess}, A. 2019, \apj, 875, 145,
  \dodoi{10.3847/1538-4357/ab0ebf}

\bibitem[{{Khadka} \& {Ratra}(2020{\natexlab{a}})}]{2020MNRAS.492.4456K}
{Khadka}, N., \& {Ratra}, B. 2020{\natexlab{a}}, \mnras, 492, 4456,
  \dodoi{10.1093/mnras/staa101}

\bibitem[{{Khadka} \& {Ratra}(2020{\natexlab{b}})}]{2020MNRAS.497..263K}
---. 2020{\natexlab{b}}, \mnras, 497, 263, \dodoi{10.1093/mnras/staa1855}

\bibitem[{{Khadka} \& {Ratra}(2021)}]{2021MNRAS.502.6140K}
---. 2021, \mnras, 502, 6140, \dodoi{10.1093/mnras/stab486}

\bibitem[{{Khadka} \& {Ratra}(2022)}]{2022MNRAS.510.2753K}
---. 2022, \mnras, 510, 2753, \dodoi{10.1093/mnras/stab3678}

\bibitem[{{Khadka} {et~al.}(2023){Khadka}, {Zaja{\v{c}}ek}, {Prince}, {Panda},
  {Czerny}, {Mart{\'\i}nez-Aldama}, {Jaiswal}, \& {Ratra}}]{Khadka2023}
{Khadka}, N., {Zaja{\v{c}}ek}, M., {Prince}, R., {et~al.} 2023, \mnras,
  \dodoi{10.1093/mnras/stad1040}

\bibitem[{{Kroupa} {et~al.}(2020){Kroupa}, {Subr}, {Jerabkova}, \&
  {Wang}}]{2020MNRAS.498.5652K}
{Kroupa}, P., {Subr}, L., {Jerabkova}, T., \& {Wang}, L. 2020, \mnras, 498,
  5652, \dodoi{10.1093/mnras/staa2276}

\bibitem[{Larribe \& Fearnhead(2011)}]{larribe2011composite}
Larribe, F., \& Fearnhead, P. 2011, Statistica Sinica, 43

\bibitem[{{Lenart} {et~al.}(2023){Lenart}, {Bargiacchi}, {Dainotti},
  {Nagataki}, \& {Capozziello}}]{biasfreeQSO2022}
{Lenart}, A.~{\L}., {Bargiacchi}, G., {Dainotti}, M.~G., {Nagataki}, S., \&
  {Capozziello}, S. 2023, \apjs, 264, 46, \dodoi{10.3847/1538-4365/aca404}

\bibitem[{{Levine} {et~al.}(2022){Levine}, {Dainotti}, {Zvonarek}, {Fraija},
  {Warren}, {Chandra}, \& {Lloyd-Ronning}}]{Levine2022ApJ...925...15L}
{Levine}, D., {Dainotti}, M., {Zvonarek}, K.~J., {et~al.} 2022, \apj, 925, 15,
  \dodoi{10.3847/1538-4357/ac4221}

\bibitem[{{Li} {et~al.}(2022){Li}, {Huang}, \& {Wang}}]{2022MNRAS.517.1901L}
{Li}, Z., {Huang}, L., \& {Wang}, J. 2022, \mnras, 517, 1901,
  \dodoi{10.1093/mnras/stac2735}

\bibitem[{{Liao} {et~al.}(2019){Liao}, {Shafieloo}, {Keeley}, \&
  {Linder}}]{2019ApJ...886L..23L}
{Liao}, K., {Shafieloo}, A., {Keeley}, R.~E., \& {Linder}, E.~V. 2019, \apjl,
  886, L23, \dodoi{10.3847/2041-8213/ab5308}

\bibitem[{Lindsay {et~al.}(2011)Lindsay, Yi, \& Sun}]{lindsay2011issues}
Lindsay, B.~G., Yi, G.~Y., \& Sun, J. 2011, Statistica Sinica, 71

\bibitem[{{Lusso} \& {Risaliti}(2016)}]{lr16}
{Lusso}, E., \& {Risaliti}, G. 2016, \apj, 819, 154,
  \dodoi{10.3847/0004-637X/819/2/154}

\bibitem[{{Lusso} {et~al.}(2010){Lusso}, {Comastri}, {Vignali}, {Zamorani},
  {Brusa}, {Gilli}, {Iwasawa}, {Salvato}, {Civano}, {Elvis}, {Merloni},
  {Bongiorno}, {Trump}, {Koekemoer}, {Schinnerer}, {Le Floc'h}, {Cappelluti},
  {Jahnke}, {Sargent}, {Silverman}, {Mainieri}, {Fiore}, {Bolzonella}, {Le
  F{\`e}vre}, {Garilli}, {Iovino}, {Kneib}, {Lamareille}, {Lilly}, {Mignoli},
  {Scodeggio}, \& {Vergani}}]{2010A&A...512A..34L}
{Lusso}, E., {Comastri}, A., {Vignali}, C., {et~al.} 2010, \aap, 512, A34,
  \dodoi{10.1051/0004-6361/200913298}

\bibitem[{{Lusso} {et~al.}(2020){Lusso}, {Risaliti}, {Nardini}, {Bargiacchi},
  {Benetti}, {Bisogni}, {Capozziello}, {Civano}, {Eggleston}, {Elvis},
  {Fabbiano}, {Gilli}, {Marconi}, {Paolillo}, {Piedipalumbo}, {Salvestrini},
  {Signorini}, \& {Vignali}}]{2020A&A...642A.150L}
{Lusso}, E., {Risaliti}, G., {Nardini}, E., {et~al.} 2020, \aap, 642, A150,
  \dodoi{10.1051/0004-6361/202038899}

\bibitem[{{Netzer}(2013)}]{qsophysics}
{Netzer}, H. 2013, Cambridge University Press,
  \dodoi{https://doi.org/10.1017/CBO9781139109291}

\bibitem[{Park \& Ratra(2019)}]{Park:2017xbl}
Park, C.-G., \& Ratra, B. 2019, Astrophys. J., 882, 158,
  \dodoi{10.3847/1538-4357/ab3641}

\bibitem[{{Peebles}(1984)}]{peebles1984}
{Peebles}, P.~J.~E. 1984, \apj, 284, 439, \dodoi{10.1086/162425}

\bibitem[{{Perlmutter} {et~al.}(1999){Perlmutter}, {Aldering}, {Goldhaber},
  {Knop}, {Nugent}, {Castro}, {Deustua}, {Fabbro}, {Goobar}, {Groom}, {Hook},
  {Kim}, {Kim}, {Lee}, {Nunes}, {Pain}, {Pennypacker}, {Quimby}, {Lidman},
  {Ellis}, {Irwin}, {McMahon}, {Ruiz-Lapuente}, {Walton}, {Schaefer}, {Boyle},
  {Filippenko}, {Matheson}, {Fruchter}, {Panagia}, {Newberg}, {Couch}, \&
  {Project}}]{perlmutter1999}
{Perlmutter}, S., {Aldering}, G., {Goldhaber}, G., {et~al.} 1999, \apj, 517,
  565, \dodoi{10.1086/307221}

\bibitem[{{Petrosian} {et~al.}(2022){Petrosian}, {Singal}, \&
  {Mutchnick}}]{2022ApJ...935L..19P}
{Petrosian}, V., {Singal}, J., \& {Mutchnick}, S. 2022, \apjl, 935, L19,
  \dodoi{10.3847/2041-8213/ac85ac}

\bibitem[{{Planck Collaboration} {et~al.}(2020){Planck Collaboration},
  {Aghanim}, {Akrami}, {Ashdown}, {Aumont}, {Baccigalupi}, {Ballardini},
  {Banday}, {Barreiro}, {Bartolo}, {Basak}, {Battye}, {Benabed}, {Bernard},
  {Bersanelli}, {Bielewicz}, {Bock}, {Bond}, {Borrill}, {Bouchet}, {Boulanger},
  {Bucher}, {Burigana}, {Butler}, {Calabrese}, {Cardoso}, {Carron},
  {Challinor}, {Chiang}, {Chluba}, {Colombo}, {Combet}, {Contreras}, {Crill},
  {Cuttaia}, {de Bernardis}, {de Zotti}, {Delabrouille}, {Delouis}, {Di
  Valentino}, {Diego}, {Dor{\'e}}, {Douspis}, {Ducout}, {Dupac}, {Dusini},
  {Efstathiou}, {Elsner}, {En{\ss}lin}, {Eriksen}, {Fantaye}, {Farhang},
  {Fergusson}, {Fernandez-Cobos}, {Finelli}, {Forastieri}, {Frailis},
  {Fraisse}, {Franceschi}, {Frolov}, {Galeotta}, {Galli}, {Ganga},
  {G{\'e}nova-Santos}, {Gerbino}, {Ghosh}, {Gonz{\'a}lez-Nuevo}, {G{\'o}rski},
  {Gratton}, {Gruppuso}, {Gudmundsson}, {Hamann}, {Handley}, {Hansen},
  {Herranz}, {Hildebrandt}, {Hivon}, {Huang}, {Jaffe}, {Jones}, {Karakci},
  {Keih{\"a}nen}, {Keskitalo}, {Kiiveri}, {Kim}, {Kisner}, {Knox},
  {Krachmalnicoff}, {Kunz}, {Kurki-Suonio}, {Lagache}, {Lamarre}, {Lasenby},
  {Lattanzi}, {Lawrence}, {Le Jeune}, {Lemos}, {Lesgourgues}, {Levrier},
  {Lewis}, {Liguori}, {Lilje}, {Lilley}, {Lindholm}, {L{\'o}pez-Caniego},
  {Lubin}, {Ma}, {Mac{\'\i}as-P{\'e}rez}, {Maggio}, {Maino}, {Mandolesi},
  {Mangilli}, {Marcos-Caballero}, {Maris}, {Martin}, {Martinelli},
  {Mart{\'\i}nez-Gonz{\'a}lez}, {Matarrese}, {Mauri}, {McEwen}, {Meinhold},
  {Melchiorri}, {Mennella}, {Migliaccio}, {Millea}, {Mitra},
  {Miville-Desch{\^e}nes}, {Molinari}, {Montier}, {Morgante}, {Moss}, {Natoli},
  {N{\o}rgaard-Nielsen}, {Pagano}, {Paoletti}, {Partridge}, {Patanchon},
  {Peiris}, {Perrotta}, {Pettorino}, {Piacentini}, {Polastri}, {Polenta},
  {Puget}, {Rachen}, {Reinecke}, {Remazeilles}, {Renzi}, {Rocha}, {Rosset},
  {Roudier}, {Rubi{\~n}o-Mart{\'\i}n}, {Ruiz-Granados}, {Salvati}, {Sandri},
  {Savelainen}, {Scott}, {Shellard}, {Sirignano}, {Sirri}, {Spencer},
  {Sunyaev}, {Suur-Uski}, {Tauber}, {Tavagnacco}, {Tenti}, {Toffolatti},
  {Tomasi}, {Trombetti}, {Valenziano}, {Valiviita}, {Van Tent}, {Vibert},
  {Vielva}, {Villa}, {Vittorio}, {Wandelt}, {Wehus}, {White}, {White},
  {Zacchei}, \& {Zonca}}]{planck2018}
{Planck Collaboration}, {Aghanim}, N., {Akrami}, Y., {et~al.} 2020, \aap, 641,
  A6, \dodoi{10.1051/0004-6361/201833910}

\bibitem[{{Postnikov} {et~al.}(2014){Postnikov}, {Dainotti}, {Hernandez}, \&
  {Capozziello}}]{Postnikov14}
{Postnikov}, S., {Dainotti}, M.~G., {Hernandez}, X., \& {Capozziello}, S. 2014,
  \apj, 783, 126, \dodoi{10.1088/0004-637X/783/2/126}

\bibitem[{Prabhu(1988)}]{prabhu1988statistical}
Prabhu, N.~U. 1988, Statistical Inference from Stochastic Processes:
  Proceedings of the AMS-IMS-SIAM Joint Summer Research Conference Held August
  9-15, 1987, with Support from the National Science Foundation and the Army
  Research Office, Vol.~80 (American Mathematical Soc.)

\bibitem[{{Rea} {et~al.}(2015){Rea}, {Gull{\'o}n}, {Pons}, {Perna}, {Dainotti},
  {Miralles}, \& {Torres}}]{Rea2015ApJ...813...92R}
{Rea}, N., {Gull{\'o}n}, M., {Pons}, J.~A., {et~al.} 2015, \apj, 813, 92,
  \dodoi{10.1088/0004-637X/813/2/92}

\bibitem[{{Riess} {et~al.}(2019){Riess}, {Casertano}, {Yuan}, {Macri}, \&
  {Scolnic}}]{2019ApJ...876...85R}
{Riess}, A.~G., {Casertano}, S., {Yuan}, W., {Macri}, L.~M., \& {Scolnic}, D.
  2019, \apj, 876, 85, \dodoi{10.3847/1538-4357/ab1422}

\bibitem[{{Riess} {et~al.}(1998){Riess}, {Filippenko}, {Challis},
  {Clocchiatti}, {Diercks}, {Garnavich}, {Gilliland}, {Hogan}, {Jha},
  {Kirshner}, {Leibundgut}, {Phillips}, {Reiss}, {Schmidt}, {Schommer},
  {Smith}, {Spyromilio}, {Stubbs}, {Suntzeff}, \& {Tonry}}]{riess1998}
{Riess}, A.~G., {Filippenko}, A.~V., {Challis}, P., {et~al.} 1998, \aj, 116,
  1009, \dodoi{10.1086/300499}

\bibitem[{{Riess} {et~al.}(2022){Riess}, {Yuan}, {Macri}, {Scolnic}, {Brout},
  {Casertano}, {Jones}, {Murakami}, {Anand}, {Breuval}, {Brink}, {Filippenko},
  {Hoffmann}, {Jha}, {D'arcy Kenworthy}, {Mackenty}, {Stahl}, \&
  {Zheng}}]{2022ApJ...934L...7R}
{Riess}, A.~G., {Yuan}, W., {Macri}, L.~M., {et~al.} 2022, \apjl, 934, L7,
  \dodoi{10.3847/2041-8213/ac5c5b}

\bibitem[{{Risaliti} \& {Lusso}(2015)}]{rl15}
{Risaliti}, G., \& {Lusso}, E. 2015, \apj, 815, 33,
  \dodoi{10.1088/0004-637X/815/1/33}

\bibitem[{{Risaliti} \& {Lusso}(2019)}]{rl19}
---. 2019, Nature Astronomy, 195, \dodoi{10.1038/s41550-018-0657-z}

\bibitem[{{Rowlinson} {et~al.}(2014){Rowlinson}, {Gompertz}, {Dainotti},
  {O'Brien}, {Wijers}, \& {van der Horst}}]{Rowlinson2014MNRAS.443.1779R}
{Rowlinson}, A., {Gompertz}, B.~P., {Dainotti}, M., {et~al.} 2014, \mnras, 443,
  1779, \dodoi{10.1093/mnras/stu1277}

\bibitem[{{Salvestrini} {et~al.}(2019){Salvestrini}, {Risaliti}, {Bisogni},
  {Lusso}, \& {Vignali}}]{salvestrini2019}
{Salvestrini}, F., {Risaliti}, G., {Bisogni}, S., {Lusso}, E., \& {Vignali}, C.
  2019, \aap, 631, A120, \dodoi{10.1051/0004-6361/201935491}

\bibitem[{Schiavone {et~al.}(2022)Schiavone, Montani, \&
  Bombacigno}]{Schiavone:2022wvq}
Schiavone, T., Montani, G., \& Bombacigno, F. 2022.
\newblock \doarXiv{2211.16737}

\bibitem[{{Scolnic} {et~al.}(2022){Scolnic}, {Brout}, {Carr}, {Riess}, {Davis},
  {Dwomoh}, {Jones}, {Ali}, {Charvu}, {Chen}, {Peterson}, {Popovic}, {Rose},
  {Wood}, {Brown}, {Chambers}, {Coulter}, {Dettman}, {Dimitriadis},
  {Filippenko}, {Foley}, {Jha}, {Kilpatrick}, {Kirshner}, {Pan}, {Rest},
  {Rojas-Bravo}, {Siebert}, {Stahl}, \& {Zheng}}]{pantheon+}
{Scolnic}, D., {Brout}, D., {Carr}, A., {et~al.} 2022, \apj, 938, 113,
  \dodoi{10.3847/1538-4357/ac8b7a}

\bibitem[{{Scolnic} {et~al.}(2018){Scolnic}, {Jones}, {Rest}, {Pan},
  {Chornock}, {Foley}, {Huber}, {Kessler}, {Narayan}, {Riess}, {Rodney},
  {Berger}, {Brout}, {Challis}, {Drout}, {Finkbeiner}, {Lunnan}, {Kirshner},
  {Sand ers}, {Schlafly}, {Smartt}, {Stubbs}, {Tonry}, {Wood-Vasey}, {Foley},
  {Hand}, {Johnson}, {Burgett}, {Chambers}, {Draper}, {Hodapp}, {Kaiser},
  {Kudritzki}, {Magnier}, {Metcalfe}, {Bresolin}, {Gall}, {Kotak}, {McCrum}, \&
  {Smith}}]{scolnic2018}
{Scolnic}, D.~M., {Jones}, D.~O., {Rest}, A., {et~al.} 2018, \apj, 859, 101,
  \dodoi{10.3847/1538-4357/aab9bb}

\bibitem[{{Sharov}(2016)}]{2016JCAP...06..023S}
{Sharov}, G.~S. 2016, \jcap, 2016, 023, \dodoi{10.1088/1475-7516/2016/06/023}

\bibitem[{{Sharov} \& {Vasiliev}(2018)}]{2018arXiv180707323S}
{Sharov}, G.~S., \& {Vasiliev}, V.~O. 2018, arXiv e-prints, arXiv:1807.07323.
\newblock \doarXiv{1807.07323}

\bibitem[{{Singal} {et~al.}(2011){Singal}, {Petrosian}, {Lawrence}, \&
  {Stawarz}}]{2011ApJ...743..104S}
{Singal}, J., {Petrosian}, V., {Lawrence}, A., \& {Stawarz}, {\L}. 2011, \apj,
  743, 104, \dodoi{10.1088/0004-637X/743/2/104}

\bibitem[{{Srianand} \& {Gopal-Krishna}(1998)}]{1998A&A...334...39S}
{Srianand}, R., \& {Gopal-Krishna}. 1998, \aap, 334, 39.
\newblock \doarXiv{astro-ph/9803183}

\bibitem[{Srinivasaragavan {et~al.}(2020)Srinivasaragavan, Dainotti, Fraija,
  Hernandez, Nagataki, Lenart, Bowden, \& Wagner}]{Srinivasaragavan2020}
Srinivasaragavan, G.~P., Dainotti, M.~G., Fraija, N., {et~al.} 2020, The
  Astrophysical Journal, 903, 18, \dodoi{10.3847/1538-4357/abb702}

\bibitem[{{Steffen} {et~al.}(2006){Steffen}, {Strateva}, {Brandt}, {Alexander},
  {Koekemoer}, {Lehmer}, {Schneider}, \& {Vignali}}]{steffen06}
{Steffen}, A.~T., {Strateva}, I., {Brandt}, W.~N., {et~al.} 2006, \aj, 131,
  2826, \dodoi{10.1086/503627}

\bibitem[{{Tananbaum} {et~al.}(1979){Tananbaum}, {Avni}, {Branduardi}, {Elvis},
  {Fabbiano}, {Feigelson}, {Giacconi}, {Henry}, {Pye}, {Soltan}, \&
  {Zamorani}}]{1979ApJ...234L...9T}
{Tananbaum}, H., {Avni}, Y., {Branduardi}, G., {et~al.} 1979, \apjl, 234, L9,
  \dodoi{10.1086/183100}

\bibitem[{{Tripp}(1998)}]{1998A&A...331..815T}
{Tripp}, R. 1998, \aap, 331, 815

\bibitem[{Varin(2008)}]{varin2008composite}
Varin, C. 2008, Asta advances in statistical analysis, 92, 1

\bibitem[{{Visser} \& {Barcel{\'o}}(2000)}]{2000ppeu.conf...98V}
{Visser}, M., \& {Barcel{\'o}}, C. 2000, in COSMO-99, International Workshop on
  Particle Physics and the Early Universe, ed. U.~{Cotti}, R.~{Jeannerot},
  G.~{Senjanovi}, \& A.~{Smirnov}, 98, \dodoi{10.1142/9789812792129_0014}

\bibitem[{{Wang} {et~al.}(2022){Wang}, {Liu}, {Yuan}, {Liang}, {Yu}, \&
  {Wu}}]{2022arXiv221014432W}
{Wang}, B., {Liu}, Y., {Yuan}, Z., {et~al.} 2022, arXiv e-prints,
  arXiv:2210.14432.
\newblock \doarXiv{2210.14432}

\bibitem[{{Weinberg}(1989)}]{1989RvMP...61....1W}
{Weinberg}, S. 1989, Reviews of Modern Physics, 61, 1,
  \dodoi{10.1103/RevModPhys.61.1}

\bibitem[{{Willingale} {et~al.}(2007){Willingale}, {O'Brien}, {Osborne},
  {Godet}, {Page}, {Goad}, {Burrows}, {Zhang}, {Rol}, {Gehrels}, \&
  {Chincarini}}]{2007ApJ...662.1093W}
{Willingale}, R., {O'Brien}, P.~T., {Osborne}, J.~P., {et~al.} 2007, \apj, 662,
  1093, \dodoi{10.1086/517989}

\bibitem[{{Wong} {et~al.}(2020){Wong}, {Suyu}, {Chen}, {Rusu}, {Millon},
  {Sluse}, {Bonvin}, {Fassnacht}, {Taubenberger}, {Auger}, {Birrer}, {Chan},
  {Courbin}, {Hilbert}, {Tihhonova}, {Treu}, {Agnello}, {Ding}, {Jee},
  {Komatsu}, {Shajib}, {Sonnenfeld}, {Blandford}, {Koopmans}, {Marshall}, \&
  {Meylan}}]{2020MNRAS.498.1420W}
{Wong}, K.~C., {Suyu}, S.~H., {Chen}, G. C.~F., {et~al.} 2020, \mnras, 498,
  1420, \dodoi{10.1093/mnras/stz3094}

\bibitem[{{Yang} {et~al.}(2021){Yang}, {Pan}, {Di Valentino}, {Mena}, \&
  {Melchiorri}}]{2021JCAP...10..008Y}
{Yang}, W., {Pan}, S., {Di Valentino}, E., {Mena}, O., \& {Melchiorri}, A.
  2021, \jcap, 2021, 008, \dodoi{10.1088/1475-7516/2021/10/008}

\bibitem[{{Zamorani} {et~al.}(1981){Zamorani}, {Henry}, {Maccacaro},
  {Tananbaum}, {Soltan}, {Avni}, {Liebert}, {Stocke}, {Strittmatter},
  {Weymann}, {Smith}, \& {Condon}}]{1981ApJ...245..357Z}
{Zamorani}, G., {Henry}, J.~P., {Maccacaro}, T., {et~al.} 1981, \apj, 245, 357,
  \dodoi{10.1086/158815}

\bibitem[{{Zhang} \& {M{\'e}sz{\'a}ros}(2001)}]{2001ApJ...552L..35Z}
{Zhang}, B., \& {M{\'e}sz{\'a}ros}, P. 2001, \apjl, 552, L35,
  \dodoi{10.1086/320255}

\end{thebibliography}
\bibliographystyle{aasjournal}

\end{document}